\RequirePackage{fix-cm}
\documentclass[smallcondensed,natbib]{svjour3_u20}     % onecolumn (ditto)
\smartqed  % flush right qed marks, e.g. at end of proof
\usepackage{graphicx}
\usepackage{amsmath}  % align equations
\usepackage{amssymb}
\usepackage[breaklinks,colorlinks=true,linkcolor=blue,anchorcolor=blue,citecolor=blue,urlcolor=blue,draft=false]{hyperref}
\usepackage{color}
\usepackage{mathrsfs}
\usepackage{placeins}
\usepackage{xcolor}
\usepackage{xspace}
\newcommand{\aap}{\mbox{A\&A}\xspace}
\newcommand{\aapr}{\mbox{A\&A\,Rev}\xspace}

\newcommand{\apj}{\mbox{ApJ}\xspace}
\newcommand{\ssr}{\mbox{SSRv}\xspace}
\newcommand{\apjl}{\mbox{ApJL}\xspace}

\newcommand{\pasj}{\mbox{PASJ}\xspace}
\newcommand{\apjs}{\mbox{ApJS}\xspace}

\newcommand{\mnras}{\mbox{MNRAS}\xspace}
\newcommand{\araa}{\mbox{ARA\&A}\xspace}
\newcommand{\nat}{\mbox{Nature}\xspace}
\newcommand{\prd}{\mbox{PRD}\xspace}
\newcommand{\aj}{\mbox{AJ}\xspace}
\newcommand{\physrep}{\mbox{Phys.~Rep.}\xspace}

\newcommand{\jcap}{\mbox{JCAP}\xspace}
\newcommand{\prl}{\mbox{PRL}\xspace}

\newcommand{\HST}{\textit{HST}\xspace}
\newcommand{\XMMNewton}{\textit{XMM-Newton}\xspace}

\newcommand{\WMAP}{\textit{WMAP}\xspace}
\newcommand{\Planck}{\textit{Planck}\xspace}
\newcommand{\Euclid}{\textit{Euclid}\xspace}
\newcommand{\WFIRST}{\textit{WFIRST}\xspace}

\newcommand{\LCDM}{{$\mathrm{\Lambda}$}CDM\xspace}
\newcommand{\Lterm}{$\mathrm{\Lambda}$\xspace}

\newcommand{\Om}{\mathrm{\Omega}_\mathrm{m}}
\newcommand{\OL}{\mathrm{\Omega}_\mathrm{\Lambda}}
\newcommand{\Ob}{\mathrm{\Omega}_\mathrm{b}}
\newcommand{\Or}{\mathrm{\Omega}_\mathrm{r}}
\newcommand{\Oz}{\mathrm{\Omega}_{0}}
\newcommand{\OX}{\mathrm{\Omega}_{X}}

\newcommand{\simgt}{\lower.5ex\hbox{$\; \buildrel > \over \sim \;$}}
\newcommand{\simlt}{\lower.5ex\hbox{$\; \buildrel < \over \sim \;$}}
\newcommand{\llangle}{\langle\!\langle}
\newcommand{\rrangle}{\rangle\!\rangle}
\newcommand{\percent}{\ensuremath{\%}}
\newcommand{\PhiN}{\Phi_\mathrm{N}}
\newcommand{\Sigmacr}{\Sigma_\mathrm{cr}}
\newcommand{\Dl}{D_l}
\newcommand{\Ds}{D_s}
\newcommand{\Dls}{D_{ls}}
\newcommand{\Nbin}{N_\mathrm{bin}}

\newcommand{\balpha}{\mbox{\boldmath $\alpha$}}
\newcommand{\bbeta}{\mbox{\boldmath $\beta$}}
\newcommand{\boldeta}{\mbox{\boldmath $\eta$}}
\newcommand{\boldxi}{\mbox{\boldmath $\xi$}} 
\newcommand{\btheta}{\mbox{\boldmath $\theta$}}
\newcommand{\bnabla}{\mbox{\boldmath $\nabla$}}
\newcommand{\bDSigma}{\mbox{\boldmath $\Delta\Sigma$}}
\newcommand{\bk}{\mbox{\boldmath $k$}}
\newcommand{\bm}{\mbox{\boldmath $m$}}
\newcommand{\bn}{\mbox{\boldmath $n$}}
\newcommand{\bd}{\mbox{\boldmath $d$}}
\newcommand{\bs}{\mbox{\boldmath $s$}}
\newcommand{\bp}{\mbox{\boldmath $p$}}
\newcommand{\bchi}{\mbox{\boldmath $\chi$}}
\newcommand{\geqs}{\geqslant}
\newcommand{\leqs}{\leqslant}
\newcommand{\kpch}{h^{-1}\mathrm{kpc}}
\newcommand{\Mpch}{h^{-1}\mathrm{Mpc}}
\newcommand{\Gpch}{h^{-1}\mathrm{Gpc}}
\newcommand{\Msunh}{h^{-1}M_\odot}
\newcommand{\Msun}{M_\odot}

\newcommand{\FIG}{./Figs}
\journalname{The Astrononmy and Astrophysics Review}

\begin{document}

\title{Cluster--galaxy weak lensing}

\titlerunning{Cluster--galaxy weak lensing}

\author{Keiichi Umetsu} 

\institute{K. Umetsu \at
Academia Sinica Institute of Astronomy and Astrophysics (ASIAA),
No.\ 1, Section 4, Roosevelt Road, Taipei 10617, Taiwan\\
\email{keiichi@asiaa.sinica.edu.tw}     
}

\date{Received: 2 July 2020 / Accepted: 7 October 2020}

\maketitle

\begin{abstract}
Weak gravitational lensing of background galaxies provides a direct
 probe of the projected matter distribution in and around galaxy
 clusters. Here we present a self-contained pedagogical review of
 cluster--galaxy weak lensing, covering a range of topics relevant to
 its cosmological and astrophysical applications.  
We begin by reviewing the theoretical foundations of gravitational
 lensing from first principles, with special attention to the basics
 and advanced techniques of weak gravitational lensing.
We summarize and discuss key findings from recent cluster--galaxy
 weak-lensing studies on both observational and theoretical grounds,
 with a focus on cluster mass profiles, the
 concentration--mass relation, the splashback radius,
 and implications from extensive mass calibration efforts for cluster
 cosmology.      
\end{abstract}

%\newpage
\setcounter{tocdepth}{2}
\tableofcontents
%\vspace{1cm}

\section{Introduction}
\label{sec:intro}

The propagation of light rays from a distant source to the observer 
is governed by the gravitational field of local inhomogeneities,
as well as by the global geometry of the universe
\citep{1992grle.book.....S}.
Hence the images of background sources carry the imprint of
gravitational lensing by intervening cosmic structures. 
Observations of gravitational lensing phenomena 
can thus be used to study the mass distribution of cosmic objects
dominated by dark matter and to test models of cosmic structure
formation \citep{1992ARA&A..30..311B}.

Galaxy clusters represent the largest class of self-gravitating systems
formed in the universe, with typical masses of
$M\sim 10^{14-15}\Msunh$. 
In the context of the standard structure formation scenario,
cluster halos correspond to rare massive local peaks in the primordial
density perturbations \citep[e.g.,][]{Tinker+2010}.
Galaxy clusters act as powerful cosmic lenses, producing a
variety of detectable lensing effects from strong to weak lensing
\citep{Kneib+Natarajan2011},
including deflection, shearing, and magnifying of the images of
background sources \citep[e.g.,][]{Umetsu2016clash}.
The critical advantage of cluster gravitational lensing is its ability
to study the mass distribution of individual and ensemble
systems independent of assumptions about their physical and dynamical
state \citep[e.g.,][]{Clowe2006Bullet}. 

Weak gravitational lensing is responsible for the weak shape
distortion, or shear, and magnification of the images of background
sources due to the gravitational field of intervening massive objects
and large scale structure \citep{Bartelmann2001,Schneider2005,Umetsu2010Fermi,Hoekstra2013,Mandelbaum2018}.
Weak shear lensing by galaxy clusters gives rise to levels of up to a
few 10 percent of elliptical distortions in images of background
sources. Thus, the weak shear lensing signal, as measured from small but
coherent image distortions in galaxy shapes, can provide a direct
measure of the projected mass distribution of galaxy clusters
\citep[e.g.,][]{1993ApJ...404..441K,1994ApJ...437...56F,Okabe+Umetsu2008}.  
On the other hand, lensing magnification can influence the observed
surface number density of background galaxies seen behind clusters,
by enhancing their apparent fluxes and expanding the area of sky
\citep[e.g.,][]{1995ApJ...438...49B,BTU+05,1998ApJ...501..539T,Umetsu+2011,Chiu2020hsc}. 
The former effect increases the source counts above the
limiting flux, whereas the latter reduces the effective observing area in
the source plane, thus decreasing the observed number of sources per
unit solid angle.
The net effect, known as magnification bias, depends on the intrinsic
faint-end slope of the source luminosity function.

In this paper, we present a self-contained pedagogical review of 
weak gravitational lensing of background galaxies by galaxy clusters
(cluster--galaxy weak lensing),
highlighting recent advances in our theoretical and observational
understanding of the mass distribution in galaxy clusters. 
We shall begin by reviewing the theoretical foundations of gravitational
lensing (Sect.~\ref{sec:theory}), with special attention to the basics
and advanced techniques of cluster--galaxy weak lensing (Sects.~\ref{sec:basics},
 \ref{sec:method}, and \ref{sec:magbias}). 
Then, we highlight and discuss key findings from recent
 cluster--galaxy weak-lensing studies (Sects.~\ref{sec:obsreview}),
 with a focus on cluster mass distributions (Sect.~\ref{subsec:massprofile}),
 the concentration--mass relation (Sect.~\ref{subsec:cM}),
 the splashback radius (Sect.~\ref{subsec:rsp}), 
 and implications from extensive mass
 calibration efforts for cluster cosmology (Sect.~\ref{subsec:mcal}).   
Finally, conclusions are given in Sect.~\ref{sec:summary}. 
 
There have been a number of reviews of relevant subjects
\citep[e.g.,][]{1992ARA&A..30..311B,1996astro.ph..6001N,1999ARA&A..37..127M,Hattori1999,Umetsu1999,vanWaerbeke+Mellier2003,Schneider2005,Kneib+Natarajan2011,Hoekstra2013,Futamase2015,Mandelbaum2018}. 
For general treatments of gravitational lensing, we refer the reader to 
\citet{1992grle.book.....S}.
For a general review of weak gravitational lensing, see
\citet{2001PhR...340..291B} and \citet{Schneider2005}.
For a comprehensive review of cluster lensing, see
\citet{Kneib+Natarajan2011}. 
For a pedagogical review on strong lensing in galaxy clusters,
see \citet{Hattori1999}.

Throughout this paper, we denote the present-day density parameters of
matter, radiation, and \Lterm (a cosmological constant) in critical
units as $\Om, \Or$, and $\OL$, respectively 
\citep[see, e.g.,][]{Komatsu+2009WMAP5}. 
Unless otherwise noted, we assume a concordance \Lterm cold dark matter 
(\LCDM) cosmology with 
$\Om=0.3$,
$\OL=0.7$, 
and a Hubble constant of  
$H_0 = 100\,h$\,km\,s$^{-1}$\,Mpc$^{-1}$ with $h=0.7$.
We denote the mean matter density of the universe at a particular
redshift $z$ as $\overline{\rho}(z)$ and the critical density as 
$\rho_\mathrm{c}(z)$.
The present-day value of the critical density is
$\rho_\mathrm{c,0}=3H_0^2/(8\pi G)\approx 1.88\times 10^{-29}h^2$\,g\,
cm$^{-3}\approx 2.78\times 10^{11}h^2 M_\odot$\,Mpc$^{-3}$,
with $G$ the gravitational constant.
We use the standard notation 
$M_{\Delta_\mathrm{c}}$ or $M_{\Delta_\mathrm{m}}$ to
denote the mass enclosed within a sphere of radius
$r_{\Delta_\mathrm{c}}$ or $r_{\Delta_\mathrm{m}}$,
within which the mean overdensity equals
$\Delta_\mathrm{c} \times \rho_\mathrm{c}(z)$ or
$\Delta_\mathrm{m} \times \overline{\rho}(z)$ at a particular redshift
$z$.
That is,
$M_{\Delta_\mathrm{c}}=(4\pi/3)\Delta_\mathrm{c}\rho_\mathrm{c}(z)r_{\Delta_\mathrm{c}}^3$
and
$M_{\Delta_\mathrm{m}}=(4\pi/3)\Delta_\mathrm{m}\overline{\rho}(z)r_{\Delta_\mathrm{m}}^3$.
We generally denote three-dimensional radial distances as $r$ and
reserve the symbol $R$ for projected radial distances.
Unless otherwise noted, we use projected densities (e.g., $\Sigma(R)$)
and distances (e.g., $R$) both defined in physical (not comoving) units.   
All quoted errors are $1\sigma$ confidence levels (CL) unless otherwise 
stated.

\section{Theory of gravitational lensing}
\label{sec:theory}

The local universe appears to be highly inhomogeneous on a wide range of
scales from stars, galaxies, through galaxy groups and clusters, to forming
superclusters, large-scale filaments, and cosmic voids.
The propagation of light from a far-background source is thus influenced
by the gravitational field caused by such local inhomogeneities along
the line of sight.
In general, a complete description of the light propagation in an
arbitrary curved spacetime is a complex theoretical problem.
However, a much simpler description is possible under a wide range of
astrophysically relevant circumstances,
which is referred to as the gravitational lensing 
theory
\citep[e.g.,][]{1992grle.book.....S,2001PhR...340..291B,Kneib+Natarajan2011}. 
This section reviews the basics of gravitational lensing theory
to provide a basis and framework for cluster lensing studies, with
an emphasis on weak gravitational lensing.

\subsection{Bending of light in an asymptotically flat spacetime}
\label{subsec:gl} 

To begin with, let us consider the bending of light in a weak-field
regime of an asymptotically flat spacetime in the framework of general
relativity. Specifically, we assume an isolated and stationary mass
distribution \citep[][]{1992grle.book.....S}. 
Then, the metric tensor $g_{\mu\nu}$ ($\mu,\nu=0,1,2,3$) of the
perturbed spacetime can be written as:
\begin{equation}
 \label{eq:metric_sw}
  \begin{aligned}
ds^2 &= g_{\mu\nu} dx^\mu dx^\nu\\
   &= -(1+2\PhiN/c^2)c^2 dt^2 +(1-2\PhiN/c^2)
   \left[(dx^1)^2+(dx^2)^2+(dx^3)^2\right], 
  \end{aligned}
\end{equation}
where 
$(x^\mu)=(ct,x^1,x^2,x^3)$ are the four spacetime coordinates, 
$\PhiN$ is the Newtonian gravitational potential in a
weak-field regime  
$|\PhiN/c^2|\ll 1$, and $c$ is the speed of light in vacuum.
We consider the metric given by Eq.~(\ref{eq:metric_sw}) 
to be the sum of a background metric $g_{\mu\nu}^{(\mathrm{b})}$ and a
small perturbation denoted by $h_{\mu\nu}$, that is,
$g_{\mu\nu}=g_{\mu\nu}^{(\mathrm{b})} + h_{\mu\nu}$ with
$|h_{\mu\nu}|\ll 1$. 

To the first order in $\PhiN/c^2$, we have
$g_{\mu\nu}^{(\mathrm{b})}=\eta_{\mu\nu}=\mathrm{diag}(-1,1,1,1)$ and \linebreak
$h_{\mu\nu}=\mathrm{diag}(-2\PhiN,-2\PhiN,-2\PhiN,-2\PhiN)/c^2$,
where $g^{\mu\nu}$ and $g^{(\mathrm{b})\mu\nu}$ are defined by 
$g^{\mu\rho} g_{\rho\nu}= \delta^{\mu}_{\nu}$ and
$g^{(\mathrm{b})\mu\rho} g^{(\mathrm{b})}_{\rho\nu}= \delta^{\mu}_{\nu}$,
with $\delta^\mu_\nu$ the Kronecker delta symbol in four dimensions.
%In the linear formalism raising and lowering of indices are understood
%using the Minkowski metric $\eta$, so that
Then, to the first order of $h$, we have
$g^{\mu\nu}= g^{(\mathrm{b})\mu\nu}-h^{\mu\nu}$, where $h^{\mu\nu}$ is defined by
$h^{\mu\nu}\equiv g^{(\mathrm{b})\mu\rho} g^{(\mathrm{b})\nu\sigma} h_{\rho\sigma} =\eta^{\mu\rho}\eta^{\nu\sigma}h_{\rho\sigma}$.

The propagation of light is described by null geodesic equations:
\begin{equation}
\label{eq:4mom}
 \begin{aligned}
k^\mu &\equiv \frac{dx^\mu(\lambda)}{d\lambda},\\
0&=g_{\mu\nu}k^\mu k^\nu,\\
\frac{dk^\mu}{d\lambda} &=-\Gamma^{\mu}_{\nu\lambda}k^\nu k^\lambda,
 \end{aligned}
\end{equation}
where $k^\mu$ is the four-momentum,
$\lambda$ is the affine parameter, and
$\Gamma^\mu_{\nu\lambda}$ denotes the Christoffel symbol,
$\Gamma^{\mu}_{\nu\rho} = (1/2) g^{\mu\lambda}\left(g_{\lambda \nu,\rho} + g_{\lambda \rho, \nu} - g_{\nu\rho,\lambda}\right)$,
with
$g_{\mu\nu}^{(\mathrm{b})}=\eta_{\mu\nu}$ and $\Gamma^{(\mathrm{b})\mu}_{\nu\rho}=0$
in the background Minkowski spacetime.
For a light ray propagating along the $x^3$-direction in the background
metric, the photon four-momentum $k^{(\mathrm{b})\mu}$ and the 
unperturbed orbit $x^{(\mathrm{b})\mu}$ are given by
$k^{(\mathrm{b})\mu}=dx^{(\mathrm{b})\mu}/d\lambda=(1,0,0,1)$ and
$x^{(\mathrm{b})\mu}=(\lambda,0,0,\lambda)$.
%where we have chosen our coordinate system such that the $x^1$-axis is
%parallel to the direction of the incident photon momentum.

Now we consider the light ray propagation in a perturbed spacetime.
To this end, we express the perturbed orbit $x^\mu(\lambda)$ as a 
sum of the unperturbed path $x^{(\mathrm{b})\mu}(\lambda)$ and the
deviation vector $\delta x^\mu(\lambda)$:
\begin{equation}
 x^\mu(\lambda) = x^{(\mathrm{b})\mu}(\lambda) + \delta x^\mu(\lambda).
\end{equation}
Without loss of generality, we can take the deflection angle to
lie in the $x^3 x^1$ plane with $x^2=0$,
and we denote $(x^1, x^3) = (x^\perp, x^{||})$.
In the weak-field limit of $|\PhiN/c^2|\ll 1$, 
the impact parameter $b$ of the incoming light ray is much greater than
the Schwarzschild radius of the deflector with mass $M$, that is,
$b\gg 2GM/c^2$.
%with $G$ the gravitational constant.
Then, the linearized null geodesic equations are written as:\footnote{See
\citet{Pyne+Birkinshaw1993} for a detailed discussion of the consistency
conditions for the truncation of perturbed null geodesics.}  
\begin{equation}
\label{eq:geodesic_pert}
 \begin{aligned}
 k^\mu(\lambda) &= k^{(\mathrm{b})\mu}(\lambda) + \delta k^\mu(\lambda),\\
0&=h_{\mu\nu}k^{(\mathrm{b})\mu} k^{(\mathrm{b})\nu}+ 2g^{(\mathrm{b})}_{\mu\nu}
k^{(\mathrm{b})\mu}\delta k^\nu,\\ 
\frac{d(\delta k^\mu)}{d\lambda} &=
 -2\Gamma^{(\mathrm{b})\mu}_{\nu\lambda} k^{(\mathrm{b})\nu} \delta k^\lambda
  -\delta \Gamma^{\mu}_{\nu\lambda} k^{(\mathrm{b})\nu} k^{(\mathrm{b})\lambda}.
 \end{aligned}
\end{equation}
The perturbed Christoffel symbol is
$\delta  \Gamma^{\mu}_{\nu\rho}=
(1/2)
\eta^{\mu\lambda}\left(
h_{\lambda \nu,\rho} +
h_{\lambda \rho, \nu}-
h_{\nu\rho,\lambda}
\right) + O(h^2)$.
%\begin{eqnarray}
%\frac{d(\delta k^{0})}{d\lambda} &=&
%-(\delta \Gamma^0_{01} +\delta \Gamma^0_{10})=-2\PhiN_{,1}\\
%\frac{d(\delta k^{||})}{d\lambda} &\equiv&
%\frac{d(\delta k^{1})}{d\lambda} =
%-(\delta \Gamma^1_{10} +\delta \Gamma^1_{01})=+2\PhiN_{,0}\\
%\frac{d(\delta k^{\perp})}{d\lambda} &\equiv&
%\frac{d(\delta k^{2})}{d\lambda} =
%-(\delta \Gamma^2_{00} +\delta \Gamma^2_{11})=-2\PhiN_{,2}
%\end{eqnarray}
Choosing the boundary conditions in the in-state ($\lambda\to -\infty$) as
$\delta k^\mu(-\infty)=0$,
we integrate the linearized null geodesic equations (Eq.~(\ref{eq:geodesic_pert}))
to obtain the following equations for the spatial components in the
out-state ($\lambda\to +\infty$): 
\begin{equation}
 \delta k^\perp(+\infty) =
  -\frac{2}{c^2}\int_{-\infty}^{+\infty}\frac{\partial\PhiN(\lambda)}{\partial
  x^\perp}\,d\lambda, \ \ \ \delta k^{||}(+\infty) = 0, 
\end{equation}
where $\lambda=x^{||}(\lambda)+O(h)$.
Taking the unperturbed path, we obtain the bending angle $\hat\alpha$ in
the small angle scattering limit ($|\hat\alpha|\ll 1$) as:\footnote{With
this sign convention, the bending angle $\hat\alpha$ has the same sign   
as the change in photon propagation direction
\citep[e.g.,][]{2000ApJ...530..547J}. 
An alternative sign convention is often used in the literature
\citep[e.g.,][]{2001PhR...340..291B},
in which $\hat{\balpha}\to -\hat{\balpha}$
(or $\balpha \to -\balpha$; see Eq.~(\ref{eq:lenseq})).}   
\begin{equation}
 \label{eq:bending}
\hat\alpha \simeq \frac{k^{\perp}(+\infty)}{k^{||}(+\infty)}
 \simeq
 -\frac{2}{c^2}\int_{-\infty}^{+\infty}\!
  \frac{\partial\PhiN(x^{||},x^\perp)}{\partial x^\perp}\,dx^{||},
\end{equation}
which is known as the Born approximation. 
This yields an explicit expression for the bending angle  
of $|\hat\alpha|\simeq 4GM/(rc^2) = 1.75'' (M/M_\odot)(r/R_\odot)^{-1}$.
General relativity predicts a deflection angle
twice as large as that Newtonian physics would provide.
%In 1919, Eddington conducted an experiment during a solar
%eclipse to test the prediction of Einstein's general relativity.
Einstein's prediction for the solar deflection of light is verified
within $\sim 0.1\percent$ \citep[e.g.,][]{1995PhRvL..75.1439L}. 

The null geodesic condition leads to
$\delta k^0(\lambda)=-2\PhiN(\lambda)/c^2 + O(h^2)$, 
or $cdt/d\lambda=1-2\PhiN(\lambda)/c^2 + O(h^2)>1$.
The gravitational time delay $\Delta t_\mathrm{grav}$,
with respect to the unperturbed light
propagation, is thus given by:
\begin{equation}
c\Delta t_\mathrm{grav} = -2\int_{-\infty}^{+\infty}\!d\lambda\,
 \PhiN(\lambda)/c^2. 
\end{equation}
Note that there is an additional time delay due to a change in the
geometrical path length caused by gravitational deflection
(see Sect.~\ref{subsubsec:thin}).

\subsection{Lens equation}
\label{subsec:lenseq}

%%%%%%%%%%%%%%%%%%%%%%%%%%%%%%%%%%%%%%%%%%%%%%%%%%%%%%%%%%%%%%%%%%%%%%%%%%%%%%%
%%% Figure 1
%%%%%%%%%%%%%%%%%%%%%%%%%%%%%%%%%%%%%%%%%%%%%%%%%%%%%%%%%%%%%%%%%%%%%%%%%%%%%%%

\begin{figure*}[!htb] %!htb
  \begin{center}
   \includegraphics[scale=0.6, angle=0, clip]{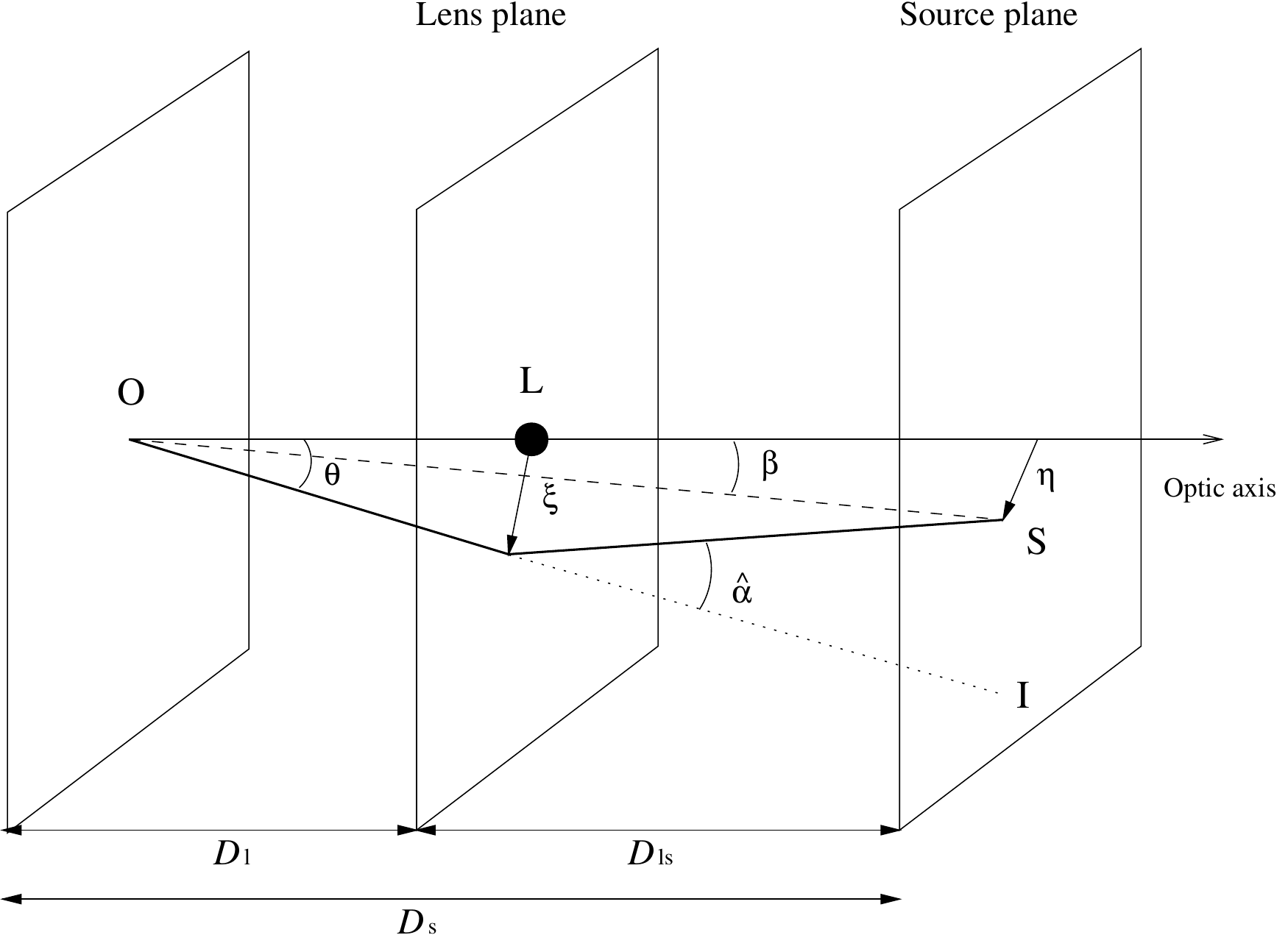} 
  \end{center}
\caption{
\label{fig:lens} 
Illustration of a typical lens system.
The light ray propagates from the source (S) at the position
 $\boldeta$ in the source plane to the observer (O), 
 passing the position $\boldxi$ in the lens plane (L),
 resulting in a bending angle $\hat{\balpha}$. 
The angular position of the source (S) relative to the optical
 axis is denoted by $\bbeta$, and that of the image (I) relative to the 
 optical axis is denoted by $\btheta$.
The $D_l$, $D_s$, and $D_{ls}$ are the observer--lens, observer--source, 
 and lens--source angular diameter distances, respectively.  
% The figure is adapted from \citet{Umetsu2010Fermi}.
 }
\end{figure*}

Let us consider the situation illustrated in Fig.~\ref{fig:lens}.
A light ray propagates from a far-distant source (S) at the position
$\boldeta$  in the source plane to an observer (O), passing the position
$\boldxi$ in the lens plane, in which the light is deflected by a
bending angle $\hat{\balpha}$.
Here the source and lens planes are defined as planes perpendicular to
the optical axis at the distance of the source and the lens,
respectively. The exact definition of the optical axis does not matter,
because the angular scales involved are very small.
The angle between the optical axis and the unlensed source (S)
position is $\bbeta$, and the angle between the optical axis and the
image (I) is $\btheta$. The angular diameter distances between the
observer and the lens, the observer and the source, and the lens and the 
source, are denoted by
$\Dl, \Ds$, and $\Dls$, respectively.

As illustrated in Fig.~\ref{fig:lens}, we have the following
geometrical relation: 
$\boldeta=(\Ds/\Dl)\boldxi+\Dls\hat{\balpha}(\boldxi)$.
Equivalently, this is translated into the relation between the angular
source and image positions,
$\bbeta=\boldeta/\Ds$ and $\btheta=\boldxi/\Dl$, as:
\begin{equation} 
\label{eq:lenseq}
\bbeta = \btheta +\frac{\Dls}{\Ds} \hat{\balpha}\equiv
 \btheta + \balpha(\btheta),
\end{equation}
where we defined the reduced bending angle, or the deflection field
\citep{2005ApJ...621...53B}, 
$\balpha(\btheta)=(\Dls/\Ds)\hat{\balpha}$ in
the last equality.   
%%%
Equation~(\ref{eq:lenseq}) is referred to as the lens equation, or the
ray-tracing equation.

%%%%%%%%%%%%%%%%%%%%%%%%%%%%%%%%%%%%%%%%%%%%%%%%%%%%%%%%%%%%%%%%%%%%%%%%%%%%%%%
%%% Figure 2
%%%%%%%%%%%%%%%%%%%%%%%%%%%%%%%%%%%%%%%%%%%%%%%%%%%%%%%%%%%%%%%%%%%%%%%%%%%%%%%

\begin{figure*}[!htb] %!htb
  \begin{center}
   \includegraphics[scale=0.7, angle=0, clip]{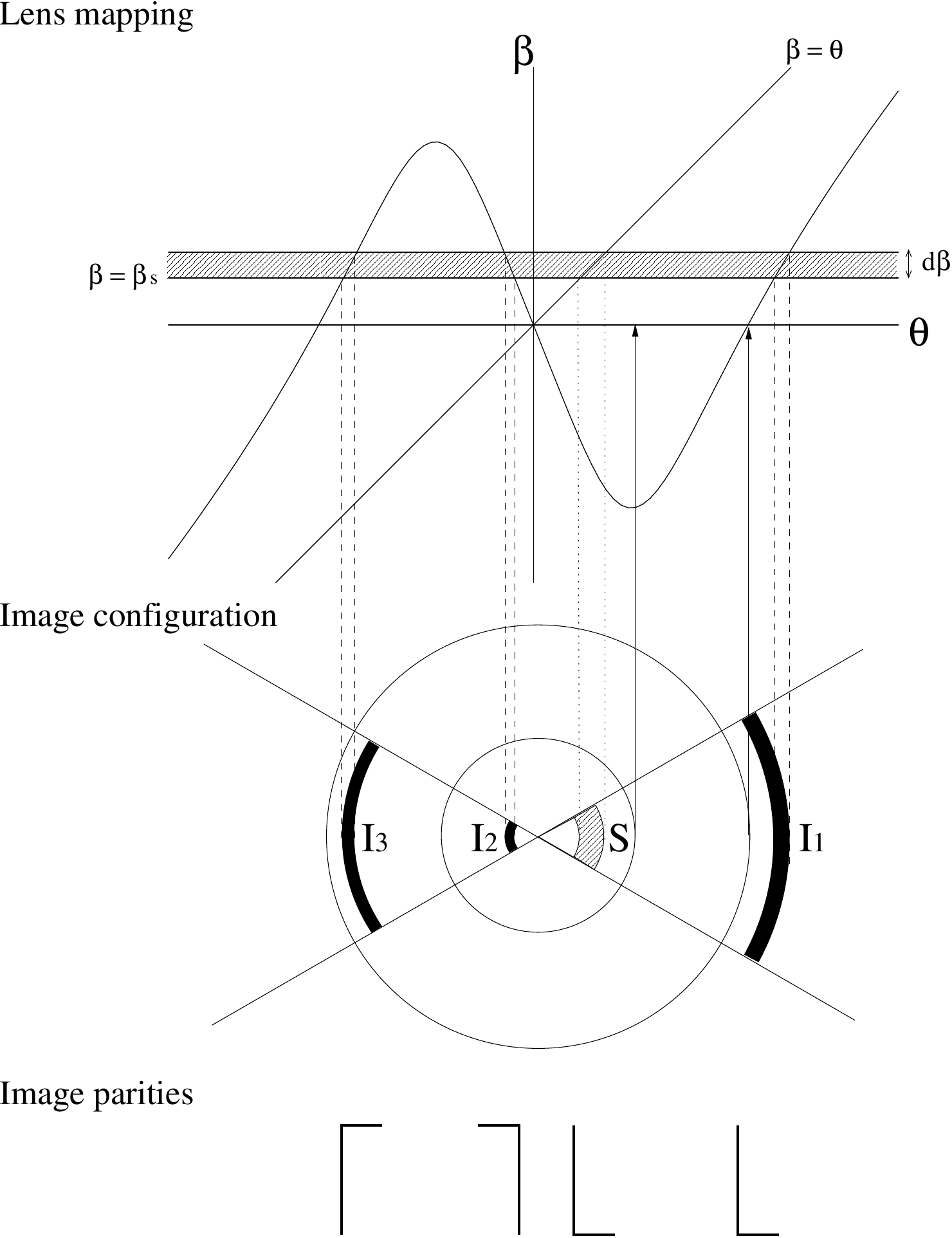} 
  \end{center}
\caption{
 \label{fig:lenseq}
Illustration of the lens mapping $\beta=\beta(\theta)$ (upper panel),
 image configuration (middle panel), and image parities (lower panel)
 for a typical axis-symmetric lens system. Three images
 ($I_1,I_2,I_3$) are produced for a source ($S$) at the location
 $\beta=\beta_s$ with a radial width $\delta\beta$. The images are
 formed at the  
 intersections of $\beta=\beta(\theta)$ and the horizontal
 line $\beta=\beta_s$. Critical curves are also shown by two solid
 concentric circles in the middle panel.
 The inner circle represents the radial critical curve where
 $d\beta(\theta)/d\theta=0$, while the outer circle represents the
 tangential critical curve where $\beta(\theta)=0$. 
 The image parities are illustrated in the lower panel. The images $I_1$
 and $I_2$ have the same total parity as the source $S$, while $I_3$ has   
 the opposite total parity to $S$.
The concepts of critical curves and image parities are given in
 Sect.~\ref{subsubsec:mu}.
 }
\end{figure*}

In general, the lens equation is nonlinear with respect to the image
position $\btheta$, so that it may have multiple solutions $\btheta$ for
a given source position $\bbeta$.  
This corresponds to multiple imaging of a background source
\citep[see][]{Hattori1999,Kneib+Natarajan2011}. An illustration of the
typical circularly symmetric lens system is shown in
Fig.~\ref{fig:lenseq}.    
We refer to \citet{Keeton2001} for a review of various families of mass 
models for gravitational lensing.

\subsection{Cosmological lens equation}
\label{subsec:gl_cosmo}

Here we turn to the cosmological lens equation that describes the light
propagation in a locally inhomogeneous, expanding universe. There are
various approaches to derive a cosmological version of the lens equation
\citep[e.g.,][]{1985A&A...143..413S,1993PThPh..90..753S,1994CQGra..11.2345S,1995PThPh..93..647F,Dodelson2003,Sereno2009}.  
We follow the approach by \citet{1995PThPh..93..647F} based on perturbed
null geodesic equations as introduced in Sect.~\ref{subsec:gl}.   
%For further details, we refer to the PhD thesis of Masahiro Takada (2000).

Consider a perturbed Friedman--Lema\'itre--Robertson--Walker (FLRW) metric
in the Newtonian gauge of the form \citep[e.g.,][]{Kodama+Sasaki1984}:
\begin{equation}
\label{eq:metric_flrw}
ds^2=
-(1+2\Psi/c^2)c^2dt^2 + (1-2\Psi/c^2) a^2(t)\left[
 d\chi^2+r^2(d\vartheta^2+\sin^2\vartheta d\varphi^2)
 \right],
\end{equation}
where
$a(t)$ is the scale factor of the universe (normalized to unity at present),
$\chi$ is the comoving distance,
$\vartheta$ and $\varphi$ are the spherical polar and azimuthal angles, respectively,
$\Psi$ is a scalar metric perturbation,
$K$ is the spatial curvature of the universe, and
$r = f_K(\chi)$ is the comoving angular diameter distance:
\begin{equation}
 \label{eq:fK}
 f_K(\chi) =
  \begin{cases} 
   (-K)^{-1/2}\sinh(\chi/\sqrt{-K}) & (K<0),\\
   \chi & (K=0),\\
   K^{-1/2}\sin(\chi/\sqrt{K}) & (K>0).
  \end{cases}
\end{equation}
The spatial curvature $K$ is expressed with
the total density parameter at the present epoch,
$\Oz\equiv\sum_X\OX=\Om+\Or+\OL$,
as $K=(\Oz-1)H_0^2/c^2$.
The evolution of $a(t)$ is determined by the Friedmann equation,
$H(a)\equiv (da/dt)/a=H_0[\Or a^{-4} + \Om a^{-3} + \OL + (1-\Oz)a^{-2}]^{1/2}$.
In the line element (\ref{eq:metric_flrw}), we have neglected all terms
of higher than $O(\Psi/c^2)$, the contributions from vector and tensor
perturbations, and the effects due to anisotropic stress.  
As we will discuss in Sect.~\ref{subsub:poisson}, $\Psi$ is
interpreted as the Newtonian gravitational potential generated by
local inhomogeneities of the matter distribution in the universe.

Since the structure of a light cone is invariant under the conformal
transformation, we work with the conformally related spacetime metric
given by
$d\tilde s^2 = a^{-2}ds^2 \equiv \tilde g_{\mu\nu} dx^\mu dx^\nu$
with $(x^\mu)=(\eta, \chi, \vartheta, \varphi)$, where
$\eta=c\int^t\!dt'/a(t')$ is the conformal time.
%We consider null geodesics given by $d\tilde s^2$.
The metric $\tilde g_{\mu\nu}$ can be
rewritten in the form of
$\tilde g_{\mu\nu}=\tilde g^{(\mathrm{b})}_{\mu\nu}+h_{\mu\nu}$, as a
sum of the background metric and a small perturbation ($|h|\ll 1$).

%%%
We follow the prescription given in Sect.~\ref{subsec:gl} to solve the
null geodesic equations in the perturbed spacetime (Eq.~(\ref{eq:metric_flrw})). 
To this end, we consider past-directed null geodesics from
the observer.
Choosing the spherical coordinate system centered on the observer,  
we have $k^{(\mathrm{b})\mu}=(-1,1,0,0)$ in the background metric with
$\Psi=0$. 
The unperturbed path is parameterized by the affine parameter $\lambda$
along the photon path as
$x^{(\mathrm{b})\mu}(\lambda)=(-\lambda,\lambda-\lambda_o,\vartheta_I,\varphi_I)$,
%with $\theta_I\ll 1$,
%where we have chosen our 
%coordinate system such that the $z$-axis is the direction to the lens
%object, and
where $\lambda_o$ is the affine parameter at the observer and
$(\vartheta_I,\varphi_I)$ denote the angular direction of the image
position on the sky. 
The comoving angular distance $r$ in the background spacetime can be
parameterized by $\lambda$ as
$r(\lambda)\equiv f_K[\chi(\lambda-\lambda_o)]$
(see Eq.~(\ref{eq:fK})). 
%%%

The perturbed null geodesic equations for the angular components
($\vartheta,\varphi$)  can be formally solved as:
\begin{equation}
\delta k^i (\lambda) =
 -\frac{2}{r^2(\lambda)}\int_{\lambda_o}^{\lambda_s}\! d\lambda'
 \partial^i \Psi(\lambda')/c^2 \ \ \ (i=\vartheta,\varphi),
\end{equation}
where
$\partial^i \Psi=(\Psi_{,\vartheta}, \sin^{-2}\vartheta\Psi_{,\varphi})$
and we have chosen
$\delta k^\vartheta(\lambda_o)=\delta k^\varphi(\lambda_o)=0$.
Inserting this result in Eq.~(\ref{eq:geodesic_pert}) and 
integrating by part yield \citep{1995PThPh..93..647F,Dodelson2003}:
\begin{equation}
 \label{eq:cosmo_lenseq}
  \begin{aligned}
\vartheta_S &= \vartheta_I
   -\frac{2}{c^2}\int_{\lambda_o}^{\lambda_s}\!
   \frac{r(\lambda_s-\lambda)}
   {r(\lambda_s)r(\lambda)}\partial^\theta \Psi(\lambda)\,d\lambda,\\
   \varphi_S &= \varphi_I
   -\frac{2}{c^2}\int_{\lambda_o}^{\lambda_s}\!
   \frac{r(\lambda_s-\lambda)}
   {r(\lambda_s)r(\lambda)}\partial^\varphi \Psi(\lambda)\,d\lambda,\\
  \end{aligned}
\end{equation}
where $\lambda_s$ is the affine parameter at the background source,
$(\vartheta_S, \varphi_S)\equiv (\vartheta(\lambda_s),\varphi(\lambda_s))$
denote the angular direction of the unlensed source position on the sky,
and we set $\delta\vartheta(\lambda_o)=\delta\varphi(\lambda_o)=0$.
Here, the integral is performed along the perturbed trajectory
$x^\mu(\lambda)=x^{(\mathrm{b})\mu}(\lambda)+\delta x^\mu(\lambda)$.
Equation~(\ref{eq:cosmo_lenseq}) relates the observed direction of the
image position $(\vartheta_I, \varphi_I)$ to the (unlensed) direction of
the source position $(\vartheta_S,\varphi_S)$ for a given background
cosmology and metric perturbation $\Psi(\bchi,\eta)$. 
This is a general expression of the cosmological lens equation obtained
by \citet{1995PThPh..93..647F}.

\subsection{Flat-sky approximation}
\label{subsub:flatsky}

Now we consider a small patch of the sky around a given line of sight
($\vartheta=0$), across which the curvature of the sky is negligible
($\vartheta\ll 1$). 
Then, we can locally define a flat plane perpendicular to the line
of sight.
By noting that $\delta\btheta\equiv(\delta\vartheta,\vartheta \delta\varphi)$
is an angular displacement vector within this sky plane,
we can express Eq.~(\ref{eq:cosmo_lenseq}) as:
\begin{equation}
 \label{eq:cosmo_lenseq2}
 \bbeta(\chi_s) = \btheta + \balpha(\chi_s),
\end{equation}
where
$\bbeta(\chi_s)$ is the (unlensed) angular position of the source, 
$\btheta$ is the apparent angular position of the source image,
and $\balpha(\chi_s)$ is the deflection field given by
\citep{1995PThPh..93..647F}: 
\begin{equation}
\balpha(\chi_s) =-\frac{2}{c^2}\int_0^{\chi_s}\!
\frac{r(\chi_s-\chi)}{r(\chi_s)} 
\bnabla_{\perp}\Psi[x^\mu(\chi)]\,d\chi,
\end{equation}
where 
$\bnabla_\perp\equiv r^{-1}(\lambda)(\partial_\vartheta,\vartheta^{-1}\partial_\varphi)$
is the transverse comoving gradient
and
the integral is performed along the perturbed trajectory
$x^\mu(\lambda)=x^{(\mathrm{b})\mu}(\lambda)+\delta x^\mu(\lambda)$
with $\lambda=\chi + O(\Psi/c^2)$.
Equation~(\ref{eq:cosmo_lenseq2}) can be applied to a range of lensing
phenomena, including 
multiple deflections of light from a background source
(Sect.~\ref{subsec:multilens}),
strong and weak gravitational lensing by
individual galaxies and clusters (Sect.~\ref{subsec:thinlens}),
and cosmological weak lensing by the intervening large-scale
structure (a.k.a., the cosmic shear).
Note that the cosmological lens equation is obtained using the
standard angular diameter distance in a background FLRW spacetime
without employing the thin-lens approximation
(see Sect.~\ref{subsec:thinlens}).

\subsection{Multiple lens equation}
\label{subsec:multilens}

We consider a discretized version of the cosmological lens equation 
(Eq.~(\ref{eq:cosmo_lenseq2})) by dividing the radial integral
between the source ($\chi=\chi_s$) and the observer ($\chi=0$) into $N$
comoving boxes ($N-1$ lens planes) separated by a constant comoving
distance of $\Delta\chi$.  
The angular position $\btheta^{(n)}$ of a light ray in the $n$th plane
($1\leqs n\leqs N$) is then given by
\citep[e.g.,][]{1992grle.book.....S,Schneider2019}: 
\begin{equation}
 \label{eq:multilens}
\bbeta^{(n)}=\btheta^{(0)}+\sum_{m=1}^{n-1} 
\frac{r(\chi_n-\chi_m)}
{r(\chi_n)}
\hat{\balpha}^{(m)},
\end{equation}
where $\btheta^{(0)} = \bbeta^{(1)}$ is the apparent angular position of
the source image and $\hat{\balpha}^{(m)}$ is the bending angle at the
$m$th lens plane ($m=1,2,\dots,n-1$):
\begin{equation}
\hat{\balpha}^{(m)}= -\frac{2}{c^2}\bnabla_\perp\Psi[\chi_m,r(\chi_m)\bbeta^{(m)}]\,\Delta\chi.
\end{equation}
The $2\times 2$ Jacobian matrix of Eq.~(\ref{eq:multilens})
($1\leqs n \leqs N$) is expressed as 
\citep[e.g.,][]{2000ApJ...530..547J}:\footnote{Note that we can write
$(D_{mn}/D_n){\cal H}_{ij}^{(m)}=-(2/c^2)g(\chi_m,\chi_n)\nabla_{\perp,i}\nabla_{\perp,j}\Psi[\chi_m,r(\chi_m)\bbeta^{(m)}]\,\Delta\chi$
with $g(\chi_m,\chi_n)=r(\chi_m)r(\chi_n-\chi_m)/r(\chi_n)$ an effective
lensing distance \citep{2000ApJ...530..547J} and $D_{mn}/D_{n}=r(\chi_n-\chi_m)/r(\chi_n)$.}
\begin{equation}
\label{eq:jacob_mult}
{\cal A}^{(n)} := \frac{\partial \bbeta^{(n)}}{\partial \btheta^{(0)}}
=
{\cal I} + \sum_{m=1}^{n-1} \frac{D_{mn}}{D_n} {\cal H}^{(m)} {\cal A}^{(m)},
\end{equation}   
where
${\cal I}$ denotes the identity matrix,
${\cal H}^{(m)}\equiv\partial\hat{\balpha}^{(m)}/\partial\bbeta^{(m)}$
is a symmetric dimensionless Hessian matrix
with 
%{\cal H}_{ij}^{(m)}\propto \partial^2\Psi/\partial\beta^{(m)}_i\partial\beta^{(m)}_j$
${\cal H}_{ij}^{(m)}=-(2/c^2)r(\chi_m)\nabla_{\perp,i}\nabla_{\perp,j}\Psi[\chi_m,r(\chi_m)\bbeta^{(m)}]\,\Delta\chi$
($i,j=1,2$),
$D_n$ is the angular diameter distance between the observer and the $n$th
lens plane, and $D_{mn}$ is the angular diameter distance between the
$m$th and $n$th lens planes ($m<n$).   
In general, the Jacobian matrix ${\cal A}^{(n)}$ can be
decomposed into the following form:
\begin{equation}
{\cal A}^{(n)}=(1-\kappa){\cal I} - \gamma_1\sigma_3 - \gamma_2\sigma_1 -i\omega \sigma_2,
\end{equation}
where
$\kappa$ is the lensing convergence,
$(\gamma_1,\gamma_2)$ are the two components of the gravitational shear
(see Sect.~\ref{subsubsec:jacobian} for the definitions and further
details of the convergence and shear), 
$\omega$ is the net rotation \citep[e.g.,][]{Cooray+Hu2002},
and
$\sigma_{a}$ $(a=1,2,3)$ are the Pauli matrices that satisfy
$\sigma_{a}\sigma_{b}=i\epsilon_{abc}\sigma_{c}$,
with $\epsilon_{abc}$ the Levi--Civita symbol in three dimensions.
The Born approximation ${\cal A}^{(m)}={\cal I}$ on the right-hand side 
of Eq.~(\ref{eq:jacob_mult}) leads to a symmetric Jacobian 
matrix with $\omega=0$.

The multiple lens equation has been widely used to study gravitational
lensing phenomena by ray-tracing through $N$-body
simulations \citep[e.g.,][]{Schneider+Weiss1988,Hamana+2000,2000ApJ...530..547J}.

\subsubsection{Cosmological poisson equation}
\label{subsub:poisson}

We assume here a spatially flat geometry with $K=0$ motivated by
cosmological observations based on cosmic microwave background (CMB) and
complementary data sets  
\citep[e.g.,][]{Hinshaw+2013WMAP9,Planck2015XIII}.
The cosmological Poisson equation relates the scalar metric perturbation
$\Psi$ (see Eq.~(\ref{eq:metric_flrw})) to the matter density
perturbation $\delta\rho$ on subhorizon scales as:
\begin{equation} 
\label{eq:poisson}
\bnabla^2 \Psi(\bchi,\eta)
=4\pi G a^2 \delta \rho =\frac{3H_0^2\Om}{2}\frac{\delta}{a},
\end{equation}
where
$\delta=\delta\rho/\overline{\rho}$ is the density contrast
with respect to the background matter density $\overline{\rho}$ of the universe,
$\overline{\rho}=a^{-3}(3H_0^2\Om)/(8\pi G)$,
and $\bnabla$ is the three-dimensional gradient operator in comoving coordinates.
A key implication of Eq.~(\ref{eq:poisson}) is that the
amplitude of $\Psi$ is related to the amplitude of $\delta$ as
$|\Psi/c^2| \sim (3\Om/2)(l/L_H)^2 (\delta/a)$ where $l$ and
$L_H=c/H_0$ 
denote the characteristic comoving scale of density perturbations and
the Hubble radius, respectively.  
Therefore, assuming the standard matter power spectrum of
density fluctuations \citep[e.g.,][]{Smith+2003halofit},
we can safely conclude that the degree of metric
perturbation is always much smaller than unity, i.e., $|\Psi/c^2|\ll 1$,
even for highly nonlinear perturbations with $|\delta|\gg 1$ on
small scales of $l\ll L_H$ ($\sim 3\Gpch$).

\subsection{Thin-lens equation}
\label{subsec:thinlens}

\subsubsection{Thin-lens approximation}
\label{subsubsec:thin}

Let us turn to the case of gravitational lensing caused by a
single cluster-scale halo.
%Galaxy clusters represent the largest class of self-gravitating systems
%formed in the universe, with typical halo masses of
%$M_\mathrm{200c} \sim 10^{14-15}\Msunh$. 
%In the context of the standard structure formation scenario,
%cluster halos correspond to rare massive local peaks in the primordial
%density perturbations \citep[e.g.,][]{Tinker+2010}.
Galaxy clusters can produce deep gravitational
potential wells, acting as powerful gravitational lenses.
In cluster gravitational lensing it is often assumed that
the total deflection angle, $\balpha(\btheta)$, is dominated by
the cluster of interest and its surrounding large-scale environment, which
becomes important beyond the cluster virial radius, $r_\mathrm{vir}$ 
\citep{Cooray+Sheth2002,Oguri+Hamana2011,Diemer+Kravtsov2014}.

Assuming that
the light propagation is approximated by a single-lens event
due to the cluster and that a light deflection occurs within a sufficiently
small region ($\chi_l-\Delta\chi/2, \chi_l+\Delta\chi/2$) 
compared to the relevant angular diameter distances, 
we can write the deflection field by a single cluster as:
\begin{equation}
\balpha(\btheta) \simeq -\frac{2}{c^2} \frac{\Dls}{\Ds}
\int_{\chi_l-\Delta\chi/2}^{\chi_l+\Delta\chi/2}\!
%a^{-1}
\bnabla_{\perp}\Psi[\chi,r(\chi_l)\btheta]\,d\chi,
\end{equation}
where $\Ds=a(\chi_s)r(\chi_s)$ and $\Dls=a(\chi_s)r(\chi_s-\chi_l)$ 
are the angular diameter distances from the
observer to the source and from the deflector to the source,
respectively, and $r(\chi_l)\btheta$ is the comoving
transverse vector on the lens plane.
In a cosmological situation, 
the angular diameter distances $D_{mn}$
between the planes $m$ and $n$ ($z_m < z_n$)
are of the order of the Hubble radius, 
$L_H\equiv c/H_0\sim 3\Gpch$, while physical extents of clusters
%\citep[say, physical halo boundaries defined by the splashback radius:
%e.g.,][]{Diemer+Kravtsov2014} 
are about $2r_\mathrm{200m} \sim (2 - 4)\Mpch$ in comoving
units. Therefore, one can safely adopt the thin-lens approximation in
cluster gravitational lensing.

We then introduce the effective lensing potential $\psi(\btheta)$
defined as: 
\begin{equation}
\psi(\btheta) \simeq
\frac{2}{c^2}
\frac{\Dls}{\Dl \Ds}
\int_{\chi_l-\Delta\chi/2}^{\chi_l+\Delta\chi/2}
\Psi[\chi,r(\chi_l)\btheta] \, a d\chi,
\end{equation}
where $\Dl$ is the angular diameter distance between the
observer and the lens, $\Dl=a(\chi_l)r(\chi_l)$.
In terms of $\psi(\btheta)$,
%the lens equation is expressed as
%\begin{equation}
% \label{eq:thinlens}
% \bbeta=\btheta + \balpha(\btheta),
%\end{equation}
%with
the deflection field $\balpha(\theta)$ is expressed as:
\begin{equation}
  \balpha(\btheta)=-\bnabla_{\theta}\psi(\btheta),
\end{equation}
where  
$\bnabla_{\theta}=r\bnabla_{\perp}=(\partial_\theta,\theta^{-1}\partial_\phi)$.

With the Fermat or time-delay potential defined by:
\begin{equation}
 \label{eq:fermat}
 \tau(\btheta; \bbeta)\equiv \frac{1}{2}(\btheta-\bbeta)^2-\psi(\btheta),
\end{equation}
the lens equation can be equivalently written as 
$\bnabla_{\theta}\tau(\btheta;\bbeta)=0$ \citep{Blandford+Narayan1986}.
Here the first term on the right hand side of Eq.~(\ref{eq:fermat})
is responsible for the geometric delay and the second term for the
gravitational time delay.
The Fermat potential
$\tau(\btheta;\bbeta)$ is related to the time delay
$\Delta t$ with respect to the unperturbed path in the observer frame by
$\Delta t(\btheta;\bbeta)=\Dl\Ds/(c\Dls)(1+z_l)\tau(\btheta;\bbeta)\equiv D_{\Delta t}\tau(\btheta;\bbeta)/c$. 
with $D_{\Delta t}=(1+z_l)\Dl\Ds/\Dls\propto H_0^{-1}$ the time-delay
distance \citep{Refsdal1964}.
According to Fermat's principle,
the images for a given source position $\bbeta$ are formed at the
stationary points of $\tau(\btheta;\bbeta)$ with respect to variations
of $\btheta$ \citep{Blandford+Narayan1986}.

Note that cluster gravitational lensing is also affected by
uncorrelated large-scale structure projected along the line of sight
\citep[e.g.,][]{1998MNRAS.296..873S,2003MNRAS.339.1155H,Umetsu+2011stack,Host2012}.
The intervening large-scale structure in the universe perturbs the 
propagation of light from distant background galaxies, producing small
but continuous transverse excursions along the light path.
For a given depth of observations, the impact of such \emph{cosmic noise}
is most important in the cluster outskirts where the cluster lensing
signal is small
\citep{2003MNRAS.339.1155H,Becker+Kravtsov2011,Gruen2015}. 

\subsubsection{Convergence and shear}
\label{subsubsec:jacobian}

Let us work with local Cartesian coordinates $\btheta=(\theta_1,\theta_2)$
centered on a certain reference point in the image plane.
The local properties of the lens mapping are described by the Jacobian
matrix defined as:
\begin{equation}
{\cal A}(\btheta):=\frac{\partial \bbeta}{\partial \btheta}
=
\left( 
	\begin{array}{cc} 
	1 - \psi_{,11}& - \psi_{,12} \\
	-\psi_{,12} & 1 - \psi_{,22}
	\end{array}\right),
\end{equation}
where we have introduced the notation,
$\psi_{,ij}=\partial^2\psi/\partial\theta_i\partial\theta_j$ ($i,j=1,2$).
Alternatively, we can write the Jacobian matrix as
${\cal A}_{ij}=\delta_{ij}-\psi_{,ij}$ ($i,j=1,2$) with $\delta_{ij}$
the Kronecker delta in two dimensions. 
This symmetric $2\times 2$ Jacobian matrix ${\cal A}$ can be decomposed
as: 
\begin{equation}
{\cal A}=(1-\kappa){\cal I}- \gamma_1\sigma_3 - \gamma_2\sigma_1,
\end{equation}
where
$\sigma_{a}$ $(a=1,2,3)$ are the Pauli matrices
(Sect.~\ref{subsec:multilens}); 
%%%
$\kappa(\btheta)$ is the lensing convergence responsible for the change
in the trace part of the Jacobian matrix ($\mathrm{tr}({\cal
A})=2(1-\kappa)$): 
\begin{equation}
\label{eq:kappa_def}
\kappa := \frac{1}{2}\left(\psi_{,11}+\psi_{,22}\right) = \frac{1}{2}\triangle\psi
\end{equation}
with $\triangle = \bnabla_\theta^2$, and ($\gamma_1,\gamma_2$) are the
 two components of the complex shear
 $\gamma(\btheta):=\gamma_1(\btheta)+i\gamma_2(\btheta)$:
\begin{equation}
 \begin{aligned}
  \gamma_1 &:= \frac{1}{2}\left(\psi_{,11}-\psi_{,22}\right),\\ 
  \gamma_2 &:= \frac{1}{2}\left(\psi_{,12}+\psi_{,21}\right)=\psi_{,12}.
 \end{aligned}
\end{equation}
Note that Eq.~(\ref{eq:kappa_def}) can be regarded as a two-dimensional Poisson
equation, $\triangle\psi(\btheta) = 2\kappa(\btheta)$.
Then, the Green function in the (hypothetical) infinite
domain is $\triangle^{-1}(\btheta,\btheta')=\ln|\btheta-\btheta'|/(2\pi)$,\footnote{Here we assume that the field size is 
sufficiently larger than the characteristic angular scale of the lensing  
clusters but small enough for the flat-sky assumption to be valid.}
so that the convergence is related to the lensing potential as:
\begin{equation}
 \psi(\btheta)=\frac{1}{\pi}
  \int\!\ln(\btheta-\btheta') \kappa(\btheta')\,d^2\theta'.
\end{equation}
The Jacobian matrix is expressed in terms of $\kappa$ and $\gamma$ as:
\begin{equation}
 \label{eq:jacob}
{\cal A}(\btheta) =
\left( 
	\begin{array}{cc} 
	1 - \kappa - \gamma_1 & - \gamma_2 \\
	- \gamma_2 & 1 - \kappa + \gamma_1
	\end{array}\right).
%(1-\kappa)\left( 
%	\begin{array}{cc} 
%	1  & 0 \\
%	0 & 1 
%	\end{array}\right)-
%\left( 
%	\begin{array}{cc} 
%	\gamma_1  & \gamma_2 \\
%	\gamma_2 &  -\gamma_1
%	\end{array}\right).
\end{equation}
The determinant of the Jacobian matrix (Eq.~(\ref{eq:jacob})) is given
as $\mathrm{det}{\cal A}=(1-\kappa)^2-|\gamma|^2$. 
In the weak-lensing limit where
$|\kappa|, |\gamma| \ll 1$, $\mathrm{det}{\cal A}\simeq 1-2\kappa$
to the first order.

The deformation of the image of an infinitesimal circular source
($d\bbeta\to 0$) behind the lens can be described by the inverse
Jacobian matrix ${\cal A}^{-1}$ of the lens equation.
In the weak-lensing limit ($|\kappa|, |\gamma|\ll 1$), we have:
\begin{equation}
\label{eq:distortion}
\left({\cal A}^{-1}\right)_{ij} 
\simeq (1 + \kappa)\delta_{ij} + {\Gamma}_{ij} \ \ \ (i,j=1,2),
\end{equation} 
where $\Gamma_{ij}$ is the symmetric trace-free shear matrix defined by 
\citep{2001PhR...340..291B, 2002ApJ...568...20C}:
\begin{equation}
\Gamma_{ij}= \left(\partial_i\partial_j-\delta_{ij}\frac{1}{2}\triangle\right)\psi(\btheta),
\end{equation}
with $\partial_i:=\partial/\partial\theta_i$ ($i=1,2$).
The shear matrix can be expressed in terms of the Pauli matrices as
$\Gamma=\sigma_3\gamma_1+\sigma_1\gamma_2$. 
The first term in Eq.~(\ref{eq:distortion}) describes the isotropic
light focusing or area distortion in the weak-lensing limit, while the
second term induces an asymmetry in lens mapping. The shear $\gamma$ is 
responsible for image distortion and can be directly observed from 
image ellipticities of background galaxies in the regime where
$|\kappa|,|\gamma|\ll 1$ (see Sect.~\ref{sec:basics}).  
%Note that both $\kappa$ and $\gamma$ contribute to the isotropic and
%anisotropic distortions in the non-weak lensing regime.
Note that both $\kappa$ and $\gamma$ contribute to the area and shape
distortions in the non-weak-lensing regime.

%%%%%%%%%%%%%%%%%%%%%%%%%%%%%%%%%%%%%%%%%%%%%%%%%%%%%%%%%%%%%%%%%%%%%%%%%%%%%%%
%%% Figure 3
%%%%%%%%%%%%%%%%%%%%%%%%%%%%%%%%%%%%%%%%%%%%%%%%%%%%%%%%%%%%%%%%%%%%%%%%%%%%%%%

\begin{figure*}[!htb] %!htb
  \begin{center}
   \includegraphics[scale=0.6, angle=0, clip]{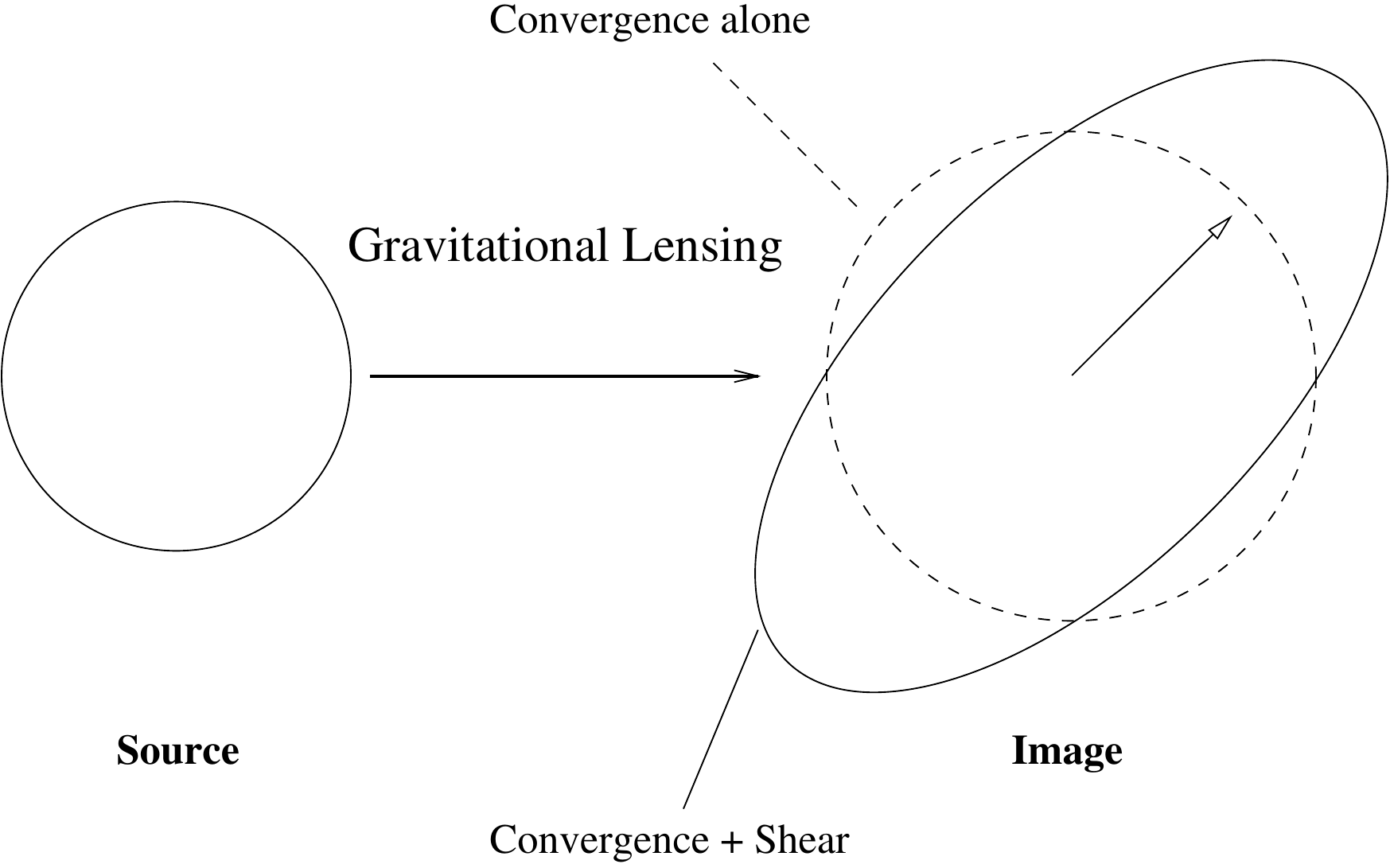} 
  \end{center}
\caption{
\label{fig:deformation} 
Illustration of the effects of the convergence $\kappa$ and the shear
 $\gamma$ on the angular shape and size of a hypothetical circular
 source. The convergence acting alone causes an isotropic focusing
 (magnification) of the image (dashed circle), while the shear deforms
 it to an ellipse. 
} 
\end{figure*}

In Fig.~\ref{fig:deformation}, we illustrate the effects of the lensing
convergence $\kappa$ and the gravitational shear $\gamma$ on the angular
shape and size of an infinitesimal circular source.
The convergence acting alone causes an isotropic magnification of the
image, while the shear deforms it to an ellipse.
Note that the magnitude of ellipticity induced by gravitational shear in
the weak-lensing regime ($|\gamma|\simlt 0.1$) is much smaller than
illustrated here.

\subsubsection{Magnification}
\label{subsubsec:mu}

Gravitational lensing describes the deflection of light by gravity.
Lensing conserves the surface brightness of a background source, a
consequence of Liouville's theorem.  
On the other hand, lensing causes focusing of light rays, resulting in
an amplification of the image flux through the local solid-angle
distortion. 
Lensing magnification $\mu$ is thus given by taking the ratio
between the lensed to the unlensed image solid angle 
as $\mu=\delta\Omega^I/\delta\Omega^S = 1/\mathrm{det}{\cal A}$, with: 
\begin{equation}
\mu(\btheta)=\frac{1}{[1-\kappa(\btheta)]^2-|\gamma(\btheta)|^2}.
\end{equation}
In the weak-lensing limit ($|\kappa|,|\gamma|\ll 1$), the magnification
factor to the first order is:
\begin{equation}
\mu(\btheta) \simeq 1+2\kappa(\btheta).
\end{equation}
The magnitude change at $\kappa(\btheta)=0.1$ is thus
$\Delta m = -(5/2)\log_{10}(\mu) \sim -0.2$.

\subsubsection{Strong- and weak-lensing regimes}
\label{subsubsec:slwl}

The ${\cal A}(\btheta)$ matrix has two local eigenvalues
$\Lambda_\pm(\btheta)$ at each image position $\btheta$:
\begin{equation}
\label{eq:eigen}
 \Lambda_{\pm}=1-\kappa\pm|\gamma|,
\end{equation}
with $\Lambda_+ \geqs \Lambda_-$.

\begin{figure*}[!htb] %!htb
  \begin{center}
   \includegraphics[scale=0.7, angle=0, clip]{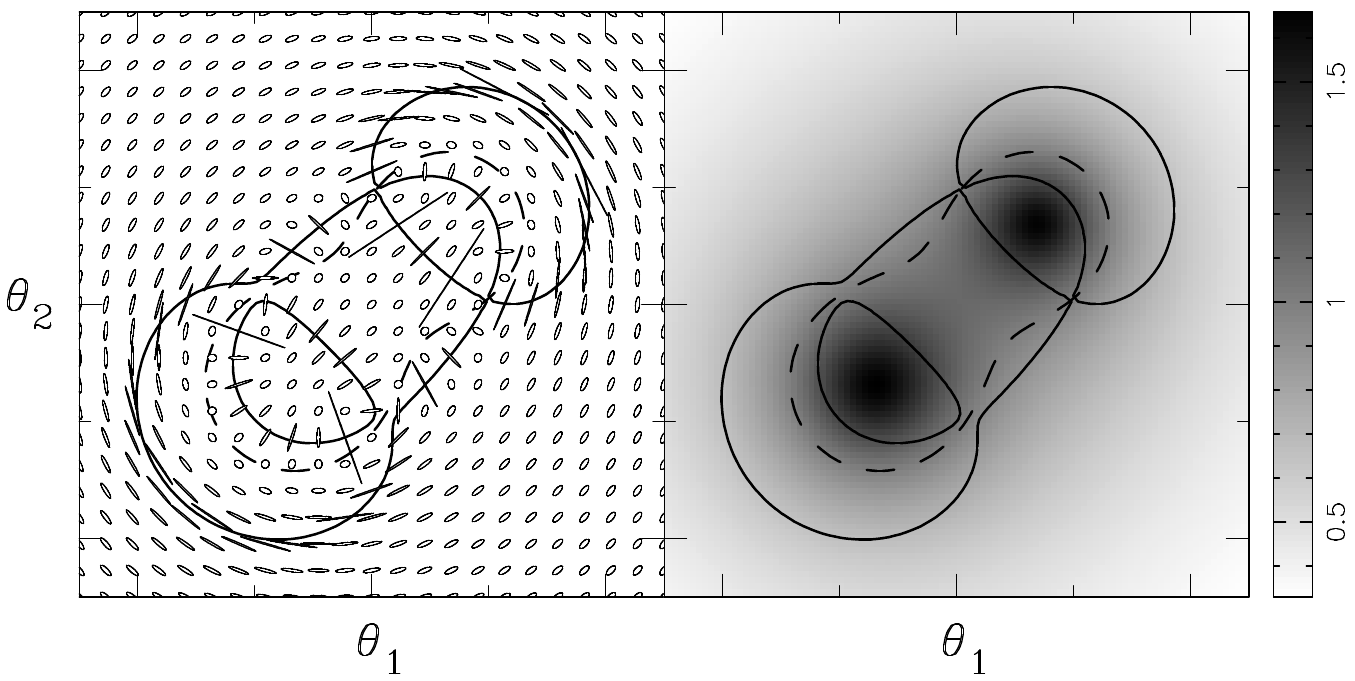} 
  \end{center}
\caption{
 \label{fig:demo}
Shape distortion field produced by a simulated lens with a bimodal mass
 distribution. 
At each grid point in the image plane (left panel), we have drawn the
 apparent shape for an intrinsically circular source using the local
 deformation factors $\Lambda_\pm$ (Eq.~(\ref{eq:eigen})). All
 ellipses have an equal area regardless of the magnification factor.
The right panel shows the $\kappa$ map of the bimodal lens.
In both panels, the solid lines indicate the critical curves.
The image distortion disappears locally along the curve $\kappa=1$
 indicated by the dashed line, which lies in the negative-parity
 region.
}
\end{figure*}

Images with $\mathrm{det}{\cal A}(\btheta)>0$
have the same parity as the source,
while those with $\mathrm{det}{\cal A}(\btheta)<0$
have the opposite parity to the source.
A set of closed curves defined by $\mathrm{det}{\cal A}(\btheta)=0$
in the image plane are referred to as {\it critical curves}, on which
lensing magnification formally diverges, and those mapped into 
the source plane are referred to as {\it caustics}
\citep[see][]{Hattori1999}. 
The critical curves separate the image plane into
even- and odd-parity regions with
$\mathrm{det}{\cal A}>0$ and
$\mathrm{det}{\cal A}<0$, respectively.

An infinitesimal circular source is transformed to an ellipse with a
minor-to-major axis ratio ($\leqs 1$) of $|\Lambda_-/\Lambda_+|$ for $\kappa<1$ and
$|\Lambda_+/\Lambda_-|$ for $\kappa > 1$,
and it is magnified by the factor $|\mu|=1/|\Lambda_+\Lambda_-|$ (see
Sect.~\ref{subsubsec:mu}).
The gravitational distortion locally disappears along the curve defined
by $\mathrm{tr}[{\cal A}(\btheta)]=0$, i.e., $\kappa(\btheta)=1$,
which lies in the odd-parity region \citep{Kaiser1995}.
This is illustrated in Fig.~\ref{fig:demo} for a simulated lens with a
bimodal mass distribution.
%These two conditions define distinct sets of curves in the lens plane, called the
%tangential and the radial critical line, respectively.
Images forming along the outer (tangential) critical curve
$\Lambda_{-}(\btheta)=0$ 
are distorted tangentially to this curve, while images forming close to the 
inner (radial) critical curve $\Lambda_{+}(\btheta)=0$ are stretched in the
direction perpendicular to the critical curve.

A lens system that has a region with $\kappa(\btheta)>1$ can produce
multiple images for certain source positions $\bbeta$, and such a system
is referred to as being {\it supercritical}.  
Note that being supercritical is a sufficient but not a necessary
condition for a general lens to produce multiple images, because the
shear can also contribute to multiple imaging.
Nevertheless, this provides us with
a simple criterion to broadly distinguish the regimes of multiple and
single imaging.  
Keeping this in mind, we refer to the region where $\kappa(\btheta) \simgt 1$
as the strong-lensing regime and the region where $\kappa(\btheta)\ll 1$
as the weak-lensing regime.

\subsubsection{Critical surface mass density}
\label{subsubsec:crit}

The lensing convergence $\kappa$ is essentially a distance-weighted
mass overdensity projected along the line of sight.
We express $\kappa(\btheta)$ due to cluster gravitational lensing as:
\begin{equation}
 \label{eq:kappa}
\kappa(\btheta)=
\int_0^{\chi_s}\! (\rho-\overline{\rho})
\left(
 \frac{c^2}{4\pi G}\frac{\Ds}{\Dl \Dls}
\right)^{-1} a d\chi
\simeq \frac{\Sigma(\btheta)}{\Sigmacr},
\end{equation}
where $\chi_s$ is the comoving distance to the source plane;
$\Sigma=\int_0^{\chi_s}\! (\rho-\overline{\rho})\,a d\chi$
is the surface mass density field of the lens projected
on the sky; and $\Sigmacr$ is the critical surface mass density of 
gravitational lensing:\footnote{
In the weak-lensing literature, projected densities and
distances are often defined to be in comoving units. For example, the
critical surface mass density for lensing in comoving units,
$\Sigmacr^\mathrm{(c)}(z_l,z_s)$, is related to that in physical units,
$\Sigmacr(z_l,z_s)$, as $\Sigmacr^\mathrm{(c)}=\Sigmacr (1+z_l)^{-2}$.
Similarly, the comoving projected separation
$R^\mathrm{(c)}$ is related to that in physical units, $R$,
as $R^\mathrm{(c)}=(1+z_l)R$.}
\begin{equation}
 \begin{aligned}
  \Sigmacr(z_l,z_s)  &= \frac{c^2}{4\pi G}\frac{\Ds}{\Dl \Dls}
  \end{aligned}
\end{equation}
for $z_s > z_l$ and
$\Sigmacr^{-1}(z_l,z_s) = 0$ (i.e., $\Dls/\Ds = 0$) for
an unlensed source with $z_s\leqs z_l$.
In the second (approximate) equality of Eq.~(\ref{eq:kappa}), we
have explicitly used the thin-lens approximation
(Sect.~\ref{subsubsec:thin}). 
The critical surface mass density $\Sigmacr$ depends on the geometric
configuration ($z_l, z_s$) of the lens--source system
and the background cosmological parameters, such as ($\Om, \OL, H_0$).
For example, for $z_l=0.3$ and $z_s=1$ in our fiducial cosmology, we have
$\Sigmacr \approx 4.0 \times 10^{15}hM_\odot$\,Mpc$^{-2}$.
For a fixed lens redshift $z_l$, the geometric efficiency of
gravitational lensing is determined by the distance ratio $\Dls/\Ds$
as a function of $z_s$ and the background cosmology.
%$\Sigmacr \sim 0.1h\,$g\,cm$^{-2}\sim 0.5h \times 10^{15}M_\odot$\,Mpc$^{-2}$ for
%$\Dls\sim \Dl\sim \Ds\sim O(c/H_0)$.

\begin{figure*}[!htb] %!htb
 \begin{center}
  \includegraphics[scale=0.5, angle=0, clip]{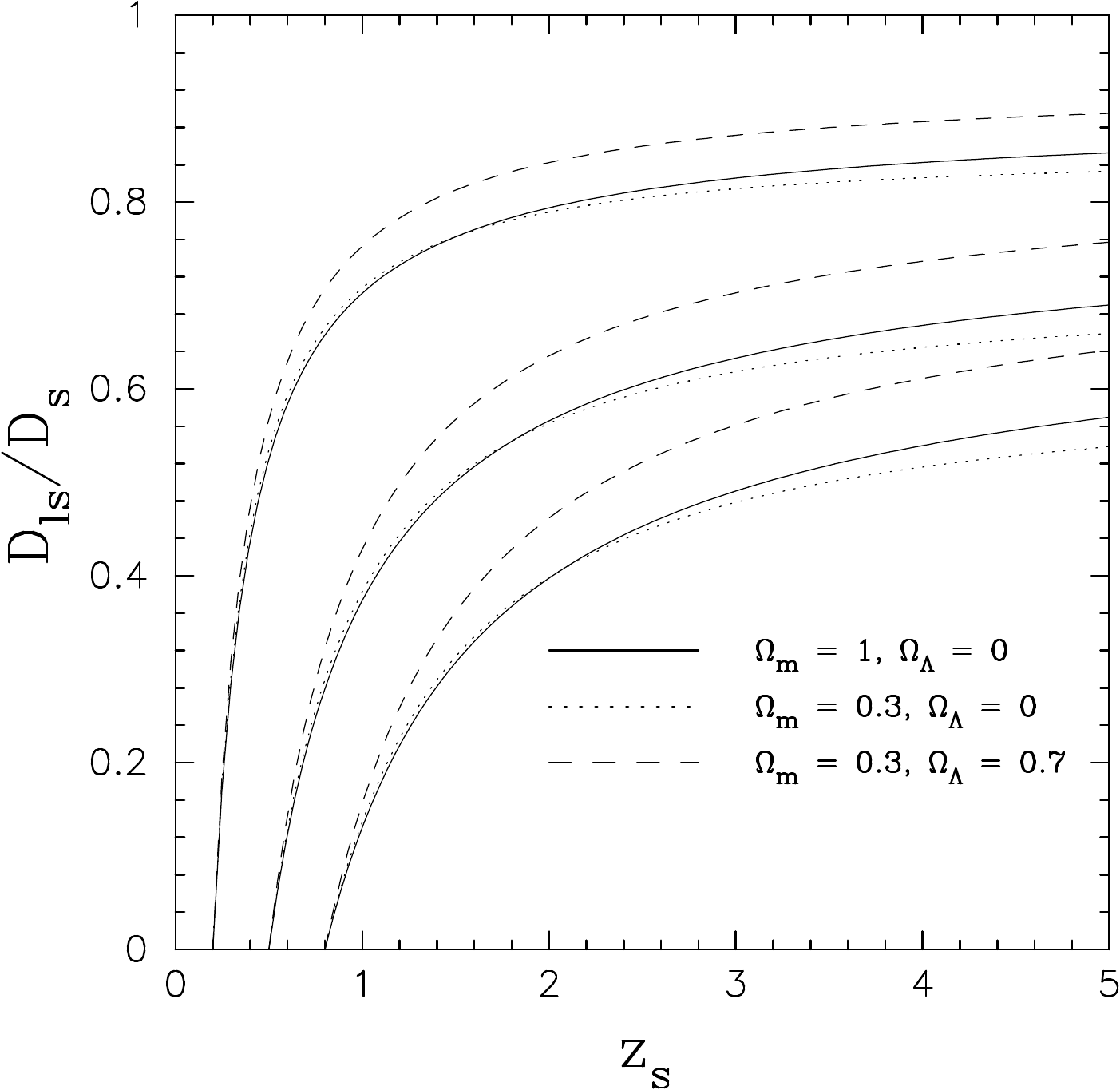} 
 \end{center}
\caption{
 \label{fig:dratio}
 Distance ratio $\Dls/\Ds$ as a function of the source redshift $z_s$ for various sets of the
 lens redshift $z_l$ and the cosmological parameters $(\Om,\OL)$.
 The ratio $\Dls/\Ds$ is shown for three different lens redshifts,
 $z_l=0.2, 0.5$, and $0.8$ (from left to right)
 and for three sets of the cosmological parameters,
 $(\Om,\OL)=(1,0), (0.3,0)$, and $(0.3,0.7)$.
}
\end{figure*}

To translate the observed lensing signal into surface mass
densities, one needs an estimate of $\Sigmacr(z_l,z_s)$ for a given
background cosmology.  
In the regime where $z_l\ll z_s$
(say, $z_l\simlt 0.2$ for background galaxy populations at $z_s\sim 1$),
$\Sigmacr$ depends weakly on the source redshift $z_s$,
so that a precise knowledge of the source-redshift distribution is less
critical \citep[e.g.,][]{Okabe+Umetsu2008,Okabe+2010WL}. 

Conversely, this distance dependence of the lensing effects can
be used to constrain the cosmological redshift--distance relation by 
examining the geometric scaling of the lensing signal as a function of
the background redshift
\citep{2007MNRAS.374.1377T,Taylor+2012,Medezinski+2011,DellAntonio2019}.  
Figure~\ref{fig:dratio} compares
$\Dls/\Ds$ as a function of $z_s$ for various sets of the lens
redshift and the 
cosmological model. 

Note that, in the limit where the lensing matter is continuously
distributed along the line of sight, the first equality of
Eq.~(\ref{eq:kappa}) can be formally rewritten as:
\begin{equation}
 \label{eq:cosmok}
 \kappa(\btheta)=
  \frac{3H_0^2}{2c^2}\Om\int_0^{\chi_s}\!d\chi\,g(\chi,\chi_s)a^{-1}(\chi)\delta[\chi,r(\chi)\btheta],
\end{equation}
 with $g(\chi,\chi_s)=r(\chi)r(\chi_s-\chi)/r(\chi_s)$
 and
$\delta=\delta\rho/\overline{\rho}$.  Equation~(\ref{eq:cosmok})
 coincides with the expression for the cosmic convergence due to
 intervening cosmic structures \citep[see][]{2000ApJ...530..547J}.
 %From the above expression, it is interesing to compare the
 It is interesting to compare the above line-of-sight integral
 (Eq.~(\ref{eq:cosmok})) to the thermal Sunyaev--Zel'dovich effect
 (SZE) in terms of the Compton-$y$ parameter 
 \citep[e.g.,][]{SZ1972,1995ARA&A..33..541R,1999PhR...310...97B}:
 \begin{equation}
  \label{eq:SZE}
 y=\frac{\sigma_\mathrm{T}}{m_\mathrm{e} c^2}\int\! n_\mathrm{e} k_\mathrm{B}
  \left(T_\mathrm{e}-T_\mathrm{CMB}\right)\,cdt\simeq
  \frac{\sigma_\mathrm{T}}{m_\mathrm{e} c^2}\int\! P_\mathrm{e}\,cdt,
 \end{equation}
 where $\sigma_\mathrm{T}$, $m_\mathrm{e}$, $k_\mathrm{B}$ are
 the Thomson scattering cross-section, the electron mass, and the Boltzmann
 constant, respectively; $T_\mathrm{CMB}= T_0(1+z)$ is the temperature
 of CMB photons with $T_0=2.725$\,K; and
 $T_e$ and $n_e$ are the electron temperature and number
 density of the intracluster gas, with
 $P_\mathrm{e}= n_\mathrm{e} k_\mathrm{B}T_\mathrm{e}$ the electron
 pressure. In the second (approximate) equality, we have used
 $T_e\gg T_0(1+z)$.
The Compton-$y$ parameter is proportional to the electron pressure
 integrated along the line of sight, thus probing the thermal energy
 content of thermalized hot plasmas residing in the gravitational
 potential wells of galaxy clusters.  The combination of the
 thermal SZE and weak lensing thus provides unique astrophysical and 
 cosmological probes \citep[e.g.,][]{Dore+2001,Umetsu+2009,Osato+2020}.

\subsubsection{Einstein radius}
\label{subsubsec:rein}

Detailed strong-lens modeling using many sets of multiple
images with measured spectroscopic redshifts allows us to determine the
location of the critical curves \citep[e.g.,][]{Zitrin2015clash,Meneghetti2017},
which, in turn, provides accurate estimates of the projected total mass
enclosed by them. 
In this context,  the term Einstein radius is often used to refer to the
size of the outer (tangential) critical curve (i.e.,
$\Lambda_{-}(\btheta)=0$; Sect.~\ref{subsubsec:slwl}). 
We note, however, that there are
several possible definitions of the Einstein radius used in the literature
\citep[see][]{Meneghetti2013arc}.  
Here we adopt the \emph{effective} Einstein radius definition
\citep{Redlich+2012,Meneghetti2013arc,Meneghetti2017,Zitrin2015clash}, 
$\vartheta_\mathrm{Ein}=\sqrt{A_\mathrm{c}/\pi}$, 
where $A_\mathrm{c}$ is the (angular) area enclosed by the outer
critical curve.  
For an axisymmetric lens, the average surface mass density within the
critical area is equal to $\Sigmacr$
\citep[see][]{Hattori1999,Meneghetti2013arc}, thus enabling us to
directly estimate the enclosed projected mass by
$M_\mathrm{2D}(<\vartheta_\mathrm{Ein})=\pi (\Dl\vartheta_\mathrm{Ein})^2\Sigmacr$.
Even for general non-axisymmetric lenses,
the projected enclosed mass profile
$M_\mathrm{2D}(<\vartheta)=\Sigmacr \Dl^2\int_{\vartheta'\le\vartheta}\!\kappa(\btheta')\,d^2\theta'$
at the location $\vartheta \sim \vartheta_\mathrm{Ein}$
is less sensitive to modeling assumptions and approaches
\citep[e.g.,][]{Umetsu+2012,Umetsu2016clash,Meneghetti2017},
thus serving as a fundamental observable quantity in the strong-lensing
regime \citep{Coe+2010}.

\section{Basics of cluster weak lensing}
\label{sec:basics}

In this section, we review the basics of cluster--galaxy weak lensing
based on the thin-lens formalism (Sect.~\ref{subsec:thinlens}). Unless
otherwise noted, we will focus on subcritical lensing (i.e., outside the
critical curves). We consider both linear ($\kappa\ll 1$) and mildly
nonlinear regimes of weak gravitational lensing.

\subsection{Weak-lensing mass reconstruction}
\label{subsec:massrec}

\subsubsection{Spin operator and lensing fields}
\label{subsubsec:spin}

For mathematical convenience, we introduce a concept of ``spin''
for weak-lensing quantities as follows
\citep{2006MNRAS.365..414B,HOLICs1,HOLICs2,Schneider+Er2008,Bacon+2009}:
a quantity is said to have spin $N$ if it has the same
value after rotation by $2\pi/N$. The product of spin-$A$ and spin-$B$
quantities has spin ($A+B$), and the product of spin-$A$ and spin-$B^*$
quantities has spin ($A-B$), where $*$ denotes the complex conjugate.

We define a complex spin-1 operator $\partial:=\partial_1+i\partial_2$ 
that transforms as a vector, $\partial'=\partial e^{i\varphi}$, with
$\varphi$  being the angle of rotation relative to the original basis.
%(see 
%Bacon et al. 2006;
%Okura, Umetsu, \& Futamase 2007;
%Okura, Umetsu, \& Futamase 2008).
Then, the lensing convergence is expressed in terms of $\psi(\btheta)$
as: 
\begin{equation}
\kappa(\btheta) = \frac{1}{2}\partial^*\partial\psi(\btheta),
\end{equation} 
where $\partial\partial^*=\nabla_\theta^2$ is a scalar or 
a spin-0 operator.
Similarly, the complex shear $\gamma=\gamma_1+i\gamma_2\equiv |\gamma|e^{2i\phi_\gamma}$
is expressed as:
\begin{equation}
\gamma(\btheta) = \frac{1}{2}\partial \partial\psi(\btheta) = \hat{\cal D} \psi(\btheta),
\end{equation} 
where
\begin{equation}
 \hat{\cal D}:=\partial\partial/2 =(\partial_1^2-\partial_2^2)/2+i\partial_1\partial_2
\end{equation}
is a spin-2 operator that transforms such that
$\hat{\cal D'}=\hat{\cal D}e^{2i\varphi}$ under a rotation of the basis axes by $\varphi$.

\subsubsection{Linear mass reconstruction}
\label{subsubsec:linrec}

Since $\gamma(\btheta)$ and $\kappa(\btheta)$ are both
linear combinations of the second derivatives of
$\psi(\btheta)$, they are related  to each other by
\citep{Kaiser1995,2002ApJ...568...20C,Umetsu2010Fermi}:
\footnote{An equivalent expression is
$\triangle\kappa(\btheta)=\partial^{i}\partial^{j}
\Gamma_{ij}(\btheta)$.}
\begin{equation}
\label{eq:local}
\triangle\kappa (\btheta)
= \partial^*\partial^*\gamma(\btheta)
= 2\hat{\cal D}^*\gamma(\btheta).
\end{equation}
The shear-to-mass inversion can thus be formally expressed as:
\begin{equation}
\label{eq:inversion}
\kappa(\btheta)
=
\triangle^{-1}(\btheta,\btheta')\left[
			       \partial^*\partial^*\gamma(\btheta')
			      \right]
=2{\hat{\cal D}^*}
\left(
\triangle^{-1}(\btheta,\btheta')\left[\gamma(\btheta')\right]
\right).
\end{equation}
%For the two-dimensional Poisson equation,
%the Green function $\triangle^{-1}(\btheta,\btheta')$
%in the infinite domain is ... so that
Using $\triangle^{-1}(\btheta,\btheta')=\ln|\btheta-\btheta'|/(2\pi)$
(Sect.~\ref{subsubsec:jacobian}),
Eq.~(\ref{eq:inversion}) in the flat-sky limit can be
solved to yield the following nonlocal relation between $\kappa$ and $\gamma$
\citep[][hereafter KS93]{1993ApJ...404..441K}:
\begin{equation}
\label{eq:ks93}
\kappa(\btheta) -\kappa_0 = 
\frac{1}{\pi}
\int\! d^2\theta'\,D^*(\btheta-\btheta')\gamma(\btheta'),
\end{equation}
where $\kappa_0$ is an additive constant and
$D(\btheta)$ is a complex kernel defined as:
\begin{equation}
\label{eq:kerneld}
D(\btheta) \equiv 2\pi \hat{\cal D}\left[\triangle^{-1}(\btheta)\right]=
\frac{\theta_2^2-\theta_1^2-2i\theta_1\theta_2}{|\theta|^4}
=-\frac{1}{(\theta_1-i\theta_2)^2}.
\end{equation}
Similarly, the complex shear field can be expressed in terms of the
convergence $\kappa$ as:
\begin{equation} 
\label{eq:kappa2gamma}
\gamma(\btheta) = 
\frac{1}{\pi}\int\!d^2\theta'\,D(\btheta-\btheta')\kappa(\btheta').
\end{equation} 
This linear mass inversion formalism is often
referred to as the KS93 algorithm.

It is computationally faster to work in Fourier domain
\citep{2000ApJ...530..547J} 
using the fast Fourier transform algorithm. By taking the Fourier
transform of Eq.~(\ref{eq:local}), we have a mass inversion
relation in the conjugate Fourier space as:
\begin{equation} 
 \begin{aligned}
  \kappa(\btheta) &= \int\!\frac{d^2k}{(2\pi)^2}\,\hat{\kappa}(\bk) e^{i\bk\cdot\btheta},\\
  \hat{\kappa}(\bk) &= \frac{k_1^2-k_2^2-2ik_1 k_2}{k_1^2+k_2^2} \hat{\gamma}(\bk) \ \ \ (\bk \ne 0),
 \end{aligned}
\end{equation}
where $\bk$ is the two-dimensional wave vector conjugate to $\btheta$,
and
$\hat{\kappa}(\bk)$ and $\hat{\gamma}(\bk)$ are the Fourier
 transforms of
 $\kappa(\btheta)$ and
 $\gamma(\btheta)=\gamma_1(\btheta)+i\gamma_2(\btheta)$,
 respectively. 
In practical applications, one may assume
$\hat{\kappa}(0)=0$ if the angular size of the observed
shear field is sufficiently large, so that the mean convergence across
the data field is approximated to zero. Otherwise, one must explicitly
account for the boundary conditions imposed by the observed shear field
to perform a mass reconstruction on a finite field
\citep[e.g.,][]{Kaiser1995,Seitz1996,1996ApJ...464L.115B,Seitz+Schneider1997,Umetsu+Futamase2000}. 
%%% 

\begin{figure*}[!htb] %!htb
 \begin{center}
  \includegraphics[scale=0.33, angle=0, clip]{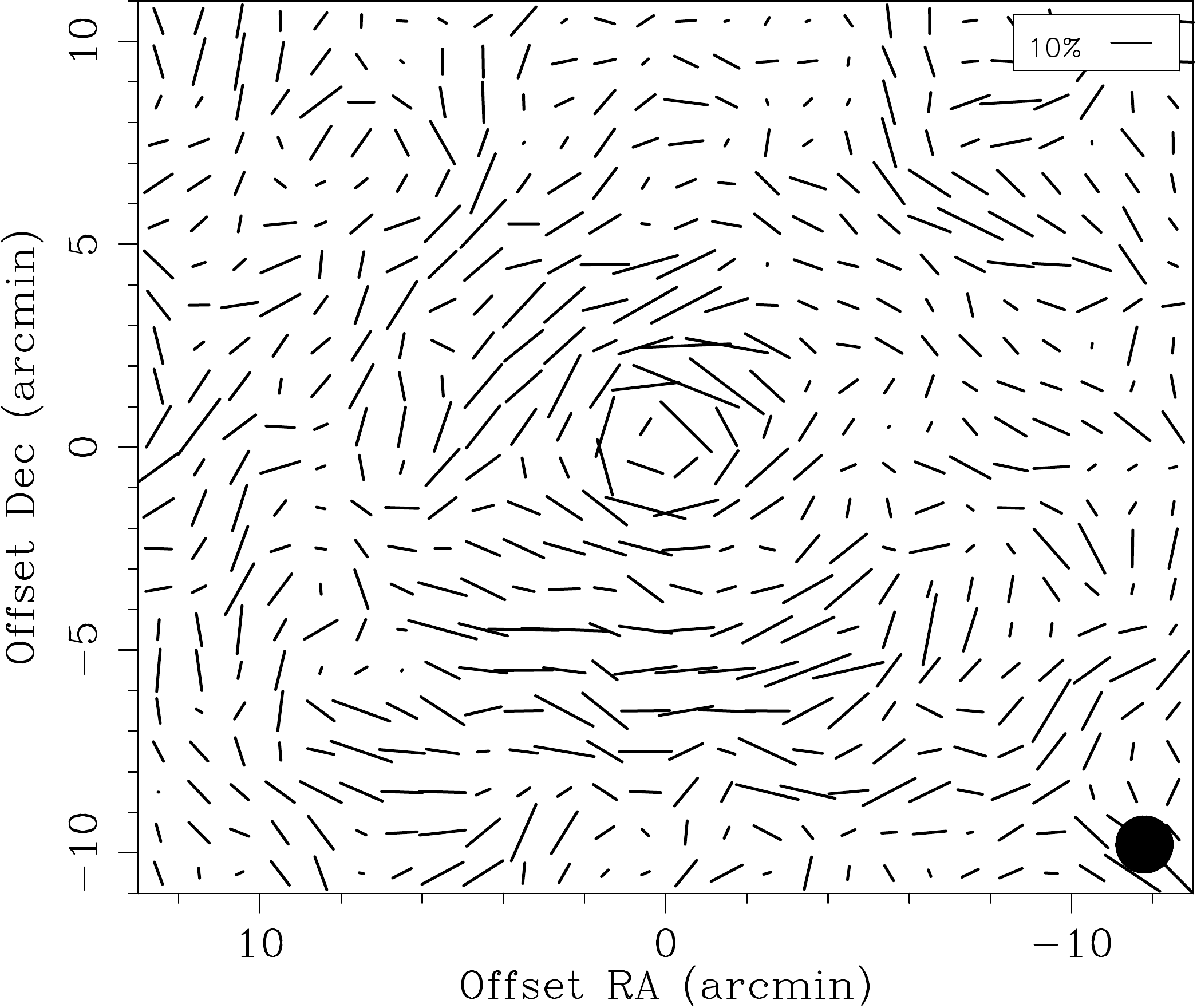}
 \end{center}
\caption{
 \label{fig:cl0024gmap}
Shape distortion field in the rich cluster Cl0024+1654
 ($z_l=0.395$) obtained from deep weak-lensing observations taken with
 Subaru/Suprime-Cam. The mean surface number density of background
 galaxies after the color--color selection
 (Sect.~\ref{subsec:dilution})
 is $n_\mathrm{g}=17.2$\,arcmin$^{-2}$.
 A stick with a length of 10\% shear is indicated in the top right
 corner. The filled circle indicates the FWHM ($1.4$\,arcmin) of the
 Gaussian smoothing. 
 The distortion field exhibits a coherent tangential pattern
 around the cluster center.
 Image reproduced with permission from \citet{Umetsu+2010CL0024}, copyright by AAS.
 }
\end{figure*}

\begin{figure*}[!htb] %!htb
 \begin{center}
  \includegraphics[scale=0.5, angle=0, clip]{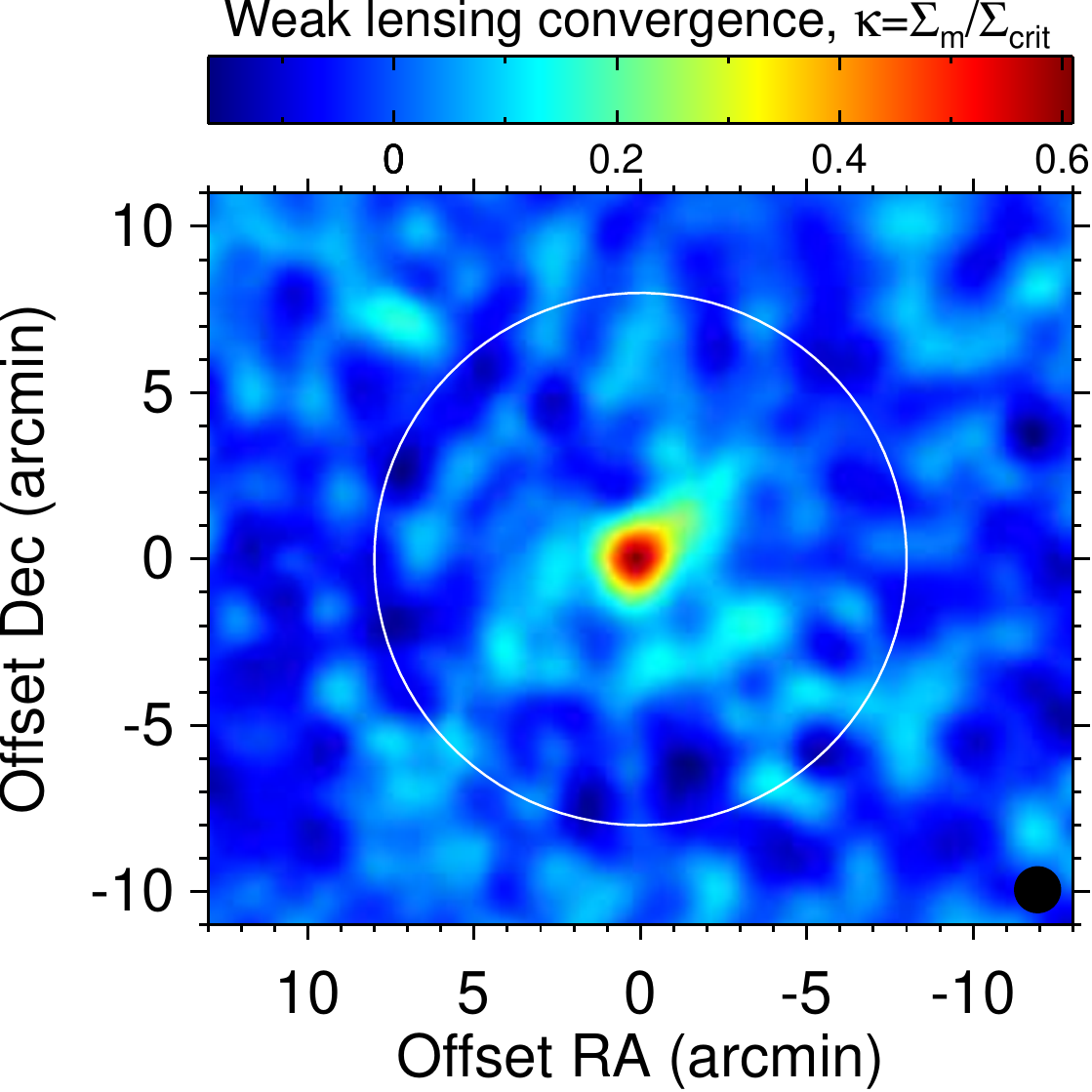}
  \includegraphics[scale=0.5, angle=0, clip]{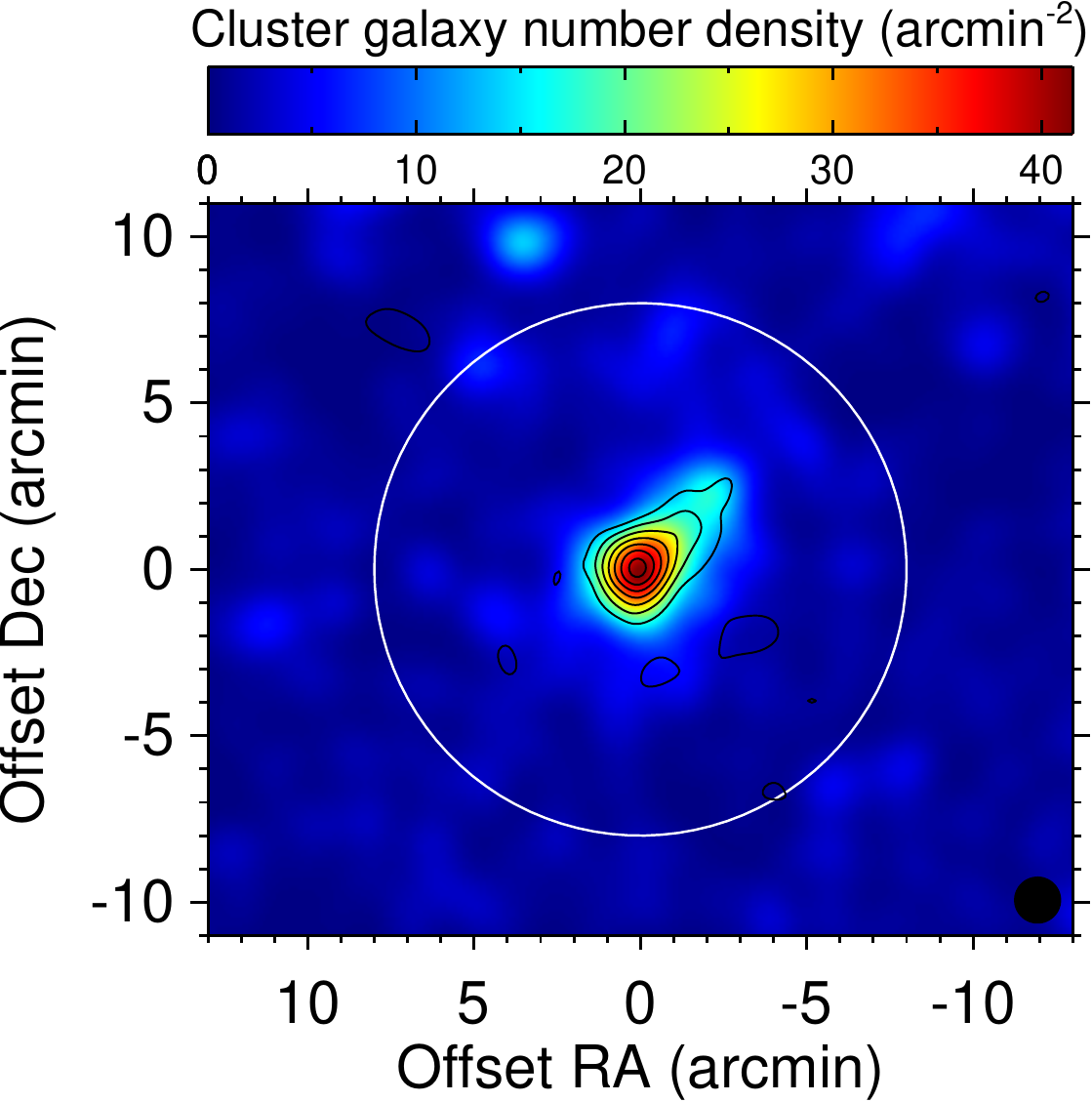}
 \end{center}
\caption{
 \label{fig:cl0024}
 Comparison of mass and galaxy distributions in the rich cluster
 Cl0024+1654 ($z_l=0.395$). 
 {\it Left panel}: projected mass distribution
 $\kappa(\btheta)=\Sigma(\btheta)/\Sigmacr$ 
 reconstructed from deep weak-lensing data (Fig.~\ref{fig:cl0024gmap})
 taken with Subaru/Suprime-Cam. 
 \emph{Right panel}: surface number density distribution $\Sigma_n(\btheta)$
 of color--color-selected cluster member galaxies.
  The solid circle in each panel indicates the cluster virial radius of 
 $r_\mathrm{vir}\approx 1.8\Mpch$.
Both maps are smoothed with a Gaussian of $1.4'$ FWHM. Also
 overlaid on the $\Sigma_n$ map is the $\kappa(\btheta)$ field shown in
 the left panel, given in units of $2\sigma$ reconstruction error from
 the lowest contour level of $3\sigma$.
% The field size is $26'\times 22'$.
 North is to the top, east to the left.
 Image reproduced with permission from \citet{Umetsu+2010CL0024}, copyright by AAS. 
}
\end{figure*}

In Fig.~\ref{fig:cl0024gmap}, we show the shape distortion field in
the rich cluster Cl0024+1654 ($z_l=0.395$) obtained by
\citet{Umetsu+2010CL0024}
from deep weak-lensing observations taken with Suprime-Cam on the 8.2\,m Subaru
telescope.
%For visualization purposes, the distortion field is smoothed with a
%Gaussian of $1.4'$ FWHM.
They accounted and corrected for the effect of the weight function used
for calculating noisy galaxy shapes, as well as for the anisotropic and 
smearing effects of the point spread function (PSF), using an improved
implementation of the modified \citet[][hereafter
KSB]{1995ApJ...449..460K} method (see Sect.~\ref{subsubsec:KSB}).
In the left panel of Fig.~\ref{fig:cl0024}, we show the
$\kappa(\btheta)$ field reconstructed from the Subaru
 weak-lensing data (see Fig.~\ref{fig:cl0024gmap}).
A prominent mass peak is visible in the cluster center, around which the
distortion pattern is clearly tangential (Fig.~\ref{fig:cl0024gmap}). 
 In this study, a variant of the linear KS93 algorithm was used to
 reconstruct the $\kappa$ map from the weak shear lensing data. 
In the right panel of Fig.~\ref{fig:cl0024}, we show the member galaxy
distribution $\Sigma_n(\btheta)$ in the cluster.
% Gaussian smoothed to the same resolution of $\mathrm{FWHM}=1.4'$.
Overall, mass and light are similarly distributed in the cluster.

\begin{figure*}[!htb] %!htb
  \begin{center}
   \includegraphics[scale=0.42, angle=0, clip]{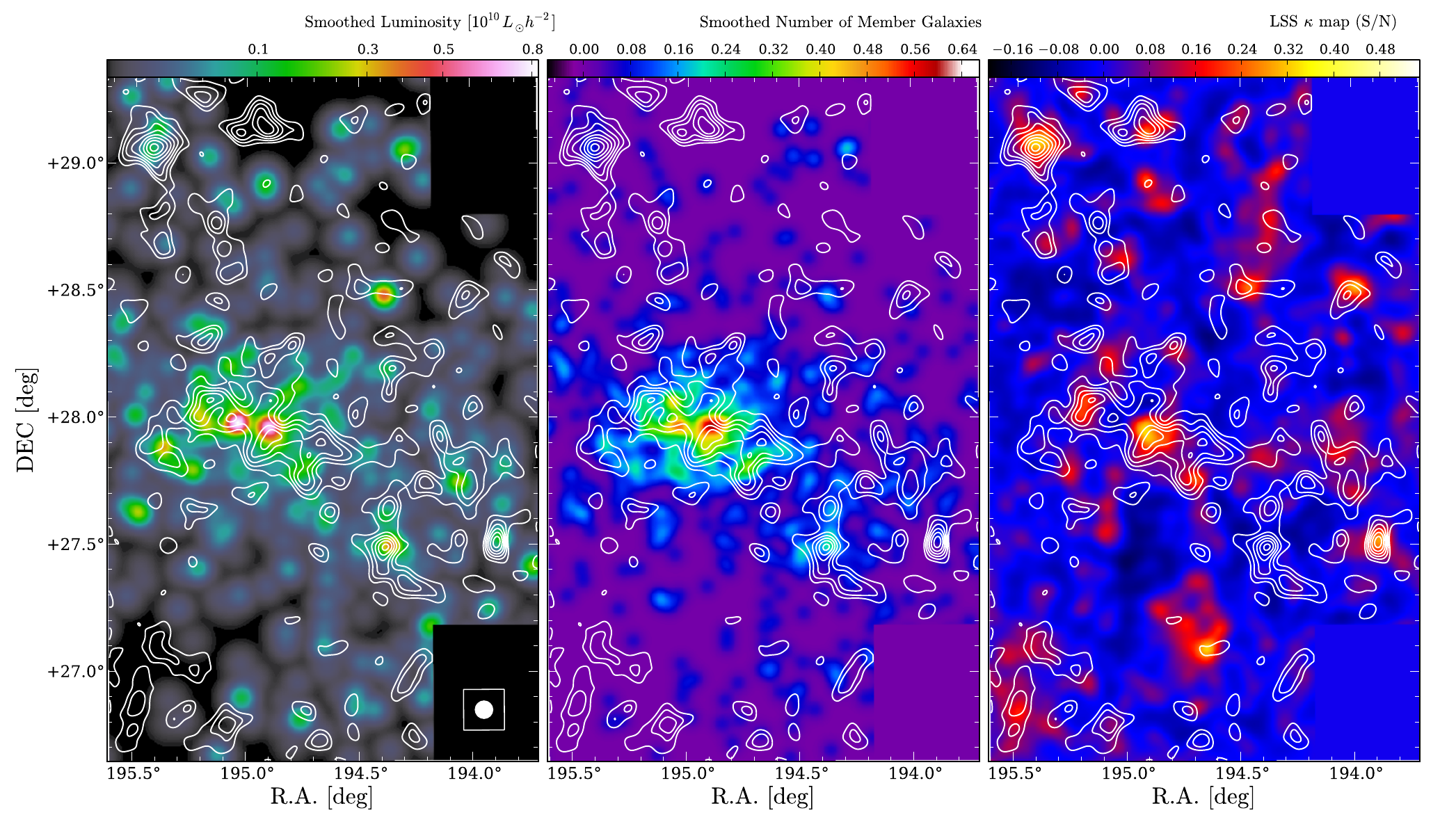}
  \end{center}
\caption{
Projected mass distribution (contours) in the Coma cluster at $z=0.0236$ 
 reconstructed from a $4$\,deg$^2$ weak-lensing survey with
 Subaru Suprime-Cam observations. \emph{Left and middle panels}:
 luminosity and number density distributions of
 spectroscopically identified cluster members, respectively. \emph{Right
 panel}: projected large-scale structure model based on galaxy--galaxy
 lensing results.
 The mean surface number density of background galaxies after the
 color--magnitude selection is $n_\mathrm{g}=41.3$\,arcmin$^{-2}$.
 Image reproduced with permission from \citet{Okabe2014coma}, copyright by AAS.
 \label{fig:coma}
 }
\end{figure*}

Figure~\ref{fig:coma} shows the projected mass distribution in the very
 nearby Coma cluster ($z_l=0.0236$) reconstructed from a $4$\,deg$^2$
 weak-lensing survey of cluster subhalos based on Subaru Suprime-Cam
 observations \citep{Okabe2014coma}. In the figure, the weak-lensing
 mass map is compared to the luminosity and number density distributions
 of spectroscopically identified cluster members, as well as to the
 projected large-scale structure model based on galaxy--galaxy
 lensing with the light-tracing-mass assumption. 
The projected mass and galaxy distributions in the Coma cluster are
 correlated well with each other. 
Thanks to the large angular extension of the Coma cluster, 
 \citet{Okabe2014coma} measured the weak-lensing masses of 32 cluster
 subhalos  down to the order of $10^{-3}$ of the cluster virial mass.

\subsubsection{Mass-sheet degeneracy} 

Adding a constant mass sheet to $\kappa(\btheta)$ in the shear-to-mass
formula (\ref{eq:kappa2gamma}) does not change the 
shear field $\gamma(\btheta)$ that is observable in the weak-lensing limit.
This leads to a degeneracy of solutions for the weak-lensing mass
inversion problem,
which is referred to as the \emph{mass-sheet degeneracy}
\citep{Falco1985,Gorenstein1988,Schneider+Seitz1995}.

As we shall see in Sect.~\ref{subsec:obs}, in general,
the observable quantity for weak shear lensing is not the  
shear $\gamma$, but the \emph{reduced} shear:
\begin{equation} 
\label{eq:redshear}
g(\btheta)=\frac{\gamma(\btheta)}{1-\kappa(\btheta)}
\end{equation}
in the subcritical regime where $\mathrm{det}{\cal A}>0$
(or $1/g^*$ in the negative-parity region with $\mathrm{det}{\cal A}<0$). 
We see that the $g(\btheta)$ field is invariant under the following 
global transformation:
\begin{equation}
\label{eq:invtrans}
\kappa(\btheta) \to \lambda \kappa(\btheta) + 1-\lambda, \ \ \ 
\gamma(\btheta) \to \lambda \gamma(\btheta)
\end{equation}
with an arbitrary scalar constant $\lambda\ne 0$
\citep{Schneider+Seitz1995}.
This transformation is equivalent to scaling 
the Jacobian matrix ${\cal A}(\btheta)$ with $\lambda$, 
$\cal {A}(\btheta) \to \lambda {\cal A}(\btheta)$.
It should be noted that this transformation leaves the location of the
critical curves ($\mathrm{det}{\cal A}(\btheta)=0$) invariant as well.
Moreover, the location of the curve defined by $\kappa(\btheta)=1$,
 on which the distortion locally disappears, is left
invariant under the transformation (Eq.~(\ref{eq:invtrans})).
%Hence, one cannot break this degeneracy as long as the only information
%regarding the image shapes is used.
A general conclusion is that all mass reconstruction methods based 
on shape information alone can determine the $\kappa$ field only up 
to a one-parameter family ($\lambda$ or $\kappa_0$) of linear
transformations (Eq.~(\ref{eq:invtrans})).

In principle this degeneracy can be broken or alleviated, for example,
by measuring the magnification factor $\mu$ in the 
subcritical regime \citep[i.e., outside the critical curves; see][]{Umetsu2013},
because $\mu$ transforms under the invariance transformation
(Eq.~(\ref{eq:invtrans})) as:
\begin{equation}
\mu(\btheta) \to \lambda^{-2} \mu(\btheta).
\end{equation}

\subsubsection{Nonlinear mass reconstruction}

Following \citet{1995A&A...297..287S},
we generalize the KS93 algorithm to include the nonlinear but subcritical  
regime (outside the critical curves).
To this end, we express the KS93 inversion formula in terms of the
observable reduced shear $g(\btheta)$. Substituting $\gamma=g(1-\kappa)$
in Eq.~(\ref{eq:ks93}), we have the following integral equation:
\begin{equation}
 \label{eq:ks93nl}
 \kappa(\vec\theta)-\kappa_0
  = \frac{1}{\pi}\int\! d^2 \theta' \,
   D^{*}(\btheta-\btheta')\, g(\btheta')\,
   [1-\kappa(\btheta')].
\end{equation}
For a given $g(\btheta)$ field, this nonlinear equation can be solved
iteratively, for example, by initially setting
$\kappa(\btheta)=0$ everywhere \citep{1995A&A...297..287S},

Equivalently,
Eq.~(\ref{eq:ks93nl}) can be formally expressed as a power series
expansion \citep{Umetsu1999}:
\begin{equation}
 \label{eq:ks93pow}
  \begin{aligned}
  \kappa(\btheta) - \kappa_0 &= (1-\kappa_0)
  \left(\hat{\cal G} -\hat{\cal G}\circ\!\hat{\cal G} +\hat{\cal G}\circ\!
   \hat{\cal G}\!\circ\! \hat{\cal G}-\cdots\right)\\
   &= (1-\kappa_0)\sum_{n=1}^{\infty}(-1)^{n-1} \hat{\cal G}^n,
  \end{aligned}
\end{equation}
where $\hat{\cal G}$ is the convolution operator defined by:
\begin{equation}
\hat{\cal G}(\btheta,\btheta')
:= \frac{1}{\pi}\int\! d^2\theta'\, D^*(\btheta-\btheta')g(\btheta')\,\times.
\end{equation}
Here $\hat{\cal G}(\vec\theta,\vec\theta')$ acts on a function of $\btheta'$.
The KS93 algorithm corresponds to the first-order approximation to this
power series expansion in the weak-lensing limit.
Note that solutions for nonlinear mass reconstructions suffer from the
generalized mass-sheet degeneracy, as explicitly shown in
Eq.~ (\ref{eq:ks93pow}).

Note that there is another class of mass inversion algorithms that uses 
maximum-likelihood and Bayesian approaches to obtain a mass map solution
and its error covariance matrix from weak-lensing data 
\citep[e.g.,][]{1996ApJ...464L.115B,Bradac2006,Merten+2009}.

\subsection{$E/B$ decomposition}
\label{subsubsec:eb}

The shear matrix $\Gamma(\btheta)=\gamma_1(\btheta)\sigma_3+\gamma_2(\btheta)\sigma_1$
that describes a spin-2 anisotropy can be expressed as a sum
of two components corresponding to the number of degrees of freedom. By
introducing two scalar fields
$\psi_E(\btheta)$ and $\psi_B(\btheta)$, we decompose the shear matrix
$\Gamma_{ij}$ ($i,j=1,2$) into two independent modes as
\citep{2002ApJ...568...20C}:
\begin{equation}
\Gamma(\btheta)=
\left( 
	\begin{array}{cc} 
	\gamma_1 & \gamma_2\\
	\gamma_2 & -\gamma_1
	\end{array}\right)
= 
\Gamma^{(E)}(\btheta) +\Gamma^{(B)}(\btheta), 
\end{equation}
with
\begin{equation}
 \begin{aligned}
\Gamma_{ij}^{(E)}(\btheta) &= 
  \left(\partial_i\partial_j-\delta_{ij}
  \frac{1}{2}\triangle\right)\psi_E(\btheta),\\
\Gamma_{ij}^{(B)}(\btheta) &= \frac{1}{2}
  \left(
  \epsilon_{kj}\partial_i\partial_k
  +
  \epsilon_{ki}\partial_j\partial_k
  \right)\psi_B(\btheta),
 \end{aligned}
\end{equation}
where $\epsilon_{ij}$ is
the Levi--Civita symbol in two dimensions,
%a $2\times 2$ antisymmetric tensor, 
defined such that
$\epsilon_{11}=\epsilon_{22}=0$, $\epsilon_{12}=-\epsilon_{21}=1$.
Here the first term associated with $\psi_E$ is a gradient or scalar $E$
component and the second term with $\psi_B$ is a curl or pseudoscalar
$B$ component.

The shear components $(\gamma_1,\gamma_2)$ are written in terms of
$\psi_E$ and $\psi_B$ as:
\begin{eqnarray}
\gamma_1&=&+\Gamma_{11}=-\Gamma_{22}=\frac{1}{2}\left(\psi_{E,11}-\psi_{E,22}\right) 
 -\psi_{B,12}\\
\gamma_2&=&\Gamma_{12}=\Gamma_{21}=\psi_{E,12}
+\frac{1}{2}\left(\psi_{B,11}-\psi_{B,22}\right).
\end{eqnarray}
As we have discussed in Sect.~\ref{subsubsec:spin}, the spin-2
$\gamma(\btheta)$ field is coordinate dependent and transforms as
$\gamma'=\gamma e^{2i\varphi}$ under a rotation of the basis axes
by $\varphi$.  
The $E$ and $B$ components can be extracted from the shear matrix as: 
\begin{equation}
 \label{eq:ebmode}
  \begin{aligned}
2\nabla^2_\theta\kappa_E &\equiv
\nabla^4_\theta\psi_E=2\partial^i\partial^j\Gamma_{ij},\\
2\nabla^2_\theta\kappa_B &\equiv
\nabla^4_\theta\psi_B=2\epsilon_{ij}
   \partial^i\partial^k\Gamma_{jk},
  \end{aligned}
\end{equation}
where we have defined the $E$ and $B$ fields,
$\kappa_E=(1/2)\triangle\psi_E$ and
$\kappa_B=(1/2)\triangle\psi_B$, respectively. 
This technique is referred to as the $E$/$B$-mode decomposition.
We see from Eq.~(\ref{eq:ebmode}) that the relations between $E$/$B$
fields and spin-2 fields are intrinsically nonlocal.
Remembering that the shear matrix due to weak lensing is given as
$\Gamma_{ij}=(\partial_i\partial_j-\delta_{ij}\triangle/2)\psi(\btheta)$
($i,j=1,2$), we identify
$\psi_E(\btheta)=\psi(\btheta)$ and $\psi_B(\btheta)=0$. Hence, for a
lensing-induced shear field, the $E$-mode signal is related to the
convergence $\kappa$, i.e., the surface mass density of the lens, 
while the $B$-mode signal is identically zero.

\begin{figure*}[!htb] %!htb
 \begin{center}
  \includegraphics[scale=0.4, angle=0, clip]{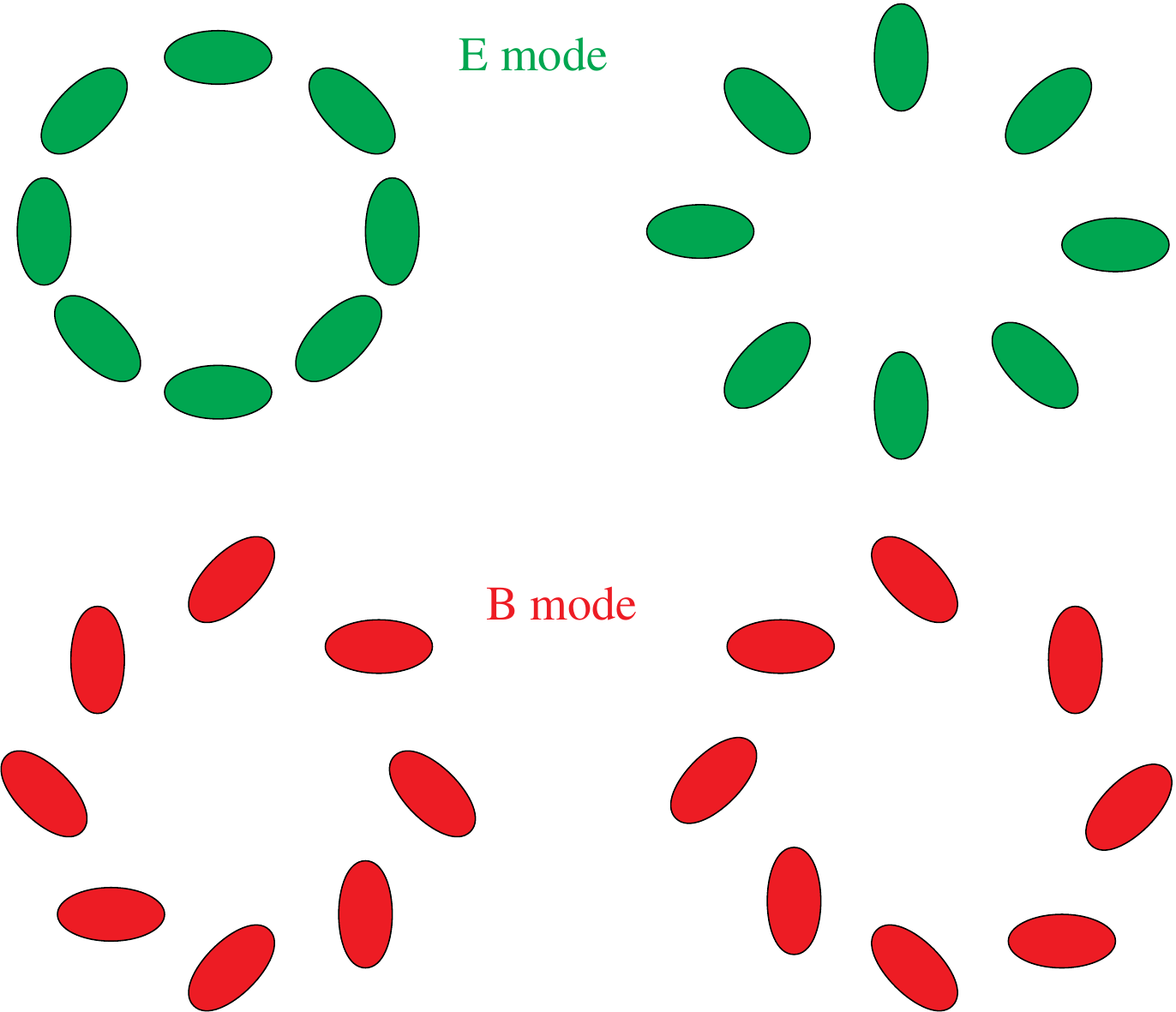}
 \end{center}
\caption{
 \label{fig:EBmode}
 Illustration of shape distortion patterns from $E$-mode and
 $B$-mode fields.
 Image reproduced with permission from \citet{vanWaerbeke+Mellier2003}. 
}
\end{figure*}

Figure~\ref{fig:EBmode} illustrates characteristic
distortion patterns from $E$-mode (curl-free) and $B$-mode
(divergence-free) fields. 
Weak lensing only produces curl-free $E$-mode signals, so that the
presence of divergence-free $B$ modes can be used as a null test for
systematic effects. In the weak-lensing regime, a tangential $E$-mode
pattern is produced by a positive mass overdensity (e.g., halos), while
a radial $E$-mode pattern is produced by a negative mass overdensity
(e.g., cosmic voids).

Now we turn to the issue of $E$/$B$-mode reconstructions from the
spin-2 shear field. Rewriting Eq.~(\ref{eq:ebmode})
in terms of the complex shear $\gamma$, we find:
\begin{equation}
\label{eq:ebmode_inv}
\begin{aligned}
\triangle\kappa_E&=
\Re\left(
2\hat{\cal D}^*\gamma\right), \\
\triangle\kappa_B&=
\Im\left(
2\hat{\cal D}^*\gamma
\right),
%=-\Re\left(
%2\hat{\cal D}^* i\gamma\right),
\end{aligned}
\end{equation}
where $\Re(Z)$ and $\Im(Z)$ denote the real part and the imaginary part
of a complex variable $Z$, respectively.
Defining $\kappa\equiv \kappa_E+i\kappa_B$, we see that 
the first of Eq.~(\ref{eq:ebmode_inv}) is identical to the mass
inversion formula
(Eq.~(\ref{eq:local})).
The $B$-mode convergence $\kappa_B$ can thus be simply obtained as
the imaginary part of Eq.~(\ref{eq:ks93}), which is expected to
vanish for a purely weak-lensing signal.
Moreover, the second of Eq.~(\ref{eq:ebmode_inv})
indicates that the transformation
$\gamma'(\btheta)=i\gamma(\btheta)$ 
($\gamma_1'=-\gamma_2, \gamma_2'=\gamma_1$)
is equivalent to an interchange operation of the $E$ and $B$ modes 
of the original maps by
$\kappa_E'(\btheta)=-\kappa_B(\btheta)$
and
$\kappa'_B(\btheta)=\kappa_E(\btheta)$.   
Since $\gamma$ is a spin-2 field that transforms as $\gamma'=\gamma e^{2i\varphi}$,
this operation is also equivalent to a rotation of each
ellipticity by $\pi/4$ with each position vector fixed.

Note that gravitational lensing can induce $B$ modes, for example, when
multiple deflections of light are involved
(Sect.~\ref{subsec:multilens}). However, these $B$ modes can be
generated at higher orders and the $B$-mode contributions coming from
multiple deflections are suppressed by a large factor compared to the
$E$-mode contributions \citep[see, e.g.,][]{Krause+Hirata2010}. 
In real observations, intrinsic ellipticities of background
galaxies also contribute to weak-lensing shear estimates.
Assuming that intrinsic ellipticities have random orientations in
projection space, such an isotropic ellipticity distribution will yield
statistically identical contributions to the $E$ and $B$ modes.
Therefore, the $B$-mode signal provides a useful null test for
systematic effects in weak-lensing observations
 (Fig.~\ref{fig:EBmode}).

\subsection{Flexion}
\label{subsubsec:flexion}

Flexion is introduced as the next higher-order lensing effects 
responsible for an arc-like and weakly skewed appearance of lensed
galaxies \citep{2005ApJ...619..741G,2006MNRAS.365..414B}
observed in a regime between weak and strong lensing (i.e., a nonlinear
but subcritical regime).
Such higher-order lensing effects occur when $\kappa(\btheta)$ and
$\gamma(\btheta)$ are not spatially constant across a source galaxy image. 
By taking higher-order derivatives of the lensing potential
$\psi(\btheta)$, we can work with higher-order transformations of galaxy
shapes by weak lensing
\citep[e.g.,][]{Massay+2007,HOLICs1,HOLICs2,2007ApJ...660.1003G,Schneider+Er2008,Viola+2012}.

\begin{figure*}[!htb] %!htb
 \begin{center}
  \includegraphics[scale=0.4, angle=0, clip]{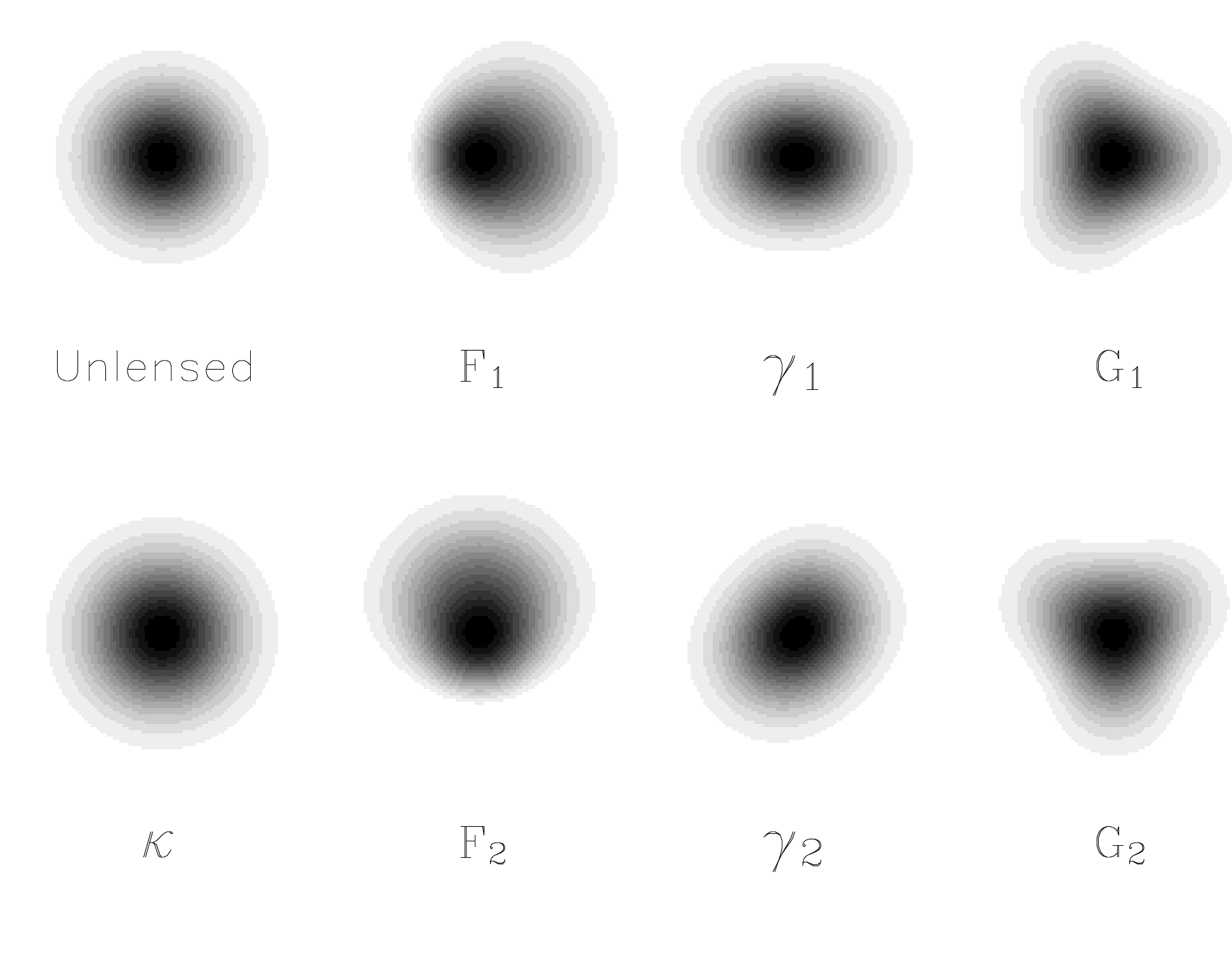}
 \end{center}
\caption{
 \label{fig:flexion}
Weak-lensing distortions with different spin values increasing from left
 to right. The convergence $\kappa$ is a spin-0 quantity, the first
 flexion $F=F_1+iF_2$ is spin-1, the shear $\gamma=\gamma_1+i\gamma_2$
 is spin-2, and the second flexion $G=G_1+iG_2$ is spin-3. 
 Image reproduced with permission from \citet{2006MNRAS.365..414B}, copyright by the authors.}
\end{figure*}

The third-order derivatives of $\psi(\btheta)$ can be combined to
form a pair of complex flexion fields as \citep{2006MNRAS.365..414B}:
\begin{equation}
 \label{eq:flexion}
  \begin{aligned}
   F&:= F_1+i{F}_2 = \frac{1}{2}\partial\partial\partial^*\psi=\partial\kappa,\\
   G&:= G_1+i{G}_2 = \frac{1}{2}\partial\partial\partial\psi=\partial\gamma.
  \end{aligned}
\end{equation}
The first flexion ${F}$ has spin-1 and the second flexion ${G}$ has
spin-3. The two complex flexion fields satisfy the following consistency
relation: 
\begin{equation}
\label{eq:consistency}
\partial^* \partial G(\btheta) = \partial\partial F(\btheta).
\end{equation}
Figure~\ref{fig:flexion} illustrates the characteristic weak-lensing
distortions with different spin values for  an intrinsically circular
Gaussian source \citep{2006MNRAS.365..414B}.

If the angular size of an image is small compared to the characteristic
scale over which $\psi(\btheta)$ varies, we can locally expand
Eq.~(\ref{eq:cosmo_lenseq2}) to the next higher order as:
\begin{eqnarray}
\label{eq:dbetaij}
 \delta\beta_i = {\cal A}_{ij}\delta\theta_j +
 \frac{1}{2}{\cal A}_{ij,k}\delta\theta_j \delta\theta_k +
 O(\delta\theta^3),
\end{eqnarray}
where ${\cal A}_{ij,k}=-\psi_{,ijk}$ ($i,j,k=1,2$).
%%%%
The ${\cal A}_{ij,k}$ matrix can be expressed with a sum of two terms:
\begin{equation}
{\cal A}_{ij,k}={F}_{ijk}+{G}_{ijk},
\end{equation}
with the spin-1 part ${F}_{ijk}$ and the spin-3 part ${G}_{ijk}$ defined
by:
\begin{eqnarray}
	{F}_{ij1} = -\frac{1}{2}\left(
\begin{array}{@{\,}cc@{\,}}
		3{F}_1  & {F}_2 \\
		 {F}_2  & {F}_1 
	\end{array}
\right) &,& \ \ 
{F}_{ij2} = -\frac{1}{2}\left(
\begin{array}{@{\,}cc@{\,}}
		{F}_2  &  {F}_1 \\
		{F}_1  & 3{F}_2
	\end{array}
\right), \\
{G}_{ij1} = -\frac{1}{2}\left(
\begin{array}{@{\,}cc@{\,}}
		{G}_1 &  {G}_2 \\
		{G}_2 & -{G}_1
	\end{array}
\right)&,& \ \ 
%\\
{G}_{ij2} = -\frac{1}{2}\left(
\begin{array}{@{\,}cc@{\,}}
		 {G}_2 & -{G}_1 \\
		-{G}_1 & -{G}_2 
	\end{array}
\right).
\end{eqnarray}
Flexion has a dimension of inverse (angular) length, so that the flexion 
effects depend on the angular size of the source image.
That is, the smaller the source image, the larger the amplitude of
intrinsic flexion contributions \citep{HOLICs2}.
The shape quantities affected by the first flexion ${F}$ alone have
spin-1 properties, while those by the second flexion ${G}$
alone have spin-3 properties.
%These third-order lensing fields naturally
%appear in the transformation equations of HOLICs between the lens and
%source planes.

Note that, as in the case of the spin-2 shear field, what is directly
observable from higher-order image distortions are the \emph{reduced}
flexion effects, $F/(1-\kappa)$ and $G/(1-\kappa)$
\citep{HOLICs1,HOLICs2,2007ApJ...660.1003G,Schneider+Er2008}, a
consequence of the mass-sheet degeneracy.

From Eq.~(\ref{eq:flexion}),
the inversion equations from flexion to $\kappa$ can be
obtained as follows \citep{2006MNRAS.365..414B}:
\begin{eqnarray}
(\kappa+iB)_{F}&=&\triangle^{-1} \partial^*{F},\\
(\kappa+iB)_{G}&=&\triangle^{-2}
 \partial^*\partial^*\partial^*{G},
\end{eqnarray}
where the complex part $iB$ describes the $B$-mode component that can be
used to assess the noise properties of weak-lensing data
\citep[e.g.,][]{HOLICs2}. 
An explicit representation for the inversion equations 
is obtained in Fourier space as:
\begin{equation}
 \begin{aligned}
\label{eq:F2k}
  \widehat\kappa_{F}(\bk) &= -i 
  \frac{k_1\widehat {F}_1 + k_2\widehat {F}_2}{k_1^2+k_2^2},\\
  \widehat\kappa_{G}(\bk) &=
  -i \frac{ 
  \widehat {G}_1 (k_1^3 - 3k_1 k_2^2)+
  \widehat {G}_2 (3k_1^2 k_2 - k_2^3) 
  }
  {(k_1^2+k_2^2)^2},
\end{aligned}
\end{equation}
for $\bk\neq 0$.
%%%

In principle one can combine independent mass reconstructions
$\widehat\kappa_a(\bk)$ $(a=\gamma,{F}, {G})$ linearly in Fourier space 
to improve the statistical significance 
with minimum noise variance weighting as \citep{HOLICs1}:
\begin{equation}
\label{eq:cmap}
\widehat\kappa(\bk)=\frac{\sum_a \widehat W_{\kappa|a}(\bk)\widehat{\kappa}_a(\bk)}
{\sum_a \widehat W_{\kappa|a}(\bk) }, 
\end{equation}
where 
$\widehat W_{\kappa|a}(\bk) = 1/P^{(N)}_{\kappa a}(\bk)$
with $P^{(N)}_{\kappa|a}(\bk)$ the two-dimensional noise power spectrum  
of $\kappa$ reconstructed using the observable $a$:
\begin{equation}
 \label{eq:power}
  \begin{aligned}
P^{(N)}_{\kappa|\gamma}(\bk) &= \frac{P_{\gamma}^{(N)}(\bk)}{2} = \frac{\sigma^2_{\gamma}}{2 n_\gamma},\\
P^{(N)}_{\kappa|F}(\bk) &= \frac{P^{(N)}_F(\bk)}{2k^2}=\frac{\sigma^2_{F}}{2 n_F \bk^2},\\
P^{(N)}_{\kappa|G}(\bk) &= \frac{P^{(N)}_G(\bk)}{2k^2}=\frac{\sigma^2_{G}}{2 n_G \bk^2},
 \end{aligned}
\end{equation}
with
$P^{(N)}_a(\bk)$ the shot noise power,
$\sigma_a$ the shape noise dispersion, and
$n_a$ the mean surface number density of background source galaxies, for
the observable $a$ ($a=\gamma, F, G$).
%%%
Assuming that errors in $\widehat{\kappa}_a(\bk)$ between different
observables are independent, 
the noise power spectrum 
for the estimator (Eq.~(\ref{eq:cmap})) is obtained as
\citep{HOLICs1}: 
\begin{equation}
P^{(N)}_{\kappa}(\bk)=\frac{1}{\sum_a \widehat{W}_a(\bk)}=
\frac{1}{\sum_a 1/P^{(N)}_{\kappa|a}(\bk)}.
\end{equation}

\begin{figure*}[!htb] %!htb
 \begin{center}
  \includegraphics[scale=0.38, angle=0, clip]{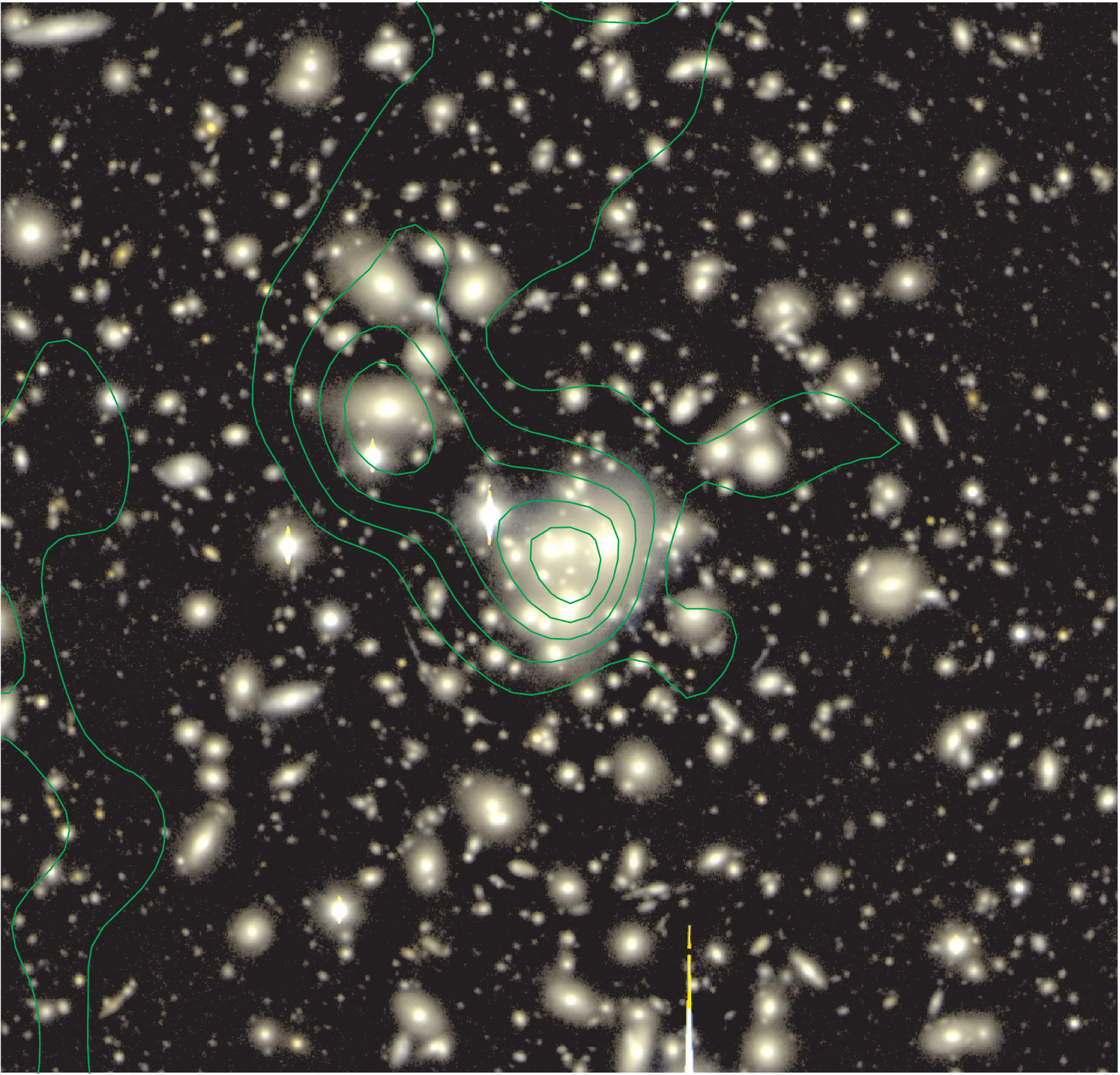}
 \end{center}
\caption{
 \label{fig:a1689flexion}
Mass contours of the rich cluster Abell 1689 ($z_l=0.183$)
 reconstructed from spin-1 flexion measurements based on Subaru
 Suprime-Cam observations, superposed on the Suprime-Cam $Vi'$ color
 image. The contours are spaced in units of $1\sigma$ reconstruction
 error estimated from the rms of the $B$-mode reconstruction.
The field size is $4'\times 4'$. North is to the top and east to the
 left.
Image reproduced with permission from \citet{HOLICs2}, copyright by AAS. 
}
\end{figure*}

Figure~\ref{fig:a1689flexion} shows the $\kappa$ field in the
central region of the rich cluster Abell 1689
($z_l=0.183$) reconstructed from the spin-1 flexion alone
\citep{HOLICs2} measured with Subaru Suprime-Cam data.
\citet{HOLICs2} used measurements of higher-order lensing image
 characteristics (HOLICs) introduced by \citet{HOLICs1}.
Their analysis accounted for the effect of the weight function used for
calculating noisy shape moments, as well as for higher-order PSF effects.
One can employ the assumption of random orientations for intrinsic
HOLICs of background galaxies to obtain a direct estimate of flexion,
in a similar manner to the usual prescription for weak shear lensing.
\citet{HOLICs2} utilized the Fourier-space relation
(Eq.~(\ref{eq:F2k})) between $F(\btheta)$ and $\kappa(\btheta)$ with the 
linear weak-lensing approximation.  
The $B$-mode convergence field was used to monitor the reconstruction
error in the $\kappa$ map.  
The reconstructed $\kappa$ map exhibits a bimodal feature in the central
region of the cluster. The pronounced main peak is associated with the
brightest cluster galaxy (BCG) and central cluster members, while the
secondary mass peak is associated with a local concentration of bright
galaxies.

Note that, as discussed in \citet{Viola+2012}, there is a
cross-talk between shear and flexion arising from shear--flexion
coupling terms, which makes quantitative measurements of flexion
challenging.

\subsection{Shear observables}
\label{subsec:obs}

Since the pioneering work of \citet[][]{1995ApJ...449..460K}, numerous
methods have been proposed and implemented in the literature to accurately  
extract the lensing signal from noisy pixelized images of background 
galaxies
\citep[e.g.,][]{Kuijken1999,Bridle+2002,Bernstein+Jarvis2002,2003MNRAS.338...35R,Hirata+Seljak2003,Miller+2007}. On
the other hand, considerable progress has been made in 
understanding and controlling systematic biases in noisy shear
estimates by relying on realistic galaxy image simulations
\citep[e.g.,][]{2006MNRAS.368.1323H,2007MNRAS.376...13M,Refregier+2012,Kacprzak+2012,GREAT3,GREAT3results,Mandelbaum2018sim}. 

Here we will review the basic idea and essential aspects of the
moment-based KSB formalism. We refer the reader to
\citet{Mandelbaum2018} for a recent exhaustive review on the subject.

\subsubsection{Ellipticity transformation by weak lensing}

In a moment-based approach to weak-lensing shape measurements, 
we use quadrupole moments $Q_{ij}$ ($i,j=1,2$) of the surface brightness
distribution $I(\btheta)$ of background galaxy images 
to quantify the shape of the images as \citep{1995ApJ...449..460K}:
\begin{equation}
\label{eq:Qij}
Q_{ij} \equiv \frac{\int\! d^2\theta\, 
q_I[I(\btheta)]\Delta\theta_i \Delta\theta_j}
{\int d^2\theta\,q_I[I(\btheta)]},
\end{equation}
where $q_I[I(\btheta)]$ is a weight function
and $\Delta\theta_i = \theta_i-\overline{\theta}_i$ denotes the offset vector
from the image centroid. Here we assume that the weight $q_I$ does not
explicitly depend on $\btheta$ but is set by the local value of the
brightness $I(\btheta)$ \citep{2001PhR...340..291B}.
The trace of $Q_{ij}$ describes the angular size of the image, while the
traceless part describes the shape and orientation of the image.
%\footnote{A practical
%implementation of the KSB method is achieved by the IMCAT package
%developed by Nick Kaiser. Note that IMCAT measures the shape moments
%with respect to the peak position rather than the centroid.}
With the quadrupole moments $Q_{ij}$,
we define the complex ellipticity $e=e_1+ie_2$ as:\footnote{Note that there are different definitions of ellipticity in the
literature, which lead to different transformation laws between the
image ellipticity and the shear \citep{2001PhR...340..291B}.}
\begin{equation}
\label{eq:cellip}
e \equiv \frac{Q_{11} - Q_{22} + 2iQ_{12}}{Q_{11} + Q_{22}}.
\end{equation}
For an ellipse with a minor-to-major axis ratio of
$q$ $(\leqs 1)$, $|e|=(1-q^2)/(1+q^2)$.

The spin-2 ellipticity $e$ (Eq.~(\ref{eq:cellip})) transforms under the
lens mapping as: 
\begin{equation}
\label{eq:chis2chi}
e^{(s)}=\frac{e-2g+e^* g^2}{1+|g|^2-2\Re(e^* g)},
\end{equation}
%where quantities with subscript ``$(s)$'' represent 
%those of (unlensed) intrinsic background sources,
where $e^{(s)}$ is the unlensed intrinsic ellipticity
and $g = \gamma/(1 - \kappa)$ is the spin-2 reduced shear.
Since $e$ is a nonzero spin quantity with a direction dependence, the
expectation value of the intrinsic 
source ellipticity $e^{(s)}$ is assumed to vanish, i.e.,
${\cal E}(e^{(s)})=0$, where ${\cal E}(X)$ denotes the expectation value
of $X$.
\citet{Schneider+Seitz1995}
showed that Eq.~(\ref{eq:chis2chi})
with the condition ${\cal E}(e^{(s)}) = 0$ is equivalent to: 
\begin{equation}
0=\sum_n w_n \frac{e_n - \delta_g}{1-\Re(e^*_n \delta_g)},
\end{equation}
where 
$e_n$ is the image ellipticity for the $n$th object,
$w_n$ is a statistical weight for the $n$th object,
and $\delta_g$ is the spin-2 complex distortion 
\citep{Schneider+Seitz1995}:
\begin{equation}
\delta_g \equiv \frac{2g}{1+|g|^2}.
\end{equation}
Note that the complex distortion $\delta_g$ is invariant under the
transformation $g\to 1/g^*$.

For an intrinsically circular source with $e^{(s)}=0$, we have:
\begin{equation}
\label{eq:e2delta}
e = \delta_g = \frac{2g}{1+|g|^2}.
\end{equation}
On the other hand,
in the weak-lensing limit ($|\kappa|, |\gamma|\ll 1$),
Eq.~(\ref{eq:chis2chi}) reduces to  
$e^{(s)} \simeq e-2g \simeq e-2\gamma$.
Assuming random orientations of source galaxies,
we average observed ellipticities over a local ensemble of source
galaxies to obtain:
\begin{equation}
\label{eq:chis2chiap}
\gamma \simeq g \simeq \frac{{\cal E}(e)}{2}.
\end{equation} 
For an input signal of $g=0.1$, Eq.~(\ref{eq:e2delta}) yields
$e\approx 0.198$. Hence, the weak-lensing approximation
(Eq.~(\ref{eq:chis2chiap})) gives a reduced-shear estimate of
$g^\mathrm{(est)}\approx 0.099$, corresponding to a negative bias of
$1\%$.  
For $g=0.2$ in the mildly nonlinear regime, 
Eq.~(\ref{eq:chis2chiap}) 
gives $g^\mathrm{(est)}\approx 0.192$, corresponding to a negative bias 
of $4\%$.

In real observations, the reduced shear $g$ may be estimated from a
local ensemble average of background galaxies as 
$\langle g \rangle\simeq \langle e\rangle/2$.
The statistical uncertainty in the reduced-shear estimate $\langle g\rangle$
decreases with increasing the number of background galaxies $N$ (see
Sect.~\ref{subsec:gest} for more details)
as $\propto \sigma/\sqrt{N}$,
with $\sigma$ the dispersion of background image ellipticities
(dominated by the intrinsic shape noise).
Weak-lensing analysis thus requires a large number
of background galaxies to increase the statistical significance of the
shear measurements.

\subsubsection{The KSB algorithm: a moment-based approach}
\label{subsubsec:KSB}

For a practical application of weak shear lensing,
we must account for various observational and instrumental effects,
such as the impact of noise on the galaxy shape measurement (both
statistical and systematic uncertainties),
the isotropic smearing component of the PSF,
and the effect of instrumental PSF anisotropy.
Therefore, one cannot simply use
Eq.~(\ref{eq:chis2chiap}) to measure the shear signal from observational
data.

A more robust estimate of the shape moments (Eq.~(\ref{eq:Qij})) is 
obtained by using a weight function $W(|\btheta|)$ that depends
explicitly on the separation $|\btheta|$ from the image centroid.
In the KSB approach, a circular Gaussian that is matched to the size of
each object is used as a weight function \citep{1995ApJ...449..460K}.
%I use a Gaussian of size h as their weight function =, i.e., the size of
%their xx the scale on which the
%object was detected at highest signicance.
The quadrupole moments obtained with such a weight function
$W(|\btheta|)$ suffer from an additional smearing and do not obey the
transformation law (Eq.~(\ref{eq:chis2chi})). Therefore,  the
expectation value ${\cal E}(e)$ of the image ellipticity is different
from the distortion $\delta_g = 2g/(1+|g|^2)$
(see Eq.~(\ref{eq:e2delta})).

The KSB formalism \citep{1995ApJ...449..460K,1998ApJ...504..636H}
accounts explicitly for the Gaussian weight function used for measuring 
noisy shape moments, the effect of spin-2 PSF anisotropy,
and the effect of isotropic PSF smearing.
The KSB formalism and its variants
assume that the PSF can be described
as an isotropic function convolved with a small anisotropic kernel.
In the limit of linear
response to lensing and instrumental anisotropies, KSB derived the
transformation law between intrinsic (unlensed) and observed
(lensed) complex ellipticities, $e^{(s)}$ and  $e$, respectively.
The linear transformation between intrinsic and observed complex 
ellipticities can be formally expressed as
\citep{1995ApJ...449..460K,1998ApJ...504..636H,2001PhR...340..291B}: 
\begin{equation}
e_i  =  e_i^{(s)} + \left(C^{g}\right)_{ij} g_j + \left(C^q\right)_{ij}
 q_j \ \ (i,j=1,2),
\end{equation}
where $q_i$ denotes the spin-2 PSF anisotropy kernel,
$(C^q)_{ij}$ is a linear response matrix for the PSF anisotropy $q_i$,
$(C^g)_{ij}$ is a linear response matrix for the reduced shear $g_i$.
%are linear response matrices for the spin-2 anisotropy fields
%($g_\alpha$ and $q_\alpha$),
The PSF anisotropy kernel and the response matrices can be calculated
from observable weighted shape moments of galaxies and stellar objects
\citep{1995ApJ...449..460K,2001PhR...340..291B,Erben2001}.
In real observations, the PSF anisotropy kernel $q(\btheta)$ can be
estimated from image ellipticities $e^*$ observed for a sample of
foreground stars for which $e^{(s)}$ and $g$ vanish, so that
$q_i(\btheta)=(C^q)^{-1}_{ij}e^*_{j}$.

Assuming that the expectation value of the intrinsic source ellipticity
vanishes ${\cal E}[e^{(s)}]=0$, we find the following linear
relation between the reduced shear and the ensemble-averaged image
ellipticity: 
\begin{equation}
 \label{eq:gest}
 g_i  = {\cal E}\left[
		 \left(C^{g}\right)^{-1}_{ij} (e-C^q q)_j
		\right] \ \ \ (i,j=1,2).
\end{equation}
In the KSB formalism, the shear response matrix 
$C^g$ is denoted as $P^g$ (or $P^\gamma$) and dubbed \emph{pre-seeing
shear polarizability}. Similarly, $C^q$ is denoted as $P^\mathrm{sm}$
and dubbed \emph{smear polarisability}.

A careful calibration of the signal response $P^g$ is essential for
any weak shear lensing analysis that relies on accurate shape
measurements from galaxy images
\citep[see][]{Mandelbaum2018}.
The levels of shear calibration bias are often quantified in terms of a
multiplicative bias factor $m$ and an additive calibration offset $c$
through the following relation
between the true input shear signal,
$g^\mathrm{true}$, and the recovered signal, $g^\mathrm{obs}$
\citep{2006MNRAS.368.1323H,2007MNRAS.376...13M,GREAT3}:
%We quantify deviations from perfect shear recovery via a
%linear fit that incorporates a multiplicative ¡Æcalibration bias¡Ç m and an
%additive ¡Æresidual shear offset¡Ç c
\begin{equation}
 g_i^{\mathrm{obs}} = (1+m_i) g_i^{\mathrm{true}} +  c_i \ \ (i=1,2),
\end{equation}
%where $m_i$ and $c_i$ are the multiplicative and additive bias factors,
%respectively \citep{}
The original KSB formalism, when applied to noisy 
observations, is known to suffer from systematic biases that depend
primarily on the size and the detection signal-to-noise ratio (S/N) of
galaxy images
\citep[e.g.,][]{Erben2001,Refregier+2012}.
Different variants of the \citet{1995ApJ...449..460K} method (KSB+) have been developed and
implemented in the literature primarily to study mass distributions of 
high-mass galaxy clusters
\citep[e.g.,][]{1998ApJ...504..636H,Hoekstra2015CCCP,2004ApJ...604..596C,Umetsu+2010CL0024,Umetsu2014clash,Oguri+2012SGAS,WtG1,Okabe+Smith2016,Schrabback2018spt}.
Note that KSB+ pipelines calibrated against realistic image
simulations of crowded fields can achieve a $\sim 2\percent$ shear 
calibration accuracy even in the cluster lensing regime
\citep[e.g.,][]{Herbonnet+2020,Hernandez-Martin2020wl}. 
%Hern«¡ndez-Mart«¿n et al. 2020). 

\subsection{Tangential- and cross-shear components}
\label{subsec:gammat}

As we have seen in Sect.~\ref{subsec:massrec}, the spin-2 shear
components $\gamma_1$ and $\gamma_2$ are coordinate dependent,
defined relative to a reference Cartesian coordinate frame (chosen by
the observer). 
It is useful to consider components of the shear that are coordinate
independent with respect to a certain reference point, such as the
cluster center. 
%in a rotated frame so as to measure them with respect to a certain
%reference point, such as the cluster center.

We define a polar coordinate system ($\vartheta, \varphi$) 
centered on an arbitrary point $\btheta_\mathrm{c}$ on the sky, such that
$\btheta=(\vartheta\cos\varphi,\vartheta\sin\varphi)+\btheta_\mathrm{c}$.
The convergence $\overline{\kappa}(\vartheta)$ averaged within a circle
of radius $\vartheta$ about $\btheta_\mathrm{c}$ is then expressed as:
\begin{equation}
 \label{eq:avkappa}
 \begin{aligned}
  \overline{\kappa}(\vartheta) &:=
  \frac{2}{\vartheta^2}\int_0^\vartheta\! d\vartheta'\,
  \vartheta' \kappa(\vartheta') \equiv \frac{\overline{\Sigma}(\vartheta)}{\Sigmacr},\\
  \kappa(\vartheta) &:=\oint\!
  \frac{d\varphi}{2\pi}\,\kappa(\vartheta,\varphi)\equiv\frac{\Sigma(\vartheta)}{\Sigmacr},
 \end{aligned}
\end{equation}
where
$\overline{\Sigma}(\vartheta)$ is the surface mass density averaged
within a circle of radius $\vartheta$ about $\btheta_\mathrm{c}$ and 
$\Sigma(\vartheta)$ is the surface mass density averaged over a
circle of radius $\vartheta$ about $\btheta_\mathrm{c}$. 
The reference point $\btheta_\mathrm{c}$ can be taken to be the cluster center, which
can be determined from symmetry of the strong-lensing pattern, the X-ray
centroid position, or the BCG position.

Let us introduce the tangential and $45^\circ$-rotated cross shear
components,
$\gamma_+(\btheta)$ and $\gamma_\times(\btheta)$, respectively, defined
relative to the position $\btheta_\mathrm{c}$ as:
\begin{equation}
 \label{eq:gtgx}
 \begin{aligned}
  \gamma_+(\btheta)         & = -\gamma_1(\btheta)\cos(2\varphi) - \gamma_2(\btheta)\sin(2\varphi),\\
  \gamma_\times(\btheta)    & = +\gamma_1(\btheta)\sin(2\varphi) - \gamma_2(\btheta)\cos(2\varphi),
 \end{aligned}
\end{equation}
which are directly observable in the weak-lensing limit where
$|\kappa|,|\gamma|\ll 1$ (see Sect.~\ref{subsec:obs}).
Using the two-dimensional version of Gauss' theorem, we find the
following identity for an arbitrary choice of $\btheta_\mathrm{c}$
\citep{Kaiser1995}: 
\begin{equation}
 \label{eq:identity}
  \begin{aligned}
   \gamma_+(\vartheta) &:= \oint\!
  \frac{d\varphi}{2\pi}\,\gamma_+(\vartheta,\varphi) = \overline{\kappa}(\vartheta) -\kappa(\vartheta)\equiv\frac{\Delta\Sigma(\vartheta)}{\Sigmacr},\\ 
   \gamma_\times(\vartheta) &:= \oint\!\frac{d\varphi}{2\pi}\,\gamma_\times(\vartheta,\varphi)= 0,
  \end{aligned}
\end{equation}
where we have defined the excess surface mass density
$\Delta\Sigma(\vartheta)$ around $\btheta_\mathrm{c}$ as a function of 
$\vartheta$ by \citep{1991ApJ...370....1M}:
\begin{equation}
 \Delta\Sigma(\vartheta) = \overline{\Sigma}(\vartheta) -\Sigma(\vartheta).
\end{equation}
From Eqs.~(\ref{eq:avkappa}) and (\ref{eq:identity}), we find:
\begin{equation}
 \label{eq:k2gt}
\frac{d\overline{\kappa}(\vartheta)}{d\ln\vartheta}
=-2\gamma_+(\vartheta).
\end{equation}
Equation~(\ref{eq:identity}) shows that,
given an arbitrary circular loop of radius $\vartheta$ around the chosen
center $\btheta_\mathrm{c}$, 
the tangential and cross shear components
averaged around the loop extract $E$-mode and $B$-mode distortion
patterns (Sect.~\ref{subsubsec:eb}).

An important implication of the first of Eq.~(\ref{eq:identity}) is
that, with tangential shear measurements from individual source galaxies (see
Sect.~\ref{subsec:obs}), one can directly determine the azimuthally
averaged $\Delta\Sigma(\vartheta)$ profile around lenses in the
weak-lensing regime, even if the mass distribution $\Sigma(\btheta)$ is
not axis-symmetric around $\btheta_\mathrm{c}$.  
Moreover, the second of Eq.~(\ref{eq:identity}) tells us that the
azimuthally averaged $\times$ component, or the $B$-mode signal, is
expected to be statistically consistent with zero if the signal is 
due to weak lensing.  Therefore, a measurement of the $B$-mode signal
$\langle g_\times(\vartheta)\rangle$ provides a useful null test against
systematic errors.

\subsection{Reduced tangential shear}
\label{subsec:gt}

\subsubsection{Azimuthally averaged reduced tangential shear}
\label{subsubsec:az_avg}

The reduced tangential shear $g_+(\vartheta)$ averaged over a circle of
radius $\vartheta$ about an arbitrary reference point
$\btheta_\mathrm{c}$ is expressed as:
\begin{equation}
 \begin{aligned}
  g_+(\vartheta) &:=  
  \oint\!\frac{d\varphi}{2\pi}\,g_+(\vartheta,\varphi)
  =\oint\!\frac{d\varphi}{2\pi}\,\frac{\gamma_+(\vartheta,\varphi)}{1-\kappa(\vartheta,\varphi)}.
 \end{aligned}
\end{equation}
If the projected mass distribution around a cluster has quasi-circular
symmetry (e.g., elliptical symmetry), then
the azimuthally averaged reduced tangential shear
$\langle g_+(\vartheta)\rangle$ around the cluster center can be
interpreted as:
\begin{equation}
 \label{eq:gtr}
 g_+(\vartheta) \simeq \frac{\gamma_+(\vartheta)}{1-\kappa(\vartheta)},
\end{equation}
where $\gamma_+(\vartheta)$ and $\kappa(\vartheta)$ are the tangential
shear and the convergence, respectively, averaged over a circle of radius
$\vartheta$ about $\btheta_\mathrm{c}$.

According to $N$-body simulations in hierarchical \LCDM models of cosmic
structure formation, dark matter halos exhibit aspherical mass
distributions that can be well approximated by triaxial mass models
\citep[e.g.,][]{2002ApJ...574..538J,Limousin2013,Despali2014}.
Since triaxial halos have elliptical isodensity
contours in projection on the sky \citep{Stark1977},
Eq.~(\ref{eq:gtr}) can give a good approximation to describe the
weak-lensing signal for regular clusters with a modest degree of
perturbation. However, the approximation is likely to break down for
merging and interacting lenses having complex, 
multimodal mass distributions. To properly model the weak-lensing signal
in such a complex merging system, one needs to directly model the
two-dimensional reduced-shear field $(g_1(\btheta),g_2(\btheta))$ 
with a lens model consisting of multi-component halos
\citep[e.g.,][]{Watanabe+2011,Okabe+2011A2163,Medezinski+2013}.
Alternatively, one may attempt to reconstruct the convergence field
$\kappa(\btheta)$ in a free-form manner from the observed
reduced shear field, with additional constraints or assumptions to break
the mass-sheet degeneracy
\citep[e.g.,][]{Jee+2005,Bradac2006,Merten+2009,Jauzac+2012,Umetsu2015A1689,Tam+2020wlsl}. 

On the other hand,
for a statistical ensemble of galaxy clusters, the average mass
distribution around their centers tends to be spherically symmetric if
the assumption of statistical isotropy holds
\citep[e.g.,][]{Okabe+2013}. Hence, the stacked weak-lensing signal for
a statistical ensemble of clusters can be interpreted using
Eq.~(\ref{eq:gtr}).  For more details,
see Sects.~\ref{subsubsec:avg_DSigma} and \ref{subsec:stack}.

\subsubsection{Source-averaged reduced tangential shear}
\label{subsubsec:src_avg}

With the assumption of quasi-circular symmetry in the projected mass
distribution around clusters (see Eq.~(\ref{eq:gtr})), let us
consider the nonlinear effects on 
the source-averaged cluster lensing profiles.
The reduced tangential shear for a given lens--source pair is written as:
\begin{equation}
 \begin{aligned}
  g_+(\vartheta|z_l,z_s)&=\Delta\Sigma(\vartheta)\sum_{n=0}^\infty
  \left[\Sigmacr^{-1}(z_l,z_s)\right]^{n+1}\Sigma^n(\vartheta).
 \end{aligned}
\end{equation}

To begin with, let us consider the expectation value for the reduced
tangential shear averaged over an ensemble of source galaxies. 
For a given cluster, the source-averaged reduced tangential shear is
expressed as: 
 \begin{equation}
  \label{eq:gt_avg}
   \langle g_+(\vartheta|z_l)\rangle =
   \Delta\Sigma\left[
		\langle\Sigma_{\mathrm{cr},l}^{-1}\rangle +
		\langle\Sigma_{\mathrm{cr},l}^{-2}\rangle\Sigma +
		\langle\Sigma_{\mathrm{cr},l}^{-3}\rangle\Sigma^2+\cdots\right],
 \end{equation}
where $\langle\cdots\rangle$ denotes the averaging over all sources,
defined such that:
\begin{equation}
 \label{eq:ISigmacr_avg}
  \langle\Sigma_{\mathrm{cr},l}^{-n}\rangle =
  \left(\sum_{s}w_{s}\Sigma_{\mathrm{cr},ls}^{-n}\right)\left(\sum_s w_{s}\right)^{-1},
\end{equation}
where the index $s$ runs over all source galaxies around the lens ($l$)
and $w_s$ is a statistical weight for each source galaxy.
An optimal choice for the statistical weight is
$w_{s} = 1 / \sigma_{g_{+},s}^{2}$,
with $\sigma_{g_{+},s}$ the statistical uncertainty of 
$g_+(\vartheta|z_l,z_s)$ estimated for each source galaxy.
Note that this choice for the weight assumes that $\sigma_{g_{+},s}$ is
independent of the lensing shear signal
\citep[see][]{Schneider+Seitz1995,1995A&A...297..287S}. 
In the continuous limit, Eq.~(\ref{eq:ISigmacr_avg}) is written as:  
\begin{equation} 
 \langle\Sigma_{\mathrm{cr},l}^{-n}\rangle =
  \left[\int_0^\infty\!dz\,\frac{dN(z)}{dz}w(z)\Sigmacr^{-n}(z_l,z)\right]
  \left[\int_0^\infty\!dz\,\frac{dN(z)}{dz}w(z)\right]^{-1},
\end{equation}
with
$dN(z)/dz$ the redshift distribution function of the source sample and
$w(z)$ a statistical weight function.
For a given cluster lens, $\Sigmacr^{-1}(z_l,z_s)\propto \Dls/\Ds$,
so that
$\langle\Sigma_{\mathrm{cr},l}^{-n}\rangle\propto \langle (\Dls/\Ds)^{n}\rangle$.

In the weak-lensing limit, Eq.~(\ref{eq:gt_avg}) gives
$\langle g_+\rangle\simeq \langle\gamma_+\rangle$.
The next order of approximation is given by \citep{Seitz+Schneider1997}:
\begin{equation}
 \label{eq:gt_avg_app}
 \begin{aligned}
 \langle g_+\rangle &\simeq
  \frac{\langle\gamma_+\rangle}{1-f_l\langle\kappa\rangle},\\
  f_l&= \frac{\langle
  \Sigma_{\mathrm{cr},l}^{-2}\rangle}{\langle\Sigma_{\mathrm{cr},l}^{-1}\rangle^2}.\\
 \end{aligned}
\end{equation}
From Eq.~(\ref{eq:gt_avg_app}), we see that an interpretation of
the averaged weak-lensing signal $\langle g_+(\vartheta|z_l)\rangle$ does
not require knowledge of individual source redshifts. Instead, it
requires ensemble information regarding the statistical redshift
distribution $dN(z)/dz$ of background source galaxies used for
weak-lensing measurements. 
%%%

For a lens at sufficiently low redshift
(see Sect.~\ref{subsubsec:crit}), $f_l\approx 1$,
thus leading to the single source-plane approximation:
$\langle g_+\rangle\simeq \langle\gamma_+\rangle/(1-\langle\kappa\rangle)$.
The level of bias introduced by this approximation is
$\Delta\langle g_+\rangle/\langle g_+\rangle\simeq (f_l-1)\langle\kappa\rangle$.
In typical ground-based deep observations of $z_l\sim 0.4$ clusters,
$\Delta f_l=f_l-1$ is found to be of the order of several percent
\citep{Umetsu2014clash}, so that the relative error
$\Delta\langle g_+\rangle/\langle g_+\rangle$ is negligibly
small in the mildly nonlinear regime of cluster lensing.

\subsubsection{Source-averaged excess surface mass density}
\label{subsubsec:src_avg_DSigma}

Next, let us consider the following estimator for the excess surface
mass density $\Delta\Sigma(\vartheta)$ for a given lens--source pair:
\begin{equation}
\label{eq:DSigma_plus}
\Delta\Sigma_+(\vartheta|z_l,z_s) :=
 \Sigmacr(z_l,z_s)g_+(\vartheta|z_l,z_s).
\end{equation}
This assumes that an estimate of $\Sigmacr^{-1}(z_l,z_s)$ for each
individual source galaxy is available, for example, from
photometric-redshift (photo-$z$) measurements.
This estimator is widely used in recent cluster weak-lensing
studies thanks to the availability of multi-band imaging data and the
advances in photo-$z$ techniques
(see Sect.~\ref{subsec:dilution}). 

In real observations, if the photo-$z$ probability distribution
function (PDF), $P_s(z)$, is available for individual source galaxies
($s$), one can calculate:
\begin{equation}
\Sigma^{-1}_{\mathrm{cr},ls}\equiv
 \left[\int\!dz\,P_s(z)\Sigmacr^{-1}(z_l,z)\right]\left[\int\!dz\,P_s(z)\right]^{-1}
\end{equation}
averaged over the PDF for each source galaxy.
Similarly to Eq.~(\ref{eq:gt_avg}),
$\Delta\Sigma_+(\vartheta|z_l,z_s)$ averaged over all sources is
expressed as:
\begin{equation}
 \label{eq:DSigma_avg}
 \langle\Delta\Sigma_+\rangle =
  \Delta\Sigma\left[1 +
	       \langle\Sigma_{\mathrm{cr},l}^{-1}\rangle\Sigma +
	       \langle\Sigma_{\mathrm{cr},l}^{-2}\rangle\Sigma^2+\cdots\right]
\end{equation}
with
\begin{equation}
 \label{eq:ISigmacr_avg_sum}
 \langle\Sigma_{\mathrm{cr},l}^{-n}\rangle =
  \left(\sum_{s}w_{ls}\Sigma_{\mathrm{cr},ls}^{-n}\right)\left(\sum_s w_{ls}\right)^{-1},
\end{equation}
where the index $s$ runs over all source galaxies around the lens ($l$)
and $w_{ls}$ is a statistical weight for each source galaxy. 
An optimal choice for the statistical weight is:
\begin{equation}
 \label{eq:wls_stack}
 w_{ls} = \Sigma_{\mathrm{cr},ls}^{-2} / \sigma_{g_{+},s}^{2},
\end{equation}
where $\sigma_{g_{+},s}$ is the statistical uncertainty of 
$g_+(\vartheta|z_l,z_s)$ estimated for each source galaxy
(Sect.~\ref{subsubsec:src_avg}).

In the weak-lensing limit, we thus have
$\langle\Delta\Sigma_+\rangle\simeq \Delta\Sigma$.
The next order of approximation is:
\begin{equation}
 \label{eq:DSigma_avg_app}
 \langle \Delta\Sigma_+\rangle \simeq
  \frac{\Delta\Sigma}{1-\langle\Sigma_{\mathrm{cr},l}^{-1}\rangle\Sigma}.
\end{equation}

\subsubsection{Lens--source-averaged excess surface mass density}
\label{subsubsec:avg_DSigma}

Finally, we consider an ensemble of galaxy clusters. Now, let
$\Delta\Sigma$ be the ensemble mass distribution of these clusters.
Then, $\Delta\Sigma_+$ averaged over all lens--source
($ls$) pairs is expressed as   \citep{Johnston+2007b}:
\begin{equation}
 \llangle\Delta\Sigma_+\rrangle =
  \Delta\Sigma\left[1 +
	       \llangle\Sigma_{\mathrm{cr}}^{-1}\rrangle\Sigma +
	       \llangle\Sigma_{\mathrm{cr}}^{-2}\rrangle\Sigma^2+\cdots\right]
\end{equation}
with
\begin{equation}
 \label{eq:DSigma_stack}
 \llangle\Sigma_{\mathrm{cr}}^{-n}\rrangle =
  \left(\sum_{l,s}w_{ls}\Sigma_{\mathrm{cr},ls}^{-n}\right)\left(\sum_{l,s} w_{ls}\right)^{-1},
\end{equation}
where
$\llangle\cdots\rrangle$ denotes the averaging over all lens--source pairs,
the double summation is taken over all clusters ($l$) and all source
galaxies ($s$), and $w_{ls}$ is a statistical weight for each
lens--source pair ($ls$). 
An optimal choice for the statistical weight is given by
Eq.~(\ref{eq:wls_stack}).

Again, the weak-lensing limit yields
$\llangle\Delta\Sigma_+\rrangle\simeq \Delta\Sigma$ and
the next order of approximation is given by \citep{Umetsu2014clash,Umetsu2020xxl}:
\begin{equation}
 \label{eq:DSigma_stack_app}
 \llangle \Delta\Sigma_+\rrangle \simeq
  \frac{\Delta\Sigma}{1-\llangle\Sigmacr^{-1}\rrangle\Sigma}.
\end{equation}
Equation~(\ref{eq:DSigma_stack_app}) can be used to interpret the
stacked weak-lensing signal including the nonlinear regime of cluster
lensing. In Sect.~\ref{subsec:stack}, we provide more details on
the stacked weak-lensing methods.

\subsection{Aperture mass densitometry}
\label{subsubsec:map}

In this subsection, we introduce a nonparametric technique to infer a
projected total mass estimate from weak shear lensing observations.
Integrating Eq.~(\ref{eq:k2gt}) between two concentric radii
$\vartheta_\mathrm{in}$ and $\vartheta_\mathrm{out}$
$(>\vartheta_\mathrm{in})$, 
we obtain an expression for the $\zeta$ statistic as
\citep{1994ApJ...437...56F,Kaiser1995,1996ApJ...473...65S}:
\begin{equation}
 \label{eq:zeta}
  \begin{aligned}
\zeta(\vartheta_\mathrm{in},\vartheta_\mathrm{out})
&:=\overline{\kappa}(\vartheta_\mathrm{in})-
   \overline{\kappa}(\vartheta_\mathrm{in},\vartheta_\mathrm{out})\\
&=\frac{2}{1-(\vartheta_\mathrm{in}/\vartheta_\mathrm{out})^2}
\int_{\vartheta_\mathrm{in}}^{\vartheta_\mathrm{out}}\!
d\ln\vartheta'
\,\gamma_+(\vartheta'),
  \end{aligned}
\end{equation}
where $\overline{\kappa}(\vartheta_\mathrm{in},\vartheta_\mathrm{out})$
 is the convergence averaged within a concentric annulus between 
  $\vartheta_\mathrm{in}$ and $\vartheta_\mathrm{out}$:
\begin{equation}
\overline{\kappa}(\vartheta_\mathrm{in},\vartheta_\mathrm{out})
:=\frac{1}{\pi(\vartheta_\mathrm{out}^2-\vartheta_\mathrm{in}^2)}
\int_{\vartheta_\mathrm{in}}^{\vartheta_\mathrm{out}\!}d\vartheta'\,\vartheta'
\kappa(\vartheta').
\end{equation}
In the weak-lensing regime where
$\gamma_+(\vartheta)\simeq g_+(\vartheta)$,
$\zeta$ can be determined uniquely from the shape distortion field in a
finite annular region at
$\vartheta_\mathrm{in}\leqs\theta\leqs\vartheta_\mathrm{out}$, because
additive constants $\kappa_0$ 
from the invariance transformation (Eq.~(\ref{eq:invtrans}))
cancel out in Eq.~(\ref{eq:zeta}).
Note that this technique is also referred to as \emph{aperture mass
densitometry}. 

Since galaxy clusters are highly biased tracers of the cosmic mass
distribution, 
$\overline{\kappa}(\vartheta_\mathrm{in},\vartheta_\mathrm{out})$ around a cluster is
expected to be positive, so that
$\zeta(\vartheta_\mathrm{in},\vartheta_\mathrm{out})$ yields a lower limit to
$\overline{\kappa}(\vartheta_\mathrm{in})$. That is, the quantity 
$M_\zeta \equiv \pi(\Dl \vartheta_\mathrm{in})^2 \Sigmacr\zeta(\vartheta_\mathrm{in},\vartheta_\mathrm{out})$
yields a lower limit to the projected mass inside a circular aperture of  
radius $\vartheta_\mathrm{in}$,
$M_\mathrm{2D}=\pi(\Dl\vartheta_\mathrm{in})^2\overline{\Sigma}(\vartheta_\mathrm{in})$.
%When we attempt to estimate the mass inside a circular boundary where 
%the weak-lensing limit holds (i.e., $\kappa \ll 1$ and $|\gamma|\ll 1$), 
%Thus, in the weak-lensing limit,
%we can directly infer the lensing mass inside $\vartheta_\mathrm{in}$ 
%from weak shear data outside this boundary.
This technique provides a powerful means to estimate the total cluster
mass from shear data in the weak-lensing regime $|\gamma|\ll 1$.
%We note that the shape of an aperture need not be restricted to a 
%circle. Aperture masses for arbitrary aperture shapes are 
%dealt with in Ref.~\cite{ScB97}.
%in Schneider and Bartelmann (1997). 

We now introduce a variant of aperture mass densitometry
defined as \citep{2000ApJ...539..540C}:
\begin{equation}
 \label{eq:zetac}
 \begin{aligned}
  \zeta_\mathrm{c}(\vartheta|\vartheta_\mathrm{in},\vartheta_\mathrm{out})
  &:=
  \overline{\kappa}(\vartheta)-\overline{\kappa}(\vartheta_\mathrm{in},\vartheta_\mathrm{out})\\
  &=
  2\int_{\vartheta}^{\vartheta_\mathrm{in}}\!d\ln\vartheta'\,\gamma_+(\vartheta')
  +
  \frac{2}{1-(\vartheta_\mathrm{in}/\vartheta_\mathrm{out})^2}
\int_{\vartheta_\mathrm{in}}^{\vartheta_\mathrm{out}}\!
d\ln\vartheta'\,\gamma_+(\vartheta'),
  \end{aligned}
\end{equation}
where the aperture radii
$(\vartheta,\vartheta_\mathrm{in},\vartheta_\mathrm{out})$ satisfy
$\vartheta < \vartheta_\mathrm{in} < \vartheta_\mathrm{out}$,
and the first and second terms in the second line of
Eq.~(\ref{eq:zetac}) are equal to
$\overline{\kappa}(\vartheta)-\overline{\kappa}(\vartheta_\mathrm{in})$
and
$\overline{\kappa}(\vartheta_\mathrm{in})-\overline{\kappa}(\vartheta_\mathrm{in},\vartheta_\mathrm{out})$,
respectively.
In the weak-lensing limit,
the quantity
\begin{equation}
M_{\zeta_\mathrm{c}}(<\vartheta)\equiv \pi
(\Dl\vartheta)^2\Sigmacr\zeta_\mathrm{c}(\vartheta|\vartheta_\mathrm{in},\vartheta_\mathrm{out})
\end{equation}
yields a lower limit to the projected mass inside a circular aperture of
radius $\vartheta$, that is:
\begin{equation}
M_\mathrm{2D}(<\vartheta)=\pi(\Dl\vartheta)^2\overline{\Sigma}(\vartheta).
\end{equation}
 
We can regard
$\zeta_\mathrm{c}(\vartheta|\vartheta_\mathrm{in},\vartheta_\mathrm{out})$
as a function of $\vartheta$ for fixed values of $(\vartheta_\mathrm{in},\vartheta_\mathrm{out})$
and measure
$\zeta_\mathrm{c}(\vartheta|\vartheta_\mathrm{in},\vartheta_\mathrm{out})$
at several independent aperture radii $\vartheta$.
As in the case of the standard $\zeta$ statistic (Eq.~(\ref{eq:zeta})),
one may choose the inner and outer annular radii 
($\vartheta_\mathrm{in}, \vartheta_\mathrm{out}$) 
%of the annular background region
to lie in the weak-lensing regime
where $g_+\simeq \gamma_+$. 
In general, however, $\vartheta$ may lie in the nonlinear regime where
$\gamma_+$ is not directly observable. 
In the subcritical regime, we can express $\gamma_+(\vartheta)$ in terms
of the observed reduced tangential shear $g_+(\vartheta)$ as:
\begin{equation}
\gamma_+(\vartheta) = g_+(\vartheta) [1-\kappa(\vartheta)],
\end{equation}
when assuming a quasi-circular symmetry in the projected mass
distribution (Sect.~\ref{subsec:gt}).
If these conditions are satisfied,
for a given boundary condition $\overline{\kappa}_0\equiv\overline{\kappa}(\vartheta_\mathrm{in},\vartheta_\mathrm{out})$,
Eq.~(\ref{eq:zetac}) can be solved iteratively as
\citep{Umetsu+2010CL0024}:
\begin{equation}
 \label{eq:zetac_kappa}
    \begin{aligned}
  \zeta_\mathrm{c}^{(n+1)}(\vartheta|\vartheta_\mathrm{in},\vartheta_\mathrm{out})
  &=
     2\int_{\vartheta}^{\vartheta_\mathrm{in}}\!d\ln\vartheta'\,
     g_+(\vartheta') \left[1-\kappa^{(n)}(\vartheta')\right]\\
  &+
  \frac{2}{1-(\vartheta_\mathrm{in}/\vartheta_\mathrm{out})^2}
\int_{\vartheta_\mathrm{in}}^{\vartheta_\mathrm{out}}\!
     d\ln\vartheta'\, g_+(\vartheta'),\\
    \kappa^{(n)}(\vartheta) &=
     \hat{\cal L}\zeta_\mathrm{c}^{(n)}(\vartheta|\vartheta_\mathrm{in},\vartheta_\mathrm{out})
     + \overline{\kappa}_0,
     %\ \ \ \ (\vartheta<\vartheta'<\vartheta_\mathrm{in}),
    \end{aligned}
\end{equation}
where we have introduced a differential operator defined as
$\hat{\cal L}(\vartheta) = \frac{1}{2\vartheta^2}\frac{d}{d\ln\vartheta}\vartheta^2$
that satisfies
$\hat{\cal L}\overline{\kappa}(\vartheta)=\kappa(\vartheta)$
and
$\hat{\cal L}1 = 1$, and the quantities indexed by $(n)$
refer to those in the $n$th iteration ($n=0,1,2,\dots$).

We solve a discretized version of Eq.~(\ref{eq:zetac_kappa}).
See Appendix~A of \citet{Umetsu2016clash} for discretized
expressions for $g_+(\vartheta)$ and $\overline{\kappa}(\vartheta)$.
One can start the iteration process with an initial guess of 
$\kappa^{(0)}(\vartheta)=0$ for all $\vartheta$ bins
and repeat it until convergence is reached in all $\vartheta$ bins.
This procedure will yield a solution for the
binned mass profile:
\begin{equation}
\overline{\kappa}(\vartheta) =\zeta_\mathrm{c}(\vartheta|\vartheta_\mathrm{in},\vartheta_\mathrm{out})+\overline{\kappa}_0,
\end{equation}
for a given value of $\overline{\kappa}_0$.
Note that the errors for the mass profile solution
in different radial bins are correlated.
The bin-to-bin error covariance matrix 
$C_{bb'}\equiv \mathrm{Cov}[\overline\kappa(\vartheta_b),\overline\kappa(\vartheta_{b'})]$
($b,b'=1,2,\dots$)
can be calculated with the linear approximation $\kappa(\vartheta)\ll 1$
in Eq.~(\ref{eq:zetac_kappa}),
by propagating the errors for the binned $g_+(\vartheta)$
profile \citep[e.g.,][]{Okabe+Umetsu2008,Umetsu+2010CL0024,Okabe+2010WL}.

Alternatively, one can attempt to determine the boundary term
$\overline{\kappa}_0$ from shear data by incorporating  
additional iteration loops.
Starting with an initial guess of 
$\overline{\kappa}_0=0$,  one can update the value of
$\overline{\kappa}_0$ in each iteration by using a specific mass model
(e.g., a power-law profile) that best fits the binned
$\overline{\kappa}(\vartheta)$ profile. This iteration procedure is
repeated until convergence is obtained
\citep[see][]{Umetsu+2010CL0024}. 
%%%

\section{Standard shear analysis methods}
\label{sec:method}
 
In this section, we outline procedures to obtain cluster mass estimates 
from azimuthally averaged reduced tangential shear measurements for a
given galaxy cluster.

\subsection{Background source selection}
\label{subsec:dilution}

\begin{figure*}[!htb] %!htb
  \begin{center}
   \includegraphics[scale=0.3, angle=0, clip]{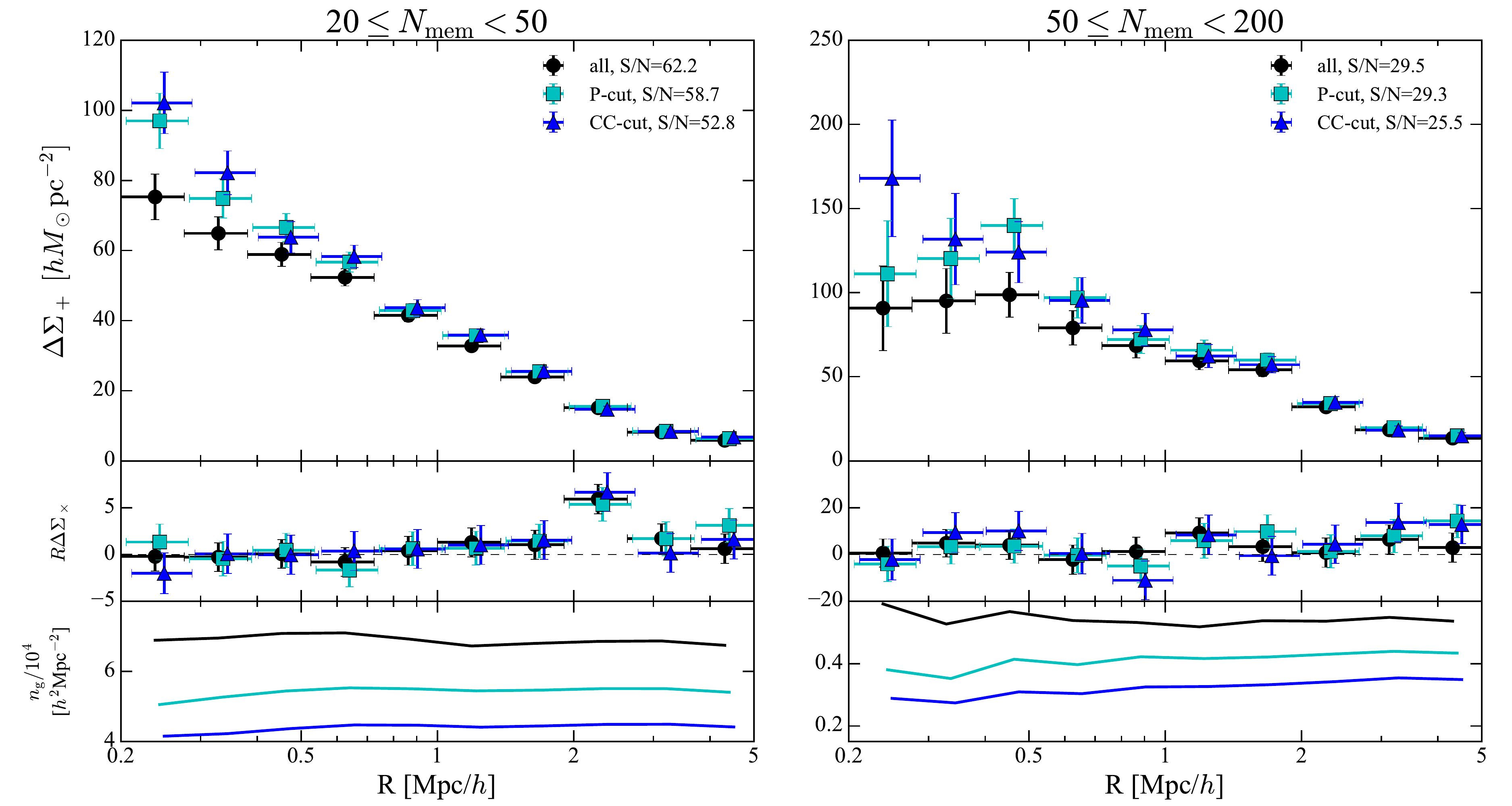}
  \end{center}
\caption{
 \label{fig:dilution}
 Stacked weak-lensing profiles around CAMIRA clusters from the
 Subaru HSC survey, shown as a function of cluster-centric comoving
 radius $R$. 
 This figure compares different source-selection methods for two
 different richness bins 
 (left for $20\leqs N_ <50$ and
  right or $50\leqs N_ <200$).
Upper panels show the excess surface mass density profile
 $\llangle\Delta\Sigma_+\rrangle$; middle panels show the
 $45^\circ$-rotated component, expected to be consistent with zero;
 and lower panels show the effective number density of source galaxies.
All quantities in this figure were calculated using photo-$z$ PDFs of 
 individual source galaxies, $P(z)$
 (see also Sect.~\ref{subsubsec:src_avg_DSigma}).
Different lines in each panel show different source-selection schemes:
 using all galaxies incorporating their $P(z)$
 (black), using $P$-cut selected galaxies (for which 98\% of $P(z)$ lies
 behind $z_l+0.2$; cyan), or CC-cut-selected galaxies
 (blue). Weak-lensing S/N values for each selection are given in the
 legend of each panel.
 Image reproduced with permission from \citet{Medezinski2018src}, copyright by the authors.
 }
\end{figure*}

A critical source of systematics in weak lensing comes from accurately 
estimating the redshift distribution of background source galaxies,
which is needed to convert the lensing signal into physical mass units
\citep{Medezinski2018src}.
Contamination of background samples by unlensed foreground and cluster
galaxies with $\Sigmacr^{-1}(z_l,z_s)=0$, when not accounted for, leads
to a systematic underestimation of the true lensing signal.
Inclusion of foreground galaxies produces a dilution of the lensing
signal that does not depend on the cluster-centric radius. 
In contrast, the inclusion of cluster galaxies significantly dilutes the
lensing signal at smaller cluster radii and causes an
underestimation of the concentration of the cluster mass profile
\citep{BTU+05}, as  well as of the halo mass $M_\Delta$ especially at
higher overdensities $\Delta$. 
The level of contamination by cluster galaxies increases with the
cluster mass or richness (see Fig.~\ref{fig:dilution}).
%%%
A secure selection of background galaxies is thus key for obtaining
accurate cluster mass estimates from weak gravitational lensing
\citep{2007ApJ...663..717M,Medezinski+2010,Medezinski2018src,UB2008,Okabe+2013,Gruen2014}.

In real observations, acquiring spectroscopic redshifts for individual
source galaxies is not feasible, particularly to the depths of
weak-lensing observations. 
Instead of spectroscopic redshifts, photo-$z$'s can be used when
multi-band imaging is available. Cluster weak-lensing studies, however,
often rely on two to three optical bands for deep imaging
\citep[e.g.,][]{BTU+05,Medezinski+2010,Oguri+2012SGAS,Okabe+Smith2016},
so that reliable photo-$z$'s could not be obtained.
Instead, well-calibrated field photo-$z$ catalogs, such
as COSMOS \citep{Ilbert+2009COSMOS,Laigle2016cosmos},
were used to determine the redshift distribution $dN(z)/dz$
of background galaxies for a given color-magnitude selection
\citep{Medezinski+2010,Okabe+2010WL}.
Such field surveys are often limited to deep but small areas and thus
subject to cosmic variance.
%Furthermore, this approach does not account for contamination of the
%background sample by cluster members. 

Dedicated wide-area optical surveys, such as
the Hyper Suprime-Cam Subaru Strategic Program
\citep[HSC-SSP;][]{Miyazaki2018hsc,hsc2018ssp,hsc2018dr1},
the Dark Energy Survey \citep[DES;][]{DESdr1},
 and the upcoming Large Synoptic Survey Telescope \citep[LSST;][]{LSST2008},
 are designed to observe in several broad bands, so that photo-$z$'s are
 better determined.
 These photo-$z$ estimates will still suffer from a large fraction of outliers  
%(20\%--30\%)
due to inherent color--redshift degeneracies,
as limited by a finite number of broad optical bands.
 The photo-$z$ uncertainties are folded in by incorporating the full PDF
 for each source galaxy \citep{WtG3}. 
 However, photo-$z$ PDFs are often sensitive
 to the assumed priors. Moreover, the accuracy of photo-$z$ PDFs will be  
 limited by the representability of spectroscopic-redshift samples used
 for calibration. Alternative approaches 
 rely on more stringent color cuts to reject objects with biased
 photo-$z$'s
 \citep{Medezinski+2010,Medezinski+2011,Umetsu+2010CL0024,Umetsu+2012,Umetsu2014clash,Okabe+2013}, 
 which however lead to lower statistical power because they result in lower
 source galaxy densities.

Using the first-year CAMIRA \citep[Cluster finding Algorithm based on
Multiband Identification of Red-sequence gAlaxies;][]{Oguri2018camira}
catalog of $\sim 900$ clusters 
($0.1<z_l<1.1$) with richness $N \geqs 20$ found in
$\sim 140$\,deg$^2$ of HSC-SSP survey data, 
\citet{Medezinski2018src} investigated robust source-selection methods
for cluster weak lensing.
%that separate background galaxies for weak-lensing shape
%measurements and alleviate dilution of the lensing signal.
They compared three different source-selection schemes:
(1) relying on photo-$z$'s and their full PDFs $P(z)$ to correct for
dilution (all),
(2) selecting background galaxies in color--color space (CC-cut),
and (3) selection of robust photo-$z$'s by applying constraints on their 
cumulative PDF ($P$-cut).
All three methods use photo-$z$ PDFs of individual source galaxies,
$P(z)$, to convert the lensing signals into physical mass units.
%%%
With perfect $P(z)$ information, all these methods should thus yield 
consistent, undiluted $\llangle\Delta\Sigma_+(R)\rrangle$ profiles.
After applying basic quality cuts, \citet{Medezinski2018src} found the
typical mean unweighted galaxy number density in the HSC shape catalog 
to be $n_\mathrm{g}=21.7$\,arcmin$^{-2}$.
Similarly, they found
$n_\mathrm{g}=11.6$\,arcmin$^{-2}$ and
$n_\mathrm{g}=13.8$\,arcmin$^{-2}$
for cluster lenses at $z_l<0.4$
using the CC-cut and $P$-cut methods, respectively.

\citet{Medezinski2018src} showed that simply relying on the photo-$z$
PDFs (all) results in a $\llangle\Delta\Sigma_+\rrangle$ profile that
suffers from dilution due to residual contamination by cluster galaxies.
%This in turn leads to an underestimation of the cluster mass
%$M_\mathrm{200c}$ ($13\percent \pm 4\percent$) and
%the concentration $c_\mathrm{200c}$ of the stacked mass profile 
%($24\percent \pm 11\percent$). 
Using proper limits, the CC- and $P$-cut methods give
consistent $\llangle\Delta\Sigma_+\rrangle$ profiles with
minimal dilution.
Differences are only seen for rich clusters with $N \geqs 50$,
where the $P$-cut method produces a slightly 
diluted signal in the innermost radial bin
compared to the CC-cuts (see Fig.~\ref{fig:dilution}). 
Employing either the $P$-cut or CC-cut selection results in cluster
contamination consistent with zero to within the 0.5\% uncertainties. 
For more details on the source-selection methods, we refer the
reader to \citet{Medezinski2018src} and references therein.
An alternative approach to correct for dilution of the lensing signal is
to statistically estimate the level of contamination and subtract it off
\citep[e.g.,][]{Varga2019des}, in which the effect of magnification bias 
must be properly taken into account (see Sect.~\ref{sec:magbias}).

\subsection{Tangential shear signal}
\label{subsec:gest}

Here we describe a procedure to derive azimuthally averaged radial 
profiles of the tangential ($+$) and cross ($\times$) shear components
around a given cluster lens at a certain redshift, $z_l$. 
Specifically, we calculate for each cluster the lensing profiles,  
$\{\langle g_+(\vartheta_b)\rangle\}_{b=1}^{\Nbin}$
and
$\{\langle g_\times(\vartheta_b)\rangle\}_{b=1}^{\Nbin}$, 
in $\Nbin$ discrete cluster-centric bins
spanning the range $\vartheta\in[\vartheta_\mathrm{min},\vartheta_\mathrm{max}]$.

Since weak shear measurements of individual background galaxies
(Eq.~(\ref{eq:gest})) are very noisy, we calculate the weighted
average of the source ellipticity components as:    
\begin{equation}
\begin{aligned}
\langle g_+(\vartheta_b\rangle &=
 \frac{\sum_{s\in b}w_{s}\,g_{+,s}}{\sum_{s\in b} w_{s}},\\
\langle g_\times(\vartheta_b)\rangle &=\frac{\sum_{s\in b} w_{s}\,g_{\times,s}}{\sum_{s\in b}w_{s}}, \\
\end{aligned}
\end{equation}
where the summation is taken over all source galaxies ($s$) that lie in
the bin ($b$);
$g_{+,s}$ and $g_{\times,s}$ represent the tangential and
$45^\circ$-rotated cross components of the reduced shear
(Eq.~(\ref{eq:gtgx})), respectively, estimated for each source galaxy;
and $w_s$ is its statistical weight. The azimuthally averaged cross
component,
$\langle g_\times(\vartheta)\rangle$,
is expected to be statistically consistent with zero
(see Sect.~\ref{subsubsec:az_avg}). 

The statistical uncertainty per shear component per source galaxy is
denoted by
$\sigma_{g_{+},s} = \sigma_{g_{\times},s} \equiv \sigma_{g,s}$, which is 
dominated by the shape noise. Here $\sigma_{g,s}$ includes both
contributions from the shape measurement uncertainty and the intrinsic
dispersion of source ellipticities \citep[e.g.,][]{Mandelbaum2018}.
In general, an optimal choice for weighting is to apply an
inverse-variance weighting with $w_s = 1/\sigma_{g,s}^2$
(Sect.~\ref{subsubsec:src_avg}). 
%%%
However, using inverse-variance weights from noisy variance estimates
may result in an unbalanced weighting scheme (e.g., sensitive to
extreme values). To avoid this, one can employ a variant of
inverse-variance weighting, 
$w_s = 1/(\sigma_{g,s}^2+\alpha_g^2)$, with $\alpha_g$ a properly
chosen softening constant
\citep[see, e.g.,][]{2003ApJ...597...98H,Umetsu+2009,Umetsu2014clash,Okabe+2010WL,Oguri2010LoCuSS,Okabe+Smith2016}. 
The error variance per shear component for
$\langle g_{+,\times}(\vartheta_b)\rangle$ 
%and $\langle g_\times(\vartheta_b)\rangle$
is given by:
\begin{equation}
 \label{eq:var_shape}
\sigma^2_\mathrm{shape}(\vartheta_b) =
\frac{\sum_{s\in b} w_s^2\sigma_{g,s}^2}
  {\left(\sum_{s\in b} w_s\right)^2},
\end{equation}
where we have assumed isotropic, uncorrelated shape noise, 
${\cal E}(\Delta g_{i,s}\Delta g_{j,s'})=\sigma_{g,s}^2\delta_{s s'}\delta_{ij}$
($i,j=+,\times$) with $s$ and $s'$ running over all source
galaxies.

To quantify the significance of the tangential shear profile measurement 
$\{g_+(\vartheta_b)\}_{b=1}^{\Nbin}$,
we define a linear S/N estimator by \citep{Sereno2017psz2lens,Umetsu2020xxl}:
\begin{equation}
\label{eq:SNR}
 \left(\mathrm{S/N}\right)_\mathrm{L} = \frac{\sum_{b=1}^{\Nbin}
 g_+(\vartheta_b)/\sigma^2_\mathrm{shape}(\vartheta_b)}
 {\left[\sum_{b=1}^{\Nbin}1/\sigma^2_\mathrm{shape}(\vartheta_b)\right]^{1/2}}.
% \left(\mathrm{S/N}\right)_\mathrm{L} = \frac{\sum_{b=1}^{\Nbin}
% \Delta\Sigma_+(\vartheta_b)/\sigma^2_\mathrm{shape}(\vartheta_b)}
% {\left[\sum_{b=1}^{\Nbin}1/\sigma^2_\mathrm{shape}(\vartheta_b)\right]^{1/2}}.
\end{equation}
This estimator gives a weak-lensing S/N integrated over the radial range
of the data. Equation~(\ref{eq:SNR}) assumes that the total uncertainty
is dominated by the shape noise and ignores the covariance between
different radial bins. Note that we shall use the full covariance matrix
for cluster mass measurements (Sect.~\ref{subsec:Lg}).
This S/N estimator is different from the conventional quadratic
estimator defined by
\citep[e.g.,][]{UB2008,Okabe+Smith2016}:
\begin{equation}
\label{eq:SNRq}
\mathrm{S/N} = \left[
		\sum_{b=1}^{\Nbin}g_{+}^2(\vartheta_b)/\sigma_{\mathrm{shape}}^2(\vartheta_b)
	       \right]^{1/2}.
\end{equation}
As discussed by \citet{Umetsu2016clash,Umetsu2020xxl}, this quadratic 
definition breaks down and leads to an overestimation of significance in 
the noise-dominated regime where the actual per-bin S/N is less than
unity.

The observed $\langle g_+(\vartheta)\rangle$ profile can be
interpreted according to Eq.~(\ref{eq:gt_avg_app}).
Then, it is important to define the corresponding bin radii
$\vartheta_b$ so as to minimize systematic bias in cluster mass measurements.
We define the effective clutter-centric bin radius $\vartheta_b$
 ($b=1,2,\dots,\Nbin$) using the weighted harmonic mean of lens--source
 transverse separations as \citep{Okabe+Smith2016,Sereno2017psz2lens}:
 \begin{equation}
  \label{eq:rcent}
 \vartheta_{b} = \frac{\sum_{s\in b}w_s}{\sum_{s\in b} w_s \vartheta_s^{-1}}.
 \end{equation}
%which allows for a nearly unbiased determination of the underlying
%cluster lensing profile.
 If source galaxies are uniformly distributed in the image
 plane and $w_s$ is taken to be constant,
 Eq.~(\ref{eq:rcent}) in the continuous limit yields
 $\vartheta_b = (\vartheta_{b1}+\vartheta_{b2})/2$
% $\vartheta_b =
% [\int_{\vartheta_{b1}}^{\vartheta_{b2}}\!d\vartheta\vartheta
% w(\vartheta))][\int_{\vartheta_{b1}}^{\vartheta_{b2}}\!d\vartheta w(\vartheta)]^{-1}
% = (\vartheta_{b1}+\vartheta_{b2})/2$
for a single radial bin defined in the range
 $\vartheta\in[\vartheta_{b1},\vartheta_{b2}]$.\footnote{In general, the 
 weighted bin center is defined by $\vartheta_b=
 [\int_{\vartheta_{b1}}^{\vartheta_{b2}}\!d\vartheta\vartheta^2
 w(\vartheta))][\int_{\vartheta_{b1}}^{\vartheta_{b2}}\!d\vartheta\vartheta
 w(\vartheta)]^{-1}$ with $w(\vartheta)$ a weight function. Assuming
 a power-law form for the weight function
 $w(\vartheta)\propto\vartheta^{-n}$, we see that Eq.~(\ref{eq:rcent})
 corresponds to the case where   
 $w(\vartheta)\propto \vartheta^{-1}$, which is optimal for an
 isothermal density profile with
 $\gamma_+(\vartheta)=\kappa(\vartheta)\propto 1/\vartheta$ 
 \citep{Okabe+Smith2016}.}

\subsection{Lens mass modeling}
\label{subsec:lensmodel}

\subsubsection{NFW model}
\label{subsubsec:nfw}

The radial mass distribution of galaxy clusters is often modeled with 
a spherical Navarro--Frenk--White \citep[][hereafter
NFW]{1996ApJ...462..563N} profile,
which has been motivated by cosmological $N$-body simulations
\citep[][]{1996ApJ...462..563N,Navarro+2004}. 
%which has been motivated by cosmological $N$-body simulations  
%\citep[][]{1996ApJ...462..563N,1997ApJ...490..493N,Oguri+Hamana2011,Child2018cm}. 
The radial dependence of the two-parameter NFW density profile is given 
by:
\begin{equation}
 \label{eq:NFW}
 \rho(r)=\frac{\rho_\mathrm{s}}{(r/r_\mathrm{s})(1+r/r_\mathrm{s})^2},
\end{equation}
where $\rho_\mathrm{s}$ is the characteristic density parameter and
$r_\mathrm{s}$ is the characteristic scale radius at which the
logarithmic density slope,
$\gamma_\mathrm{3D}(r)\equiv d\ln{\rho(r)}/d\ln{r}$, equals $-2$.
The logarithmic gradient of the NFW profile is
$\gamma_\mathrm{3D}(r)=-[1+3(r/r_\mathrm{s})]/[1+(r/r_\mathrm{s})]$.
For $r/r_\mathrm{s}\ll 1$,
$\gamma_\mathrm{3D}\to-1$, whereas for $r/r_\mathrm{s}\gg 1$,
$\gamma_\mathrm{3D}\to-3$.
The radial range where the logarithmic density slope is close to the
``isothermal'' value of $-2$ is particularly important, given that such a
mass distribution is needed to explain the flat rotation curves observed
in galaxies.

The overdensity mass $M_\Delta$ is given by integrating
Eq.~(\ref{eq:NFW}) out to the corresponding overdensity radius
$r_\Delta$ at 
which the mean interior density is $\Delta\times \rho_\mathrm{c}(z_l)$ 
(Sect.~\ref{sec:intro}).
For a physical interpretation of the cluster lensing signal, it is
useful to specify the NFW model by the halo mass, 
$M_\mathrm{200c}$, and the concentration parameter,
$c_\mathrm{200c}=r_\mathrm{200c}/r_\mathrm{s}$.  
The characteristic density $\rho_\mathrm{s}$ is then given by:
\begin{equation}
 \label{eq:rhos_NFW}
 \rho_\mathrm{s}=
\frac{\Delta}{3}
 \frac{c_{\Delta}^3}{\ln(1+c_{\Delta})-c_{\Delta}/(1+c_{\Delta})}\rho_\mathrm{c}(z_l).
\end{equation}

Analytic expressions for the radial dependence of the
projected NFW profiles,
$\Sigma_\mathrm{NFW}(R)=2\rho_\mathrm{s}r_\mathrm{s}\times f_\mathrm{NFW}(R/r_\mathrm{s})$
and
$\overline{\Sigma}_\mathrm{NFW}(R)=2\rho_\mathrm{s}r_\mathrm{s}\times g_\mathrm{NFW}(R/r_\mathrm{s})$
with $R=\Dl\vartheta$,
are given by
\citet[][see also \citealt{1996A&A...313..697B}]{2000ApJ...534...34W}:
%%\citep[][see also \citealt{1996A&A...313..697B}]{2000ApJ...534...34W}:
\begin{equation}
 f_\mathrm{NFW}(x)=
  \begin{cases}
  \frac{1}{1-x^2}\left(-1+\frac{2}{\sqrt{1-x^2}}\mathrm{arctanh}{\sqrt{\frac{1-x}{1+x}}}\right)  & (x <1),\\
  \frac{1}{3}  & (x =1),\\
  \frac{1}{x^2-1}\left[1-\frac{2}{\sqrt{x^2-1}}\arctan{\sqrt{\frac{x-1}{x+1}}}\right)  & (x>1),
   \end{cases}
\end{equation}
and
\begin{equation}
 g_\mathrm{NFW}(x)=
  \begin{cases}
   \frac{2}{x^2}\left[\frac{2}{\sqrt{1-x^2}}\mathrm{arctanh}\sqrt{\frac{1-x}{1+x}}+\ln\left(\frac{x}{2}\right)\right]  & (x <1),\\
   2\left[1+\ln\left(\frac{1}{2}\right)\right]  & (x =1),\\
   \frac{2}{x^2}\left[\frac{2}{\sqrt{x^2-1}}\arctan\sqrt{\frac{x-1}{x+1}}+\ln\left(\frac{x}{2}\right)\right]  & (x>1).
   \end{cases}
\end{equation}
The excess surface mass density for an NFW halo is then obtained as
$\Delta\Sigma_\mathrm{NFW}(R)=\overline{\Sigma}_\mathrm{NFW}(R)-\Sigma_\mathrm{NFW}(R)$.
These projected NFW functionals provide a good approximation for the
projected matter distribution around cluster-size halos
\citep{Oguri+Hamana2011}.

\begin{figure*}[!htb] %!htb
  \begin{center}
   \includegraphics[scale=0.45, angle=0, clip]{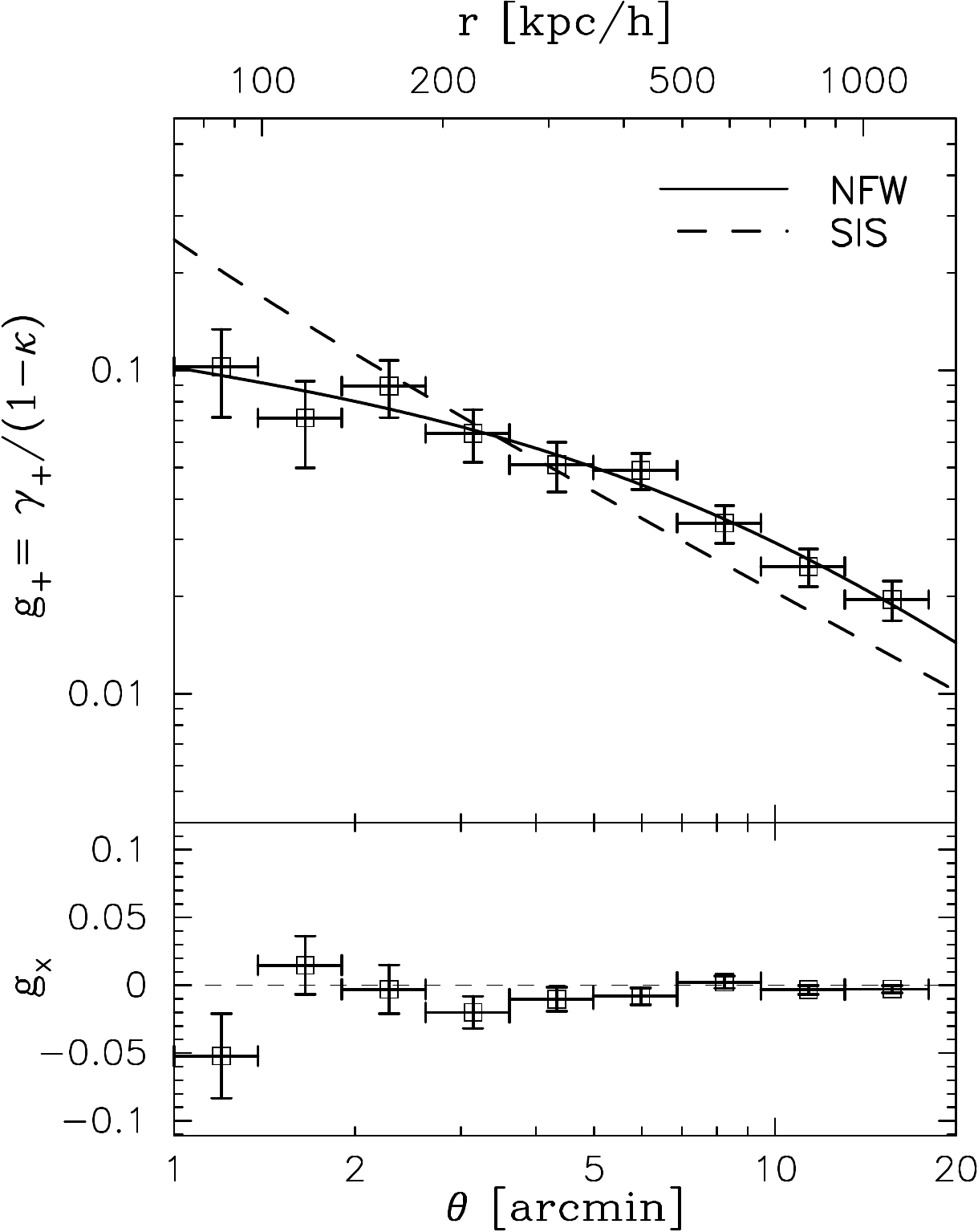}
   \includegraphics[scale=0.45, angle=0, clip]{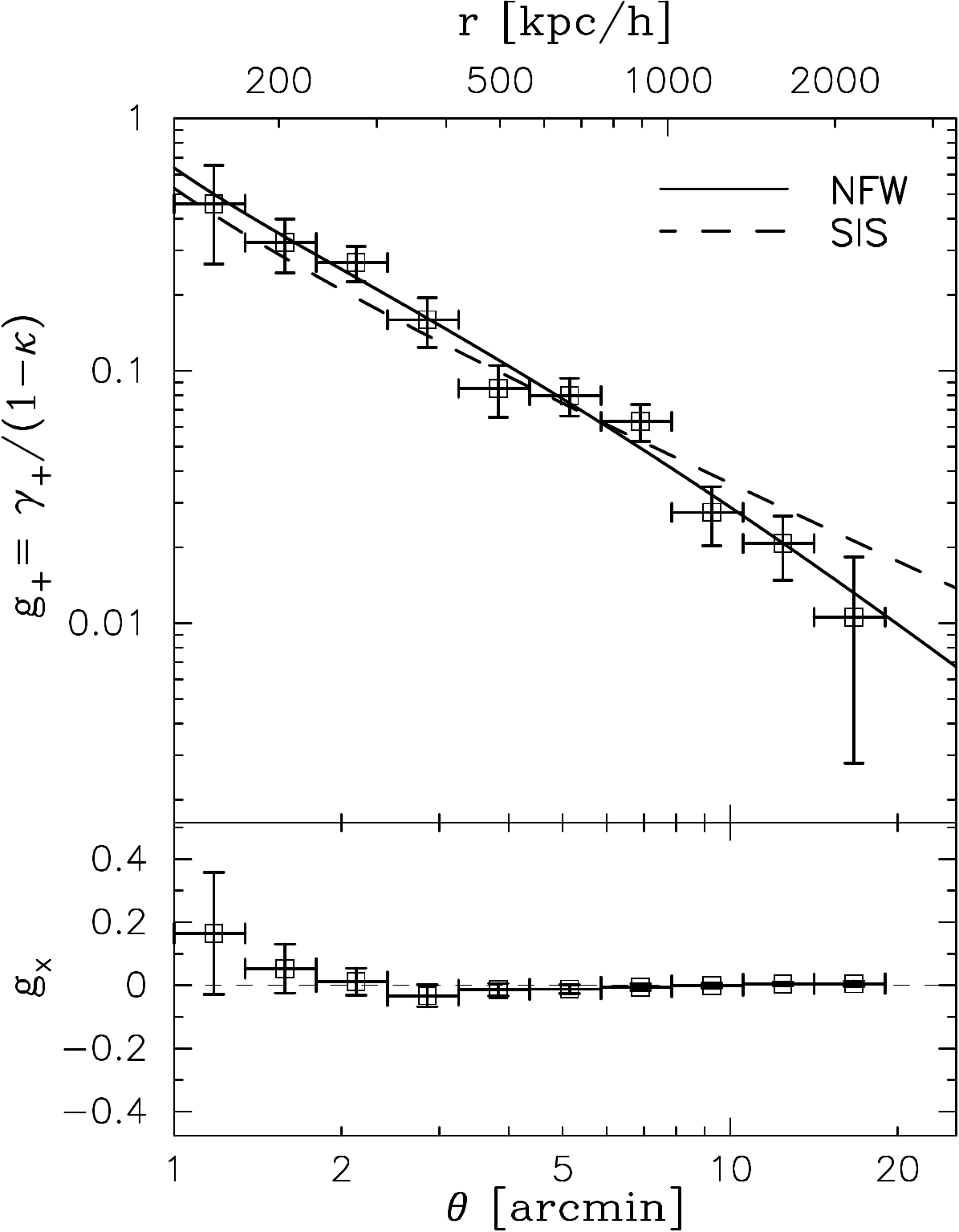}
  \end{center}
\caption{
 \label{fig:gt}
 Azimuthally averaged radial profiles of the reduced tangential shear
 $g_+$ (top panels) and the cross-shear component $g_\times$ (bottom
 panels) for Abell 2142 (left) and Abell 1689 (right)
 based on Subaru Suprime-Cam data. The solid and dashed lines show the
 best-fit NFW and SIS profiles for each cluster, respectively.
 The $45^\circ$-rotated cross-shear component is expected to be
 consistent with zero.
 Image reproduced with permission from \citet{Umetsu+2009}, copyright by AAS.
}
\end{figure*}

As an example, we show in Fig.~\ref{fig:gt} the reduced tangential and  
$45^\circ$-rotated shear profiles
$\langle g_+(\vartheta)\rangle$ and $\langle
g_\times(\vartheta)\rangle$, respectively,
for two high-mass clusters, Abell 2142 and Abell 1689,
obtained from Subaru Suprime-Cam data \citep{Umetsu+2009}.
The $\langle g_+(\vartheta)\rangle$ profiles are compared with
their best-fit NFW and singular isothermal sphere (SIS) models.
The SIS density profile is given by $\rho(r)=\sigma_v^2/(2\pi G r^2)$,
 with $\sigma_v$ the one-dimensional velocity dispersion.
For both clusters, the observed $\langle g_+(\vartheta)\rangle$ profiles are
better fitted by the NFW model having an outward-steepening density profile.  
Abell 2142 is a nearby cluster at $z_l=0.091$ perturbed by merging substructures
\citep[e.g.,][]{Okabe+Umetsu2008,Umetsu+2009,Liu2018a2142}. The radial
curvature observed in the $\langle g_+\rangle$ profile of Abell 2142 is highly
pronounced, so that the power-law SIS model is strongly disfavored by
the Subaru weak-lensing data. From the best-fit NFW model, the mass and
concentration parameters of Abell 2142 are constrained as
$M_\mathrm{200c}=(9.1\pm 1.9)\times 10^{14}\Msunh$
%$M_\mathrm{200c}=(13.0\pm2.7)\times 10^{14}\Msun$
and 
$c_\mathrm{200c}=4.1\pm 0.8$ \citep{Umetsu+2009,Liu2018a2142}.

In contrast, Abell 1689 ($z_l=0.183$) is among the best-studied clusters
and the most powerful known lenses to date 
\citep[e.g.,][]{2005ApJ...621...53B,2007ApJ...668..643L,UB2008,Lemze+2009,Kawaharada+2010,Coe+2010,Diego2015a1689,Umetsu+2011,Umetsu2015A1689},
characterized by a large Einstein radius (Sect.~\ref{subsubsec:rein}) of
$\vartheta_\mathrm{Ein}=(47.0\pm 1.2)$\,arcsec for a fiducial source at
$z_s=2$ \citep{Coe+2010}. This indicates a high 
degree of mass concentration in projection of the sky.
In fact, the observed $\langle g_+(\vartheta)\rangle$ profile of Abell
1689 is well fitted by an NFW profile with a high concentration of
$c_\mathrm{200c}\sim 10$ 
\citep{BTU+05,UB2008,Umetsu+2009,Umetsu2015A1689,Coe+2010}, compared to 
the theoretical expectations, $c_\mathrm{200c}\sim 4$  
\citep[e.g.,][]{Bhatt+2013,Diemer+Kravtsov2015}.
%even after accounting for substantial intrinsic scatter (see
%Sect.~\ref{subsec:cM}).
%%
From full triaxial modeling of two-dimensional weak-lensing, X-ray, and
SZE observations, \citet{Umetsu2015A1689} obtained
$M_\mathrm{200c}=(12.1\pm 1.9)\times 10^{14}\Msunh$
%$M_\mathrm{200c}=(17.3\pm 2.7)\times 10^{14}\Msun$
and
$c_\mathrm{200c}=7.91\pm 1.41$,
which overlaps with the $1\sigma$ tail of the predicted distribution of
halo concentration. Moreover, the multi-probe data set is in favor of a
triaxial geometry with a minor-to-major axis ratio of $c/a=0.39\pm 0.15$ and a
major axis closely aligned with the line of sight by $(22\pm 10)^\circ$. 
Therefore, the superb lensing efficiency of Abell 1689 is likely caused by
its intrinsically high mass concentration combined with a chance
alignment of its major axis with the line-of-sight direction \citep[see
also][]{Oguri2005}.

\subsubsection{Halo model}
\label{subsubsec:halomodel}

At large cluster-centric distances,
the correlated matter around the cluster, 
namely the 2-halo term \citep{Cooray+Sheth2002},
contributes to the lensing signal.
In the standard halo-model prescription
\citep{Oguri+Takada2011,Oguri+Hamana2011}, the total lensing signal
$\Delta\Sigma(R)$ is 
 given as the sum of the 1-halo and 2-halo terms. The 1-halo term
 $\Delta\Sigma_\mathrm{1h}$ 
 accounts for the mass distribution within the main cluster halo,
which can be described by a smoothly truncated NFW profile
\citep[][hereafter BMO]{BMO}:
\begin{equation}
 \rho(r)=\frac{\rho_\mathrm{s}}{(r/r_\mathrm{s})(1+r/r_\mathrm{s})^2}  \left[1+\left(\frac{r}{r_\mathrm{t}}\right)^2\right]^{-2},
\end{equation}
where the truncation parameter $r_\mathrm{t}$ is set to a fixed
multiple of the halo outer radius
\citep[e.g., $r_\mathrm{t}\approx 2.6r_\mathrm{vir}$ or
$r_\mathrm{t}\approx 3r_\mathrm{200c}$; see][]{Oguri+Hamana2011,Covone2014,Umetsu2014clash}.
Analytic expressions for the radial dependence of the projected BMO
profiles are given by \citet{BMO} and \citet{Oguri+Hamana2011}.
%%%

The 2-halo term contribution $\Delta\Sigma_\mathrm{2h}$ to the
tangential shear signal is expressed as \citep[][see
\citealt{dePutter+Takada2010} for the 
 full-sky expression]{Oguri+Takada2011,Oguri+Hamana2011}:\footnote{The
 corresponding 2-halo term in comoving density units is obtained by
 replacing $\Dl(z_l)$ in Eq.~(\ref{eq:2ht}) with the comoving
 angular diameter distance 
 $(1+z_l)\Dl(z_l)=f_K[\chi(z_l)]$.}
\begin{equation}
 \label{eq:2ht}
 \Delta\Sigma_\mathrm{2h}(R|M_\mathrm{200c},z_l) =
  \frac{\overline{\rho}(z)
  b_\mathrm{h}(M_\mathrm{200c};z_l)}{(1+z_l)^3 \Dl^2(z_l)}
  \int\!\frac{\ell d\ell}{2\pi}\,J_2(\ell\vartheta)P(k_\ell; z_l),
\end{equation}
where
$b_\mathrm{h}(M_\mathrm{200c}; z_l)$ is the linear halo bias
\citep[e.g.,][]{Tinker+2010}, 
$k_\ell\equiv \ell/[(1+z_l)\Dl(z_l)]$,
$P(k;z_l)$ is the linear matter power spectrum as a function of the
comoving wavenumber $k$ evaluated at the cluster redshift $z_l$, and
$J_n(x)$ is the Bessel function of the first kind and the $n$th order.
We can
compute the corresponding radial profile
$\Sigma_\mathrm{2h}(R|M_\mathrm{200c},z_l)$ of the lensing convergence
by replacing $J_2(x)$ in Eq.~(\ref{eq:2ht}) with the zeroth-order Bessel
function $J_0(x)$. The 2-halo term is proportional to the product
$b_\mathrm{h}\sigma_8^2$, with $\sigma_8$ the rms amplitude of linear
mass fluctuations in a sphere of comoving radius $8\Mpch$.
In the standard \LCDM model, the 2-halo term contribution to
$\Delta\Sigma$ (or $\Sigma$) becomes important, on average, at $R\simgt$
several (or two) virial radii
\citep[][]{Oguri+Hamana2011,Becker+Kravtsov2011}.     
In particular cases where clusters are residing in extremely dense
environments, the 2-halo contribution to the lensing signal
could become  more significant \citep{Sereno+2018natas}.

\subsubsection{DK14 model}
\label{subsubsec:DK14}

\citet[][hereafter DK14]{Diemer+Kravtsov2014} provide a useful fitting
function for the spherically averaged density profile $\rho(r)$ around
dark matter halos calibrated for a suite of $N$-body simulations in
\LCDM cosmologies. 
The DK14 density profile is given by:
\begin{equation}
 \label{eq:DK14}
 \begin{aligned}
\Delta\rho &\equiv \rho(r)-\overline{\rho} =\rho_\mathrm{inner}\times f_\mathrm{trans}+\rho_\mathrm{outer},\\
\rho_\mathrm{inner} &= \rho_{-2}
 \exp\left\{-\frac{2}{\alpha_\mathrm{E}}\left[
\left(\frac{r}{r_{-2}}\right)^{\alpha_\mathrm{E}}-1\right]\right\},\\
f_\mathrm{trans} &= \left[1+
		      \left(\frac{r}{r_\mathrm{t}}\right)^\beta\right]^{-\frac{\gamma}{\beta}},\\
\rho_\mathrm{outer}&=\frac{b_\mathrm{e}\overline{\rho}}{\Delta_\mathrm{max}^{-1}
  + (r/r_\mathrm{piv})^{s_\mathrm{e}}},\\
 \end{aligned}
\end{equation}
with $r_\mathrm{piv}=5r_\mathrm{200m}$ and $\Delta_\mathrm{max} = 10^3$, 
which is introduced as a maximum cutoff density of the outer term to
avoid a spurious contribution at small halo radii \citep{colossus}.  
The inner profile $\rho_\mathrm{inner}(r)$ describes the intra-halo
mass distribution in a multi-stream region,
which is modeled by an Einasto profile
\citep{Einasto1965}
with $\rho_{-2}$ and $r_{-2}$ the scale density and radius at which the
logarithmic slope is $-2$ and
$\alpha_\mathrm{E}$ the shape parameter describing the degree of profile
curvature. 
The transition term $f_\mathrm{trans}(r)$ characterizes the steepening  
around a truncation radius, $r_\mathrm{t}$.  
The outer term $\rho_\mathrm{outer}$, given by a softened power law, is
responsible for the material infalling toward the cluster in a
single-stream region at large halo radii. 
%the 2-halo term.
DK14 found that this fitting function provides a precise description 
($\simlt 5\%$) of their simulated DM density profiles at
$r\simlt 9r_\mathrm{vir}$.
At larger radii ($r\simgt 9r_\mathrm{vir}$), the outer term is expected
to follow a shape proportional to the matter correlation function. 
%\citep[e.g.,][]{Oguri+Takada2011}.
As in the case of the NFW profile,
it is useful to define the halo concentration by
$c_\Delta=r_\Delta/r_{-2}$.

The DK14 profile is described by eight parameters, 
$(\rho_{-2}, r_{-2}, \alpha_\mathrm{E},
\beta, \gamma, r_\mathrm{t}, b_\mathrm{e}, s_\mathrm{e})$,
and
is sufficiently flexible to reproduce a range of fitting functions, such
as the halo model \citep{Oguri+Hamana2011,Hikage2013} and density
profiles with a sharp steepening feature associated with
the splashback radius (see Sect.~\ref{subsec:rsp}).
Equation~(\ref{eq:DK14}) can be used as a fitting
function in conjunction with generic priors for some of the shape and
structural parameters 
\citep[see][]{Diemer+Kravtsov2014,More2015splash,Umetsu+Diemer2017,Chang2018sp}. 
By projecting $\Delta\rho(r)$ along the line of sight, we
obtain the surface mass density responsible for gravitational lensing as:
\begin{equation}
\Sigma(R)=2\int_R^\infty\frac{\Delta\rho(r)rdr}{\sqrt{r^2-R^2}},
\end{equation}
where the line-of-sight integral is evaluated numerically.
The publicly available code,
\textsc{colossus} 
\citep{colossus}, implements a range of calculations relating to
three-dimensional and projected halo profiles including the NFW,
Einasto, and DK14 models.

\subsection{Shear likelihood function}
\label{subsec:Lg}

The likelihood function ${\cal L}$ of a mass model for weak shear
observations $\bd\equiv\{\langle g_+(\vartheta_b)\rangle\}_{b=1}^{\Nbin}$ is
written as: 
\begin{equation}
 \label{eq:lnLg}
  \begin{aligned}
 -2\ln{\cal L(\bp}) &= \sum_{b,b'=1}^{\Nbin}
  \left[\langle g_+(\vartheta_b)\rangle-\widehat{g}_+(\vartheta_b|\bp)\right]
  (C^{-1})_{bb'}
  \left[\langle g_+(\vartheta_{b'})\rangle-\widehat{g}_+(\vartheta_{b'}|\bp)\right]\\
   &+ \ln\left[(2\pi)^{\Nbin}\mathrm{det}(C)\right],
  \end{aligned}
\end{equation}
where $C$ is the $\Nbin\times\Nbin$ error covariance matrix for the
binned reduced tangential shear profile $\bd$ and
$\widehat{g}_+(\vartheta_b|\bp)$ represents the theoretical
expectation for 
$\langle g_+(\vartheta_b)\rangle$ (Eq.~(\ref{eq:gt_avg_app}))
predicted by the model parameterized by a set of parameters $\bp$.
Note that modeling of the cluster lensing signal $\widehat{g}_+(\vartheta|\bp)$
requires information of the lensing depth
$\langle\Sigmacr^{-1}\rangle$ averaged over the source-redshift
distribution (Sect.~\ref{subsubsec:src_avg}). 
Similarly, one can define a likelihood function for the lensing
convergence profile $\kappa(\vartheta)$, which can be reconstructed from
combined shear and magnification measurements
\citep[e.g.,][]{Umetsu+2011,Umetsu2014clash}.  

A well-characterized inference of the model parameters $\bp$ can be  
obtained within the Bayesian framework by properly choosing the priors
\citep{Umetsu2020xxl}. 
In this context,
when interpreting the cluster lensing signal with an NFW profile
(Sect.~\ref{subsubsec:nfw}),
it is useful to take
$\bp=(M_\mathrm{200c},c_\mathrm{200c})$ as fitting parameters.\footnote{In general, it is
appropriate to assume a log-uniform prior, instead of a uniform prior,
for a positive-definite quantity  
such as $M_\mathrm{200c}$ and $c_\mathrm{200c}$,
especially when the quantity spans a wide dynamic range
\citep{Umetsu2014clash,Umetsu2020xxl,Sereno2017psz2lens,Okabe2019hsc}.
Such a treatment is also self-consistent with mass-scaling relation
analysis, where one often works with logarithmic quantities (e.g.,
$\ln{M_\Delta}, \ln{c_\Delta}$). 
Since the corresponding prior distributions in 
$M_\mathrm{200c}$ and $c_\mathrm{200c}$ scale as $1/M_\mathrm{200c}$ and
$1/c_\mathrm{200c}$, the choice of their lower bounds is relatively
important.}
%%%
Tangential shear fitting with a spherical NFW profile is a standard
approach for measuring individual cluster masses 
from weak lensing
\citep[e.g.,][]{Okabe+2010WL,WtG3,Hoekstra2015CCCP}. 
Numerical simulations suggest that mass estimates from tangential shear
fitting can be biased low (by $\sim 5\percent-10\percent$;
\citealt{Meneghetti+2010a,Becker+Kravtsov2011,Rasia+2012}) 
because local substructures that are abundant in the
outskirts of massive clusters dilute the shear tangential to the cluster
center. 
Moreover, systematic deviations of the lensing signal from the assumed 
NFW profile form in projection can lead to a substantial level of mass bias,
even if the spherically averaged density profiles $\rho(r)$
in three dimensions are well described by the NFW form
\citep[e.g.,][]{Sereno2016einasto,Umetsu2020xxl}.
Therefore, it is highly important to optimize the radial range for
tangential shear fitting so as to alleviate the mass bias
\citep{WtG1,WtG3,Pratt2019}.

\begin{figure*}[!htb] %!htb
  \begin{center}
   \includegraphics[scale=0.5, angle=0, clip]{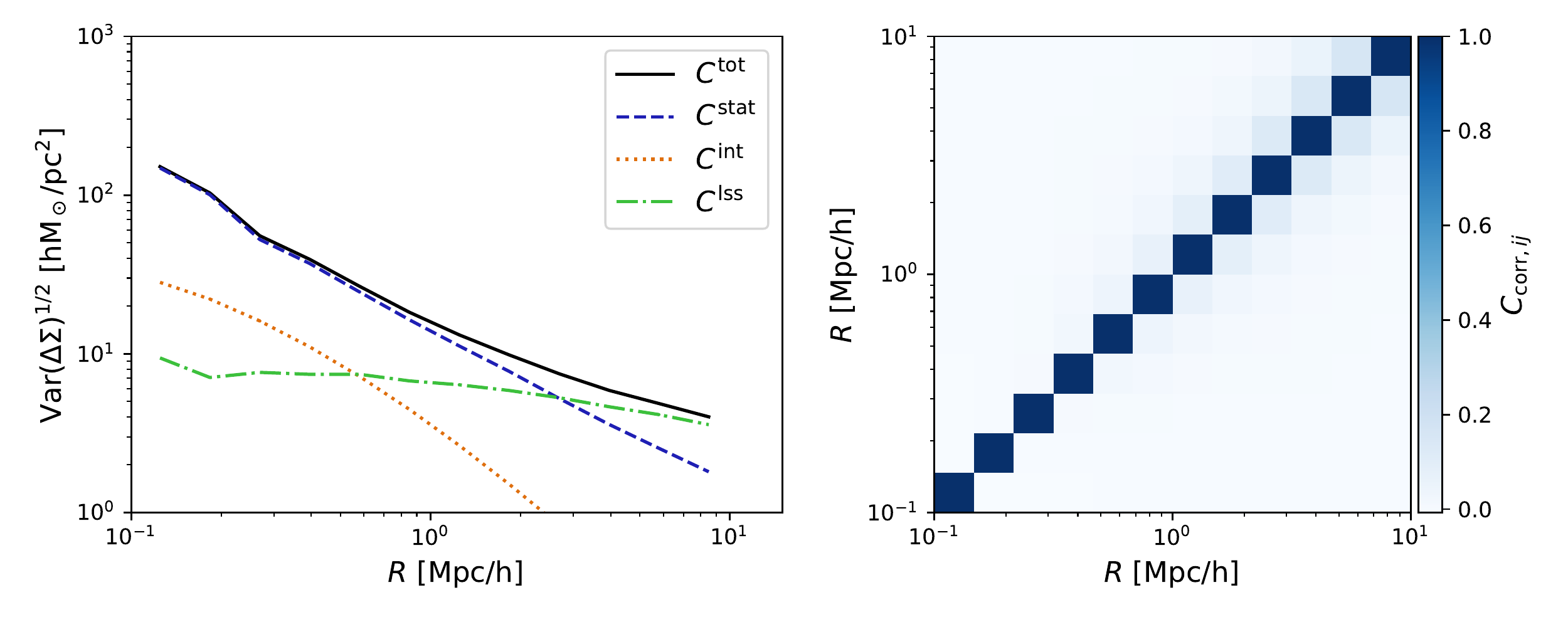}
  \end{center}
\caption{
 \label{fig:cmat}
\emph{Left panel}: diagonal components of the covariance 
 matrix as a function of cluster-centric radius $R$ obtained from a 
 stacked shear analysis based on Subaru Hyper Suprime-Cam observations. 
 The black solid line shows the total covariance $C$; the blue  
 dashed line is the uncertainty due to intrinsic
 shapes of source galaxies ($C^\mathrm{stat}$, denoted as  
 $C^\mathrm{shape}$ in this paper), the orange dotted line is the
 covariance due to intrinsic variations of the projected cluster lensing
 signal ($C^\mathrm{int}$), and the green dashed-dotted line is the
 cosmic noise covariance due to large-scale structure uncorrelated with
 clusters ($C^\mathrm{lss}$).
 Shape noise is dominant for $R\simlt 3\Mpch$ (comoving), while the
 cosmic noise dominates at larger separations. \emph{Right panel}:
 correlation matrix of the stacked total covariance as a function of the 
 radial bin. The correlation between 
 radial bins appears at large separation due to the cosmic noise
 covariance.
 Image reproduced with permission from \citet{Miyatake2019actpol}, copyright by AAS.
}
\end{figure*}

To obtain robust constraints on the underlying cluster mass
distribution, we need to ensure that the shear likelihood function
(Eq.~(\ref{eq:lnLg})) 
includes all relevant sources of uncertainty
\citep{Gruen2015}.
Following \citet{Umetsu2016clash,Umetsu2020xxl},
we decompose the error covariance matrix $C$ for
$\bd=\{\langle g_+(\vartheta_b)\rangle\}_{b=1}^{\Nbin}$ as:
\begin{equation}
 \label{eq:Cmat}
 C = C^\mathrm{shape} + C^\mathrm{lss} + C^\mathrm{int},
\end{equation}
where
$(C^\mathrm{shape})_{bb'}=\sigma^2_\mathrm{shape}(\vartheta_b)\delta_{bb'}$
 is the diagonal statistical uncertainty due to the shape noise
(Eq.~(\ref{eq:var_shape}));
 $(C^\mathrm{lss})_{bb'}$ is the cosmic noise covariance
due to uncorrelated large-scale structure projected along the
line of sight \citep{1998MNRAS.296..873S,2003MNRAS.339.1155H};
and $(C^\mathrm{int})_{bb'}$ accounts for statistical fluctuations of 
the projected cluster lensing signal due to intrinsic variations
associated with assembly bias and cluster asphericity
\citep{Gruen2015}.

Figure~\ref{fig:cmat} shows the diagonal elements of the
covariance matrix used in a stacked weak-lensing analysis of
\citet{Miyatake2019actpol}
and the correlation matrix, defined with the total covariance matrix as  
$C_{bb'}/\sqrt{C_{bb}C_{b'b'}}$. 
A similar figure but for the $\kappa$ profile was presented in
\citet{Umetsu2016clash}, which presents a joint weak and strong lensing
analysis of 20 high-mass clusters targeted by the CLASH survey
\citep[Cluster Lensing And Supernova survey with
Hubble;][]{Postman+2012CLASH}.

The elements of the $C^\mathrm{lss}$ matrix are given by 
\citep{2003MNRAS.339.1155H,Oguri+Takada2011}:
\begin{equation}
 \label{eq:Clss}
 (C^\mathrm{lss})_{bb'}=\int\!\frac{\ell d\ell}{2\pi}P_\kappa(\ell)
  \hat{J_2}(\ell\vartheta_b)\hat{J_2}(\ell\vartheta_{b'}),
\end{equation}
where $\hat{J}_2(\ell\vartheta_b)$ is the Bessel function of the first kind
and second order averaged over the $b$th annulus \citep[for the case of
the binned convergence profile,
see][]{Umetsu+2011stack,Umetsu2016clash,Gruen2015}; and $P_\kappa(\ell)$
is the two-dimensional convergence power spectrum (see Eq.~(\ref{eq:cosmok}))
as a function of angular multipole $\ell$
calculated using the flat-sky and
Limber's approximation as \citep{Limber1953,Kaiser1992}:
\begin{equation}
 \label{eq:Pkappa}
  P_\kappa(\ell) = \frac{9H_0^4\Om^2}{4c^4}\int_0^{\chi_s}\!d\chi\,
%  \left[\frac{r(\chi_s-\chi)}{r(\chi_s)}\right]^2
  W^2(\chi,\chi_s)
 a^{-2}(\chi)P_\mathrm{NL}\left[k=\frac{\ell}{r(\chi)};\chi\right],
\end{equation}
with
$\chi$ the comoving coordinate along the line of sight,
$W(\chi,\chi_s)=r(\chi_s-\chi)/r(\chi_s)$ the ratio of angular diameter
distances $\Dls/\Ds$,
and
$P_\mathrm{NL}(k;\chi)$ the nonlinear matter power spectrum.
The convergence power spectrum $P_\kappa(\ell)$ can be evaluated for a
given source population and a cosmological model. In Eq.~(\ref{eq:Pkappa}),
we have assumed a single comoving distance $\chi_s$ corresponding to the
effective single-plane redshift of source galaxies (i.e., all source
galaxies lying at $\chi=\chi_s$).
Provided that $\Delta\vartheta_b/\vartheta_b\ll 1$ with $\Delta\vartheta_b$
the radial bin width,
we have $\hat{J}_2(\ell\vartheta_b)\simeq J_2(\ell\vartheta_b)$ (without
bin-averaging) in Eq.~(\ref{eq:Clss}). 

The $C^\mathrm{int}$ matrix describes statistical fluctuations of
the projected cluster lensing signal at fixed halo mass due to intrinsic
variations in halo concentration, triaxiality and orientation, and 
correlated secondary structures around the cluster, as well as to
deviations from the assumed NFW form
\citep{Becker+Kravtsov2011,Gruen2015}.\footnote{When simultaneously
determining the mass and concentration for a given individual cluster,
strictly speaking, the contribution from the intrinsic variance in
concentration should be excluded from $C^\mathrm{int}$. Nevertheless, 
the effect of the concentration variance becomes important only at small
cluster radii \citep{Gruen2015}.}
\citet{Gruen2015} constructed a semi-analytical model of
$C^\mathrm{int}$ that is calibrated to cosmological numerical
simulations. 
\citet{Umetsu2016clash} found that
the diagonal part of the intrinsic covariance for the convergence
$\kappa$ can be well approximated by:\footnote{Following 
\citet{Umetsu2016clash,Umetsu2020xxl}, this formalism excludes the
external contribution from $C^\mathrm{lss}$ (Eq.~(\ref{eq:Clss})),
which was formally included in the $C^\mathrm{int}$ covariance by
\citet{Gruen2015}.} 
\begin{equation}
 (C^\mathrm{int}_{\kappa})_{bb} \simeq \alpha_\mathrm{int}^2\kappa^2(\vartheta_b)
\end{equation}
with $\alpha_\mathrm{int}=0.2$ in the intracluster (1-halo) regime at
$R=\Dl\vartheta\simlt r_\mathrm{200m}$. This suggests that the intrinsic
variance is self-similar in the sense that
$\sqrt{(C^\mathrm{int}_\kappa)_{bb}}/\kappa(\vartheta_b)\sim \mathrm{const}$.
A further simplification can be made by setting the off-diagonal elements
of $C^\mathrm{int}_\kappa$ to zero, i.e.,
$(C^\mathrm{int}_{\kappa})_{bb'}=\alpha^2_\mathrm{int}\kappa^2(\vartheta_b)\delta_{bb'}$. 
In general, the diagonal approximation to $C^\mathrm{int}_\kappa$ can
lead to an underestimation of parameter uncertainties
 \citep[][see their Fig.~5]{Gruen2015}, where the degree of
 underestimation depends on the binning scheme, depth of weak-lensing
 observations, and halo mass. 
The impact of the diagonal approximation is more severe for deeper
observations (or higher S/N weak-lensing data).\footnote{Adopting a
constant logarithmic binning with $\Delta\ln\vartheta\sim0.3$,
  \citet{Umetsu2016clash} found that
  the lensing S/N estimated using the
  diagonal approximation to $C^\mathrm{int}_\kappa$ is accurate to
  $\sim 10\percent$ for their ground-based weak-lensing observations of
  high-mass clusters with $M_\mathrm{200c}\sim 10^{15}\Msunh$.}
Assuming a representative mass profile, it is possible to convert the intrinsic
covariance matrix $C^\mathrm{int}_\kappa$ for the convergence into that
for the tangential shear.
This can be done by assuming an NFW density profile
together with the concentration--mass relation
$c_\mathrm{200c}(M_\mathrm{200c},z)$ for a given
cosmological model \citep{Miyatake2019actpol,Umetsu2020xxl}.  
The covariance $C^\mathrm{int}$ for the $g_+$ profile obtained in this
way thus depends on halo mass. 
\citet{Miyatake2019actpol} found that, however, the intrinsic
covariances with different halo masses remain nearly self-similar in their
shapes once scaled by $R\to R/r_\mathrm{200m}$.

\subsection{Stacked weak-lensing estimator}
\label{subsec:stack}

Stacking an ensemble of galaxy clusters helps average out large
statistical fluctuations inherent in noisy weak-lensing observations of
individual clusters. The statistical precision can be largely improved
by stacking together a large number of clusters, allowing for tighter
and more robust constraints on the ensemble properties of the cluster
mass distribution. 
A stacked lensing analysis is thus complementary to an alternative
approach that relies on individual cluster mass measurements
(Sects.~\ref{subsec:gest} and \ref{subsec:Lg}).
In particular, a comparison of the two approaches provides a useful
consistency check in different S/N regimes  
\citep[e.g.,][]{Okabe+2010WL,Umetsu2014clash,Umetsu2016clash,Umetsu2020xxl,Okabe+Smith2016}.

Let us consider an ensemble of $N$ galaxy clusters. 
We model the ensemble mass distribution of these clusters in terms of
the excess surface mass density profile as:
\begin{equation}
 \bm = \left\{\Delta\Sigma(R_b)\right\}_{b=1}^{\Nbin}.
\end{equation}
Specifically, our model is a vector of $\Nbin$ parameters containing the 
 binned $\Delta\Sigma(R)$ profile as a function of the projected
 cluster-centric radius $R$ (see Sect.~\ref{subsec:gammat}).
Here we aim to construct an unbiased estimator for the 
 model $\bm$, or the ensemble $\Delta\Sigma(R)$ profile,
 given weak-lensing observations of $N$ individual clusters.

We assume that these clusters are distributed in redshift, having
different geometric responses to the lensing signal through
$\Sigmacr^{-1}(z_l,z_s)$.  
We express weak-lensing observations
$\bd_l=\{\langle g_{+}(R_b|z_l)\rangle\}_{b=1}^{\Nbin}$ 
for a given cluster ($l$) as a sum of the signal vector $\bs_l$ and the  
noise vector $\bn_l$ as:
\begin{equation}
  \bd_{l} =\bs_{l} + \bn_{l} \ \ \ (l=1,2,\dots,N),
\end{equation}
with
\begin{equation}
 \bs_{l} = a_{l} \bm,
\end{equation}
%the signal vector of the $l$th cluster as
%\begin{eqnarray}
%\bs_{l} \equiv \left\{\langle g_{+}(R_b|z_l)\rangle\right\}_{b=1}^{\Nbin} =\omega_{l} \bm,
%\end{eqnarray}
%where $\langle g_{+}(R|z_l)\rangle$ denotes the
%source-averaged reduced shear profile observed for the $l$th cluster
%(Sect.~\ref{subsubsec:src_avg}) and 
%where $\bd_l=\{\langle g_{+}(R_b|z_l)\rangle\}_{b=1}^{\Nbin}$ is the
%source-averaged reduced shear profile observed for the $l$th cluster
%(Sect.~\ref{subsubsec:src_avg}) and
where the response coefficient $a_l$ represents the source-averaged
inverse critical surface mass density evaluated for the $l$th cluster
(Eq.~(\ref{eq:ISigmacr_avg})):
\begin{equation}
 a_{l} = \left\langle\Sigma_{\mathrm{cr},l}^{-1}\right\rangle.
\end{equation}
In this expression, we assume that both $\bd_l$ and $a_l$ have been
averaged over an ensemble of source galaxies to represent the 
respective source-averaged quantities for the $l$th cluster.
For simplicity, we have ignored the nonlinearity between the lensing
signal $g_+$ and the surface mass density $\Delta\Sigma$
(see Sect.~\ref{subsubsec:src_avg}).\footnote{Remember that the
observable quantity for weak shear lensing is the reduced shear, $g=\gamma/(1-\kappa)$.}
We refer to \citet{Umetsu2020xxl} for a
treatment of the stacked weak-lensing analysis accounting for the
nonlinear correction for the source-averaging effect. 

Assuming that $\bn$ obeys Gaussian statistics
and that the noise vectors between different clusters are statistically
independent,  we can write the total likelihood function of observations 
$\bd=\{\bd_1,\bd_2,\dots,\bd_N\}$ as:
\begin{equation}
  P(\bd|\bm) = \left( \prod_{l=1}^{N}{\cal N}_l\right) \exp{\left[
  -\frac{1}{2}\sum_{l=1}^N (\bd_l-a_l\bm)^t C_l^{-1} (\bd_l-a_l\bm)
  \right]},
\end{equation}
where $C_l=\langle \bn_l \bn_l^t\rangle$ is the error covariance matrix
(Sect.~\ref{subsec:Lg}) for the $l$th cluster and 
${\cal N}_l=(2\pi)^{-\Nbin/2}|C_l|^{-1/2}$ is a normalization factor.
In ground-based cluster weak-lensing observations, the shear covariance
matrix $(C_l)_{bb'}$ per cluster ($b,b'=1,2,\dots,\Nbin$) is
dominated by the statistical uncertainty due to the shape noise.
The contribution from cosmic noise (Sect.~\ref{subsec:Lg}) becomes
important at large cluster-centric distances (Fig.~\ref{fig:cmat}).
% %

The total log-likelihood function $\ln{P(\bd|\bm)}$ is expressed as:
\begin{equation}
\ln{P(\bd|\bm)} =  -\frac{1}{2}\sum_{l=1}^N (\bd_l-a_l\bm)^t C_l^{-1}
 (\bd_l-a_l\bm) + \mathrm{const}.
\end{equation}
According to Bayes' theorem,
the posterior probability distribution of $\bm$ given the data $\bd$ is:
\begin{equation}
P(\bm|\bd) = P(\bd|\bm)\frac{P(\bm)}{P(\bd)},
\end{equation}
where $P(\bm)$ is the prior probability distribution for the model $\bm$
and $P(\bd)$ is the evidence that serves as a normalization factor here.
We assume an uninformative uniform prior for our model $\bm$, such that 
$P(\bm|\bd) \propto P(\bd|\bm)$.
%%%
By maximizing $\ln{P(\bm|\bd)}$ with respect to $\bm$,
we obtain the desired expression for the stacked weak-lensing 
estimator $\widehat{\bm}$ as
\citep[e.g.,][]{Umetsu+2011stack}:
\begin{equation}
\label{eq:stack}
\widehat{\bm} = 
\left(\displaystyle\sum_{l=1}^N a_l^2 C_l^{-1} \right)^{-1}
 \,
\left(
\displaystyle\sum_{l=1}^{N}{a_l C_l^{-1} \bd_l}
\right) \equiv \llangle\bDSigma_+\rrangle.
\end{equation}
%where $\llangle...\rrangle$ denotes the averaging over all lens--source pairs.
%where $\llangle\bDSigma\rrangle$
% $\llangle\bDSigma\rrangle=\{\llangle\Delta\Sigma(R_i)\rrangle\}_{i=1}^M$
%represents a vector containing $M$ binned $\Delta\Sigma$ values of the
%ensemble mass distribution.
Note that the weight assigned to $\Delta\Sigma_+$ of each cluster is
proportional to 
$a_l^2=\left\langle\Sigma_{\mathrm{cr},l}^{-1}\right\rangle^2$  (see
also Eq.~(\ref{eq:wls_stack})) because $\bs_l\propto a_l$.
The error covariance matrix ${\cal C}$ for the stacked
estimator $\llangle\bDSigma_+\rrangle$ (Eq.~(\ref{eq:stack})) is
given by:
\begin{equation}
{\cal C}_{bb'} = \left({F}^{-1}\right)_{bb'},
\end{equation}
with ${F}$ the Fisher information matrix defined as 
\citep[e.g.,][]{Umetsu+2011stack}:
\begin{equation}
 {F}_{bb'} \equiv - {\cal E}\left[
			 \frac{\partial^2 \ln{P}(\bm|\bd)}{\partial m_b
			 \partial m_{b'}}
		     \right]_{\bm=\widehat{\bm}}
=
\left(\sum_{l=1}^N a_l^2 C_l^{-1}\right)_{bb'}.
\end{equation}
The total S/N for detection is given by \citep[e.g.,][]{UB2008}:
%\citep[e.g.,][]{Umetsu+2011stack,Umetsu2016clash}:
\begin{equation}
 (\mathrm{S/N})^2 = \sum_{b,b'=1}^{\Nbin} \widehat{m}_b \left({\cal C}^{-1}\right)_{bb'}
  \widehat{m}_{b'}
  = \widehat{\bm}^t {\cal C}^{-1}\widehat{\bm}.
\end{equation}
Again, this quadratic S/N estimator breaks down and leads to
an overestimation of significance if the actual per-bin S/N is less than
unity (see Sect.~\ref{subsec:gest}).

It is noteworthy that interpreting the effective mass from the stacked
 lensing signal (Eq.~(\ref{eq:stack})) requires caution especially
 when the cluster sample spans a wide range in mass and redshift.
This is because the
amplitude of the lensing signal is weighted by the redshift-dependent  
sensitivity and it is not linearly proportional to the cluster mass
\citep[e.g.,][]{Mandelbaum2005halomodel,Umetsu2016clash,Umetsu2020xxl,Melchior2017des,Sereno2017psz2lens}.   
We refer the reader to \citet{Miyatake2019actpol} and
\citet{Murata2019hsc} for further discussion of this issue.

\begin{figure*}[!htb] %!htb
  \begin{center}
   \includegraphics[scale=0.4, angle=0, clip]{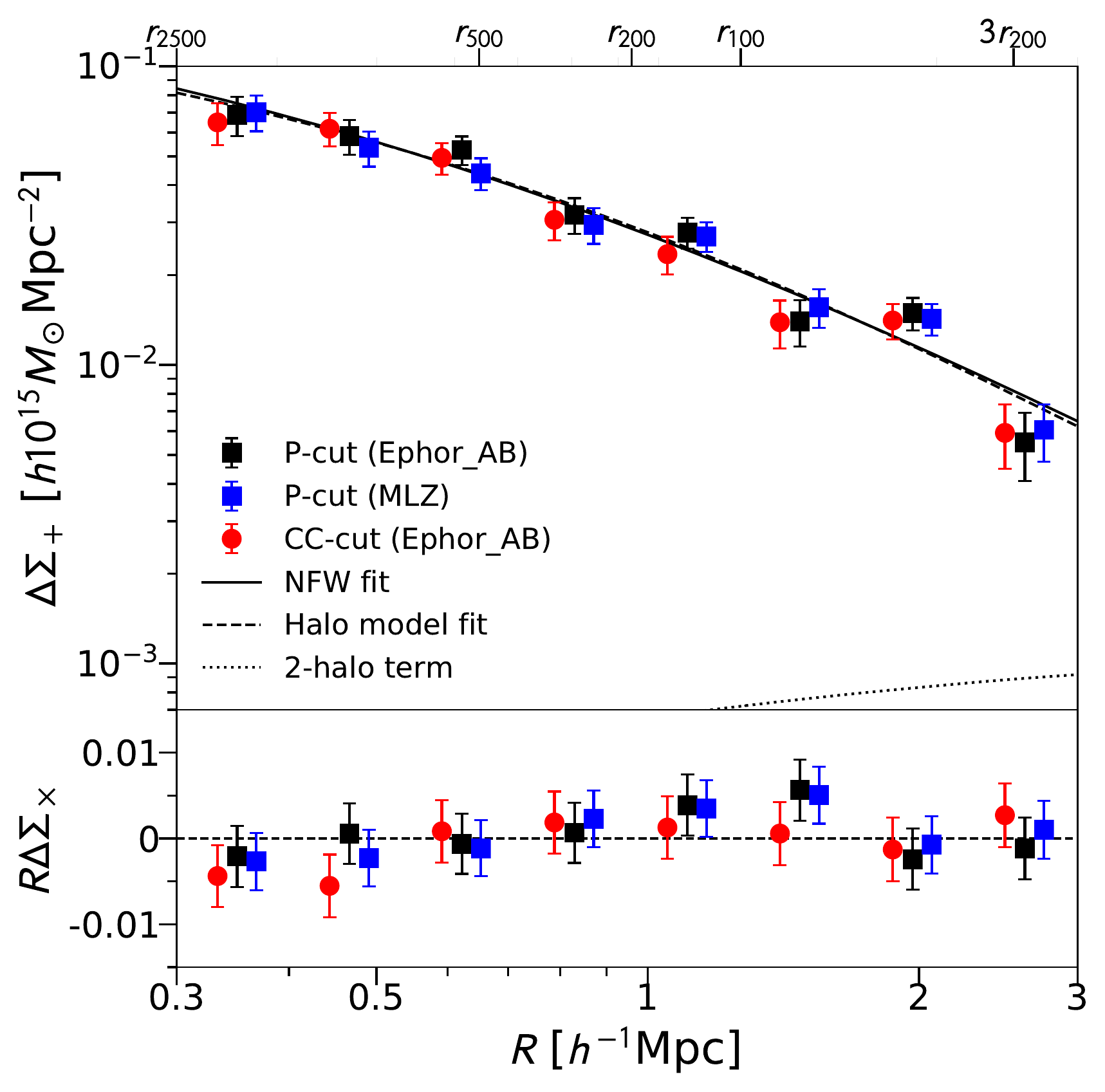}
  \end{center}
\caption{
\label{fig:U20stack}
 Stacked tangential ($\Delta\Sigma_+$: top) and cross
 ($\Delta\Sigma_\times$: bottom) shear profiles
 as a function of cluster-centric comoving radius $R$, derived
  for a sample of 136 spectroscopically confirmed X-ray-selected groups
 and clusters ($0.033\leqs z_l\leqs 1.033$) detected in the 25\,deg$^2$
 XXL-N region.  
 The results are based on Subaru HSC-SSP survey data, shown for three 
 different source-selection methods
 (black squares, blue squares, and red circles).
The data points with different selection methods are horizontally
 shifted for visual clarity. The solid line and the
 dashed line represent the best-fit NFW model and the halo model,
 respectively, derived from the fiducial $P$-cut measurements. The
 dotted line shows the 2-halo term of the best-fit halo
 model.
Ephor\_AB and MLZ refer to HSC photo-$z$ codes \citep{Tanaka2018photoz}.
 Image reproduced with permission from \citet{Umetsu2020xxl}, copyright by AAS.
}
\end{figure*}

Figure~\ref{fig:U20stack} shows the stacked weak-lensing signals around  
a sample of 136 spectroscopically confirmed X-ray groups and clusters at
$0.033\leqs z_l\leqs 1.033$ selected from the \textit{XMM}-XXL survey
\citep{Umetsu2020xxl}. Their weak-lensing analysis is based on HSC-SSP
survey data. The figure compares stacked 
$\llangle\Delta\Sigma_{+}\rrangle$ profiles of the XXL sample obtained
with different source-selection methods
(see Sect.~\ref{subsec:dilution}).   
This comparison shows no significant difference between these profiles 
within errors in all bins.
From a single-mass-bin NFW fit to the stacked shear profile, 
\citet{Umetsu2020xxl} found
$M_\mathrm{200c}=(8.7\pm 0.8)\times 10^{13}\Msunh$
and
$c_\mathrm{200c}=3.5\pm 0.9$
 at a lensing-weighted mean redshift of $z_l\approx 0.25$.
 This is in agreement with the mean concentration expected for
 dark matter halos in the standard \LCDM cosmology, 
 $c_\mathrm{200}\approx 4.1$ at
 $M_\mathrm{200c}=8.7\times 10^{13}\Msunh$ and $z=0.25$
 \citep[e.g.,][]{Diemer+Joyce2019}. 
Figure~\ref{fig:U20stack} also displays the best-fit halo model
including the effects of surrounding large-scale structure 
as a 2-halo term.
Figure~\ref{fig:U20stack} shows that the 2-halo term in the range
$R\in [0.3,3]\,\Mpch$ (comoving) is negligibly small even in low-mass
clusters and groups \citep[e.g.,][]{Leauthaud2010cosmos,Covone2014,Sereno2015s8},
for which the maximum radius corresponds to $\sim 3r_\mathrm{200c}$.
This is because the tangential shear
$\Delta\Sigma(R)=\overline{\Sigma}(R)-\Sigma(R)$
is insensitive to flattened sheet-like structures \citep{Schneider+Seitz1995}.

\subsection{Quadrupole shear}
\label{subsec:quad}

Halos formed in collisionless CDM simulations are not spherical and can
have complex shapes. A more realistic description of individual cluster
halos is as triaxial ellipsoids with minor-to-major axis ratios of order
$a/c\sim 0.5$, slowly increasing with halo-centric radius
\citep{2002ApJ...574..538J,Bonamigo2015}.
More massive halos are less spherical and more prolate, as they tend to 
form later. The projected matter distributions around clusters are thus
expected to be anisotropic, with typical axis ratios of $q\sim 0.6$
\citep[e.g.,][]{OkabeT2018}. The projected axis ratio of cluster halos
varies slowly with cluster-centric distance 
\citep[e.g.,][]{OkabeT2018}.

For sufficiently massive clusters at low redshift,
deep weak-lensing observations allow us to constrain the halo shape on
an individual cluster basis,
by forward-modeling the observed two-dimensional shear (or convergence)
field with elliptical lens models
%@@KU: Citation to Wegner+17 excluded
%\citep{Oguri2010LoCuSS,Okabe+2011A2163,Watanabe+2011,Umetsu+2012,Medezinski+2013,Medezinski2016}. 
\citep{Oguri2010LoCuSS,Okabe+2011A2163,Watanabe+2011,Umetsu+2012,Medezinski+2013,Medezinski2016,2017ApJ...844...67W}. 
This tow-dimensional fitting approach is flexible and can be readily
generalized to include multiple halo components
\citep[see the discussion in Sect.~\ref{subsubsec:az_avg}, e.g.,][]{Okabe+2011A2163,Medezinski+2013}
and triaxial halo shapes
\citep[e.g.,][]{Oguri2005,Sereno+Umetsu2011,Umetsu2015A1689,Chiu2018clump3d}. 

In this subsection, we introduce quadrupole shear estimators for
measuring the projected halo shape for a stacked ensemble of galaxy
clusters.

\subsubsection{Projected halo shape and multipole expansion}
\label{subsubsec:multipole}

Following \citet{Adhikari2015}, we introduce a formalism that allows
for modeling the effects of halo ellipticity on weak shear observables
based on an angular multipole expansion of the
lensing fields.\footnote{For example, we decompose
the $\kappa$ field into angular multipoles as
$\kappa(R,\varphi)=\sum_{m=-\infty}^{\infty}\kappa^{(m)}(R)\,e^{im\varphi}$.
Explicitly, the multipoles are
$\kappa^{(0)}(R)=(2\pi)^{-1}\oint\!\kappa(R,\varphi)\,d\varphi$ for
$m=0$ and
$\kappa^{(m)}(R)=\pi^{-1}\oint\!\kappa(R,\varphi)\cos{m\varphi}\,d\varphi$
for $m\geqs 1$.} 
Let us write the azimuthally averaged projected mass density profile as
$\Sigma^{(0)}(R)\propto R^{-\eta_0}$ with 
$\eta_0 = -d\ln{\Sigma^{(0)}(R)/d\ln{R}}>0$.
Assuming that $q$ is constant with cluster-centric radius, we can write
the surface mass density around clusters
as $\Sigma(x,y)\propto R_q^{-\eta_0}$
\citep{Adhikari2015,Clampitt+Jain2016}, where
$R_q$ is an elliptical radial coordinate defined as
\citep{Evans+Bridle2009,Oguri+2012SGAS,Umetsu+2012,Umetsu2018clump3d}:
\begin{equation} 
R_q = \left(qx^2+y^2/q\right)^{1/2},
\end{equation}
with $q$ the minor-to-major axis ratio $(0<q\leqs 1)$.
Here we have chosen the Cartesian coordinate system ($x,y$) centered on
the halo, such that the 
$x$-axis is aligned with the major axis of the projected ellipse.
We define the corresponding mass ellipticity by
$e=(1-q^2)/(1+q^2)$.

We express the multipole expansion of $\Sigma$
as \citep{Adhikari2015,Clampitt+Jain2016}:
\begin{equation}
 \label{eq:Sigma_multipole}
 \begin{aligned}
 \Sigma(R,\varphi) \propto R_q^{-\eta_0} &= R^{-\eta_0}\left[
  1+\frac{e\eta_0}{2}\cos{2\varphi} + O(e^2)
  \right],\\
  &\equiv \Sigma^{(0)}(R) + \Sigma^{(2)}(R)\cos{2\varphi} + ...,
  \end{aligned}
\end{equation}
where $\varphi$ is the azimuthal angle relative to the halo's major axis,
the multipole $\Sigma^{(m)}(R)$ is the coefficient of the
$e^{im\varphi}$ component of the azimuthal behavior, and 
we assume $e\eta_0/2\ll 1$ to justify the neglect of higher-order terms in
the expansion. 
%On group/cluster scales, several authors have constrained
%the halo ellipticity from stacked weak-lensing measurements
%(Evans & Bridle 2009; Clampitt & Jain 2016; van Uitert
%et al. 2017; Shin et al. 2018) using quadrupole shear
%estimators and their variants (e.g., Natarajan & Refre-
%gier 2000; Adhikari et al. 2015; Clampitt & Jain 2016). 
We thus model the projected mass distributions of clusters as the sum of
a monopole and a quadrupole. We further assume that:
\begin{equation}
 \label{eq:mono_quad}
 e \simeq \frac{2\Sigma^{(2)}(R)}{\eta_0(R)\Sigma^{(0)}(R)}.
\end{equation}
The quadrupole $\Sigma^{(2)}$ can thus be completely determined by
the monopole $\Sigma^{(0)}\propto R^{-\eta_0}$, up to a multiplicative
factor corresponding to the halo ellipticity $e$.

Similarly, the quadrupole moments of the tangential ($+$) and cross
($\times$) shear components are given by \citep{Adhikari2015}:
\begin{equation}
 \label{eq:quadshear_tan}
 \begin{aligned}
  \Sigmacr\gamma_+^{(2)}(R) &= \frac{e}{2}\left[-\Sigma^{(0)}(R)\eta_0(R)+I_1(R)+I_2(R)\right]\cos{2\varphi},\\
  \Sigmacr\gamma_\times^{(2)}(R) &= \frac{e}{2}\left[I_1(R)-I_2(R)\right]\sin{2\varphi},
 \end{aligned}
\end{equation}
where $I_1(R)$ and $I_2(R)$ are defined by \citep{Clampitt+Jain2016}:
\begin{equation}
 \begin{aligned}
  I_1(R) &=
  \frac{3}{R^4}\int_0^R\!R'^3\Sigma^{(0)}(R')\eta_0(R')\,dR',\\
  I_2(R) &=\int_R^\infty\!\frac{\Sigma_0(R')\eta_0(R')}{R'}\,dR'.
 \end{aligned}
\end{equation}
Equation~(\ref{eq:quadshear_tan}) suggests an optimal estimator weighted 
by $\cos{2\varphi}$ to extract the quadrupole of the excess surface mass
density, $\Delta\Sigma^{(2)}(R)$, from tangential shear measurements.

Weighted quadrupole estimators for the tangential and cross shear components
are given by
\citep{Natarajan+Refregier2000,Mandelbaum2006ellip}:\footnote{Equation~(\ref{eq:wlquad_tan}) 
corresponds to 
$2fe_\mathrm{g}\Delta\Sigma_\mathrm{iso}$ and
$2f_{45}e_\mathrm{g}\Delta\Sigma_\mathrm{iso}$ of
\citet{Mandelbaum2006ellip}, where $\Delta\Sigma_\mathrm{iso}$ is the
monopole of the excess surface mass density, $e_\mathrm{g}$ is the
observed ellipticity of the tracer distribution, and $f$ and $f_{45}$
represent the quadrupole strengths of the tangential and cross shear
components, respectively.
See also \citet{Clampitt+Jain2016}.}
\begin{equation}
 \label{eq:wlquad_tan}
 \begin{aligned}
  \llangle\Delta\Sigma^{(2)}_+\rrangle &=
  \left[\sum_{l,s}w_{ls}\Delta\Sigma_+(R|z_l,z_s)\cos{2\varphi_{ls}}\right]
  \left[\sum_{l,s}w_{ls}\cos^2{2\varphi_{ls}}\right]^{-1},\\
  \llangle\Delta\Sigma^{(2)}_\times\rrangle &=
  \left[\sum_{l,s}w_{ls}\Delta\Sigma_\times(R|z_l,z_s)\sin{2\varphi_{ls}}\right]
  \left[\sum_{l,s}w_{ls}\sin^2{2\varphi_{ls}}\right]^{-1},
 \end{aligned}
\end{equation}
where
$\Delta\Sigma_{+,\times}(R|z_l,z_s)=\Sigmacr(z_l,z_s) g_{+,\times}(R|z_l,z_s)$
(Eq.~(\ref{eq:DSigma_plus})); 
$w_{ls} = \Sigma_{\mathrm{cr},ls}^{-2} / \sigma_{g,ls}^{2}$
is the statistical weight for each lens--source pair ($ls$),
with $\sigma_{g,ls}$ the statistical uncertainty per shear component
(see Eq.~(\ref{eq:wls_stack})); and $\varphi_{ls}$ is the azimuthal
angle of each source galaxy ($s$) relative to the major axis of each
cluster lens ($l$).
In real observations, we must rely on the major axis of the distribution
of baryonic tracers (e.g., central galaxies, X-ray gas) to perform
aligned, stacked lensing measurements by Eq.~(\ref{eq:wlquad_tan})
\citep{Mandelbaum2006ellip,vanUitert2012,vanUitert+2017,Clampitt+Jain2016}. 

As discussed by \citet{2004ApJ...606...67H} and
\citet{Mandelbaum2006ellip}, in practical 
applications, Eq.~(\ref{eq:wlquad_tan}) is susceptible to a possible
systematic alignment of lens galaxy (e.g., BCGs) and source
 ellipticities. Such a spurious alignment signal can arise from an
 incomplete correction of the PSF anisotropy, which tends to affect
 neighboring objects in a similar manner.
On the other hand, when interpreting the quadrupole shear signal,
%$\llangle\Delta\Sigma^{(2)}_+\rrangle$,
one must take into account possible misalignment between the underlying
matter and tracer distributions, which will cause a dilution of the
quadrupole shear signal. 
Moreover, modeling of the quadrupole shear based on the multipole 
expansion (Eq.~(\ref{eq:mono_quad})) should only be applied to the
case with a small halo ellipticity ($e\eta/2\ll 1$), so that the
higher-order terms can be safely ignored
(see Eq.~(\ref{eq:Sigma_multipole})).

\subsubsection{Cartesian estimator}
\label{subsubsec:quad_cart}
 
\begin{figure*}[!htb] %!htb
  \begin{center}
   \includegraphics[scale=0.3, angle=0, clip]{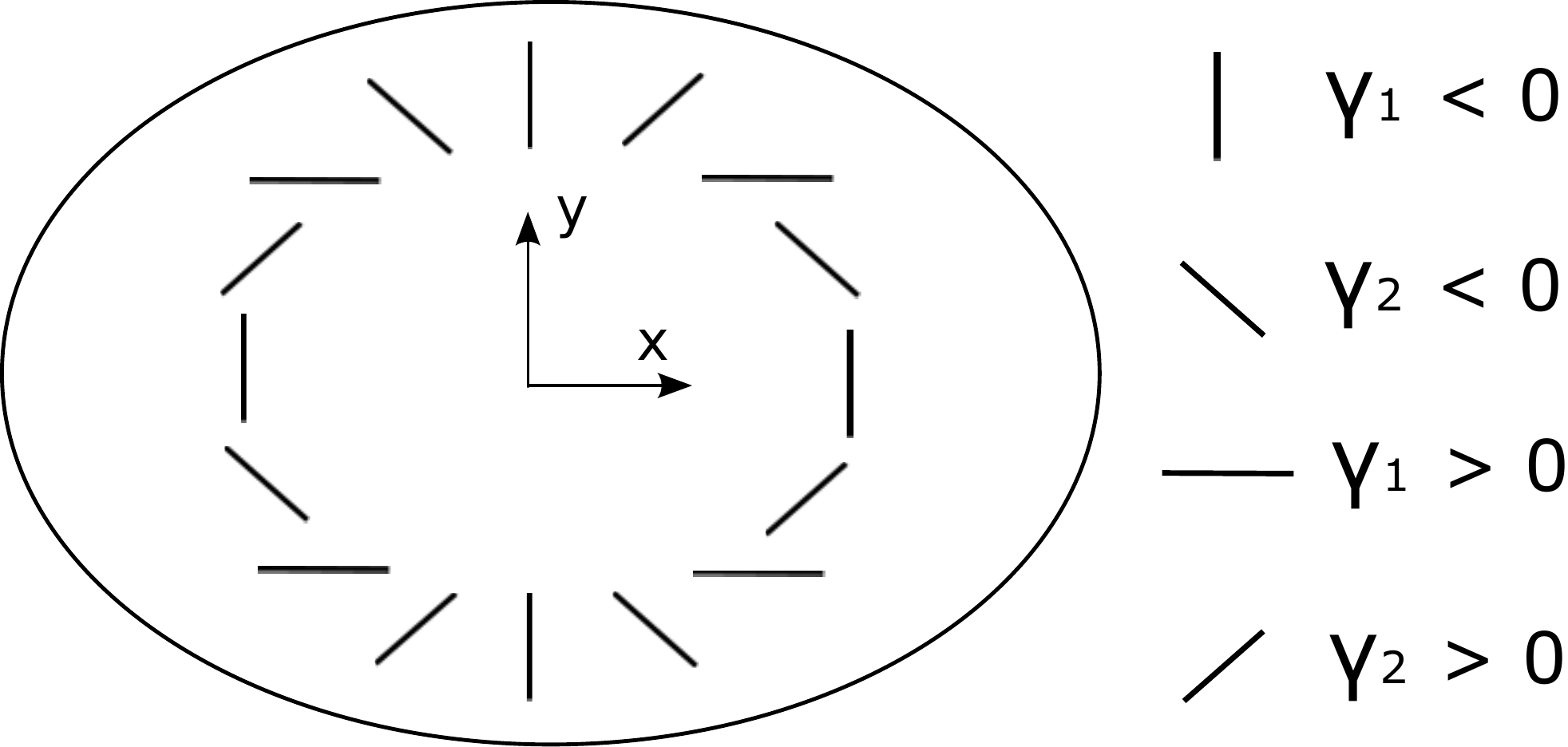}
  \end{center}
 \caption{
The quadrupole shear pattern produced by an elliptical mass distribution.
 The $x$-axis of the Cartesian coordinate system is aligned with
the major axis of the tracer distribution, which is assumed to be aligned with
 the major axis of the underlying mass distribution.
The sign convention for the Cartesian shear components
 ($\gamma_1,\gamma_2$) is shown at the right.
Note that the monopole shear (which is purely tangential) is not
 contributing to the shear pattern illustrated here. 
 Image reproduced with permission from \citet{Clampitt+Jain2016}, copyright by the authors. 
 \label{fig:quad_diagram}
}
\end{figure*}

Now we introduce the Cartesian estimator of \citet{Clampitt+Jain2016}.
%that nulls the purely tangential monopole shear contribution.
%There are several factors contributing to this alignment. First, the
 %point spread function (PSF) has a net large-scale alignment with the
 %survey scan direction,
Compared to the estimator of \citet{Natarajan+Refregier2000}, a
practical advantage of this estimator is that 
one of the two Cartesian components ($\Delta\Sigma_2^{(\pm)}$ defined
below) is insensitive to the spurious alignment of lens--source galaxy
ellipticities \citep{Clampitt+Jain2016} discussed at the end of
Sect.~\ref{subsubsec:multipole}.   
With this
estimator, we measure the stacked quadrupole shear signal with respect
to a coordinate system with the $x$-axis 
aligned with the major axis of the distribution of baryonic tracers
(e.g., central galaxies, X-ray gas).
The monopole signal is nulled with this Cartesian estimator.
We adopt the same sign convention for the Cartesian
$\gamma_1$ and $\gamma_2$ components as defined in
\citet[][]{Clampitt+Jain2016} and use $\varphi$ to denote the azimuthal
angle relative to the $x$-axis of each cluster. This is illustrated in
Fig.~\ref{fig:quad_diagram}. 

The Cartesian shear components are related to the tangential and cross
components (see Eq.~(\ref{eq:gtgx})) by:
\begin{equation}
 \begin{aligned}
  \gamma_1(R,\varphi) &= -\gamma_+(R,\varphi)\cos{2\varphi}+\gamma_\times(R,\varphi)\sin{2\varphi},\\
  \gamma_2(R,\varphi) &= -\gamma_+(R,\varphi)\sin{2\varphi}-\gamma_\times(R,\varphi)\cos{2\varphi}.
 \end{aligned}
\end{equation}
In the framework of \citet{Adhikari2015} based on the multipole
expansion, the multipole moments of the Cartesian shear components are
written as follows \citep{Clampitt+Jain2016}:
\begin{equation}
 \label{eq:quadshear_cart}
 \begin{aligned}
  \Sigmacr\gamma_1^{(2)} &=
  \frac{e}{4}\left(\left[\Sigma^{(0)}(R)\eta_0(R)-2I_1(R)\right]\cos{4\varphi}
  + \Sigma^{(0)}(R)\eta_0(R)-2I_2(R)\right),\\
  \Sigmacr\gamma_2^{(2)}&= \frac{e}{4}\left[\Sigma^{(0)}(R)\eta_0(R)-2I_1(R)\right]\sin{4\varphi}.
 \end{aligned}
\end{equation}
Equation~(\ref{eq:quadshear_cart}) shows that the azimuthal dependence
of the Cartesian shear components goes as $\cos{4\varphi}$
(except for the two terms without $\varphi$ dependence; see
\citealt{Clampitt+Jain2016} for more discussion)
and $\sin{4\varphi}$, so that there is a sign change in both
components after every angle $\pi/4$. When moving around the circle,
the shear signal from elliptical clusters transitions between regions
where $\gamma_1^{(2)}$ and then $\gamma_2^{(2)}$ alternately dominate
\citep{Clampitt+Jain2016}, as illustrated in Fig.~\ref{fig:quad_diagram}.

\begin{figure*}[!htb] %!htb
  \begin{center}
   \includegraphics[scale=0.35, angle=0, clip]{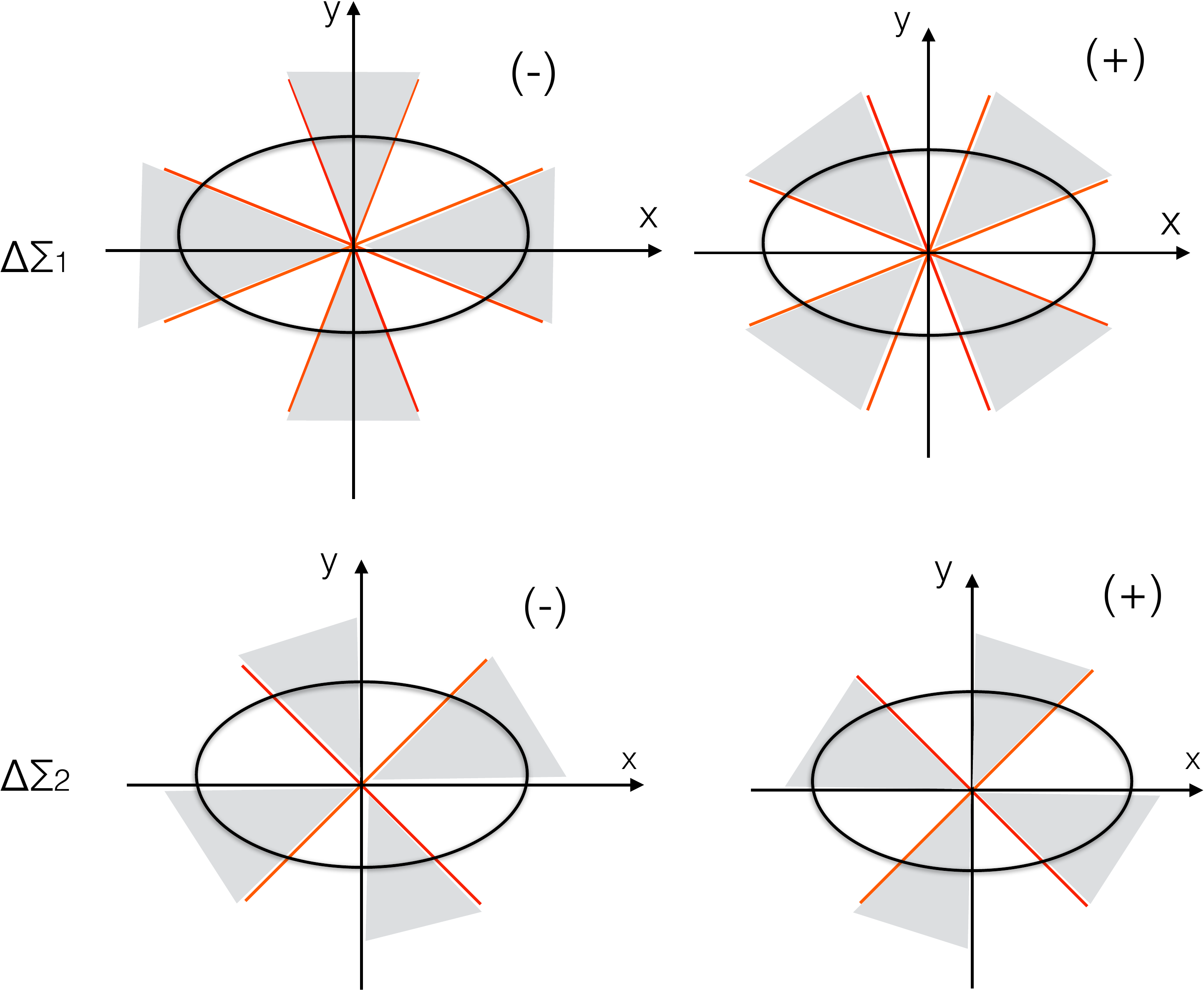}
  \end{center}
 \caption{
 Illustration of the Cartesian quadrupole shear estimator of
 \citet{Clampitt+Jain2016}. 
 The $x$-axis of the Cartesian coordinate system is aligned with the
 major axis of the tracer distribution
 % which is assumed to be aligned with the underlying mass distribution
 (black solid ellipse). 
 We group together the Cartesian first and second shear components in
 same-sign regions of $\cos{4\varphi}$ and $\sin{4\varphi}$ (gray-shaded
 regions), respectively, and define four quadrupole shear components,
 namely, 
 $\Delta\Sigma_{1}^{(-)}$ (upper left),
 $\Delta\Sigma_{1}^{(+)}$ (upper right),
 $\Delta\Sigma_{2}^{(-)}$ (lower left), and
 $\Delta\Sigma_{2}^{(+)}$ (lower right).
 Image reproduced with permission from \citet{Umetsu2018clump3d}, copyright by AAS. 
 \label{fig:quad}
}
\end{figure*}

Following \citet{Clampitt+Jain2016}, we group together the first and
second shear components $(g_1,g_2)$ of background source galaxies in the
regions where $\cos{4\varphi}$ and $\sin{4\varphi}$ have the same sign
(see Fig.~\ref{fig:quad}), respectively, and define the following
estimator: 
\begin{equation}
 \label{eq:wlquad}
   \begin{aligned}
    \llangle\Delta\Sigma_i^{(\pm)}(R)\rrangle&=\left[\sum_{l,s}
    w_{ls}\Delta\Sigma_{i}(R|z_l,z_s) \right]
    \left[\sum_{l,s} w_{ls}\right]^{-1} \ \ \ (i=1,2),
   \end{aligned}
\end{equation}
where we have introduced the notation in analogy to
the tangential shear (Eq.~(\ref{eq:DSigma_plus})):
\begin{equation}
 \Delta\Sigma_i(R|z_l,z_s) = \Sigmacr(z_l,z_s) g_i(R|z_l,z_s),
\end{equation}
%with $R=\Dl\vartheta$ the projected cluster-centric radius. 
where $w_{ls} = \Sigma_{\mathrm{cr},ls}^{-2} / \sigma_{g,ls}^{2}$
is the statistical weight for each lens--source pair ($ls$),
with $\sigma_{g,ls}$ the statistical uncertainty per shear component
(see Eq.~(\ref{eq:wls_stack}));
%for $g_1(R|z_l,z_s)$ and $g_2(R|z_l,z_s)$
and
$s$ runs over all source galaxies that fall in the specified bin
$(R,\varphi)$, different for each shear component 
$i$ and sign \citep{Clampitt+Jain2016}:
$i=1$, $\mathrm{sign}=-$, $-\pi/8\leqs \varphi<\pi/8$;
$i=1$, $\mathrm{sign}=+$, $\pi/8\leqs \varphi<3\pi/8$;
$i=2$, $\mathrm{sign}=-$, $0\leqs \varphi<\pi/4$;
$i=2$, $\mathrm{sign}=+$, $\pi/4\leqs \varphi<\pi/2$.
For each case, the summation in Eq.~(\ref{eq:wlquad}) also includes
source galaxies lying in symmetrical regions shifted by $\pi/2$,
$\pi$, and $3\pi/2$, as illustrated in Fig.~\ref{fig:quad}.
%An optimal choice for $w_{ls}$ is
%$w_{ls} = \Sigma_{\mathrm{cr},ls}^{-2} / \sigma_{g,ls}^{2}$
%where $\sigma_{g,ls}$ is the statistical uncertainty per shear component
%for estimated $g_1(R|z_l,z_s)$ and $g_2(R|z_l,z_s)$.

\begin{figure*}[!htb] %!htb
  \begin{center}
   \includegraphics[scale=0.33, angle=0, clip]{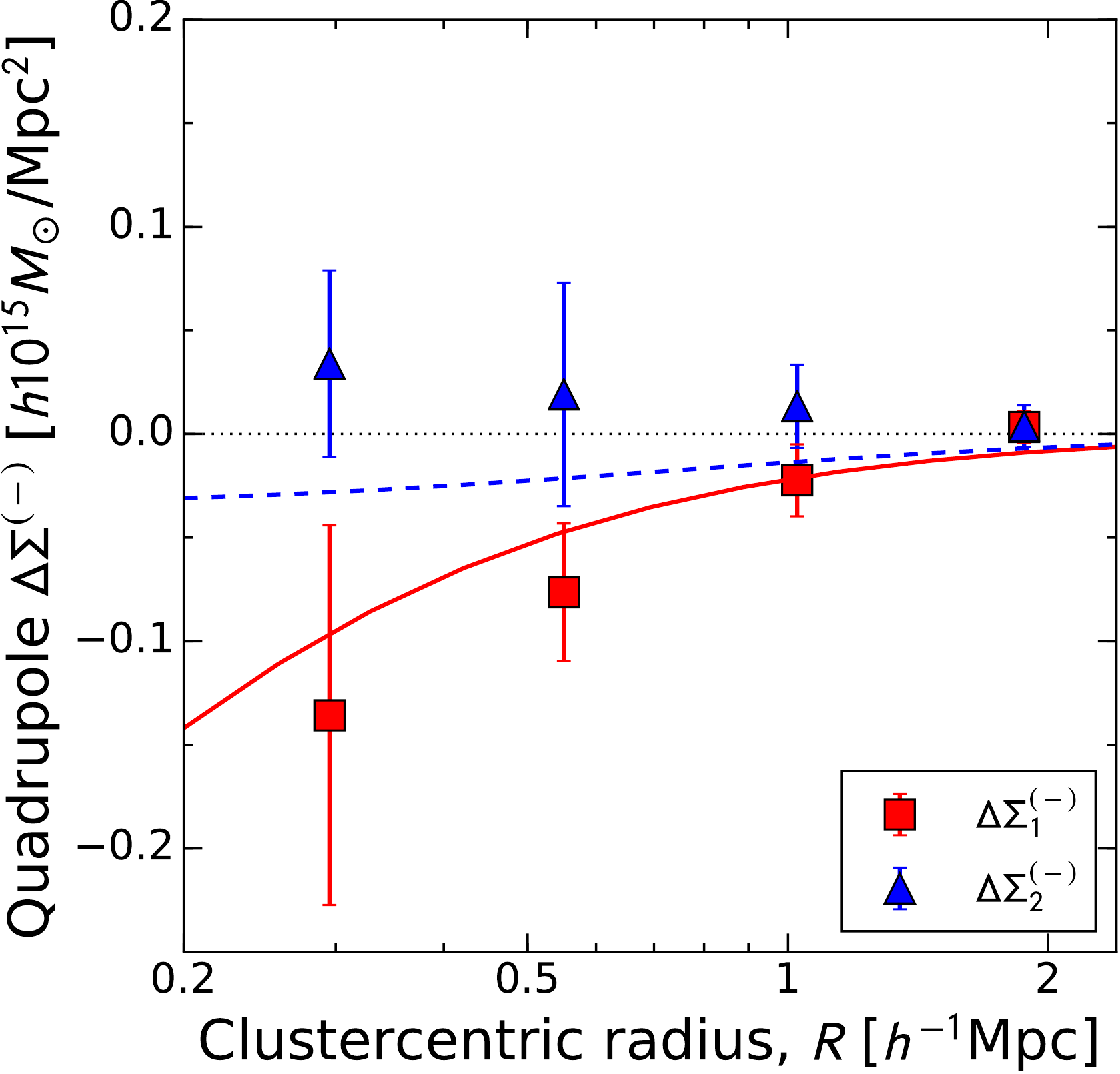}
   \includegraphics[scale=0.33, angle=0, clip]{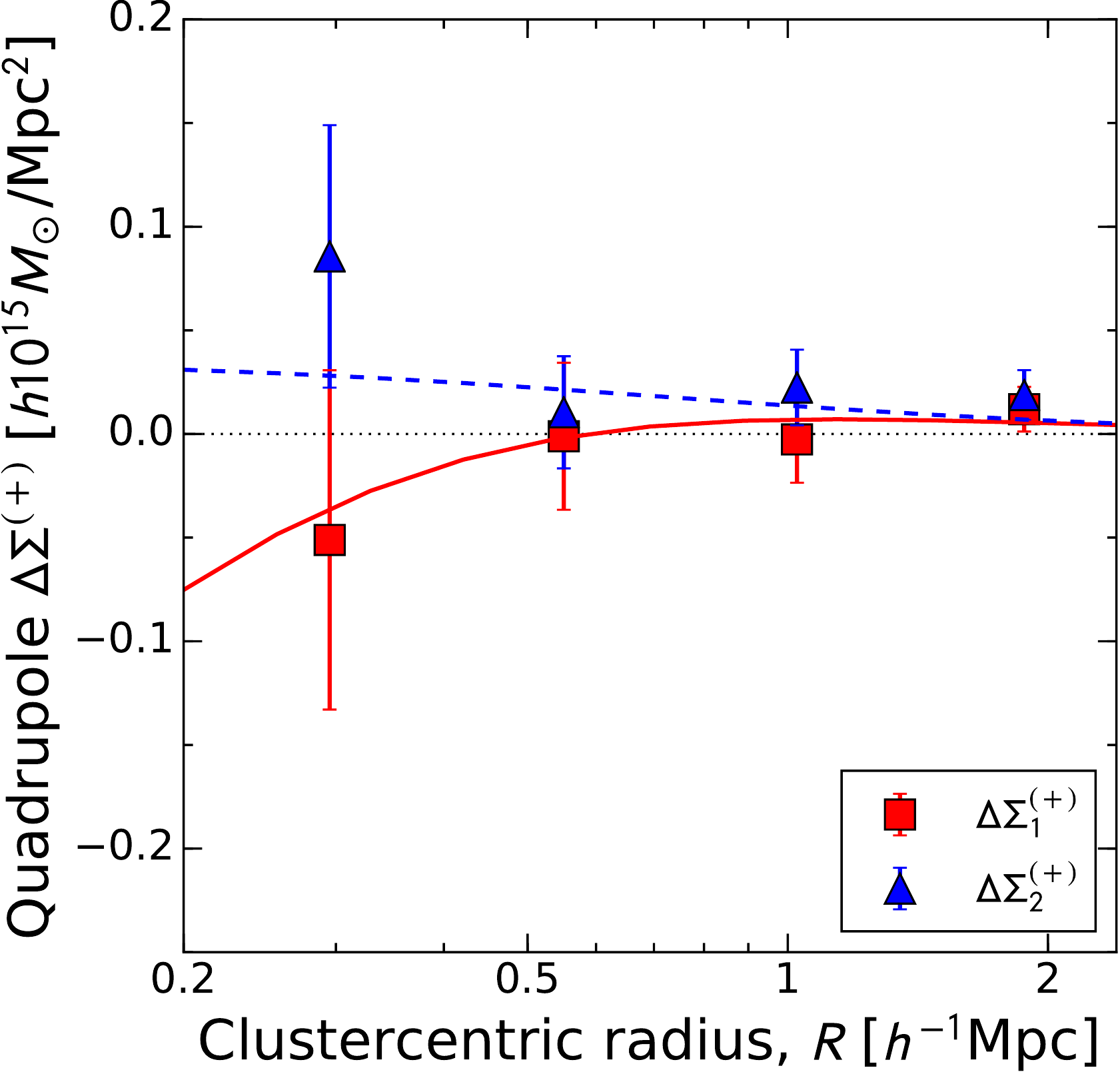}
  \end{center}
\caption{
Stacked quadrupole shear profiles for a sample of 20 CLASH
 clusters measured with respect to the X-ray major axis of each 
 cluster.
\emph{Left panel}: observed   
 $\llangle\Delta\Sigma_1^{(-)}\rrangle$ (red squares) and
 $\llangle\Delta\Sigma_2^{(-)}\rrangle$ (blue triangles)
 profiles shown along with the best-fit elliptical-NFW model.
\emph{Right panel}: same as the left panel, but showing the results 
 for the 
 $\llangle\Delta\Sigma_{1,2}^{(+)}\rrangle$ profiles. The best-fit
 model was obtained from a simultaneous elliptical-NFW fit to the four
 quadrupole shear profiles.
 Image reproduced with permission from \citet{Umetsu2018clump3d}, copyright by AAS.
 \label{fig:U18}
}
\end{figure*}

Figure~\ref{fig:U18} shows the stacked quadrupole shear profiles,
$\llangle \Delta\Sigma_1^{(+)}\rrangle$,
$\llangle \Delta\Sigma_1^{(-)}\rrangle$,
$\llangle \Delta\Sigma_2^{(+)}\rrangle$, and
$\llangle \Delta\Sigma_2^{(-)}\rrangle$,
derived for a sample of 20 high-mass CLASH clusters
\citep{Umetsu2018clump3d}.   
The quadrupole shear signal was measured with respect to the major axis
of the X-ray gas shape of each cluster. 
\citet{Umetsu2018clump3d} modeled the stacked
$\llangle\Delta\Sigma_{1,2}^{(\pm)}\rrangle$
profiles by assuming an elliptical-NFW density profile with the major
axis aligned with the X-ray major axis (for an elliptical extension of
lensing mass models, see \citealt{Keeton2001}).
Any misalignment would thus lead to a dilution of the quadrupole signal
and hence an underestimation of the halo ellipticity.
\citet{Umetsu2018clump3d} obtained stacked
constraints on the projected axis ratio of $q=0.67\pm 0.10$
(or $1-q=0.33\pm 0.10$),
which is fully consistent with the median axis ratio 
$q=0.67\pm 0.07$ of this sample
obtained from their two-dimensional shear and magnification
analysis of the 20 individual clusters.
%based on shear and magnification.   
%This consistency supports the robustness of our results
%and suggests again a tight alignment between the intracluster
%gas and DM.
Their results suggest that the total matter distribution is
closely aligned with the X-ray brightness distribution (with a median
misalignment angle of $|\Delta\mathrm{PA}|=21^\circ\pm 7^\circ$)
%(see Sect.~\ref{subsec:triax}),
as expected from cosmological hydrodynamical
simulations \citep[see][]{OkabeT2018}.

\section{Magnification bias}
\label{sec:magbias}

In addition to the shape distortions, gravitational lensing can cause
focusing of light rays, which results in an amplification of the image
flux through the solid-angle distortion (Sect.~\ref{subsubsec:mu}).    
Lensing magnification provides complementary and independent
observational alternatives to gravitational shear,
especially at high redshift where source galaxies are more difficult to
resolve
\citep[][]{vanWaerbeke+2010,Hildebrandt+2011,Ford2014cfhtlens,Chiu2016magbias,Chiu2020hsc}.

\subsection{Magnified source counts} 
\label{subsec:counts}

Let us consider source number counts $n_0(>F)$ per unit solid angle as a 
function of the limiting flux $F$ for a given population of background
objects (e.g., color--magnitude-selected galaxies, quasars, etc.). In the
absence of gravitational lensing, the intrinsic source counts can be
written as:
\begin{equation}
 \label{eq:n0}
 n_0(>F) 
  =\int\!dz\,\frac{d^2V}{dzd\Omega}\,
  \int_{L(z)}^\infty\!dL\,\frac{d^2N(L,z)}{dLdV}
  \equiv \int\!dz\,\frac{dn_0[z|>L(z)]}{dz},
\end{equation}
where
$d^2V(z)/dz/d\Omega$ is the comoving volume element per redshift
interval per unit solid angle,
$d^2N(L,z)/dL/dV$ is the luminosity function of the background population,
$L(z)=4\pi D_L^2(z)F$
is the luminosity threshold corresponding to
the flux limit $F$ at redshift $z$, with $D_L(z)$ the luminosity
distance,
and $dn_0[z|>L(z)]/dz$ is the redshift distribution function.

Here we focus on the subcritical regime with $\mu(\btheta) > 0$ (i.e.,
outside the critical curves).
Lensing magnification causes focusing of light rays while conserving the
surface brightness (Sect.~\ref{subsubsec:mu}), resulting in the
following two competing effects \citep{1995ApJ...438...49B,Umetsu2013}:
\begin{enumerate}
\item[1] Area distortion: $\delta\Omega\to \mu(\btheta)\delta\Omega$;
\item[2] Flux amplification: $F\to \mu(\btheta)F$.
\end{enumerate}
The former effect reduces the geometric area in the source plane,
thus decreasing the observed number of background sources per unit solid
angle. 
On the other hand, the latter effect amplifies the flux of background
sources, increasing the observed number of sources above the limiting
flux.

In the presence of gravitational lensing, the magnified source counts are
given as:
\begin{equation}
 \label{eq:nmu}
  \begin{aligned}
  n_\mu(>F)&=\int\!dz\,\frac{d^2V}{\mu(z)dzd\Omega}\,
  \int_{L(z)/\mu(z)}^\infty\!dL\,\frac{d^2N(L,z)}{dLdV}\\
   &= \int\!dz\,\frac{dn_0[z|>L(z)/\mu(z)]}{\mu(z)dz}\equiv\int\!dz\,\frac{dn_\mu[z|>L(z)]}{dz},
  \end{aligned}
\end{equation}
where $dn_\mu[z|>L(z)]/dz$ is the magnified redshift distribution
function of the source population.
Hence, the net change of the magnified source counts
$n_\mu(>F)/n_0(>F)$, known as magnification bias, depends on the
intrinsic (unlensed) source luminosity function, $d^2N(L,z)/dL/dV$. 
One can calculate the expectation value for the magnified source counts
$n_\mu(>F)$ for a given background cosmology and a given source
luminosity function.

In real observations, we apply different cuts
(e.g., size, magnitude, and color cuts)
in source selection for measuring the shear and magnification effects,
thus leading to different source-redshift distributions. In contrast to
the former effect, measuring the effect of magnification bias does not
require source galaxies to be spatially resolved, but it does require a 
stringent flux limit against incompleteness effects
\citep{Hildebrandt2016,Chiu2020hsc}.

Equation~(\ref{eq:nmu}) indicates that, when redshift information of
individual source galaxies is available from spectroscopic redshifts, we
can directly measure the magnified redshift distribution of background
source galaxies for a flux-limited sample \citep{1995ApJ...438...49B}:
\begin{equation}
 \label{eq:nzmag}
 \frac{dn_\mu[z|>L(z)]}{dz}
  =\frac{dn_0[z|>L(z)/\mu(z)]}{\mu(z)dz}.
\end{equation}
Hence, in principle, the lensing-induced distortion of the redshift
distribution $dn_\mu[z|>L(z)]/dz$ can be measured from
spectroscopic-redshift measurements with respect to the unlensed
distribution $dn_0[z|>L(z)]/dz$, which can be found in random  
fields. In particular, the integrated magnification-bias effect will
translate into an enhancement in mean source redshift of the background
sample (i.e., the first moment of Eq.~(\ref{eq:nzmag})).
 Using 300,000 BOSS survey galaxies with accurate spectroscopic 
 redshifts, \citet{Coupon+2013} measured their mean redshift depth
 behind four large samples of optically selected clusters from the Sloan
 Digital Sky Survey (SDSS), totaling 5000--15,000 clusters. 
 They found a $\simgt 1$ percent level of mean redshift increase
 $\delta z(R)$ toward the cluster center for SDSS-defined optical
 clusters with an effective mass of $M_\mathrm{200c}\sim (1-2)\times
 10^{14}M_\odot$.

\subsection{Magnification observables}
\label{subsec:magobs}

To simplify the calculations, we discretize Eq.~(\ref{eq:nmu}) as:
\begin{equation}
 n_\mu(>F) \simeq \sum_s  \frac{dn_\mu[z_s|>L(z_s)]}{dz}\Delta z\equiv \sum_s
 n_\mu(z_s|>F),
\end{equation}
where $n_\mu(z_s|>F)$ represents a subsample of the
background population in the redshift interval $[z_s,z_s+\Delta z]$.
If the change in flux due to magnification is small compared to the
range over which the slope of the 
luminosity function varies, the intrinsic source counts
$n_0[z_s|>L(z_s)]$ can be approximated at $L(z_s)$ by a power law with
a logarithmic slope of:\footnote{Note that, instead of $\alpha$,
$s\equiv d\log_{10}{n_0}(<m)/dm = 0.4\alpha$ in terms of the limiting
magnitude $m$ is also used in the literature
\citep[e.g.,][]{BTU+05,Umetsu+2011,Umetsu2014clash}.}
\begin{equation}
 \alpha(z_s) = -\frac{d\log n_0(z_s|>L)}{d\log L}\Big{|}_{L(z_s)}.
\end{equation}

%Note that the value of $\alpha$ depends on the luminosity function
%$dn/dL(L,z)$ of background sources (and hence the object type, such as
%late/early type galaxies and  quasars) and the observing wavelength.
The magnified source counts $n_\mu(z_s|>F)$ in the redshift
interval $[z_s,z_s+\Delta z]$ are given by:
\begin{equation}
 \begin{aligned}
  n_\mu(z_s|>F) &=
  \frac{1}{\mu}\frac{dn_0[z_s|>L(z_s)/\mu]}{dz}\Delta z
   &=  n_0(z_s|>F)\,\mu^{\alpha(z_s)-1}.
 \end{aligned}
\end{equation}
The corresponding magnification bias is given by:
\begin{equation}
 \label{eq:bmuz}
 b_\mu(z_s) := \frac{n_\mu(z_s|>F)}{n_0(z_s|>F)} = \mu^{\alpha-1}.
\end{equation}
Equation~(\ref{eq:bmuz}) implies a positive bias for $\alpha > 1$ and a
negative bias for $\alpha < 1$. The net magnification effect on the source
counts vanishes at $\alpha=1$.
For a depleted sample of background sources with $\alpha\ll 1$,
the effect of magnification bias is dominated by the geometric area
distortion ($b_\mu \to \mu^{-1}$ at $\alpha\to 0$)
and insensitive to the intrinsic source luminosity function \citep{Umetsu2013}.
In the weak-lensing limit ($|\gamma|,|\kappa|\ll 1$), we have:
\begin{equation}
b_\mu \simeq 1+2(\alpha-1)\kappa.
\end{equation}
Hence, the flux magnification bias $b_\mu$ in 
the weak-lensing limit provides a local measure of the surface mass
density field, $\kappa(\btheta)$. 
The combination of shear and magnification can thus be used
to break or alleviate mass-sheet degeneracy
\citep{Schneider+2000,BTU+05,UB2008,Umetsu+2011,Umetsu2014clash,Umetsu2018clump3d}.

In practical applications, we need to average over a broad range of source
redshifts to increase the S/N. The magnification bias averaged over the
source-redshift distribution is expressed as:
\begin{equation}
 \begin{aligned} 
 \langle b_\mu\rangle := \frac{n_\mu(>F)}{n_0(>F)} &=
  \frac{\sum_s n_0(z_s|>F) \mu^{\alpha(z_s)-1}}{\sum_s n_0(z_s|>F)}.
  \end{aligned}
\end{equation}
In the continuous limit $\sum_s n_0(z_s|>F) \to
\int\!dz\,dn_0(z|>F)/dz$, we have the following equation
\citep{Umetsu2013,Umetsu2016clash}: 
\begin{equation}
 \label{eq:bmu_avg}
\langle b_\mu\rangle=
 \left[\int\!dz\,\frac{dn_0(z|>F)}{dz}\mu^{\alpha(z)-1}\right]
 \left[\int\!dz\,\frac{dn_0(z|>F)}{dz} \right]^{-1}
 =\langle\mu^{\alpha-1}\rangle.
\end{equation}
Equation~(\ref{eq:bmu_avg}) gives a general expression for the flux
magnification bias.
Deep multi-band photometry spanning a wide wavelength
range allows us to identify distinct populations of background
galaxies
\citep[e.g.,][]{Medezinski+2010,Medezinski+2011,Medezinski2018src,Umetsu+2012,Umetsu2014clash,Umetsu2015A1689}. 
Since a given flux limit ($F$) corresponds to different intrinsic
luminosities $L(z_s)$ at different source redshifts $z_s$
(Eq.~(\ref{eq:n0})), source counts of distinctly different background 
populations probe different regimes of magnification bias. 
The bias is strongly negative for quiescent
galaxies at $\langle z_s\rangle\sim 1$,
with a faint-end slope of $\alpha\sim 0.4$
at the limiting magnitude
$z'\approx 25.6$\,ABmag \citep[e.g.,][]{Umetsu2014clash,Umetsu2015A1689}.
A net count depletion ($b_\mu<1$) results for such a source
population with $\alpha\ll 1$
\citep[e.g.,][]{Broadhurst1995,1998ApJ...501..539T,BTU+05,UB2008,Umetsu+2011,Umetsu+2012,Umetsu2014clash,Umetsu2015A1689,Ford2012,Coe+2012A2261,Medezinski+2013,Radovich2015planck,Ziparo2016locuss,Wong2017}, 
because the effect of magnification bias is dominated by 
the geometric area distortion.
In the regime of density depletion,
a practical advantage is that the effect is not
sensitive to the exact form of the source luminosity function.
The S/N for detection of $b_\mu$ increases progressively as the flux
limit $F$ decreases.

\begin{figure*}[!htb] %!htb
  \begin{center}
   \includegraphics[scale=0.45, angle=0, clip]{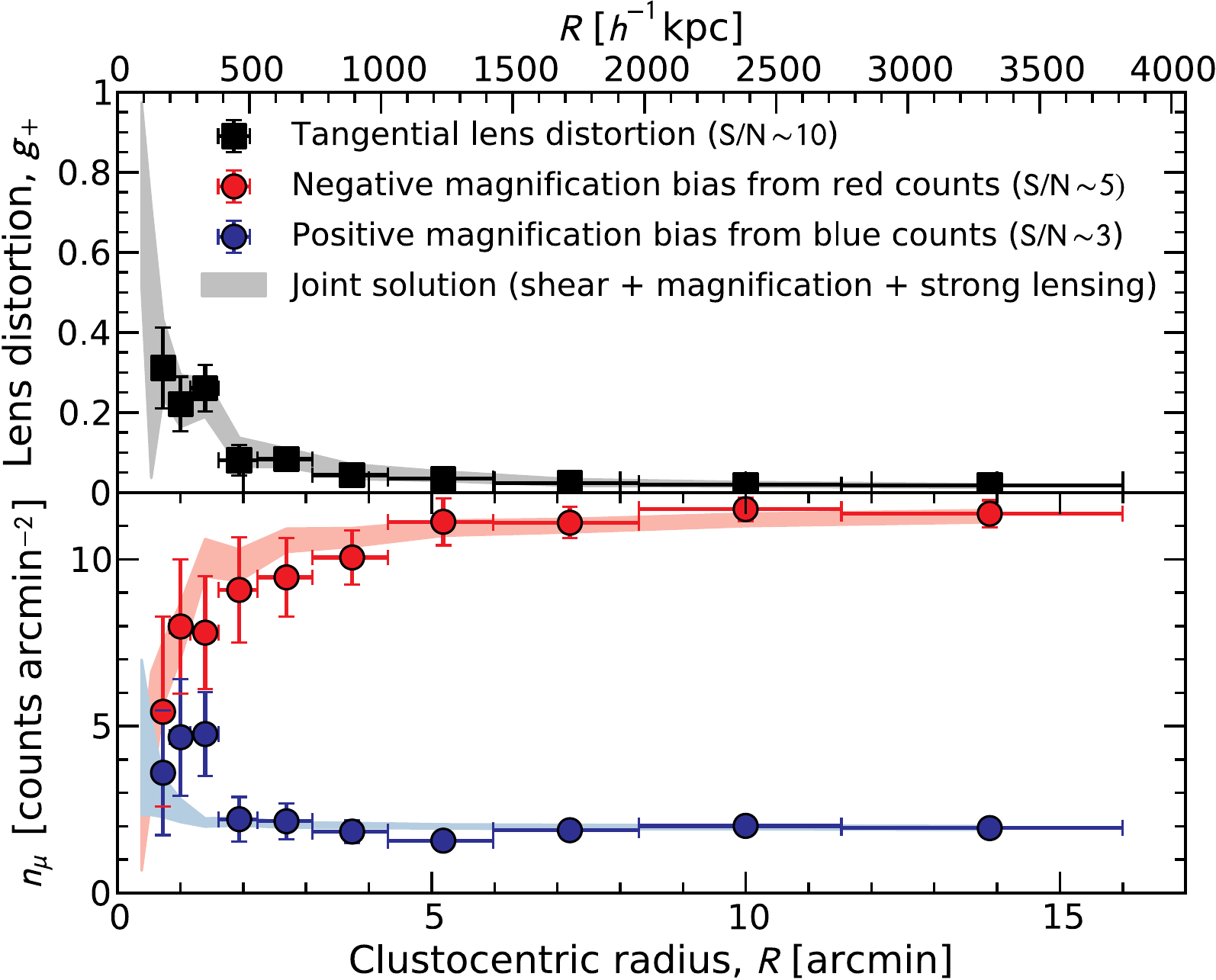} 
  \end{center}
\caption{
 \label{fig:U13}
Weak-lensing radial profiles of MACS~J1206.2$-$0847 ($z_l = 0.439$)
 from Subaru Suprime-Cam observations.
The top panel shows the reduced tangential shear profile $g_+$ (squares).
The bottom panel shows the coverage- and masking-corrected number-count
profiles $n_\mu$ for flux-limited samples of blue and red background
 galaxies (blue and red circles, respectively).
The error bars include contributions from Poisson counting uncertainties
 and those from intrinsic angular clustering of each source
 population. For the red sample, a strong depletion of the source counts
 is seen toward the cluster center 
 due to magnification of the sky area, while a slight enhancement
 of blue counts is present in the innermost radial bins due to the effect of
 positive magnification bias.
Also shown for each observed profile is the joint
Bayesian reconstruction (shaded area) from combined strong-lensing,
 weak shear lensing, and positive/negative magnification-bias
 measurements.
 Image reproduced with permission from \citet{Umetsu2013}, copyright by AAS. 
}
\end{figure*}

Figure~\ref{fig:U13} displays weak-lensing radial profiles for the
cluster MACS~J1206.2$-$0847 at $z_l = 0.439$ derived from Subaru
Suprime-Cam observations \citep{Umetsu2013}. It is a highly massive
X-ray cluster with 
$M_\mathrm{200c}=(11.1\pm 2.5)\times 10^{14}\Msunh$
%$M_\mathrm{200c}=(15.9\pm 3.6)\times 10^{14}\Msun$
\citep{Umetsu2014clash} targeted by the CLASH survey.
%%\citep{Postman+2012CLASH}. 
The black squares in the top panel show
the reduced tangential shear profile $g_+(R)$. The blue and red circles
in the bottom panel are positive and negative magnification-bias
measurements $n_\mu(R)$ showing density enhancement and depletion,
respectively, as a function of cluster-centric radius $R$.
These weak-lensing measurements yield respective S/N values of
$10.2$, $2.9$, and $4.7$.
Figure~\ref{fig:U13} also shows a joint Bayesian reconstruction of each
observed profile obtained from combined strong-lensing, weak shear
lensing, and positive/negative magnification-bias measurements.

\subsection{Nonlinear effects on the source-averaged magnification bias}

It is instructive to consider a maximally \emph{depleted} population of
source galaxies with $\alpha=-d\log{n_0(>F)}/d\log{F}=0$ at the limiting
flux $F$.
For such a population, the effect of magnification bias is purely
geometric, $\langle b_\mu\rangle=\langle\mu^{-1}\rangle$, and insensitive
to details of the intrinsic source luminosity function, $d^2N(L,z)/dL/dz$.
In the nonlinear subcritical regime, the
source-averaged  inverse magnification factor is expressed as \citep{Umetsu2013}:
\begin{equation}
 \begin{aligned}
 \langle\mu^{-1}\rangle &= \left\langle
  (1-\kappa)^2-|\gamma|^2\right\rangle\\
  &= \left(1-\langle\kappa\rangle\right)^2 - |\langle\gamma\rangle|^2
  + (f_l-1)\left(
  \langle\kappa\rangle^2 - |\langle\gamma\rangle|^2\right)\\
  &\equiv (1-\langle\kappa\rangle)^2-|\langle\gamma\rangle|^2+\Delta\langle\mu^{-1}\rangle,
 \end{aligned}
\end{equation}
where $\langle\cdots\rangle$ denotes the averaging over the source-redshift
distribution  (see Eq.~(\ref{eq:bmu_avg})),
$f_l=\langle\Sigma_{\mathrm{cr},l}^{-2}\rangle/\langle\Sigma_{\mathrm{cr},l}\rangle^2$
is a quantity of the order of unity,
and $\Delta\langle\mu^{-1}\rangle$ is the correction with respect to the
single source-plane approximation.
The error associated with the single source-plane approximation is
$\Delta\langle\mu^{-1}\rangle/\langle\mu^{-1}\rangle\simeq(f_l-1)\left(\langle\kappa\rangle^2-|\langle\gamma\rangle|^2\right)$,
which is much smaller than unity for background populations
with $\alpha\sim 0$ in the mildly nonlinear
subcritical regime where $|\langle\kappa\rangle|\sim
|\langle\gamma\rangle|\sim O(10^{-1})$. It is therefore
reasonable to use the single source-plane approximation for
calculating the magnification bias of depleted source populations with 
$\alpha\ll 0$.

In the regime of density enhancement ($\alpha>1$), on the
other hand, interpreting the observed lensing signal
requires detailed knowledge of the intrinsic source 
luminosity function \citep[see, e.g.,][]{Chiu2016magbias,Chiu2020hsc},
especially in the nonlinear regime where the flux amplification factor
is correspondingly large (say, $\mu\simgt 1.5$).
For example, a blue distant population of background galaxies
is observed to have a well-defined redshift distribution that is fairly
symmetric and peaked at a mean redshift of $\langle z_s\rangle\sim 2$ 
\citep[e.g.,][]{Lilly+2007,Medezinski+2010}. Therefore,
the majority of these faint blue galaxies are in the far background of
typical cluster lenses, so that the lensing signal has a weaker
dependence on the source redshift $z_s$. In such a case, the single
source-plane approximation may well be justified \citep{Umetsu2013}.

\subsection{Observational systematics and null tests} 
\label{subsec:nulltest}

In real observations, contamination of the background sample by unlensed
galaxies is a critical source of systematics in cluster weak lensing, as
discussed in Sect.~\ref{subsec:dilution}.
In particular,
contamination by cluster galaxies has a direct impact on the
interpretation of background source counts because it will cause an
apparent density enhancement at small cluster-centric radii.
To avoid significant contamination and alleviate this problem as much as
possible, one often relies on a
stringent color--color selection (Sect.~\ref{subsec:dilution}) to
measure the lensing magnification signal from a distinct population of
background galaxies
\citep[e.g.,][]{BTU+05,UB2008,Umetsu+2011,Umetsu2014clash,Ziparo2016locuss,Chiu2016magbias,Chiu2020hsc}.
If well-calibrated photo-$z$ PDFs are available from multi-band
observations, the impact of cluster contamination can be characterized
and assessed by statistically decomposing the photo-$z$ PDF $P(z)$ into
the cluster and random-field populations \citep[e.g.,][]{Gruen2014,Chiu2020hsc}. 

Moreover, for unbiased magnification-bias measurements,
one has to correct for the incomplete area coverage due to masked
regions and incomplete measurement annuli.
%(e.g., at large cluster-centric distances).
Specifically, masking (or blocking) of background galaxies
by foreground objects, cluster members, and saturated or bad pixels
needs to be properly accounted for
\citep[see][]{Umetsu+2011,Chiu2020hsc}. Another concern is the impact of
blending effects in the crowded regions of cluster environments and the
presence of intracluster light \citep{Gruen2019des}, which could bias
the photometry and thus photo-$z$'s. The effects of masking and
blending on the source counts can be examined and quantified by
injecting synthetic galaxies into real images from observations
\citep{Huang2018hsc,Chiu2020hsc}. 
%Another concern regarding cluster fields is the presence of
%intracluster light (ICL), which could bias the photometry, photo-
%z, and/or mass modelling. Gruen et al. (2019) 
%
% correct for the incomplete area coverage (accounting for masked
% regions and incomplete outer annuli 
%for the incomplete area coverage, accounting for masked regions due to
%bright saturated stars, w 
%

\begin{figure*}[!htb] %!htb
  \begin{center}
   \includegraphics[scale=0.31, angle=0, clip]{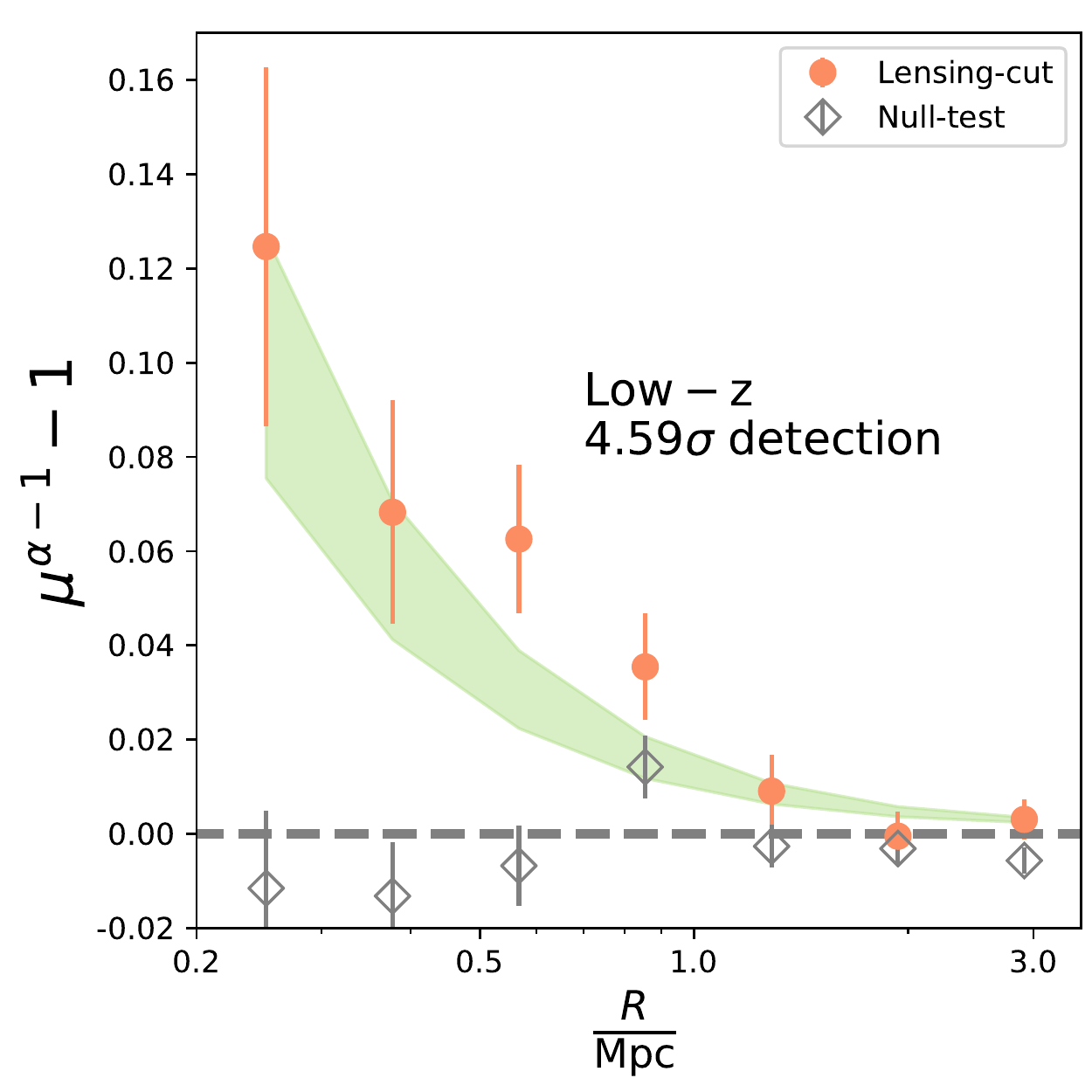}
   \includegraphics[scale=0.31, angle=0, clip]{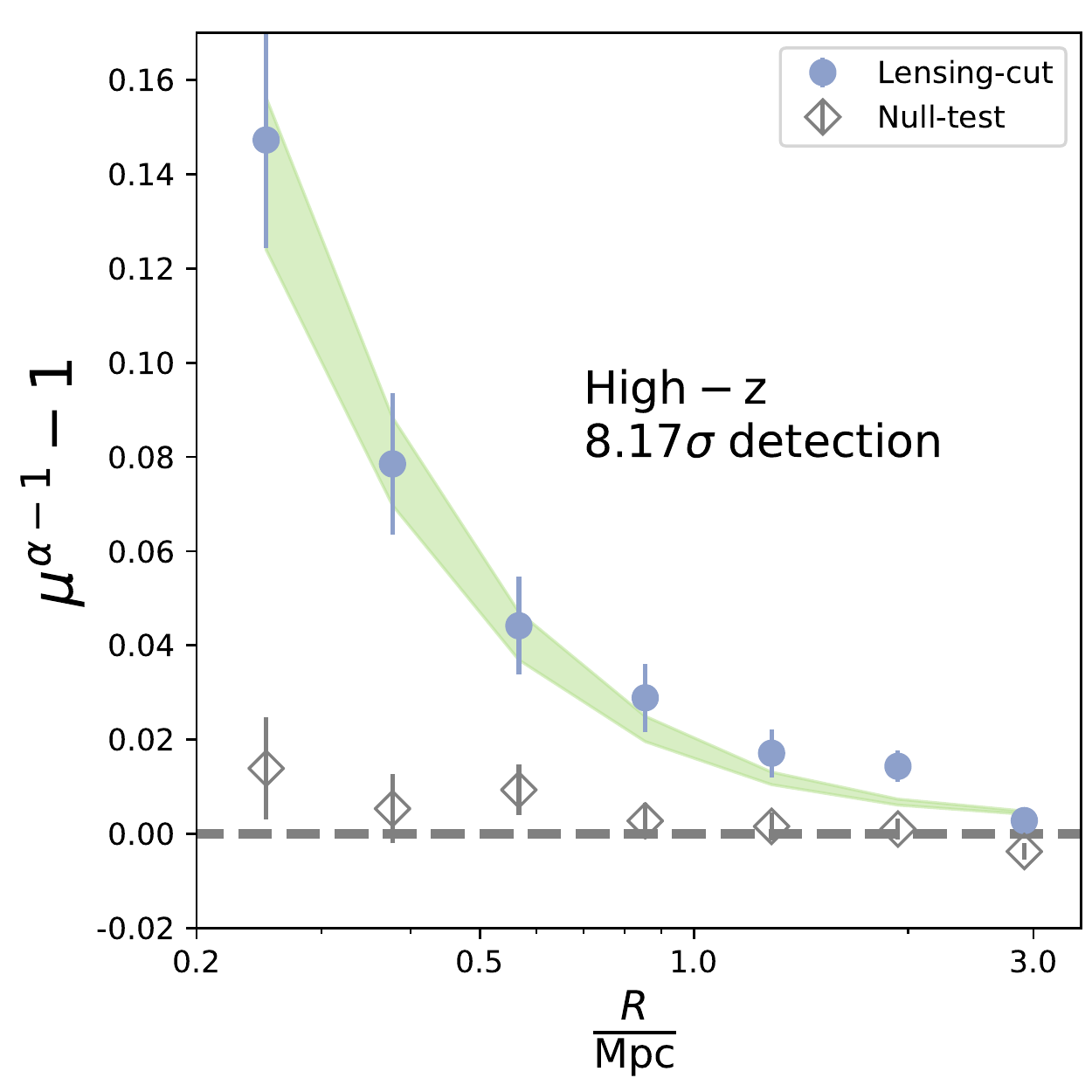}
   \includegraphics[scale=0.31, angle=0, clip]{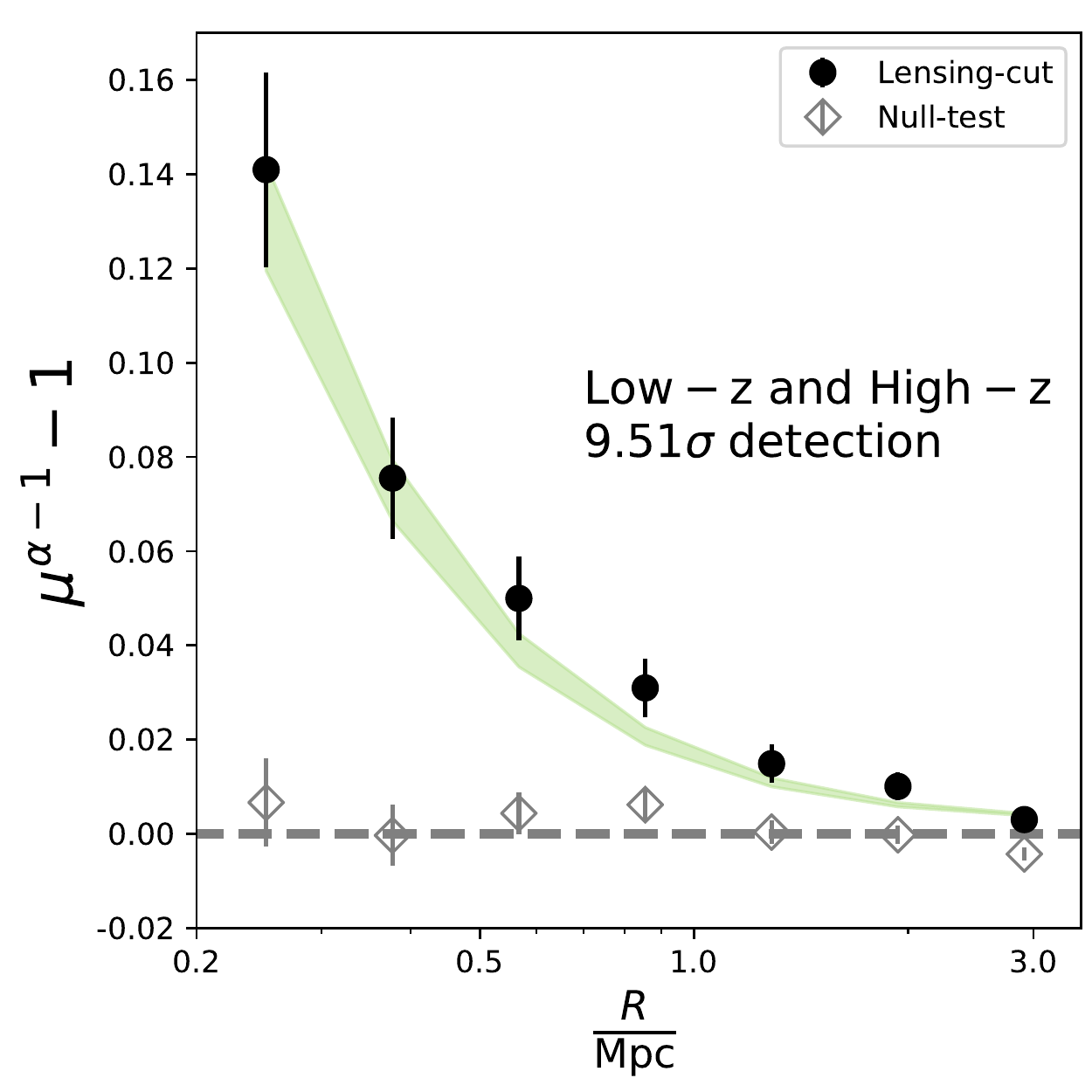}
  \end{center}
\caption{
 \label{fig:Chiu20}
Stacked magnification-bias profiles around a sample of 3029 CAMIRA
 clusters obtained using low-$z$ (left) and high-$z$
 (middle) samples, as well as the joint (right) background sample, from
 the Subaru HSC-SSP survey. The results are shown as a function of the
 projected physical cluster-centric radius $R$.
Filled circles with error bars represent the stacked magnification-bias
 measurements with the respective lensing-cut samples. The total
 detection significance for each measurement is labeled in each panel.
Open diamonds with error bars show the results for ``null-test'' samples,
 for which the net effect of magnification bias is expected to be zero.
The green shaded region in each panel represents the $1\sigma$
 confidence interval around the best-fit profile.
 Image reproduced with permission from \citet{Chiu2020hsc}, copyright by the authors.
}
\end{figure*}

Since the net effect of magnification bias is expected to vanish for a
flux-limited background sample defined at $\alpha=1$
(Sect.~\ref{subsec:magobs}), weak-lensing magnification provides a powerful
null test, similar to the cross-shear ($B$-mode) signal in the case of
weak shear lensing (Sect.~\ref{subsec:gest}).
By performing a null test,
one can empirically assess the level of residual bias
that could be present in the measurement for a ``lensing-cut'' sample
defined at $\alpha > 1$ or $\alpha < 1$.
The only assumption made in this approach is that
residual systematics are the same between the lensing-cut and null-test
samples defined at different flux (magnitude) limits.
This null test allows us to quantify the impact of deblending effects,
biased photometry in crowded regions, and any incorrect assumptions about
$P(z)$-decomposition \citep{Chiu2020hsc}.
This is demonstrated in Fig.~\ref{fig:Chiu20}.
The figure shows the stacked magnification-bias profiles around a sample
of 3029 CAMIRA clusters with richness $N>15$ in the redshift range
$0.2\leqs z < 1.1$,
obtained using flux-limited low-$z$ and high-$z$ background samples,
as well as the joint sample, selected in the $g-i$ versus $r-z$ diagram
from HSC-SSP survey data \citep{Chiu2020hsc}.  
The magnification-bias signal for the full CAMIRA sample is detected at
a significance level of $9.5\sigma$.  On the other hand, the residual
bias estimated from the null-test samples was found to be statistically
consistent with zero \citep{Chiu2020hsc}.

\section{Recent advances in cluster weak-lensing observations}
\label{sec:obsreview}

Galaxy clusters provide valuable information from the physics driving
cosmic structure formation to the nature of dark matter and dark
energy. Their content reflects that of the universe:
$\sim 85\%$ dark matter and $\sim 15\percent$ baryons 
(cf. $\Ob/\Om=(15.7\pm 0.4)\percent$; \citealt{Planck2015XIII}),
with $\sim 90\percent$ of the baryons residing in
%the gaseous intracluster medium.
a hot, X-ray-emitting phase of the intracluster medium.
%the gaseous intracluster medium.
Massive clusters dominated by dark matter are not expected to be
significantly affected by baryonic gas cooling
\citep{Blumenthal+1986,Duffy+2010},
unlike individual galaxies, because the high temperature and low density
prevent efficient cooling and gas contraction.
Consequently, for clusters in a state of quasi-equilibrium, the
form of their total mass profiles reflects closely the underlying
dark matter distribution. Hence, galaxy clusters
%Since clusters are highly massive and dominated by dark matter, they
offer fundamental tests on the assumed properties 
of dark matter, as well as on models of nonlinear structure formation.

The \LCDM paradigm assumes that dark matter is effectively cold
(nonrelativistic) and collisionless on astrophysical and cosmological
scales \citep{Bertone+Gianfranco2018}. 
In this context, the standard CDM model and its variants, 
such as self-interacting dark matter
\citep[SIDM;][]{Spergel+Steinhardt2000}
and wave dark matter
\citep[$\psi$DM;][]{Peebles2000dm,Hu2000dm,Schive2014psiDM},
can provide a range of testable predictions for the properties of
cluster-size halos 
(Sects.~\ref{subsec:massprofile} and \ref{subsec:cM}).
%\ref{subsec:selfsim}). 
A prime example is the ``Bullet Cluster'', a merging pair of clusters  
exhibiting a significant offset between the centers of the gravitational 
lensing mass and the X-ray peaks of the collisional cluster gas
\citep{2004ApJ...604..596C,Clowe2006Bullet}.
The data support that dark matter in clusters is effectively
collisionless like galaxies, placing a robust upper limit on the
self-interacting cross section for dark matter of
$\sigma_\mathrm{DM}/m<1.25$\,cm$^2$\,g$^{-1}$ 
\citep{Randall2008}.

The abundance of clusters as a function of redshift
provides a sensitive probe of the amplitude and growth rate of
primordial density perturbations, as well as of the cosmological volume
element $d^2V(z)/dz/d\Omega$. 
This cosmological sensitivity arises mainly because clusters populate
the high-mass exponential tail of the halo 
mass function \citep[e.g.,][]{2001ApJ...553..545H}.
In principle galaxy clusters can complement other cosmological probes,
such as CMB, galaxy clustering, cosmic shear, and distant supernova
observations. 
To place cosmological constraints using clusters, however, it is
essential to study large cluster samples with well-characterized
selection functions, spanning a wide range in mass and redshift
\citep{Allen+2004,Vikhlinin+2009CCC3,Mantz2010}.
Currently, the ability of galaxy clusters to provide robust cosmological
constraints is limited by systematic uncertainties in their mass
calibration \citep{Pratt2019}. 
Since the level of mass bias is sensitive to
calibration systematics of the instruments \citep{Donahue2014clash,Israel+2015}
and is likely mass dependent
\citep{CoMaLit5,Umetsu2020xxl}, a concerted effort is needed to enable
an accurate mass calibration with weak gravitational lensing (see
Sect.~\ref{subsec:mcal}).

Substantial progress has been made in constructing statistical samples 
of galaxy clusters thanks to dedicated wide-field surveys in various wavelengths
\citep{2014A&A...571A..29P,Planck2015XXVII,SPT2015sze,Miyazaki2018wl,Oguri2018camira}.
Systematic lensing studies of galaxy clusters often target X-ray-
or SZE-selected samples 
\citep[e.g.,][]{WtG1,Postman+2012CLASH,Gruen2014,Hoekstra2015CCCP,Sereno2017psz2lens}. 
This is because the hot intracluster gas provides an excellent tracer of
the cluster's gravitational potential
\citep[e.g.,][]{Donahue2014clash}, except for the cases of massive
cluster collisions caught in an ongoing phase of dissociative mergers
\citep{Clowe2006Bullet,Okabe+Umetsu2008}.  Moreover, X-ray and SZE
observations provide useful centering information of individual
clusters.  The effect of off-centered clusters is to dilute and flatten
the observed $\Sigma(R)$ profile at scales smaller than the offset scale 
$\sigma_\mathrm{off}$ \citep{Johnston+2007b,George+2012,Du+Fan2014}.
Since flattened, sheet-like mass distributions produce little shear, the
impact of miscentering on $\Delta\Sigma(R)$ is much larger.
The off-centered $\Delta\Sigma(R)$ profile is strongly suppressed by
smoothing at scales $R\simlt 2.5\sigma_\mathrm{off}$ 
\citep{Johnston+2007b}. 
A comparison of X-ray, SZE, and optical (e.g., BCGs) center positions 
allows us to empirically assess the level of halo miscentering.
It should be noted, however, that a merger can boost the X-ray and SZE
signals and make their peaks off-centered during the
compression phase \citep{Molnar2012}. Although the timescale on which
this happens is expected to be short \citep[$\sim 1$\,Gyr;][]{2001ApJ...561..621R}, 
it could induce a selection effect and contribute to the scatter in
their scaling relations \citep{Umetsu2020xxl}.

In this section, we review recent advances in our understanding of the
distribution and amount of mass in galaxy clusters based on cluster
weak-lensing observations.

\subsection{Cluster mass distribution}
\label{subsec:massprofile}

The distribution and concentration of mass in dark-matter-dominated
halos depend fundamentally on the properties of dark matter.
For the case of collisionless CDM models, cosmological $N$-body
simulations with sufficiently high resolution can provide accurate
predictions for the end product of collisionless collapse in an
expanding universe.  
Although the formation of halos is a complex, nonlinear dynamical
process and halos are evolving through accretion and mergers,
\LCDM models predict that 
the structure of quasi-equilibrium halos characterized in terms of the
spherically averaged density profile $\rho(r)$ is approximately
self-similar with a  characteristic density cusp in their centers,
$\rho(r)\propto 1/r$  
\citep{1996ApJ...462..563N,1997ApJ...490..493N}.
The density profile $\rho(r)$ of dark-matter-dominated halos steepens
continuously with radius and it is well described by the NFW
form out to the virial radius, albeit with large variance associated
with the assembly histories of individual halos \citep{Jing+Suto2000}.

%The three-dimensional shape of dark matter halos is predicted to be not
%spherical but triaxial, reflecting the collisionless nature of dark
%matter \citep{2002ApJ...574..538J}.
%%

Subsequent numerical studies with improved statistics and higher
resolution found that the spherically averaged density profiles of \LCDM
halos are better approximated by the three-parameter Einasto function
with an additional degree of freedom
\citep[][]{Merritt+2006,Gao+2008}, which is closely linked with the mass
accretion history of halos \citep[][]{Ludlow+2013}. 
The Einasto profile has a power-law logarithmic slope of
$\gamma_\mathrm{3D}(r) = -2(r/r_{-2})^{\alpha_\mathrm{E}}$
(Sect.~\ref{subsubsec:DK14}).
For a given halo concentration,
an Einasto profile with $\alpha_\mathrm{E}\approx 0.18$ closely
resembles the NFW profile over roughly two decades in radius
\citep{Ludlow+2013}.
The shape parameter $\alpha_\mathrm{E}$ of \LCDM halos
increases gradually with halo mass and redshift 
\citep[see][$0.15\simlt \alpha_\mathrm{E}\simlt 0.25$ at $z=0$]{Gao+2008,Child2018cm},
so that the density profiles of \LCDM halos are
not strictly self-similar \citep{Navarro+2010}.
By analyzing a large suite of $N$-body simulations in \LCDM, 
\citet{Child2018cm} found that
both Einasto and NFW profiles provide a good description of the stacked
mass distributions of cluster-size halos at low redshift,
implying that the two fitting functions are nearly
indistinguishable for stacked ensembles of low-redshift clusters, in
contrast to clusters at higher redshift ($z\simgt 1$).  
    
The three-dimensional shape of collisionless halos is predicted to be 
generally triaxial with a preference for prolate shapes
\citep{Warren+1992,2002ApJ...574..538J},
reflecting the collisionless nature of dark matter
\citep{Ostriker+Steinhardt2003}.
Older halos tend to be more relaxed and thus to be rounder.
Since more massive halos form later on average,
cluster-size halos are expected to be more elongated than less
massive systems \citep{Shaw+2006,Ho2006,Despali2014,Despali+2017,Bonamigo2015}.
Accretion of matter from the surrounding large-scale environment also
plays a key role in determining the shape and orientation of halos. The 
halo orientation tends to be in the preferential infall direction of
the subhalos and hence aligned along the surrounding filaments
\citep{Shaw+2006}.
The shape and orientation of galaxy clusters thus provide an independent
test of models of structure formation (see Sect.~\ref{subsec:quad}).

\begin{figure*}[!htb] %!htb
  \begin{center}
   \includegraphics[scale=0.33, angle=0, clip]{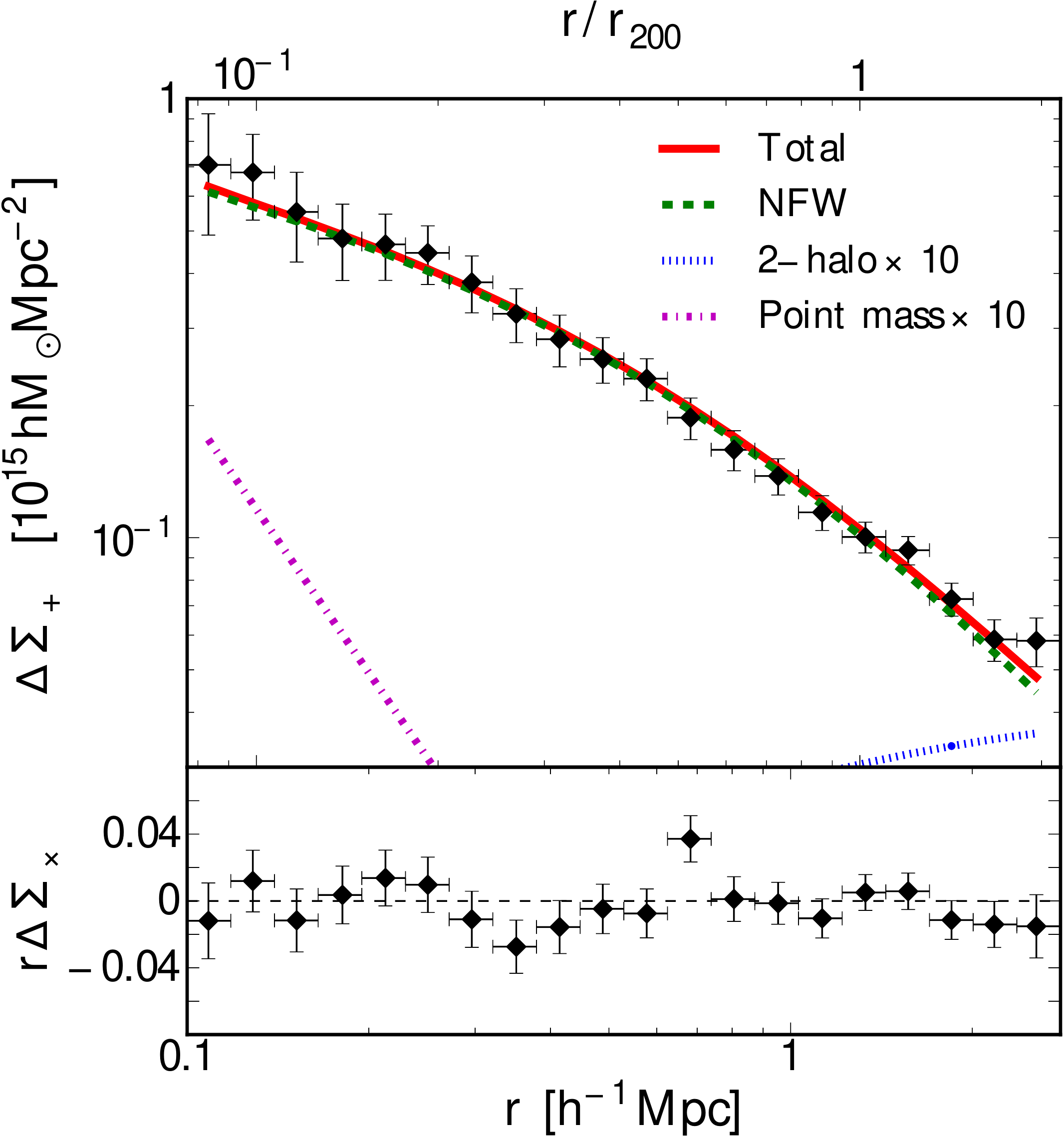}
  \end{center}
\caption{
\label{fig:okabe16gt}
Stacked tangential ($\Delta\Sigma_+$: top) and cross
 ($\Delta\Sigma_\times$: bottom) shear profiles  
  obtained for a sample of 50 X-ray luminous LoCuSS clusters.
Solid red and dashed green lines are the total mass model and the NFW
 model, respectively. The dotted blue and 
 dash-dotted magenta lines represent the 2-halo term and the central
 point source multiplied by a factor of 10, respectively.
 Image reproduced with permission from \citet{Okabe+Smith2016}, copyright by the authors. 
}
\end{figure*}

Prior to dedicated wide-field optical imaging surveys such as Subaru
HSC-SSP and DES, several cluster lensing surveys carried out deep
targeted observations toward a few tens to several tens of highly
massive galaxy clusters with
$M_\mathrm{200c}\sim 10^{15}\Msun$
\citep[e.g.,][]{Postman+2012CLASH,Okabe+2013,WtG1,Hoekstra2015CCCP}.
Since such clusters are extremely rare across the sky, targeted
weak-lensing observations with deep multi-band imaging  currently
represent the most efficient approach to study in detail the high-mass
population of galaxy clusters each with sufficiently high S/N
\citep[see][]{Contigiani+2019}. 

In the last decade, cluster--galaxy weak-lensing observations have
established that the total matter distribution within clusters in 
projection can be well described by cuspy, outward-steepening density
profiles
\citep{Umetsu+2011,Umetsu2014clash,Umetsu2016clash,Silva+2013,Newman+2013a,Okabe+2013,Sereno2017psz2lens},
%%\citep{Umetsu+2011,Umetsu2014clash,Umetsu2016clash,Newman+2013a,Okabe+2013,Sereno2017psz2lens}, 
such as the NFW and Einasto profiles
with a near-universal shape \citep{Niikura2015,Umetsu+Diemer2017}, as
predicted for collisionless halos in quasi-gravitational equilibrium
\citep[e.g.,][]{1996ApJ...462..563N,1997ApJ...490..493N,Taylor+Navarro2001,Merritt+2006,Gao+2008,Hjorth+2010DARKexp,DARKexp2}.   
Moreover, the shape and orientation of galaxy clusters as constrained by 
weak-lensing and multiwavelength data sets are found to be in agreement
with \LCDM predictions
\citep[e.g.,][]{Oguri2005,Evans+Bridle2009,Oguri2010LoCuSS,Morandi2012,Sereno2013glszx,Sereno2018clump3d,Umetsu2015A1689,Umetsu2018clump3d,Chiu2018clump3d,Shin+2018}, 
although detailed studies of individual clusters are currently limited
to a relatively small number of high-mass clusters with deep
multiwavelength observations  
\citep[see][]{Sereno2018clump3d,Umetsu2018clump3d}. 
These results are all in support of the standard explanation for dark  
matter as effectively collisionless and nonrelativistic on
sub-megaparsec scales and beyond,
with an excellent match with standard \LCDM predictions
\citep[however, see][for an excess of galaxy--galaxy strong-lensing
events in clusters with respect to \LCDM]{Meneghetti+2020}.

In Fig.~\ref{fig:okabe16gt}, we show the ensemble-averaged
$\llangle\Delta\Sigma_+\rrangle$ profile in the radial range
$R\in[0.1,2.8]\,\Mpch$ obtained for a
stacked sample of 50 X-ray clusters
\citep{Okabe+Smith2016} targeted by the LoCuSS Survey
\citep[Local Cluster Substructure Survey;][]{Smith2016locuss}.
Their weak shear lensing analysis is based on two-band imaging
observations with Subaru/Suprime-Cam. 
Their cluster sample is drawn from the ROSAT All-Sky Survey
\citep[RASS;][]{Veges1999rass} 
at $0.15<z<0.3$ and is approximately X-ray luminosity limited.
%\citep{Okabe+Smith2016}.
%The LoCuSS sample is selected based on the
%X-ray luminosity, ignoring other physical and
%relaxation properties \citep{Smith2016locuss}.
The stacked shear profile of the LoCuSS sample is in excellent agreement
with the NFW profile 
with $M_\mathrm{200c}=6.37^{+0.28}_{-0.27}\times 10^{14}\Msunh$
and $c_\mathrm{200c}=3.69^{+0.26}_{-0.24}$ at $z_l=0.23$.
%The figure also displays the best-fit halo model including the effects
%of surrounding large-scale structure as a 2-halo term.
The 2-halo term contribution to $\Delta\Sigma$ for the LoCuSS sample is
negligibly small in the radial range $\simlt 2r_\mathrm{200c}$.
From a single Einasto profile fit to the stacked
$\llangle\Delta\Sigma\rrangle$ profile,
\citet{Okabe+Smith2016} obtained the best-fit Einasto shape parameter of
$\alpha_\mathrm{E}=0.161^{+0.042}_{-0.041}$, which is consistent within
the errors with the \LCDM predictions for cluster-size
halos at $z_l=0.23$ 
\citep{Gao+2008,Dutton+Maccio2014}.

\begin{figure*}[!htb] %!htb
  \begin{center}
   \includegraphics[scale=0.32, angle=0, clip]{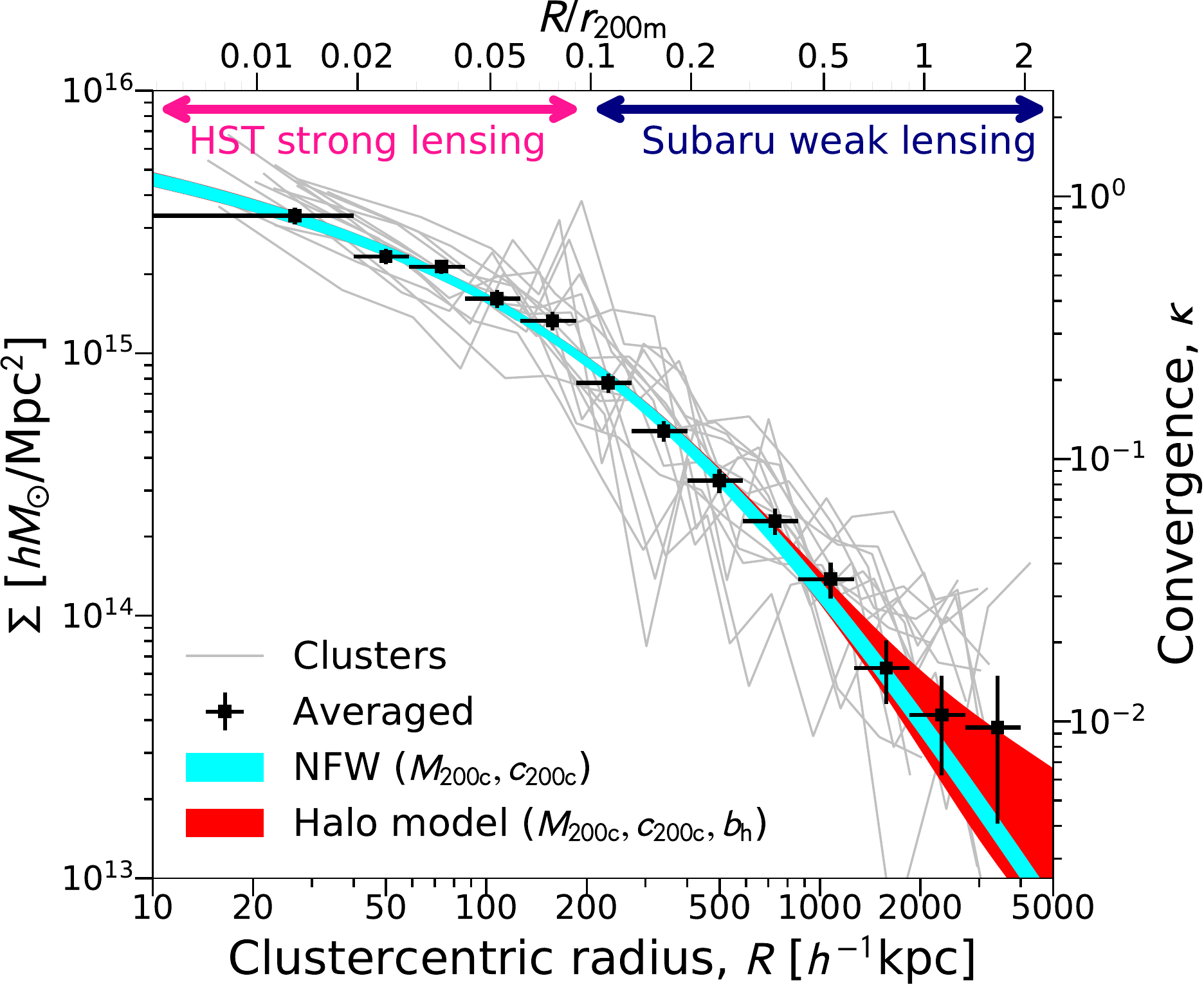} 
   \includegraphics[scale=0.30, angle=0, clip]{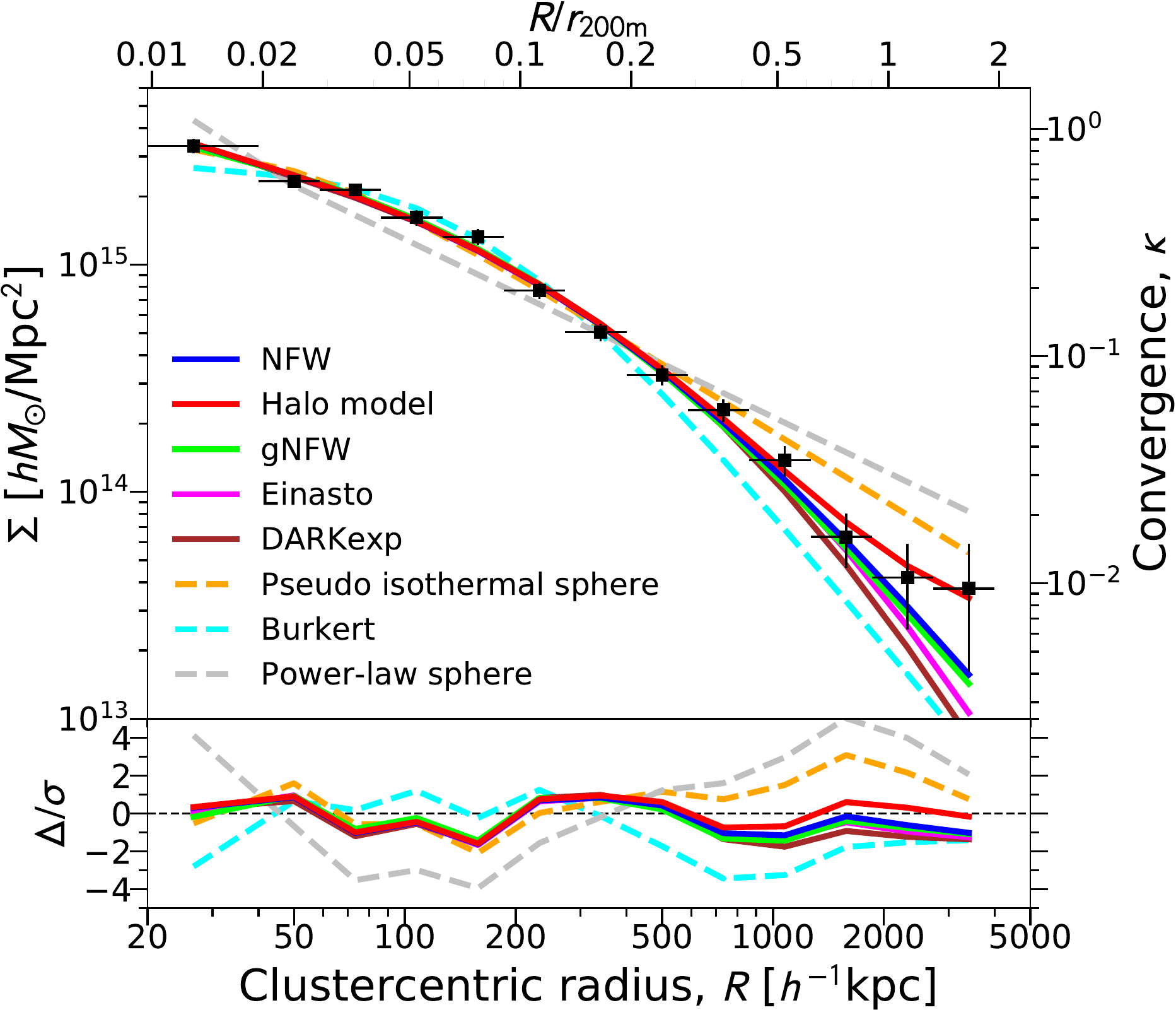} 
  \end{center}
\caption{
 \label{fig:kappa_u16}
\emph{Left panel}: ensemble-averaged projected mass density profile
 $\llangle\Sigma(R)\rrangle$ (black squares) for a sample of 16 CLASH
 X-ray-selected clusters (gray lines),
 obtained from a joint analysis of strong-lensing, weak-lensing shear
 and magnification data.
 \emph{Right panel}:
 models with $\chi^2$ probabilities to exceed (PTE) of $>0.05$ are 
 shown with solid lines, 
 while those with $\mathrm{PTE}\leqs 0.05$ are shown with dashed lines.
 The averaged mass profile is well 
 described by a family of density profiles predicted for
 dark-matter-dominated halos in gravitational equilibrium, such 
 as the NFW, Einasto, and DARKexp models.
 Cuspy halo models (red) that include the
 2-halo contribution from surrounding large-scale 
 structure provide improved agreement with the data.
 %The single power-law, cored isothermal, and Burkert density profiles
 %are statistically disfavored by the observed mass profile with a
 %pronounced radial profile curvature.
This is a slightly modified version of the figure presented in
 \citet{Umetsu2016clash}. 
}
\end{figure*} 

Figure~\ref{fig:kappa_u16} shows the ensemble-averaged
$\llangle\Sigma\rrangle$ profile of 16 CLASH X-ray-selected
clusters \citep{Umetsu2016clash}
based on a joint strong and weak lensing analysis of 16-band {\textit{Hubble Space
Telescope} (\HST) observations \citep{Zitrin2015clash} and
wide-field multicolor imaging taken primarily with Subaru/Suprime-Cam
\citep{Umetsu2014clash}. 
The CLASH survey is an \HST Multi-Cycle Treasury program designed
to study with 525 assigned orbits 
the mass distributions of 25 high-mass clusters.
In this sample, 20 clusters were selected to have regular X-ray
morphologies and X-ray temperatures above 5\,keV. Numerical
simulations suggest that this X-ray-selected subsample is mostly
composed of relaxed clusters ($\sim 70\percent$) but the rest 
($\sim 30\percent$) are unrelaxed systems
\citep{Meneghetti2014clash}. Another subset of five clusters were
selected by their high-magnification lensing properties.
%These clusters often turn out to be dynamically 
%disturbed merging systems. 
\citet{Umetsu2016clash} studied a subset of 20 CLASH clusters
(16 X-ray-selected and 4 high-magnification systems)
taken from \citet{Umetsu2014clash}, who presented a joint shear and
magnification weak-lensing analysis of these individual clusters.
The stacked $\llangle\Sigma\rrangle$ profile over two decades in radius,   
$R\in[0.02,2]r_\mathrm{200m}$,    
is well described by a family of density profiles
predicted for cuspy dark-matter-dominated 
halos in gravitational equilibrium, namely, the NFW,
Einasto, and DARKexp models
\citep{Umetsu2016clash}.\footnote{The NFW and Einasto profiles represent
phenomenological models for cuspy 
dark halos motivated by numerical simulations. The DARKexp model
describes the distribution of particle energies in collisionless
self-gravitating systems with isotropic velocity distributions
\citep{Hjorth+2010DARKexp,DARKexp2},
providing theoretical predictions for the structure of collisionless  
halos. }
In contrast, the single power-law, 
cored-isothermal, and Burkert density profiles are statistically
disfavored by the data.
Cuspy halo models that include the 2-halo term provide improved 
agreement with the data.

\citet{Umetsu2016clash} found the best-fit NFW parameters for the
stacked CLASH $\llangle\Sigma\rrangle$
profile of $M_\mathrm{200c}=10.1^{+0.8}_{-0.7}\times 10^{14}\Msunh$
and $c_\mathrm{200c}=3.76^{+0.29}_{-0.27}$ \citep{Umetsu2016clash}
at a lensing-weighted mean redshift of $z_l\approx 0.34$.
Similarly, the best-fit Einasto shape parameter for the stacked
$\llangle\Sigma\rrangle$ profile is
$\alpha_\mathrm{E}=0.232^{+0.042}_{-0.038}$,
which is in excellent agreement with
predictions from \LCDM numerical simulations,
$\alpha_\mathrm{E}=0.21 \pm 0.07$
\citep[][$\alpha_\mathrm{E}=0.24\pm 0.09$ when fitted to surface mass density profiles of
projected halos]{Meneghetti2014clash}.

Note that the innermost bin in Fig.~\ref{fig:kappa_u16} represents the
mean density interior to $R_\mathrm{min}=40\kpch$ corresponding to the
typical resolution limit of their \HST strong-lensing analysis,
$\delta\vartheta\approx 10$\,arcsec. This scale $R_\mathrm{min}$ is
about twice the typical half-light radius of the CLASH BCGs
\citep[see][]{Tian+2020}, within which the stellar baryons 
dominate the total mass of the clusters
\citep[e.g.,][]{Caminha2019}. 
Determinations of the central slope of the dark matter density
profile $\rho_\mathrm{DM}(r)$ in clusters require additional constraints
on the total mass in the innermost region, such as
from stellar kinematics of the BCG \citep{Newman+2013b}.
To constrain $\rho_\mathrm{DM}(r)$,  
one needs to carefully model the different contributions to the cluster
total mass profile, coming from the stellar mass of member  
galaxies, the hot gas component, the BCG stellar mass, and dark matter
\citep{Sartoris2020vlt}.
Moreover, it is important to take into account the velocity anisotropy
on the interpretation of the line-of-sight stellar velocity dispersion
profile of the BCG \citep{Schaller2015cl,Sartoris2020vlt,He+2020}. 
For these reasons, current measurements and interpretations of the
asymptotic central slope of $\rho_\mathrm{DM}(r)$ in galaxy clusters
appear to be controversial
\citep[e.g.,][]{Newman+2013b,Sartoris2020vlt,He+2020}. 

According to cosmological $N$-body simulations,
the spherically averaged
density profiles in the halo outskirts are most self-similar when
expressed in units of 
overdensity radii $r_{\Delta_\mathrm{m}}$ defined with respect
to the mean density of the universe, $\overline{\rho}(z)$,
especially to $r_\mathrm{200m}$  \citep{Diemer+Kravtsov2014}.
This self-similarity indicates that overdensity radii defined with
respect to the mean cosmic density are preferred to describe the
structure and evolution of the outer density profiles.
The structure and dynamics of the infall region are expected
to be universal in units of the turnaround radius, according to 
self-similar infall models
\citep{1972ApJ...176....1G,Fillmore+Goldreich1984,Bertschinger1985,Shi2016}. 
In these models, the turnaround radius is a fixed
multiple of the radius enclosing a given fixed overdensity with 
respect to the mean cosmic density.
The outer profiles can thus be expected to be self-similar in
$r/r_\mathrm{200m}$.  
In contrast, the density profiles in the intra-halo (1-halo) region are   
found to be most self-similar when they are scaled by
$r_\mathrm{200c}$
\citep[or any other critical overdensity radius with a reasonable threshold;][]{Diemer+Kravtsov2014}.  
That is, density profiles of \LCDM halos in $N$-body
simulations prefer different scaling radii in different regions of the
density profile \citep{Diemer+Kravtsov2014}.
These empirical scalings were confirmed in cosmological hydrodynamical 
simulations of galaxy clusters \citep[][see also \citealt{Shi2016}]{Lau2015}.
However, the physical explanation for the self-similarity of the inner
density profile when rescaled with $r_\mathrm{200c}$ is less clear.

\begin{figure*}[!htb] %!htb
  \begin{center}
   \includegraphics[scale=0.3, angle=0,clip]{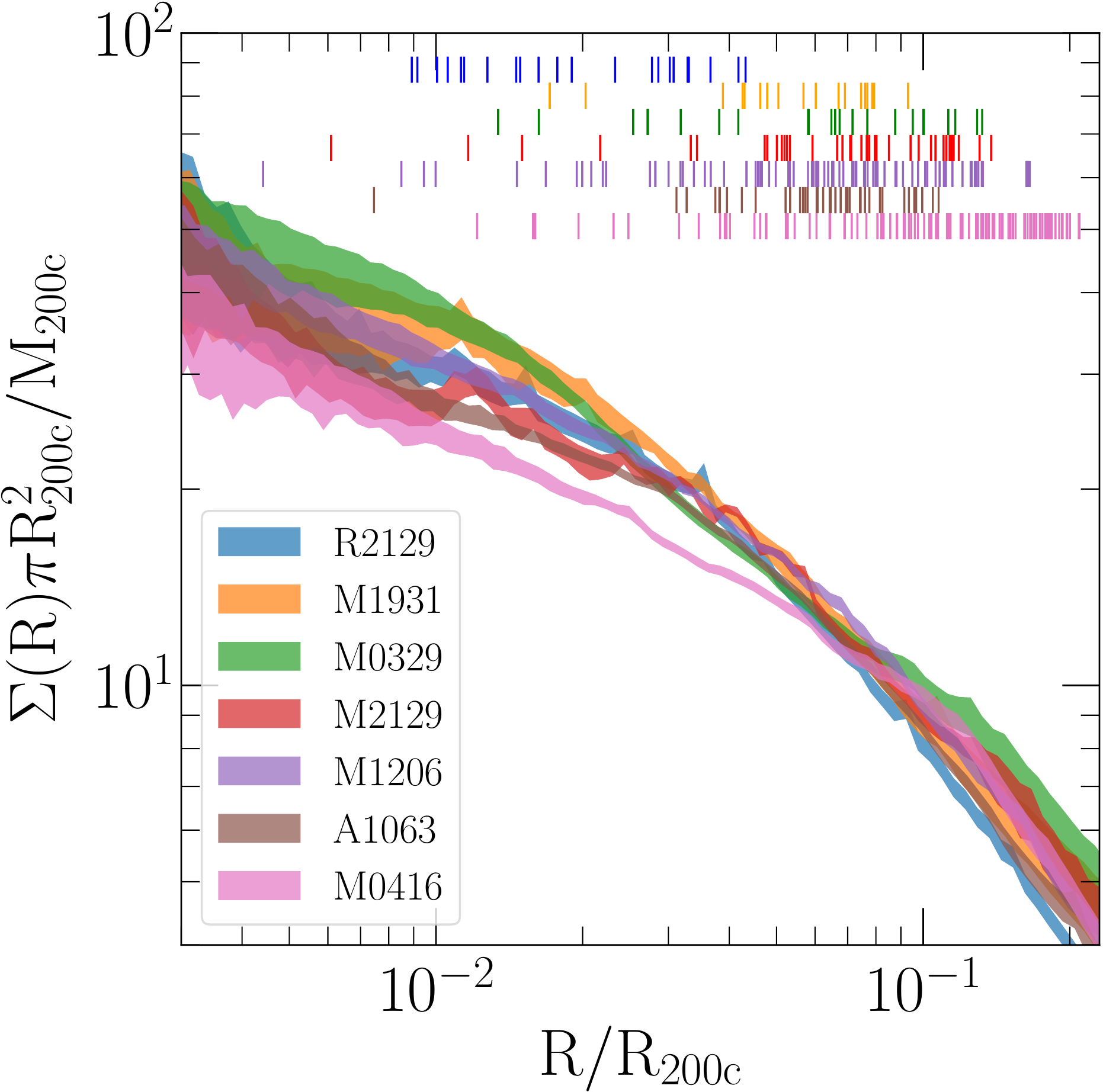}
   \includegraphics[scale=0.3, angle=0,clip]{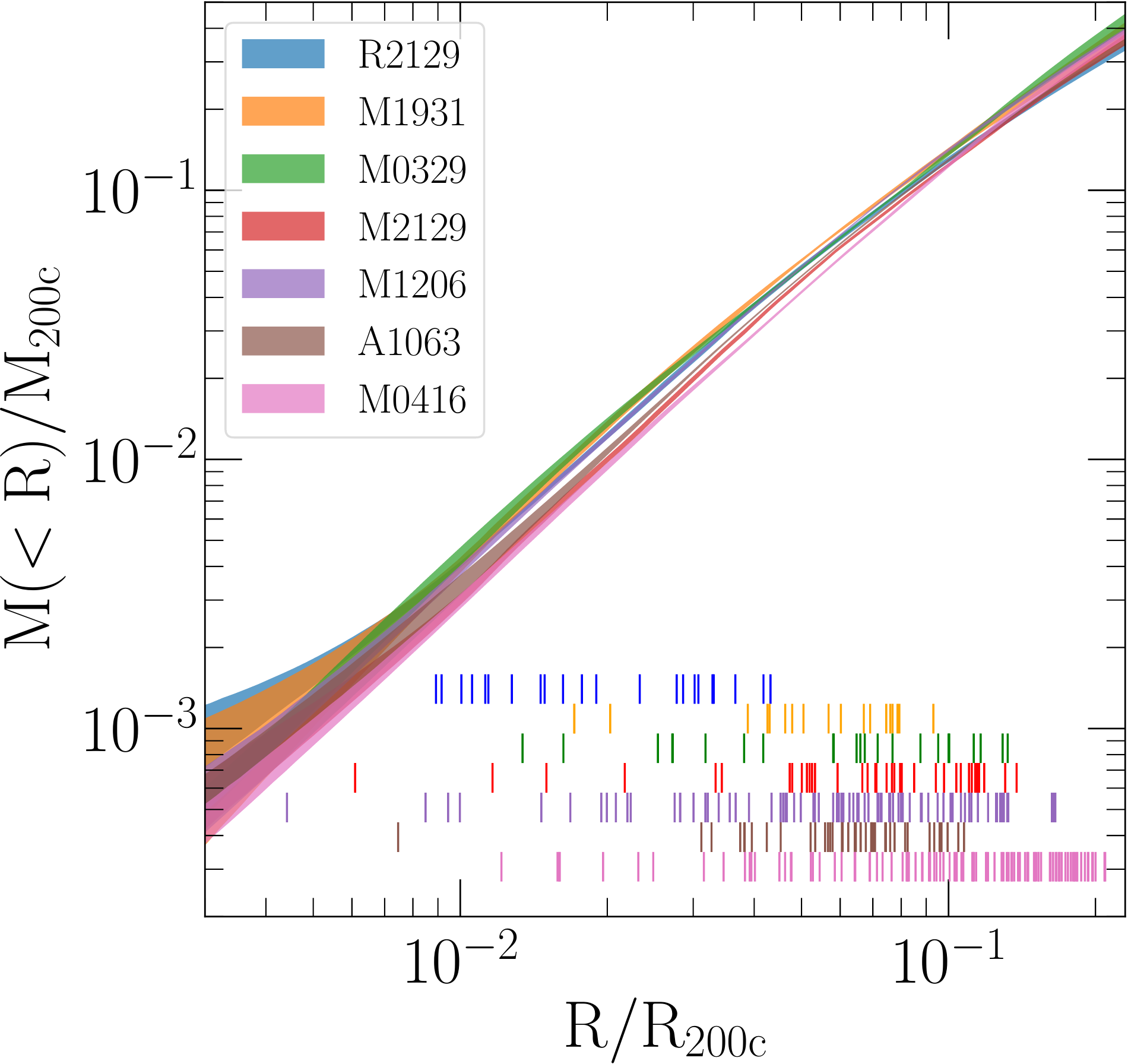}
  \end{center}
\caption{
 \label{fig:caminha19}
Projected mass density (left) and enclosed mass (right)
 profiles for seven CLASH clusters derived from a detailed
 strong-lensing analysis, 
 rescaled by $M_\mathrm{200c}$ and $r_\mathrm{200c}$ obtained from 
 NFW fits to independent weak-lensing measurements
 \citep{Umetsu2018clump3d}. 
Vertical lines color-coded for each cluster indicate the positions of
 multiple images used for the lens modeling, all belonging    
 to spectroscopic confirmed families.
MACS~J0416 with a shallower inner density slope is a highly asymmetric
 merging system.    
Image reproduced with permission from \citet{Caminha2019}, copyright by ESO.
}
\end{figure*}

In Fig.~\ref{fig:caminha19}, we show the projected total mass density
 ($\Sigma$) and enclosed mass ($M_\mathrm{2D}$) profiles for seven CLASH
 clusters derived from a detailed strong-lensing analysis of
 \citet{Caminha2019} 
based on extensive spectroscopic information, primarily from the Multi
Unit Spectroscopic Explorer (MUSE) archival data complemented with
CLASH-VLT redshift measurements \citep{Biviano2013,Rosati2014VLT}.
In the figure, the projected mass profiles of individual clusters are
rescaled using $M_\mathrm{200c}$ and $r_\mathrm{200c}$ obtained from NFW 
fits to independent ground-based weak-lensing measurements
\citep{Umetsu2018clump3d}.  
All clusters have a relatively large number of multiple-image
 constraints in the region 
$10^{-2}\simlt R/r_\mathrm{200c}\simlt 10^{-1}$,
where the shapes of the rescaled $\Sigma(R)$ and $M_\mathrm{2D}(<R)$
profiles are remarkably similar. Even MACS~J0416
\citep{Zitrin+2013M0416,Jauzac2015m0416,Grillo2015,Balestra2016m0416}, which  
is a highly asymmetric merging system, does not deviate significantly 
from the overall homologous profiles.
Within $10\%$ and  $20\%$ of $r_\mathrm{200c}$
for the seven clusters,
\citet{Caminha2019} measured a mean projected total mass value of
$0.13$ and $0.32 \times M_\mathrm{200c}$,
respectively, finding 
a remarkably small scatter of $5\percent$ and $6\percent$. At these same
radii, for the projected total mass density profiles, they found a mean value
of $9.0$ and $4.7 \times M_\mathrm{200c}/(\pi r_\mathrm{200c}^2)$,
with a slightly larger scatter of $7\%$ and $9\%$, respectively. The
observed trend is qualitatively consistent with the predictions by
\citet{Diemer+Kravtsov2014} and \citet{Lau2015}.

\subsection{The concentration--mass relation}
\label{subsec:cM}

The halo concentration $c_\Delta$ is a key quantity that characterizes
the density structure of dark matter halos,
where $c_\Delta$ is defined as the ratio of the outer halo radius
$r_\Delta$ (typically defined at an overdensity
of $\Delta_\mathrm{c}=200$)
and the inner characteristic radius $r_\mathrm{s}$ at which
the logarithmic density slope is $-2$
(Sect.~\ref{subsec:gest}).\footnote{This definition can be generalized to any
form of the density profiles other than the NFW profile, such as the
Einasto profile \citep[][Sect.~\ref{subsubsec:DK14}]{Einasto1965} and
the generalized NFW profile \citep{Zhao1996}.}
The halo concentration as a function of halo mass and redshift is
referred to as the concentration--mass ($c$--$M$) relation
\citep[e.g.,][]{2001MNRAS.321..559B,2002ApJ...568...52W}.
%\citep[e.g.,][]{Duffy+2008,2007MNRAS.381.1450N}.
For NFW halos, the $c$--$M$ relation fully specifies the structure of
halos at fixed halo mass and thus is a key ingredient of cluster
cosmology. 

In hierarchical \LCDM models, $c_\Delta$ is predicted to
depend on the accretion history. 
In the early phase of rapid mass accretion,
the scale radius $r_\mathrm{s}$ of a halo scales approximately as
the virial radius and thus $c_\Delta$ remains nearly constant
\citep{Zhao+2003}.
During the subsequent slow accretion phase,
the scale radius stays approximately constant,
whereas $r_\Delta$ continues to grow
through a mixture of physical accretion and pseudo-evolution,
resulting in an increase in halo concentration
\citep{1997ApJ...490..493N,2001MNRAS.321..559B,2002ApJ...568...52W,Diemer+2013pseudo,Correa2015cM}.
Since the mass accretion history depends on
the amplitude and shape of peaks in the initial density field,
as well as on the mass scale and the background cosmology,
%which, in turn, depend on the mass scale of the peak
%as well as on the parameters of the background cosmological model.
$c_\Delta$ depends on halo mass, redshift, and cosmological parameters
\citep{Prada2012,Dutton+Maccio2014,Diemer+Kravtsov2015,Diemer+Joyce2019}. 
There have been a number of attempts to
%express halo concentrations
%as a function of halo peak height $\nu(M_\Delta,z)$ to 
obtain a more universal representation of $c_\Delta$ as a
function of physical parameters, such as the
halo peak height $\nu(M_\Delta,z)$ and the local slope of the matter
power spectrum $d\ln{P(k)}/d\ln{k}$
\citep[see][]{Zhao+2009,Prada2012,Diemer+Kravtsov2015,Diemer+Joyce2019}.

Since galaxy clusters are, on average, dynamically young and still
growing through accretion and mergers, cluster halos are expected to
have relatively low concentrations, 
$c_\mathrm{200c}(z=0)\sim 4$, in contrast to 
individual galaxy halos that have denser central regions,
$c_\mathrm{200c}(z=0)\sim 7-8$
\citep{Bhatt+2013,Dutton+Maccio2014,Diemer+Kravtsov2015,Child2018cm,Diemer+Joyce2019}.
These general trends are complicated by diverse formation and
assembly histories of individual halos \citep{Ludlow+2013}, which
translate into substantial scatter in the $c$--$M$ relation,
with a lognormal intrinsic dispersion of 
$\sigma_\mathrm{int}(\ln{c_\mathrm{200c}})\sim 35\percent$\footnote{The
fractional scatter in natural logarithm is quoted as a percent
\citep[see, e.g.,][]{Umetsu2020xxl}.}
at fixed halo mass
\citep[e.g.,][]{Duffy+2008,Bhatt+2013,Diemer+Kravtsov2015}\footnote{According
to $N$-body simulations, the distribution of concentrations 
derived for all halos has tails at both low and high values of
$c_\mathrm{200c}$, and it is neither Gaussian nor lognormal
\citep[see Fig.~1 of][]{Diemer+Kravtsov2015}.
By selecting ``relaxed'' halos (which excludes some of the tails
resulting from poor fits), some authors found that the distribution of
concentrations is described by a lognormal distribution
\citep[e.g.,][]{Jing2000,2001MNRAS.321..559B,2007MNRAS.381.1450N},
while others found that it is better described by a Gaussian
\citep[][]{Reed+2011,Bhatt+2013}.}

\begin{figure*}[!htb] %!htb
  \begin{center}
   \includegraphics[scale=0.33, angle=0, clip]{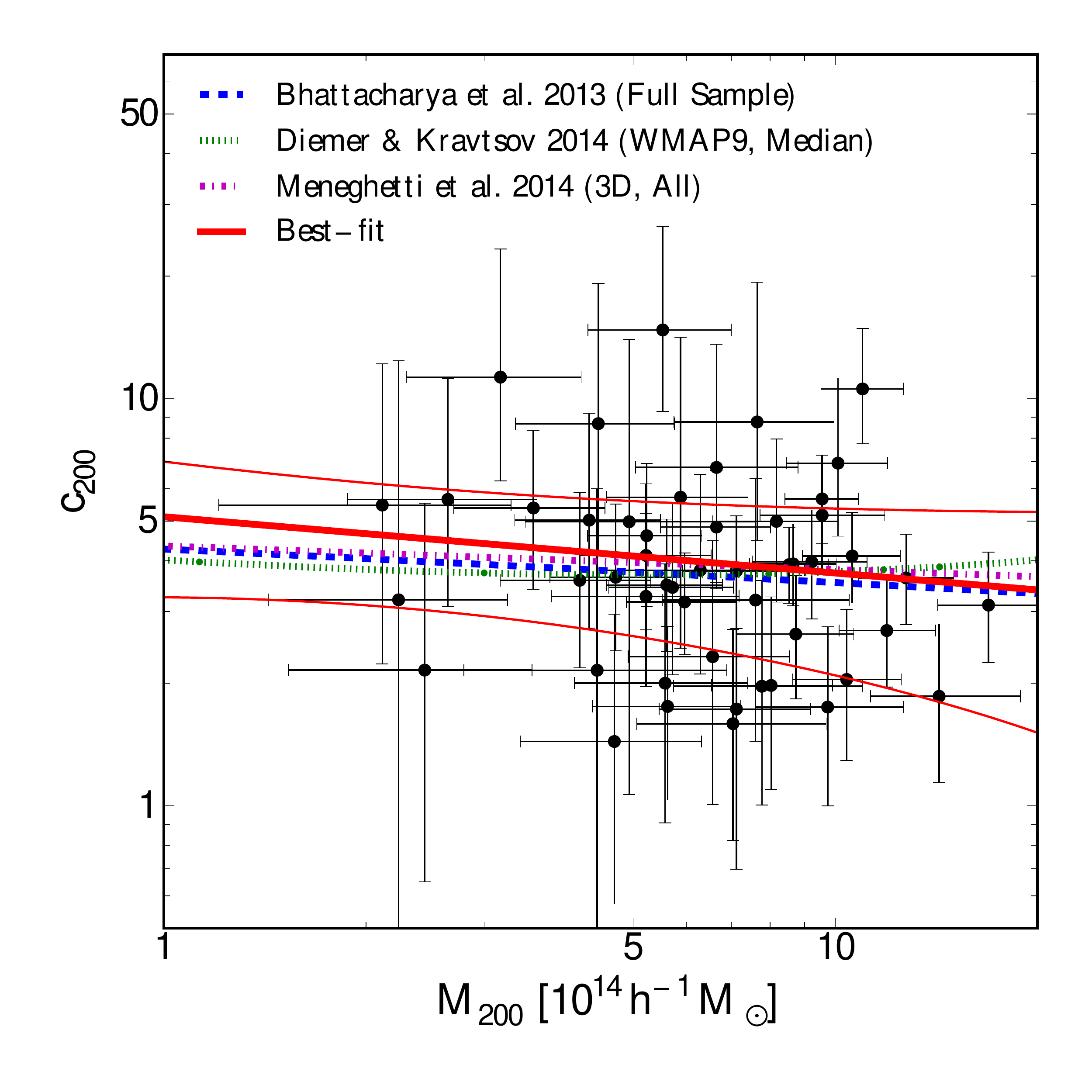} 
  \end{center}
\caption{
 \label{fig:cMplot_Okabe16}
The concentration--mass ($c_\mathrm{200c}$--$M_\mathrm{200c}$) relation
 for a sample of 50 X-ray luminous clusters (black circles with
 error bars) derived from the LoCuSS
 survey \citep{Okabe+Smith2016}. The results are based on weak-lensing
 measurements from Subaru Suprime-Cam observations. 
 The thick and thin red lines  show the best-fitting function and the
 errors, respectively. The dashed blue, dotted green, and dot-dashed
 magenta lines are the mean $c$--$M$ relation of 
 \LCDM halos derived from numerical simulations of
 \citet{Bhatt+2013}, \citet{Diemer+Kravtsov2015}, and
 \citet{Meneghetti2014clash} at $z_l = 0.23$, respectively.
Image reproduced with permission from \citet{Okabe+Smith2016}, copyright by the authors. 
}
\end{figure*}

On the observational side,
cluster lensing studies targeting lensing-unbiased samples
\citep[e.g.,][]{Merten2015clash,Du2015,Umetsu2016clash,Umetsu2020xxl,Okabe+Smith2016,Cibirka2017,Sereno2017psz2lens,Klein2019}
have found that the $c$--$M$ relations derived for these
cluster samples agree well with theoretical models calibrated for recent
\LCDM cosmologies 
\citep[e.g.,][]{Bhatt+2013,Dutton+Maccio2014,Meneghetti2014clash,Diemer+Kravtsov2015,Child2018cm,Diemer+Joyce2019}.

In Fig.~\ref{fig:cMplot_Okabe16}, we show the
 $c_\mathrm{200c}$--$M_\mathrm{200c}$ relation  derived
 from a Subaru weak-lensing analysis of 50 X-ray-selected clusters
 targeted by the LoCuSS survey \citep[see Sect.~\ref{subsec:massprofile};][]{Okabe+Smith2016}. Their results are based
 on NFW profile fits to the reduced tangential shear profiles of
 individual clusters. The best-fit $c_\mathrm{200c}$--$M_\mathrm{200c}$
 relation for the LoCuSS sample is
 $c_\mathrm{200c}=5.12^{+2.08}_{-1.44}\times (M_\mathrm{200c}/10^{14}\Msunh)^{-0.14\pm 0.16}$
 at $z_l=0.23$ \citep{Okabe+Smith2016}.
The normalization and slope of the $c_\mathrm{200c}$--$M_\mathrm{200c}$
 relation for the LoCuSS sample are in excellent agreement with
 those found for mass-selected samples of dark-matter halos in
 \LCDM numerical simulations
 \citep{Bhatt+2013,Meneghetti2014clash,Diemer+Kravtsov2015}, indicating
 no significant impact of the X-ray selection on the normalization and
 mass slope parameters.
  \citet{Okabe+Smith2016} found an intrinsic lognormal
 dispersion of $\sigma_\mathrm{int}(\ln{c_\mathrm{200c}})<20\percent$
 ($68.3\percent$ CL) at fixed halo mass, 
 which is lower than found for \LCDM halos in $N$-body
 simulations 
 \citep[$\sim 35\percent$ for the full
 population of halos including both relaxed and unrelaxed
 systems;][]{Bhatt+2013,Diemer+Kravtsov2015}.
 Similar results on the concentration scatter
 were obtained for independent X-ray cluster samples
 \citep[e.g., the CLASH and the XXL samples with
 $\sigma_\mathrm{int}(\ln{c_\mathrm{200c}})=(13\pm 6)\percent$ and
 $\sigma_\mathrm{int}(\ln{c_\mathrm{200c}})<24\percent$ at the
 $99.7\percent$ CL, respectively; see][]{Umetsu2016clash,Umetsu2020xxl}.  
This is likely caused in part by the X-ray selection bias in terms of
 the cool-core or relaxation state,
 as found by previous studies 
 \citep{Buote+2007,Ettori+2010,Eckert2011cc,Meneghetti2014clash}.

\begin{figure*}[!htb] %!htb
  \begin{center}
   \includegraphics[scale=1., angle=0, clip]{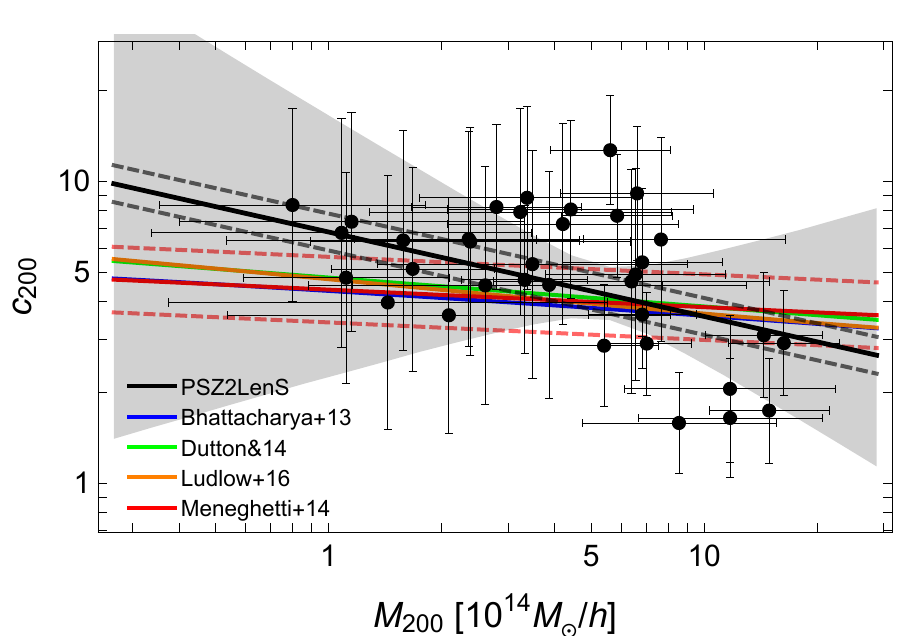} 
  \end{center}
\caption{
 \label{fig:cMplot_Sereno17}
The concentration--mass ($c_\mathrm{200c}$--$M_\mathrm{200c}$) relation
 at $z_l = 0.20$,
 derived from a weak-lensing analysis of 35 PSZ2LenS clusters
 \citep{Sereno2017psz2lens} selected from the \Planck SZE survey.
 The solid and dashed black lines show the median $c$--$M$ relation and
 its intrinsic scatter, respectively.
The shaded gray region encloses the $68.3\percent$ probability around the median
relation due to uncertainties on the scaling parameters. The blue, green,
 and orange lines show the $c$--$M$ relation of 
 \LCDM halos derived from numerical simulations of
\citet{Bhatt+2013}, \citet{Dutton+Maccio2014}, and \citet{Ludlow+2016},
 respectively. 
The solid and dashed red lines show the $c$--$M$ relation of
 \LCDM halos and its $1\sigma$ scatter, respectively,
 from \citet{Meneghetti2014clash}. 
Image reproduced with permission from \citet{Sereno2017psz2lens}, copyright by the authors.
}
\end{figure*}

Cluster samples are traditionally defined by X-ray or optical
observables, and more recently through the thermal SZE strength
\citep[e.g.,][]{Planck2015XXVII}.
The SZE is a characteristic spectral distortion in the CMB
induced by inverse Compton scattering between cold CMB photons and hot
ionized electrons
\citep[Sect.~\ref{subsubsec:crit};][]{1995ARA&A..33..541R,1999PhR...310...97B}. 
Unlike any other detection techniques,  SZE-selected cluster samples are
nearly mass-limited and have well-behaved selection functions.
This is because the SZE detection signal has a very weak dependence on the
redshift (see Eq.~(\ref{eq:SZE})) and it is also less sensitive to
the relaxation state of the cluster.
Blind SZE surveys can thus provide representative cluster samples
representative of the full population of halos out to high redshift. 
This makes SZE surveys ideal for cosmological tests based on the
evolution of the cluster abundance.   

We show in Fig.~\ref{fig:cMplot_Sereno17} the
$c_\mathrm{200c}$--$M_\mathrm{200c}$ relation at $z_l=0.20$ obtained for
a sample of \Planck clusters targeted by the PSZ2LenS
project \citep{Sereno2017psz2lens}.
The PSZ2LenS sample includes 35 optically confirmed \Planck clusters
selected from the second \Planck catalog of Sunyaev--Zel'dovich sources
\citep[PSZ2;][]{Planck2015XXVII} located in the fields of two lensing
surveys,
namely the CFHTLenS
\citep[Canada France Hawaii Telescope Lensing Survey;][]{Heymans+2012CFHTLenS}
and
RCSLenS
\citep[Red Cluster Sequence Lensing Survey;][]{Hildebrandt2016rcslens}
surveys.
%The sample is statistically complete and homogeneous in
%terms of observing facilities, data acquisition, reduction, and 
%analysis.
The PSZ2LenS sample represents a faithful subsample of
the whole population of \Planck clusters, for which homogeneous 
weak-lensing data and photometric redshifts are available from
the CFHTLenS and RCSLenS surveys \citep{Sereno2017psz2lens}.
The resulting relation between mass and concentration is in broad
agreement with theoretical predictions from $N$-body
simulations calibrated for recent \LCDM cosmologies
\citep{Bhatt+2013,Dutton+Maccio2014,Meneghetti2014clash,Ludlow+2016}. 
%which shared the same observational instrumentation
%and the same data-analysis tools. PSZ2LenS is homogeneous in
%terms of selection, observational set-up, data reduction and data
%analysis.
%\citet{Sereno2017psz2lens} performed a weak-lensing analysis of PSZ2LenS
%sample 
%were SZ selected by the Planck mission in the fields covered by the
%CFHTLenS and the RCSLenS. The surveys are not deep but the
%sample, which we named PSZ2LenS, is statistically complete and
%homogeneous in terms of observing facilities, and data acquisition,
%reduction and analysis.
%\citep{Sereno+Covone2013}

\subsubsection{Superlens clusters: are they overconcentrated?}
\label{subsubsec:superlens}

In contrast to X-ray- or SZE-selected samples, galaxy clusters
identified by the presence of strongly lensed giant arcs represent a
highly biased population.
In particular, cluster lenses selected to have large Einstein radii
(e.g., $\vartheta_\mathrm{Ein} > 30$\,arcsec for $z_s=2$)
represent the most lensing-biased population of clusters
with their major axis preferentially aligned with the observer's line of
sight 
\citep{2007ApJ...654..714H,Oguri+Blandford2009,Meneghetti+2010MARENOSTRUM,Meneghetti+2011}.
Such an extreme population of cluster lenses is referred to as
\emph{superlenses} \citep{Oguri+Blandford2009}. 
A selection bias in favor of prolate structure pointed to the observer
is expected because this orientation boosts the projected surface mass
density and hence the lensing signal.  
A population of superlens clusters is also expected to be
biased toward halos with intrinsically higher concentrations 
\citep{2007ApJ...654..714H,Sereno2010concen}.
Accordingly, in the context of \LCDM,
superlens clusters are predicted to have large apparent concentrations
in projection of the sky, compared to typical clusters with similar
masses and redshifts \citep{Oguri+Blandford2009}.
Calculations of the enhancement of the projected mass and thus boosted 
Einstein radii find a statistical bias of $\sim 34\percent$ in
concentration based on $N$-body simulations of \LCDM cosmologies
\citep{2007ApJ...654..714H}.  
Semianalytical simulations based on triaxial halos find a concentration
bias of $\sim 40\percent$--$60\percent$ for superlens clusters
\citep{Oguri+Blandford2009}.

Despite attempts to correct for potential projection and
selection biases inherent to lensing,
%ntriguingly,
initial results from combined strong- and weak-lensing
measurements assuming a spherical halo
revealed a relatively high degree of halo concentration in lensing clusters
\citep{2003A&A...403...11G,2003ApJ...598..804K,BTU+05,BUM+08,Oguri+2009Subaru,Zitrin+2011A383,Umetsu+2011stack}, 
lying above the $c$--$M$ relation calibrated for \LCDM
cosmologies \citep[e.g.,][]{2007MNRAS.381.1450N,Duffy+2008}
based on earlier
\textit{Wilkinson Microwave Anisotropy Probe} (\WMAP) releases
\citep{2003ApJS..148..175S,Komatsu+2009WMAP5}.
A possible explanation for the apparent discrepancy was that cluster
halos are highly overconcentrated than expected from
\LCDM models.
Motivated by possible implications for the overconcentration problem,
one of the key objectives of the CLASH survey is to
establish the degree of mass concentration for a lensing-unbiased sample
of high-mass clusters using combined strong- and weak-lensing
measurements with homogeneous data sets.

\begin{figure*}[!htb] %!htb
  \begin{center}
   \includegraphics[scale=0.55, angle=0, clip]{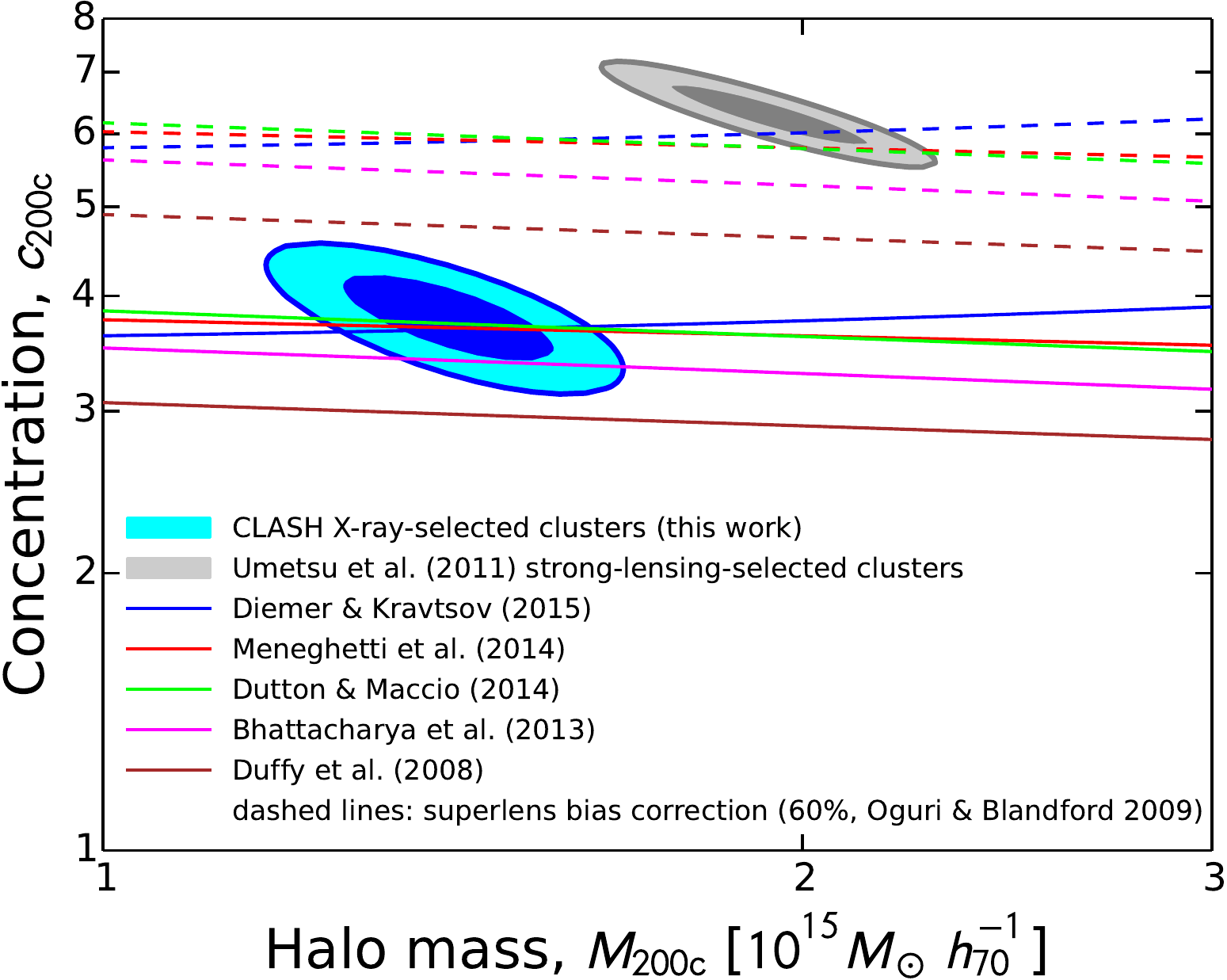}
  \end{center}
\caption{
 \label{fig:cM_superlens}
Stacked lensing constraints on the $c$--$M$ relation
 derived assuming spherical NFW halos for the CLASH X-ray-selected
 subsample
 \citep[][$c_\mathrm{200c}=3.76^{+0.29}_{-0.27}$ at $z_l\approx 0.34$;
 blue contours]{Umetsu2016clash}
 and for the superlens sample of
 \citet[][$c_\mathrm{200c}=6.31\pm 0.35$ at $z_l\approx 0,32$;
 gray contours]{Umetsu+2011stack}. 
 Both results are based on their respective strong and weak lensing
 analyses.
 For each case, the contours show the $68.3\percent$ and $95.4\percent$
 confidence levels.
The results are compared to theoretical $c$--$M$ relations (solid lines)
 from numerical simulations of \LCDM cosmologies
 \citep{Duffy+2008,Bhatt+2013,Dutton+Maccio2014,Meneghetti2014clash,Diemer+Kravtsov2015},
 all evaluated at $z=0.32$ for the full population of halos.
 The dashed lines show 60\% superlens corrections to the solid lines,  
 accounting for the effects of strong-lensing selection and orientation
 bias expected for the population of superlens clusters
 \citep[][]{Oguri+Blandford2009}.
 Image reproduced with permission from \citet{Umetsu2016clash}, copyright by AAS.
}
\end{figure*}

As a precursor study of the CLASH survey, \citet{Umetsu+2011stack}
carried out a combined strong- and weak-lensing 
analysis of four superlens clusters of similar masses \citep[A1689,
A1703, A370, and Cl0024$+$1654; see][]{Umetsu+2011} 
using strong-lensing, weak-lensing
shear and magnification measurements obtained from high-quality \HST and 
Subaru observations. 
The stacked sample has a lensing-weighted mean redshift of
$z_l\approx 0.32$.
These clusters display prominent strong-lensing features characterized
by large Einstein radii,
$\theta_\mathrm{Ein}\simgt 30$\,arcsec for $z_s=2$.
\citet{Umetsu+2011stack} found that the stacked $\llangle\Sigma\rrangle$
profile of the four clusters in the range $R\in [40,2800]\kpch$
is well described by a single NFW profile,
with an effective concentration of  
$c_\mathrm{200c}= 6.31\pm 0.35$ at 
$M_\mathrm{200c}=(13.4\pm 0.9)\times 10^{14}\Msunh$,
corresponding to an Einstein radius of 
$\vartheta_\mathrm{Ein}\approx 36$\,arcsec for $z_s=2$.
After applying a 50\% superlens correction,
\citet{Umetsu+2011stack} found a discrepancy of $\sim 2\sigma$ with 
respect to the $c$--$M$ relation of \citet{Duffy+2008} calibrated for
the \WMAP five-year cosmology.
They concluded that there is no significant tension
between the concentrations of the four clusters and those of
\LCDM halos if lensing biases are coupled to a
sizable intrinsic scatter in the $c$--$M$ relation. 

In Fig.~\ref{fig:cM_superlens}, we show in the
$c_\mathrm{200c}$--$M_\mathrm{200c}$ plane the stacked lensing
constraints obtained for the CLASH X-ray-selected subsample
\citep{Umetsu2016clash} and those for the superlens sample of
\citet{Umetsu+2011stack}. 
The stacked lensing constraints for the two cluster samples are compared
to theoretical $c$--$M$ relations of 
\citet{Duffy+2008}, \citet{Bhatt+2013}, \citet{Dutton+Maccio2014},
\citet{Meneghetti2014clash}, and 
\citet[][their mean relation]{Diemer+Kravtsov2015}, all 
evaluated for the full population of halos at $z=0.32$.
This comparison demonstrates that $c$--$M$ relations that are
calibrated for more recent simulations and
\LCDM cosmologies (\WMAP seven- and nine-year cosmologies and \Planck
cosmologies) provide better agreement with the CLASH lensing
measurements  
\citep{Umetsu2014clash,Umetsu2016clash,Merten2015clash}. 
This is also in line with the findings of \citet{Dutton+Maccio2014}, who
showed that the $c$--$M$ relation in the \WMAP five-year cosmology
has a 20\% lower normalization at $z=0$ than in the \Planck
cosmology, which has a correspondingly higher normalization in terms of
$\Om$ and $\sigma_8$.

To account for the superlens bias in the \citet{Umetsu+2011stack}
sample, Fig.~\ref{fig:cM_superlens} shows each of the $c$--$M$
predictions with a maximal 60\% correction applied \citep{Oguri+Blandford2009}.  
We see from the figure that, once the effects of 
selection and orientation bias are taken into account, the results of
\citet{Umetsu+2011stack} come into line with the models of 
\citet{Dutton+Maccio2014}, 
\citet{Meneghetti2014clash}, and
\citet{Diemer+Kravtsov2015}, the three most recent $c$--$M$ models
studied in \citet{Umetsu2016clash}.
Hence, the discrepancy found by \citet{Umetsu+2011stack} can be fully
reconciled by the higher normalization of the $c$--$M$ relation as
favored by more recent \WMAP and \Planck cosmologies
\citep{Komatsu+2011WMAP7,Hinshaw+2013WMAP9,Planck2015XIII}. 
Therefore, there appears to be no compelling evidence for
the overconcentration problem within the standard \LCDM framework
\citep[see also][]{Oguri+2012SGAS,Foex+2014,Robertson+2020}. 

Another issue when comparing to theoretical relations is that they are
generally quantified by the mean and median concentration of individual
halos \citep{Diemer+Kravtsov2015}, rather than that of stacked density
profiles \citep{Child2018cm}.
Since the lensing signal depends nonlinearly on halo concentration, even
if $c_\mathrm{200c}(M_\mathrm{200c})$ is normally distributed, there is
no guarantee that the concentration extracted from a stacked lensing
profile would be equal to the mean of the concentrations of the halos in
the stack.   
The difference can be significant especially when taking into account 
different stacking procedures used in lensing measurements
\citep[][]{Umetsu+Diemer2017}.

It is intriguing to note that, as already discussed in
Sect.~\ref{subsubsec:nfw}, full triaxial modeling of Abell 1689
($z_l=0.183$) 
shows that combined lensing, X-ray, and SZE observations of the cluster
can be consistently explained by its  intrinsically high mass
concentration combined with a chance alignment of its major axis with
the line-of-sight direction \citep{Umetsu2015A1689}.
A careful interpretation
of lensing, dynamical, and X-ray data based on $N$-body/hydrodynamical
simulations suggests that Cl0024$+$1654 ($z_l=0.395$) is the result of
a high-speed, line-of-sight collision of two massive clusters viewed
approximately $2$--$3$\,Gyr after impact when the gravitational
potential has had time to relax in the center, but before the gas has  
recovered \citep{Umetsu+2010CL0024}.
Similar to the case of Cl0024$+$1654, Abell 370 ($z_l=0.375$) is faint
in both X-ray and SZE signals and does not follow the X-ray/SZE
observable--mass-scaling relations \citep[see][]{Czakon2015}.
$N$-body/hydrodynamical simulations constrained by lensing, dynamical,
X-ray, and SZE observations suggest that Abell 370 is a post major
merger after the second core passage in the infalling phase, just before
the third core passage \citep{Molnar+2020a370}. In this post-collision
phase, the gas has not settled down in the gravitational potential well
of the cluster, which explains why A370 does not follow the mass scaling
relations.  
Note that, because of its large projected mass and high lensing
magnification capability, Abell 370 has been
selected as one of the six Hubble Frontier Fields clusters
\citep{Lotz2017hff}.

Finally, it should be noted that high-magnification-selected clusters at
$z_l>0.5$, such as those selected by the CLASH and Frontier Fields
surveys,
%\citep{Postman+2012CLASH,Lotz2017hff}, 
often turn out to be dynamically disturbed, highly massive ongoing
mergers 
\citep[e.g.,][]{Torri+2004,Zitrin+Broadhurst2009,Merten+2011,Zitrin+2013M0416,Medezinski+2013,Medezinski2016}. 
These ongoing mergers can produce substructured, highly elongated lenses
in projection of the sky \citep[e.g.,][]{2005PASJ...57..877U,2019ApJ...874..132A},
enhancing the lensing efficiency \citep{Meneghetti+2003,Meneghetti+2007,Redlich+2012} 
by boosting the number of multiple images per critical area, due to the
increased ratio of the caustic area relative to the critical area
\citep{Zitrin+2013M0416}.  The projected mass distributions of such
ongoing mergers cannot be well described by a single NFW profile.
In contrast to the superlens clusters at $z_l<0.4$
\citep{Umetsu+2011stack}, NFW fits to the lensing profiles of 
high-magnification CLASH clusters yield relatively low
concentrations \citep[see][]{Umetsu2016clash}.

\subsection{Splashback radius}
\label{subsec:rsp}

In the standard \LCDM paradigm of hierarchical
structure formation, galaxy clusters form through accretion of matter
along surrounding filamentary structures, as well as through successive
mergers of smaller objects. An essential picture of halo
assembly is that shells of matter surrounding an overdense 
region in the early universe will initially expand with the Hubble flow,
decelerate, turn around, and fall back in.
%start contracting. 
Each shell will cross previously collapsed shells that are oscillating
in the growing halo potential. In this picture, accreting particles will
pile up near the apocenter of their first orbit, thus creating a sharp
density enhancement or caustic in the halo outskirts
\citep{Fillmore+Goldreich1984,Bertschinger1985}.
%\citep{1972ApJ...176....1G,Fillmore+Goldreich1984,Bertschinger1985}.
This steepening feature depends on the slope of the initial
mass perturbation, which determines the mass accretion rate of dark
matter halos \citep{Fillmore+Goldreich1984,Lithwick+Dalal2011}. 
This is illustrated in Fig.~\ref{fig:splash}.

\begin{figure*}[!htb] %!htb
  \begin{center}
   \includegraphics[scale=0.38, angle=0, clip]{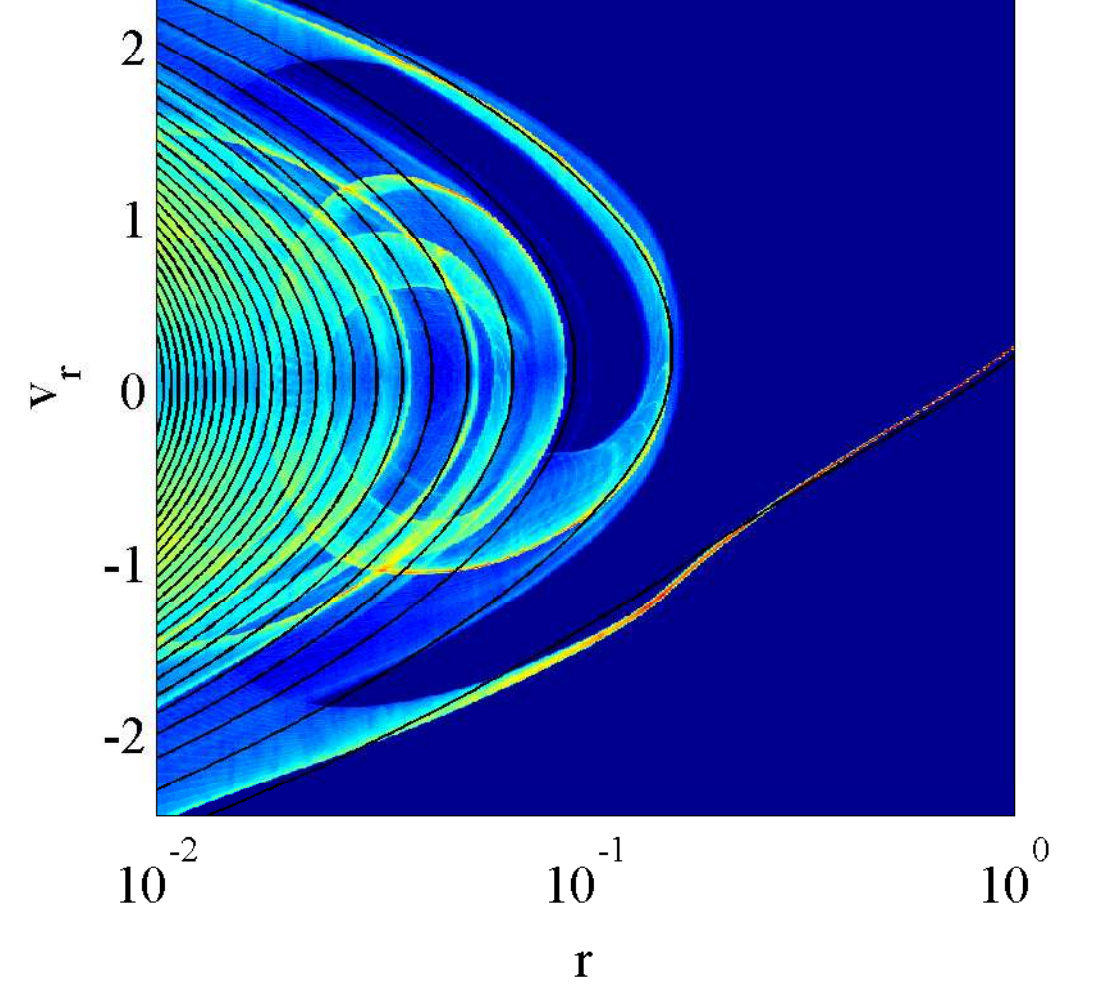}
   \includegraphics[scale=0.36, angle=0, clip]{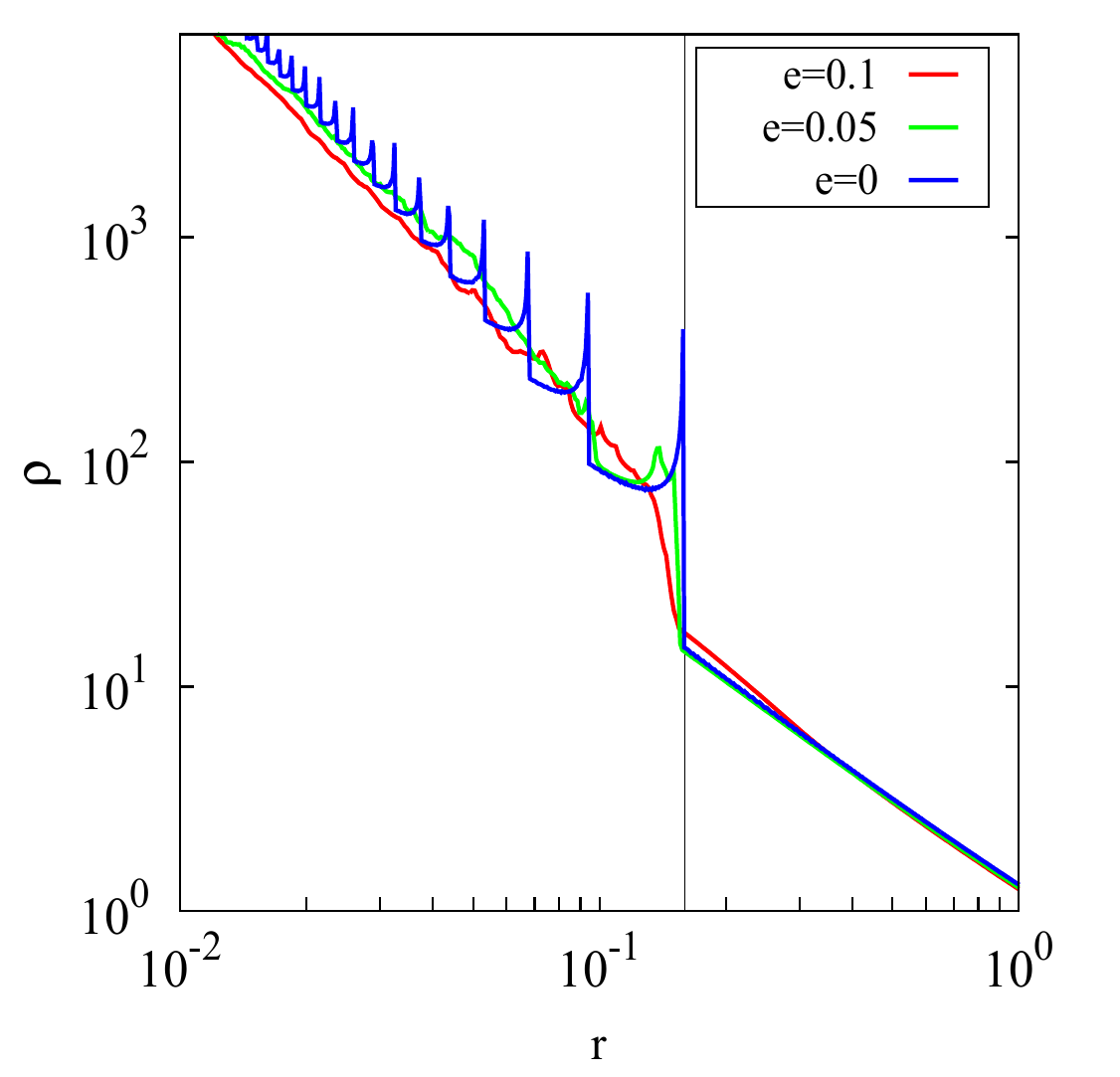} 
  \end{center}
\caption{
 \label{fig:splash}
Density caustics for self-similar collisionless halos with an accretion 
 rate of $s=d\ln{M}/d\ln{a}=3$
 \citep{Fillmore+Goldreich1984,Lithwick+Dalal2011}. 
The left panel shows particle trajectories in the phase space
 diagram for spherically symmetric collapse (solid black curve)
 and for three-dimensional triaxial collapse (color map).
The horizontal axis is the halo-centric distance ($r$) and the vertical  
 axis is the radial velocity ($V_r$). 
The right panel shows the spherically averaged density profile $\rho(r)$ 
for different values of the initial ellipticity parameter $e$
 \citep{Lithwick+Dalal2011}. 
The vertical line indicates the
 location of the outermost caustic, or the splashback radius, predicted
 by the similarity solution for spherical collisionless collapse with
 this value of accretion rate. 
The caustic location depends on the mass accretion rate of halos, while 
 the steepening caustic feature in the spherically averaged 
 $\rho(r)$ depends on $e$.
Image reproduced with permission from \citet{Adhikari2014}, copyright by IOP and SISSA.
See also
 \citet{Shi2016} for an analytical approach to modeling the outer
 profile of dark matter halos. 
 }
\end{figure*}

Recently, a closer examination of the halo density profiles in
 cosmological $N$-body simulations has revealed systematic deviations 
 from the NFW and Einasto profiles in the outskirts at  
 $r\simgt 0.5r_\mathrm{200m}$ \citep{Diemer+Kravtsov2014}. 
In particular, the halo profiles exhibit a sharp drop in density, a
feature associated with the orbital apocenter of the recently accreted
matter in the growing halo potential (see Fig.~\ref{fig:splash}).
The location of the outermost density caustic expected in collisionless 
halos is referred to as the splashback radius
\citep{Diemer+Kravtsov2014,Adhikari2014,More2015splash}.
The splashback radius constitutes a physical boundary of
halos because it sharply separates (at least in perfect spherical
symmetry) the multistream intra-halo region from the outer infall region
\citep{Diemer+Kravtsov2014,More2015splash,Mansfield+2017,Okumura+2018}. 
%It is also related to the transition scale between the 1-halo and 2-halo
%regimes to a certain extent \citep{Cooray+Sheth2002,More2015splash,Tomooka+2020}. 
The splashback radius is also related to the transition scale between
the 1-halo and 2-halo regimes of the halo--mass correlation function
\citep{Garcia+2020}.

Splashback features are determined by the orbits of dark
matter particles in the halo potential and thus fully characterized in
phase space \citep{Diemer2017,Okumura+2018}.
Hence, the steepening feature in the density profile alone  
cannot capture the full dynamical information of dark matter halos
\citep[see][]{Okumura+2018}. 
%Furthermore, the two-halo term of the density statistic is
%enhanced by the galaxy bias, 
%The steepening is less pronounced for halos with lower mass
%accretion rates.
In particular, the ``true'' location of the splashback radius based on
particle orbits is not equivalent to a particular
location in the spherically averaged density profile
\citep{Diemer2017,Diemer+2017}. 
%Keeping this in mind, it is useful
Nevertheless, it is convenient 
to define the splashback radius
$r_\mathrm{sp}$ as the halo radius where the logarithmic slope of the
three-dimensional density profile,
$\gamma_\mathrm{3D}(r)=d\ln{\rho}(r)/d\ln{r}$, is steepest
\citep{Diemer+Kravtsov2014,More2015splash}.
The splashback radius is a genuine prediction of the  standard CDM
picture of structure formation. We can possibly learn something about
the nature of dark matter and cosmology
\citep{Adhikari2018sp,Banerjee2020sp}. Moreover, this steepening feature
may be used to make the connection to the dynamical relaxation state of
halos.

In the context of \LCDM, the location of $r_\mathrm{sp}$ with respect to
$r_\mathrm{200m}$ is predicted to decrease with mass accretion rate
$s(a)\equiv d\ln{M_\mathrm{vir}(a)}/d\ln{a}$
\citep{Diemer+Kravtsov2014,Adhikari2014,More2015splash,Diemer+2017}
and to increase with $\Om(a)\equiv \overline{\rho}(a)/\rho_\mathrm{c}(a)$
\citep{More2015splash,Diemer+2017}, with some additional dependence on
peak height $\nu$ \citep[see][]{Diemer+2017}.
In \LCDM cosmologies,
fast accreting halos have $r_\mathrm{sp}\simlt r_\mathrm{200m}$ with a
sharper splashback feature,
while for slowly accreting halos $r_\mathrm{sp}$ can be as large as
$(2-3)r_\mathrm{200m}$ \citep{More2015splash,Diemer+2017}.
The steepening splashback signal is thus expected to be strongest for
massive galaxy clusters because they are, on average, fast accreting
systems \citep{Adhikari2014,Diemer+Kravtsov2014}.
Galaxy clusters are thus the best objects to look for the splashback 
feature.

The steepening feature near the splashback radius $r_\mathrm{sp}$ can be
inferred from weak lensing and density statistics of the galaxy
distribution. 
%The splashback radius $r_\mathrm{sp}$ has a well-defined feature that
%can be inferred from weak lensing and density statistics of the galaxy
%distribution. 
When using galaxies as a tracer of the mass distribution around
clusters, however, one needs to account for the effect of dynamical friction 
that acts to reduce the orbital apocenter of sufficiently massive
subhalos hosting cluster galaxies \citep[see][]{Adhikari+2016}.  
%%%%
Since the efficiency of dynamical friction increases with the ratio of subhalo
to cluster halo mass, the impact of dynamical friction on the splashback
feature is expected to depend on the luminosity of tracer galaxies
\citep[e.g.,][]{More2016splash,Chang2018sp}.
The splashback feature in stacked galaxy surface density profiles has
been routinely detected using cluster--galaxy cross correlations, thanks
to statistical samples of clusters defined from large optical or SZE
surveys
\citep{More2016splash,Baxter2017splash,Chang2018sp,Shin2019sp,Zurcher+More2019,Murata2020hsc}.

Cluster--galaxy weak lensing can be used to directly test detailed
predictions for the splashback feature in the outer density profile of
cluster halos.
Since the location of the steepest slope in three dimensions is a
trade-off between the steepening 1-halo term and the 2-halo term,
one needs to precisely measure the lensing signal in both
1-halo and 2-halo regimes spanning a wide range in cluster-centric
radius. 
The location of the three-dimensional splashback radius $r_\mathrm{sp}$
can then be inferred by forward-modeling the projected lensing profile
($\Delta\Sigma(R)$ or $\Sigma(R)$) assuming a flexible fitting function,
such as the DK14 profile (Sect.~\ref{subsubsec:DK14}).\footnote{In
projection, the 2-halo term has a substantial impact on the apparent
location of the steepest density slope $d\ln\Sigma/d\ln{R}$,
which emerges at a much smaller radius that is unrelated to the
steepening term in three dimensions \citep{Umetsu+Diemer2017}.
This highlights the importance of forward-modeling
the effects of the steepening based on the underlying three-dimensional 
density profile \citep{Umetsu+Diemer2017}.
} 
As discussed in \citet{More2016splash} and \citet{Umetsu+Diemer2017},
one would apply two requirements to claim a detection of the
splashback radius using a DK14 model (Eq.~(\ref{eq:DK14})),
namely (1) that the
location of the steepest slope in three dimensions
can be identified at high statistical significance and (2) that this  
steepening is greater than
expected from a model without a steepening feature, such as an
Einasto profile (or a DK14 model with $f_\mathrm{trans}(r)=1$).
%that expected from a DK14 model with $f_\mathrm{trans}(r)=1$ (i.e., an
%Einasto profile). 
The second criterion is important to ensure that the steepening
is associated with a density caustic rather than the transition
between the 1-halo and 2-halo terms.

\begin{figure*}[!htb] %!htb
  \begin{center}
   \includegraphics[scale=0.33, angle=0, clip]{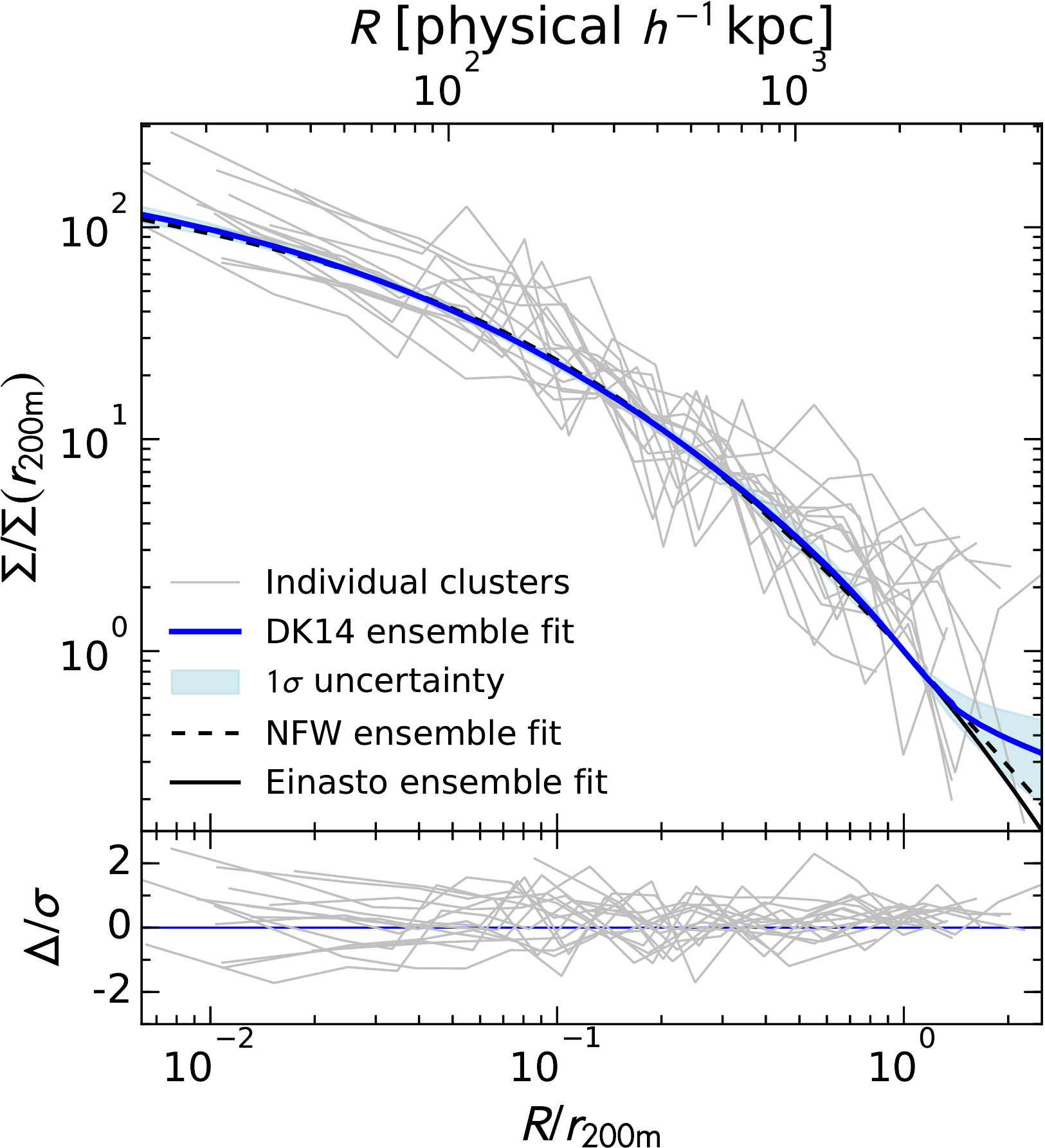}
   \includegraphics[scale=0.33, angle=0, clip]{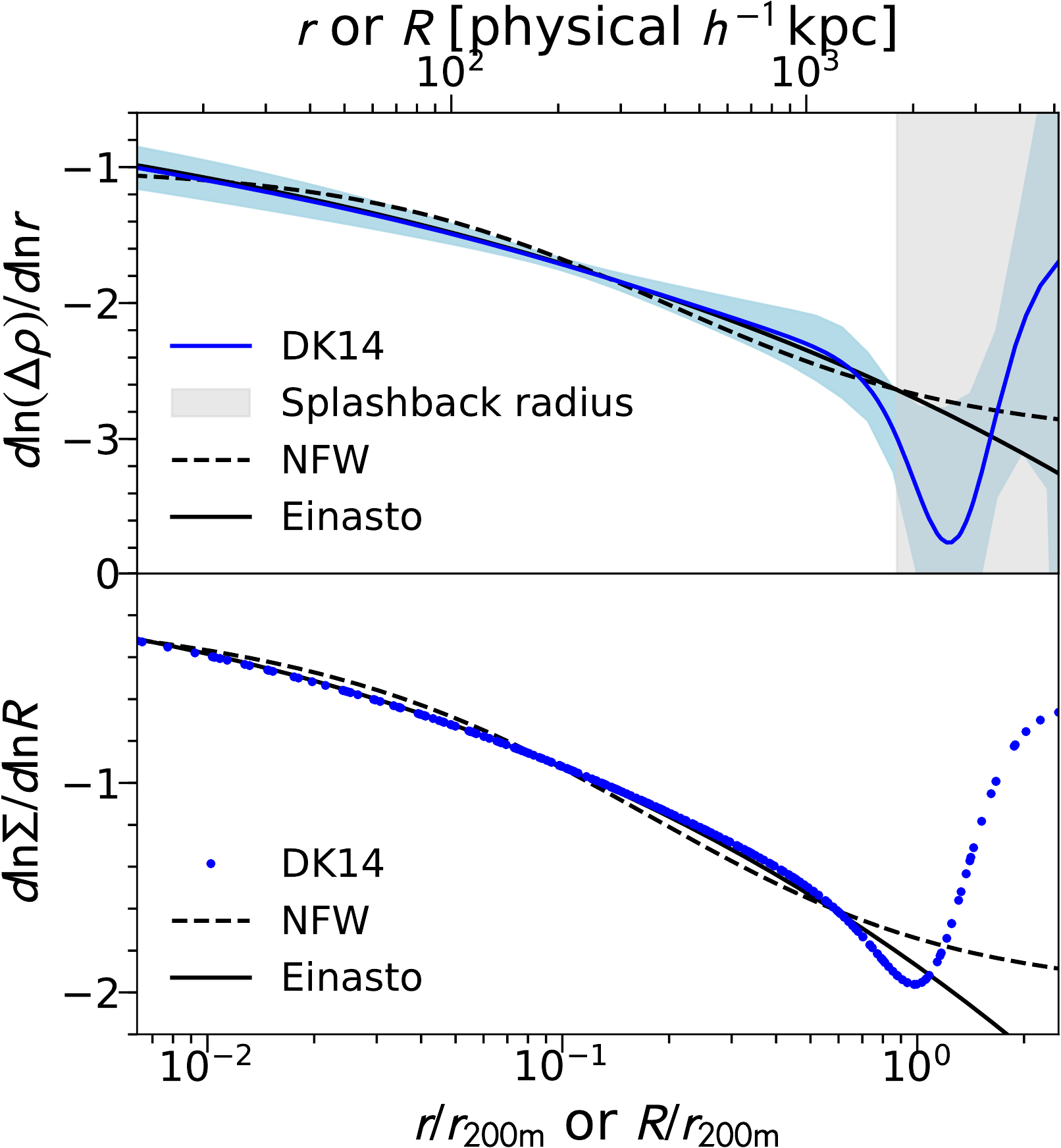} 
  \end{center}
\caption{
 \label{fig:ud17}
An overview of the first attempt to detect $r_\mathrm{sp}$ from
 lensing \citep{Umetsu+Diemer2017}. 
\emph{Left}:  The top panel shows the scaled surface mass density
 $\Sigma/\Sigma(r_\mathrm{200m})$ of the CLASH X-ray-selected sample as
 a function of $R/r_\mathrm{200m}$. The blue thick solid line and 
 the blue shaded area show the best-fit DK14 profile and its $1\sigma$
 uncertainty derived from a simultaneous ensemble fit to the scaled
 $\Sigma$ profiles of 16 individual clusters (gray lines). The
 corresponding NFW (black dashed) and Einasto (black solid) fits are
 also  shown. The lower panel shows deviations (in units of $\sigma$) of
 the observed cluster profiles from the best-fit DK14 profile.
\emph{Right}: similarly, the top panel shows the logarithmic gradient of
 the inferred three-dimensional density profiles as a function of
 $r/r_\mathrm{200m}$ for the DK14, NFW, and Einasto models.
 The gray vertical shaded area indicates the range  from the $16$th to
 the $84$th percentile of the marginalized posterior distribution of the
 splashback radius $r_\mathrm{sp}/r_\mathrm{200m}$.
 The bottom panel is the same as the top panel,
 but showing the logarithmic slope of the surface mass density
 profiles. The best-fit DK14 profile is shown as blue dots at the
 locations of the data points. 
This is a slightly modified version of the figure presented in
 \citet{Umetsu+Diemer2017}.
}
\end{figure*}

\citet{Umetsu+Diemer2017} were the first to attempt to place direct 
constraints on the splashback radius $r_\mathrm{sp}$ around clusters by
using gravitational lensing observations.  
They developed methods for modeling averaged cluster lensing profiles
scaled to a chosen halo overdensity $\Delta$, which can be optimized for 
the extraction of gradient features that are local in cluster radius, in
particular the density steepening due to the splashback radius. 
\citet{Umetsu+Diemer2017} examined the ensemble mass distribution of 16
CLASH X-ray-selected clusters with a weighted mean mass of
$M_\mathrm{200m}\approx 13\times 10^{14}\Msunh$,
by forward-modeling the $\Sigma(R)$ profiles of individual clusters
(Fig.~\ref{fig:kappa_u16}) 
obtained by \citet{Umetsu2016clash}.
The maximum radius of their ensemble analysis is
$R_\mathrm{max}\approx 5\Mpch\sim 2.5r_\mathrm{200m}$.
Their results are shown in Fig.~\ref{fig:ud17}.
They found that the CLASH ensemble mass profile in projection is
remarkably well described by an NFW or Einasto density profile out to 
$R \approx 1.2 r_\mathrm{200m}$ (Sect.~\ref{subsec:massprofile}),
beyond which the data exhibit a flattening with respect to the NFW or
Einasto profile due to the 2-halo term.
The gradient feature in the cluster outskirts is most pronounced
for a scaling with $r_\mathrm{200m}$,
which is consistent with simulation results of
\citet{Diemer+Kravtsov2014} and \citet{Lau2015}
(Sect.~\ref{subsec:massprofile}).
\citet{Umetsu+Diemer2017}, however, did not find statistically significant
evidence for the existence of $r_\mathrm{sp}$ in the CLASH lensing data,
as limited by the field of view of Suprime-Cam ($34\times
27$\,arcmin$^2$) on the Subaru telescope.
%and the statistical quality of the data.   
Assuming the DK14 profile form and generic priors calibrated with
numerical simulations, they placed an informative lower limit on the
splashback radius of the clusters, if it exists, of
$r_\mathrm{sp}/r_\mathrm{200m}>0.89$ at
$68\percent$ confidence. This constraint is in agreement with
the \LCDM expectation for the CLASH sample,
$r_\mathrm{sp}/r_\mathrm{200m}\approx 0.97$ \citep{More2015splash}.
%evaluated at the effective peak height of the CLASH
%sample, $\nu_\mathrm{200m}=4.0\pm 0.1$.
%Note that we expect the average mass accretion
%rate of the CLASH X-ray-selected sample to be low due to a high fraction
%of relaxed objects \citep{Meneghetti2014clash}. This selection effect
%could cause a bias toward higher values of
%$r_\mathrm{sp}/r_\mathrm{200m}$ in this sample
%\citep[see][]{Umetsu+Diemer2017}. 

\begin{figure*}[!htb] %!htb
  \begin{center}
   \includegraphics[scale=0.45, angle=0, clip]{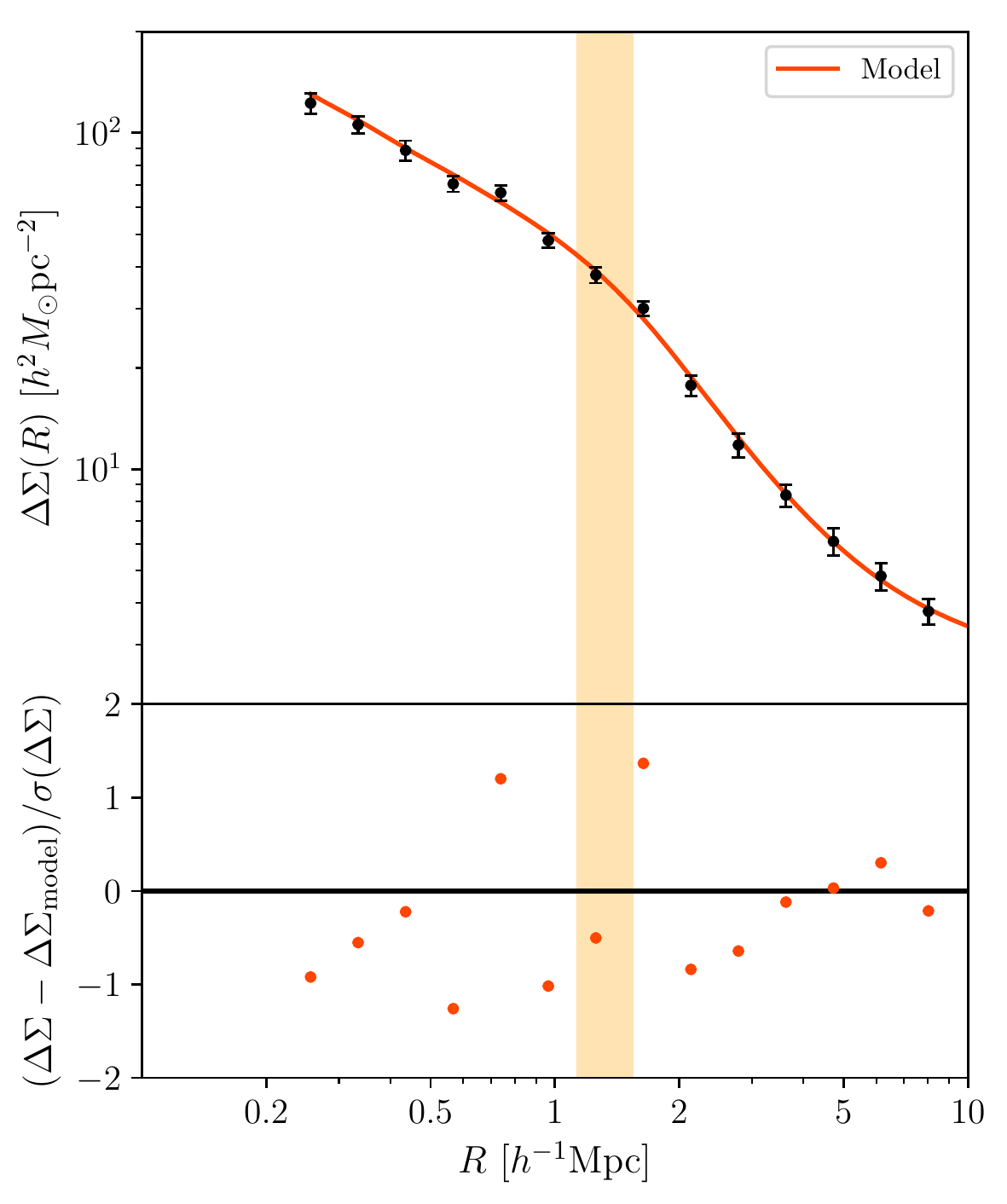}
   \includegraphics[scale=0.32, angle=0, clip]{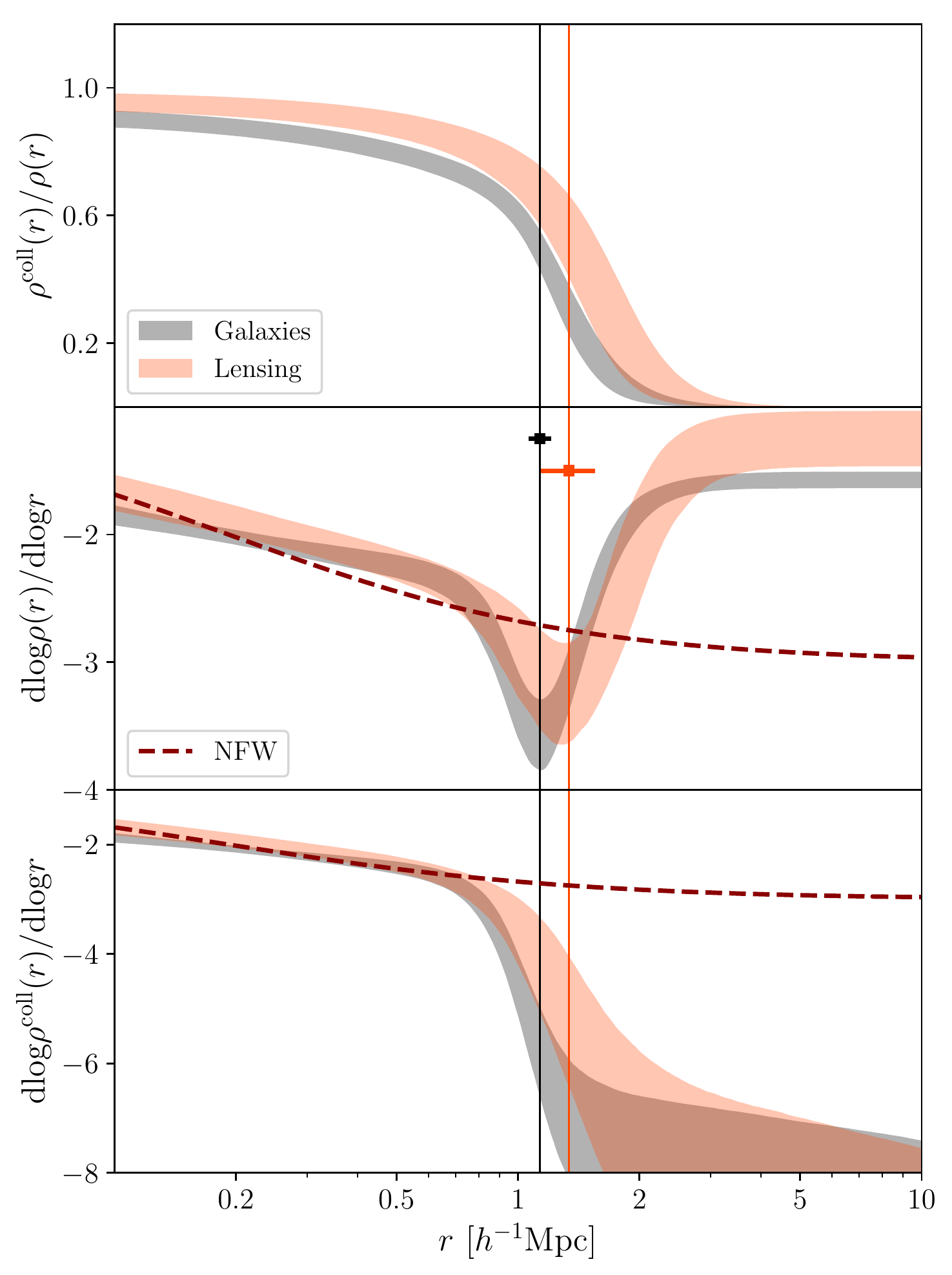} 
  \end{center}
\caption{
\label{fig:chang18}
The first successful $r_\mathrm{sp}$ determination from weak lensing by
 \citet{Chang2018sp}. 
\emph{Left}: The top panel shows the stacked
 $\llangle\Delta\Sigma_+\rrangle$ profile as a function of
 (comoving) cluster-centric radius $R$
 (black points with error bars) derived for a sample of
 $3684$ redMaPPer clusters ($0.2<z_l<0.55$; $20<\lambda<100$)
 in the first-year DES data.
The red line shows the model fit to the lensing measurements. The
 inferred location of the splashback radius $r_\mathrm{sp}$ with its
 $1\sigma$ uncertainty is marked by
 the vertical orange band.
The bottom panel shows the residual in the fits
 divided by the uncertainty of the measurement.
\emph{Right}: comparison of model-fits from the projected galaxy
 distribution (gray) and weak lensing (red).
The upper panel shows the fraction of   
 the density profile $\rho^\mathrm{coll}(r)/\rho(r)$
 for the collapsed material
 $\rho^\mathrm{coll}(r)\equiv \rho_\mathrm{E}(r)f_\mathrm{trans}(r)$
 over the total density profile $\rho(r)$ (see Eq.~(\ref{eq:DK14})).
The middle panel shows the logarithmic gradient of the
 total density profile compared to that of an
NFW profile (dashed line). The lower panel shows the logarithmic
 gradient of the profile for the collapsed material compared to 
that of an NFW profile. The vertical
lines mark the mean values of $r_\mathrm{sp}$ inferred from the model
 fits for both galaxy and lensing measurements, while the horizontal bars
in the middle panel indicate the uncertainties on $r_\mathrm{sp}$.
Image reproduced with permission from \citet{Chang2018sp}, copyright by AAS. 
}
\end{figure*}

%\citet{Chang2018sp}
The first successful determination of $r_\mathrm{sp}$ from lensing was
achieved by \citet{Chang2018sp}, who
measured both galaxy number density
($\Sigma_\mathrm{g}$) and tangential shear ($\Delta\Sigma_+$) profiles
around a statistical sample of clusters detected by the red-sequence
Matched-filter Probabilistic Percolation \citep[redMaPPer;][]{Rykoff2016}
algorithm in the first year DES data.
Their fiducial sample of 3684 clusters is defined by a redshift
selection of $0.2<z_l<0.55$ and a richness selection of
$20<\lambda<100$.
The sample is characterized by an effective mass of 
$M_\mathrm{200m}\approx 1.8\times 10^{14}\Msunh$ at a mean redshift of
$z_l\approx 0.41$.
The left panels of Fig.~\ref{fig:chang18} show the stacked
$\llangle\Delta\Sigma_+(R)\rrangle$ profile around their fiducial sample
along with the best-fit DK14 profile.
For DK14 modeling of the weak-lensing signal (see Sect.~\ref{subsubsec:DK14}),
\citet{Chang2018sp} assumed uniform priors on
$(\rho_{-2}, r_{-2}, b_\mathrm{e}\overline{\rho},s_\mathrm{e})$
and
Gaussian priors on the shape parameters of 
$\log_{10}{\alpha_\mathrm{E}}=\log_{10}(0.19)\pm 0.1$,
$\log_{10}{\beta}=\log_{10}(6.0)\pm 0.2$, and
$\log_{10}{\gamma}=\log_{10}(4.0)\pm 0.2$
\citep{More2016splash,Umetsu+Diemer2017},
allowing a representative range of values calibrated by $N$-body
simulations \citep{Diemer+Kravtsov2014,More2015splash}. 
They also marginalized over two
additional parameters describing the cluster miscentering effect
expected for optically selected clusters \citep{Johnston+2007b}.

With the stacked DES weak-lensing measurements,
\citet{Chang2018sp} constrained the location of the
steepest slope in the three-dimensional density profile to lie in the
range $r_\mathrm{sp}/r_\mathrm{200m}=0.97\pm 0.15$.
%%%
The location and steepness of this gradient feature inferred
from the weak-lensing signal agrees within the errors with those
inferred from the stacked galaxy density measurements
($r_\mathrm{sp}/r_\mathrm{200m}=0.82\pm 0.05$), as shown in the right
panels of Fig.~\ref{fig:chang18}.
Note that, as mentioned above, the $r_\mathrm{sp}$ determined by the
galaxy density profile is expected to be smaller than that of the 
underlying matter distribution because of dynamical friction, depending
on the mass of the galaxies used.
%%%
From the weak-lensing (or galaxy density) profile, 
\citet{Chang2018sp} found the total cluster density profile at the 
location of $r_\mathrm{sp}$ to be 
steeper than the NFW profile form at a significance level of $2.0\sigma$
(or $3.0\sigma$).
Similarly, they found the 1-halo term of the DK14 profile
$\rho_\mathrm{E}(r)f_\mathrm{trans}(r)$
at the location of $r_\mathrm{sp}$ to be steeper than the NFW form at
the $2.9\sigma$ (or $4.6\sigma$) level.

\citet{Chang2018sp} found that $r_\mathrm{sp}$ measured from weak
lensing is smaller than but consistent with the expectation from
$N$-body simulations within the large errors.  The $r_\mathrm{sp}$
measured from the galaxy density profile with a higher precision is
significantly smaller than that  determined from the corresponding
population of subhalos in $N$-body simulations \citep[see
also][]{Shin2019sp}, which is in agreement with the previous results based
on the SDSS redMaPPer samples \citep{More2016splash,Baxter2017splash}.
Using $N$-body simulations,  \citet{Chang2018sp} found that the effect
of dynamical friction is significant only for very massive subhalos and
the fiducial galaxy sample used in their analysis is likely not
significantly affected by dynamical friction.
This discrepancy is likely due to the systematic effects associated with
the optical cluster finding algorithm
\citep{Zu2017,Busch+White2017,Murata2020hsc}.
By analyzing synthetic galaxy catalogs created from $N$-body
simulations, \citet{Murata2020hsc} found that the level 
of systematic bias in the inferred location of $r_\mathrm{sp}$ can be 
significantly alleviated when increasing the aperture size of optical
cluster finders beyond the splashback feature.

%\citet{Chang2018sp} found that $r_\mathrm{sp}$ measured from weak
%lensing is in agreement with the expectation from $N$-body simulations
%within the large errors. In contrast, the $r_\mathrm{sp}$ measured from
%the galaxy density 
%profile with a higher precision is significantly smaller than that 
%determined from the corresponding population of subhalos in 
%$N$-body simulations, which is in agreement with
%previous results from SDSS data \citep{More2016splash,Baxter2017splash}.
%This discrepancy is likely due in part to the systematic effects
%associated with the optical cluster finding algorithm
%\citep{Zu2017,Busch+White2017}. 

More recently, \citet{Contigiani+2019} placed a stacked lensing
constraint on the splashback feature for their sample of 27 high-mass
clusters at $0.15<z_l<0.55$  targeted by the CCCP survey \citep[Cluster Canadian Comparison
Project;][]{Hoekstra2015CCCP}.
The cluster sample is characterized by a weighted mean mass of
$M_\mathrm{200m}\approx 12\times 10^{14}\Msunh$ at a mean redshift of
$z_l\sim 0.2$. Their analysis is based on wide-field weak-lensing data
taken with CFHT/MegaCam with a $1$\,deg$^2$ field of view.
Their data set is very similar in nature to the CLASH sample of
\citet{Umetsu+Diemer2017} because both studies are based on targeted
lensing observations of similarly high-mass clusters.
Although they did not detect a significant steepening,
\citet{Contigiani+2019} constrained the splashback radius for their
stacked sample as $r_\mathrm{sp}=3.6^{+1.2}_{-0.7}$\,Mpc (comoving)
assuming a DK14 profile with generic priors calibrated with numerical
simulations. Although the sample size of clusters in
\citet{Contigiani+2019} is not significantly larger than that of
\citet{Umetsu+Diemer2017}, the large field-of-view of CFHT/MegaCam
allowed them to better constrain the location and steepness of the
splashback feature in the cluster outskirts.

These studies represent a first step toward using cluster--galaxy weak
lensing and density statistics of the galaxy distribution to examine
well-defined predictions for the splashback features from cosmological
$N$-body simulations \citep[e.g.,][]{Diemer+2017} and to explore the physics
associated with the splashback radius of collisionless halos.
A significant improvement in the statistical quality of data is expected
from ongoing and upcoming wide-field surveys.
On the other hand, improved understanding of systematic effects, such as
selection bias of observable-selected clusters and projection effects,
will be needed to harness the full potential of such high-precision
measurements. 
Furthermore, the use of phase-space statistics will be extremely
useful to explore the properties and signatures of dark matter, in
particular of dark matter self interactions
\citep{More2016splash,Okumura+2018}.

\subsection{Mass calibration for cluster cosmology}
\label{subsec:mcal}

Determining the abundance of rare massive galaxy clusters above a given
mass threshold provides a powerful probe of cosmological parameters,
especially $\Om$ and $\sigma_8$ \citep[e.g.,][]{Rosati+2002}.
This constraining power is primarily due to the fact that clusters
constitute the high-mass tail of hierarchical 
structure formation, which is exponentially sensitive to the growth of
cosmic structure \citep[][]{2001ApJ...553..545H,Watson+2014}.
Conversely, obtaining an accurate calibration of the
mass scale for a given cluster sample is key to harness the power of   
cluster cosmology \citep{Pratt2019}.
In this context, large statistical samples of
clusters spanning a wide range in mass and redshift, with a
well-characterized selection function, provide an independent means of
examining any viable cosmological model  
\citep[e.g.,][]{Allen+2004,Allen2011cosmo,Vikhlinin+2009CCC3,Mantz2010,Weinberg2013cosmo,deHaan+2017}. 
In principle, clusters can complement other cosmological probes if the
systematics are well understood and controlled.

Thanks to dedicated blind SZE surveys that are made possible in recent
years, large homogeneous samples of clusters have been obtained
through the SZE selection, out to and beyond a redshift of unity over 
a wide area of the sky
\citep[e.g.,][]{2014A&A...571A..29P,Planck2015XXVII,SPT2015sze,Hilton2018actpol}. 
In particular, the \Planck satellite mission
produced representative catalogs of galaxy clusters detected via the
SZE signal from its all-sky survey.
The \Planck PSZ2 catalog contains 1653 SZE detections of cluster
candidates from the 29 month full-mission \Planck data
\citep{Planck2015XXVII}. 
The full-mission \Planck cosmology sample contains $439$ clusters down
to S/N of 6, representing the most massive population of clusters with a
well-behaved selection function
\citep{Planck2015XXVII,Planck2015XXIV}. 

Despite these recent advances, we must reiterate that accurate cluster
mass measurements are essential in the cosmological interpretation of
the cluster abundance \citep{Pratt2019}.
Since clusters are detected by the observable baryonic signature, an
external calibration of the corresponding observable--mass scaling
relation is necessary for a cosmological interpretation of the cluster
sample, by accounting for inherent statistical effects and selection  
bias \citep[e.g.,][]{Battaglia+2016,Miyatake2019actpol,Sereno2020xxl}. 
This was acutely demonstrated as an internal tension  
in the \Planck analyses \citep{Planck2013XX,Planck2015XXIV}, which
revealed a non-negligible level of discrepancy between the cosmological
parameters $(\Om,\sigma_8)$ derived from the \Planck cluster counts
and those from combining the \Planck primary CMB measurements with
non-cluster data sets, both within the framework of the standard \LCDM
cosmology.    
Here it should be noted that \citet{Planck2013XX} employed X-ray
observations with \XMMNewton to calibrate the scaling 
relation between the SZE signal strength and cluster mass for their
\Planck cluster
cosmology sample.  However, the determination of cluster mass relied on
the assumption that the intracluster gas is in hydrostatic equilibrium
(hereafter HSE) with the gravitational potential dominated by dark
matter.  

To characterize the overall level of mass bias (assumed to be constant
in cluster mass and redshift) for their SZE-selected clusters,
\citet{Planck2013XX} introduced a parameter defined as \citep[see
also][]{Planck2015XXIV}: 
\begin{equation}
 1-b = \left\langle\frac{M_\mathrm{SZE}}{M_\mathrm{true}}\right\rangle,
\end{equation}
where $M_\mathrm{SZE}$ denotes the SZE mass proxy and
 $M_\mathrm{true}$ is the true mass \citep[see][]{PennaLima2017}, both
 defined at an overdensity of $\Delta_\mathrm{c}=500$. 
Note that this factor includes not only astrophysical biases but also
 all systematics encoded in the statistical relationship between the 
 \Planck-based mass and the true mass
 \citep[see][]{Planck2013XX,Donahue2014clash}.
The \Planck team initially adopted $1-b=0.8$
 as a fiducial value with a flat prior in the range $[0.7,1.0]$
 \citep[][]{Planck2013XX}, which is about the level
 expected due to deviations from the assumed HSE,
 $b \sim (10-20)\percent$
 \citep[e.g.,][]{Nagai2007,Meneghetti+2010a,Angelinelli+2020}. 
 If the bias were zero, the \Planck CMB cosmology
 would predict far more massive clusters than observed. 
By analyzing the \Planck cosmology sample from the full PSZ2 cluster
 catalog, \citet{Planck2015XXIV} found that the level of mass bias
 required to bring the \Planck cluster counts and \Planck primary CMB
 into full agreement in their base \LCDM cosmology is
 $1-b = 0.58 \pm 0.04$.  
Intriguingly, this would imply that \Planck masses underestimate the
 true values by  
 $b = (42\pm 4)\percent$, compared to
 $b \sim 20\percent$ initially adopted by \citet{Planck2013XX}.
The level of disagreement between the \Planck cluster counts and \Planck
 primary CMB is about $2\sigma$.
 
While this tension could potentially reflect a higher-than-expected sum
of neutrino masses or more exotic physics, the confidence in such a
scenario would be limited by systematic uncertainties arising from 
both astrophysical and observational effects
\citep[e.g.,][]{Planck2013XX,Donahue2014clash,CoMaLit5}.  
In fact, the level of disagreement appears to slightly decrease after
accounting for updated lower values of the reionization optical depth   
\citep{Planck2016tau}.
Nevertheless, this tension has attracted considerable
attention in the cluster community and led to deeper investigations into
the mass calibration for this representative cosmology sample of \Planck
clusters. 

Weak gravitational lensing offers a direct probe of the cluster mass
distribution.
Cluster--galaxy weak lensing can provide an unbiased mass calibration of
galaxy clusters, if one can carefully control systematic effects, such as shear
calibration bias (Sect.~\ref{subsubsec:KSB}), photo-$z$ bias, residual
cluster contamination (Sect.~\ref{subsec:dilution}), and
mass modeling bias (Sect.~\ref{subsec:lensmodel}). This has become
possible thanks to concerted efforts by various groups over the last few
decades 
\citep[e.g.,][]{1995ApJ...449..460K,1998ApJ...504..636H,Hoekstra2015CCCP,Okabe+Umetsu2008,Okabe+2010WL,Okabe+2013,Medezinski+2010,Medezinski2018src,Oguri2010glafic,Umetsu+2011,Umetsu2014clash,Rosati2014VLT,WtG1,WtG2,WtG3,Merten2015clash,CoMaLit1,Melchior2017des,Mandelbaum2018sim,Mandelbaum2018shear,Gruen2019des}.

\begin{figure*}[!htb] %!htb
  \begin{center}
   \includegraphics[scale=0.45, angle=0, clip]{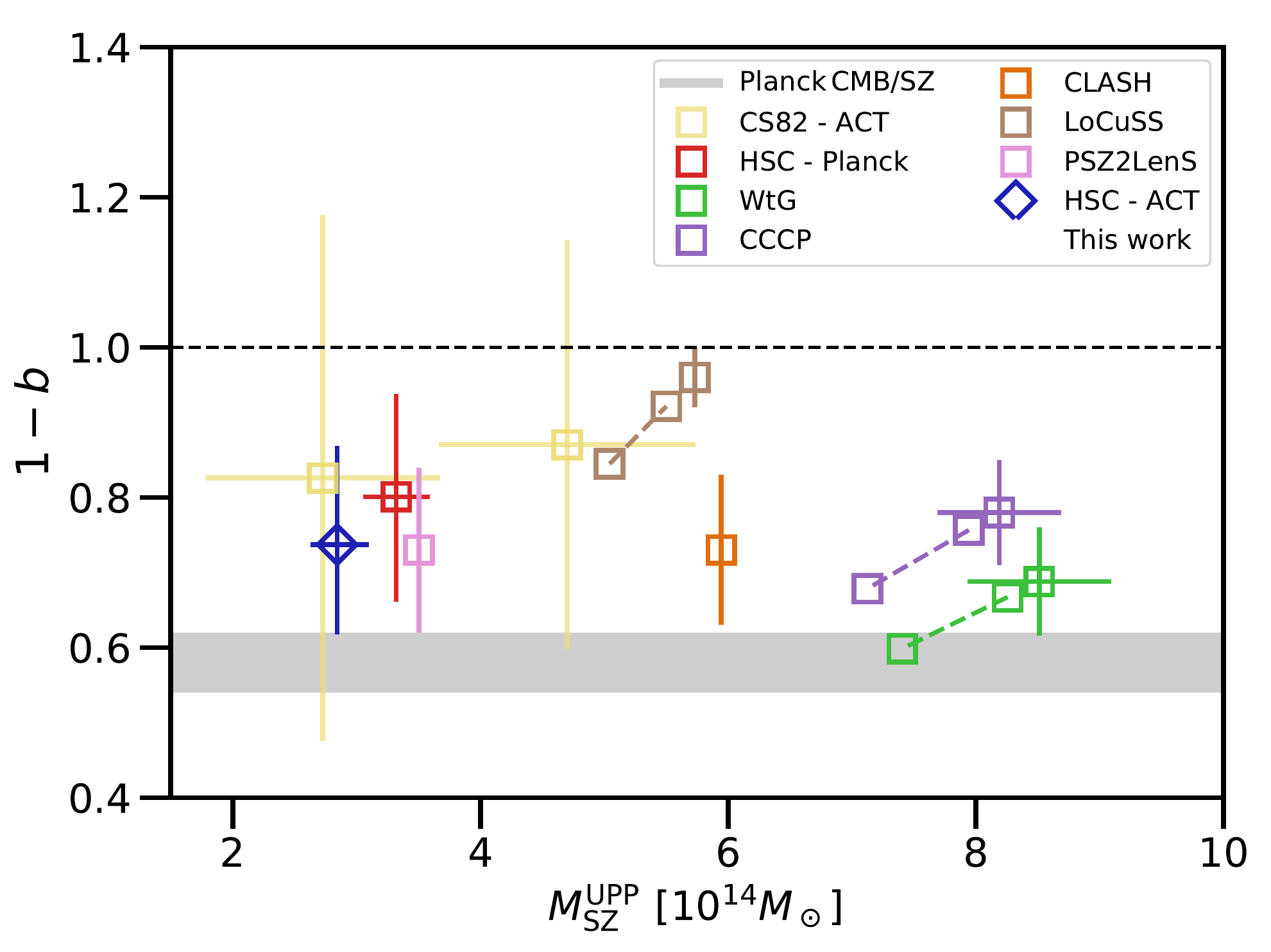}
  \end{center}
\caption{
 \label{fig:bsz}
Comparison of $1-b$ as a function of the SZE mass proxy $M_\mathrm{SZE}$
 for \Planck clusters and for clusters detected by the Atacama Cosmology
 Telescope Polarimeter (ACTpol) experiment.
The data points show the ratios of $M_\mathrm{SZE}$ to $M_\mathrm{WL}$, 
 $1-b_\mathrm{WL}=\langle M_\mathrm{SZE}/M_\mathrm{WL}\rangle$.
 These $M_\mathrm{SZE}$ masses
 (denoted as $M_\mathrm{SZ}^\mathrm{UPP}$ in the figure)
 are derived using the universal pressure profile (UPP) and X-ray
 mass-scaling relation from \citet{Arnaud+2010}, which assumes that the 
 intracluster gas is in hydrostatic equilibrium. The gray band indicates
 the values of $(1-b)$ required to reconcile the \Planck cluster counts
 \citep{Planck2015XXIV} with cosmological parameters from \Planck
 primary CMB \citep{Planck2015XIII}.
 The blue diamond shows the HSC weak-lensing mass calibration of ACTpol
 clusters by \citet{Miyatake2019actpol}.
 Previous $(1-b)$ measurements by
 CS82-ACT \citep{Battaglia+2016},
 LoCuSS \citep{Smith2016locuss},
 CLASH \citep{PennaLima2017},
 PSZ2LenS \citep{Sereno2017psz2lens},
 and HSC-\Planck \citep{Medezinski2018planck}
  are shown in yellow, brown, orange, pink, and red
  squares, respectively.
 The green and purple squares with error bars show the
 original measurements from the Weighing the Giants project
 \citep[WtG;][]{vonderLinden2014calib}
 and CCCP \citep{Hoekstra2015CCCP}, respectively,
 and the same colored squares connected by
 the dashed lines show the $3\percent$--$15\percent$ range for the
 Eddington-bias-corrected measurements calculated in
 \citet{Battaglia+2016}. 
Image reproduced with permission from \citet{Miyatake2019actpol}, copyright by AAS.
 }
\end{figure*}

Several recent studies used weak-lensing mass estimates,
$M_\mathrm{WL}$, to recalibrate cluster masses for subsets of the 
\Planck cosmology sample assuming that the average weak-lensing mass is
unbiased 
\citep{vonderLinden2014calib,Hoekstra2015CCCP,Smith2016locuss,Sereno2017psz2lens,PennaLima2017}.
The samples used in these studies typically contain a few tens of
\Planck clusters.
%and have only marginal overlap in their redshift and mass ranges.
The results of mass calibrations are controversial in terms of
the inferred level of bias,
with some studies finding relatively high values of mass bias,
$1-b_\mathrm{WL}\equiv \langle M_\mathrm{SZE}/M_\mathrm{WL}\rangle\sim 0.6-0.8$
with typical uncertainties of $\pm 0.1$
\citep[][]{vonderLinden2014calib,Hoekstra2015CCCP,Sereno2017psz2lens,PennaLima2017},
and some finding little bias for low-$z$ \Planck clusters,
$1-b_\mathrm{WL}=0.95\pm 0.04$
\citep[at $0.15<z<0.3$;][]{Smith2016locuss}.
%Systematic differences between these weak-lensing studies are at the
%$\sim 10\percent$ level \citep{Hoekstra2015CCCP,Planck2015XXIV}.
%Therefore, it is unclear  whether these differences are due to
%the systematics in their weak-lensing 
%measurements or that $(1-b)$ has a mass or redshift dependence.  
However, it should be noted that some of these mass calibrations did
not include the correction for Eddington bias and their inferred
values of $(1-b_\mathrm{WL})$ are thus likely overestimated
\citep[see][]{Battaglia+2016,Medezinski2018planck,Miyatake2019actpol}. 
The results of mass-calibration efforts for SZE-selected cluster samples 
are summarized in Fig.~\ref{fig:bsz}
\citep[for details, see][]{Miyatake2019actpol}.

Recently, \citet{Medezinski2018planck} performed a weak-lensing
analysis of five \Planck clusters located within $\sim
140$\,deg$^2$ of full-depth and full-color HSC-SSP data. 
With its unique combination of area and depth,
the HSC Wide layer will provide uniformly determined 
weak-lensing mass measurements for thousands of clusters over the total
sky area of $\sim 1000$\,deg$^2$. This is different from previous
studies in which weak-lensing measurements are based on targeted
observations and/or archival data. 
Using the high-quality HSC weak-lensing data and accounting for
Eddington bias,
\citet{Medezinski2018planck} determined the mean level of mass bias to be
$1-b_\mathrm{WL}=\langle M_\mathrm{SZE}/M_\mathrm{WL}\rangle=0.80\pm 0.14$
at a mean weak-lensing mass of
$M_\mathrm{WL}=(4.15\pm 0.61)\times 10^{14}M_\odot$.
Since \citet{Medezinski2018planck} analyzed only five \Planck clusters in a
lower-mass regime than previous weak-lensing studies,
as shown in Fig.~\ref{fig:bsz},
this relatively low bias, $b_\mathrm{WL}=(20\pm 14)\percent$, does not
stand in tension with previous higher values of $b_\mathrm{WL}$ 
nor with the level needed to explain
the high value of $\sigma_8$ found from \Planck primary CMB,
$b = (42\pm 4)\percent$ \citep{Planck2015XXIV}.

Therefore, the mystery continues. More representative subsamples with
greater overlap with the \Planck cosmology sample are needed to draw a 
definitive conclusion about $(1-b)$ and its cosmological
implications.  
%%%
When the full HSC-Wide survey is complete, we expect to have $\sim 40$
\Planck clusters observed in the total area of $\sim 1000$\,deg$^2$.
The level of uncertainty on the mass calibration, if
assuming it is statistics dominated, will be reduced from the
$\sim 10\percent$ level achieved with five \Planck-HSC clusters
\citep{Medezinski2018planck} to reach $\sim 4\percent$.
This is below the current level of systematic uncertainty in
the cluster mass calibration, $\simlt 10\%$
\citep{Medezinski2018planck,Miyatake2019actpol} and will thus require an 
even more stringent treatment of weak-lensing systematics. 
%(Medezinski et al.\ 2018b).
%With such a high S/N measurement ($\sim 50\sigma$ expected),
%With the full HSC-Wide coverage,
%we expect to obtain a tighter mass calibration for a representative
%subsample of \Planck clusters.
%and re-derive cosmological
It will also allow us to examine in detail the level of HSE bias and
study its possible dependence on cluster mass and redshift, providing
valuable information about the thermodynamic history of intracluster gas.

\section{Conclusions}
\label{sec:summary}
 
In this paper, we presented a comprehensive review of cluster--galaxy 
weak lensing,
covering a range of topics relevant to its cosmological and 
astrophysical applications. The goals of this review were (1) to provide
a self-contained pedagogical overview of the theoretical foundations for
gravitational lensing from first principles (Sect.~\ref{sec:theory}),  with special attention 
to the basics and advanced concepts of cluster weak lensing
(Sects.~\ref{sec:basics}, \ref{sec:method}, and \ref{sec:magbias}),
%such as shear, convergence, flexion, and magnification bias, as well as
%mass reconstruction techniques and standard shear analysis methods for
%individual and stacked clusters
and (2) to summarize and highlight recent advances in our understanding
of the mass distribution in and around cluster halos based on numerical
simulations and observational results (Sect.~\ref{sec:obsreview}).  

%mass reconstruction techniques
%various detectable effects
%shear magnification flexion

Thanks to concerted community efforts, there has been substantial
progress over the last few decades in this area on both observational
and theoretical grounds. 
In this review, we focused on several key issues, namely,
the shape of the mass density profile (Sect.~\ref{subsec:massprofile}),
the $c$--$M$ relation and its intrinsic scatter
(Sect.~\ref{subsec:cM}), splashback features in the cluster outskirts
(Sect.~\ref{subsec:rsp}), and cluster mass calibrations for cluster cosmology
(Sect.~\ref{subsec:mcal}). 
%% %
Observations of cluster--galaxy weak lensing are, thus far, generally
favorable for the standard \LCDM paradigm of structure formation,
in terms of the standard explanation for dark matter as
effectively collisionless and nonrelativistic on sub-megaparsec scales
and beyond, with an excellent match between data and predictions
for cluster-size massive halos (Sect.~\ref{sec:obsreview}).

These studies constitute an encouraging step toward
using cluster--galaxy weak lensing to robustly test detailed predictions
of \LCDM and its variants, such as SIDM and $\psi$DM, calibrated from
cosmological numerical simulations.
Such predictions can be unambiguously tested across a
wide range in cluster mass and redshift, with large
statistical samples of clusters from ongoing and planned lensing surveys,
such as Subaru HSC-SSP, DES, LSST,
% satellite missions
\WFIRST, and \Euclid.

% Improvements in modeling and simulation capabilities and 
%the cluster selection will enable us to control for systemtic effects 

\begin{acknowledgements}
I am grateful to Paul T.~P. Ho and Eiichiro Komatsu for the
 invitation to present this review.
 I thank the two anonymous referees for their helpful comments.
 I thank Mauro Sereno, Mark Birkinshaw, Benedikt Diemer,
 Luis~A. D{\'{\i}}az Garcia, Nobuhiro Okabe, and Sut-Ieng Tam
 for careful reading of the manuscript and constructive comments.
I acknowledge stimulating discussions with Teppei Okumura and Toshifumi
 Futamase.
I thank Misao Sasaki for his helpful
 comments on an early version of the manuscript.
This work is supported by the Ministry of
 Science and Technology of Taiwan
 (grants MOST 106-2628-M-001-003-MY3 and MOST 109-2112-M-001-018-MY3)
 and by the Academia Sinica Investigator Award (grant AS-IA-107-M01).
\end{acknowledgements}

%%%%%%%%%%%%%%%%%%%%%%%%%%%%%%%%%%%%%%%%%%%%%%%%%%%%%%%%%%%%%%%%%%%
%%%%%%%%%%%%%%%%%%%%%%%%%%%%%%%%%%%%%%%%%%%%%%%%%%%%%%%%%%%%%%%%%%%
%%%
%%% References 
%%% 
%%%%%%%%%%%%%%%%%%%%%%%%%%%%%%%%%%%%%%%%%%%%%%%%%%%%%%%%%%%%%%%%%%%
%%%%%%%%%%%%%%%%%%%%%%%%%%%%%%%%%%%%%%%%%%%%%%%%%%%%%%%%%%%%%%%%%%%

\bibliographystyle{spbasic_FS} % default
%\bibliographystyle{unsrtnat}
%%%\bibliographystyle{spmpsci}      % mathematics and physical sciences
%%%\bibliographystyle{spphys}       % APS-like style for physics

% Keiichi's note:
% [1] For submission
% Make sure comment out the line ``bibtex $fname'' in mklatex.sh
% for arXiv submission, remove the 1st line of ms.bbl ``\newcommand{\noop}[1]{}''; Execute twice ``pdflatex ms.tex'', or Execute ``mkcopy.sh''

% [2] For a draft version
% Execute ``mklatex.sh'' to produce ``ms.bbl''
%\bibliography{KeiichiU,Lensref,CMB}

%\bibliography{Nissen-Gustafsson}   % name your BibTeX data base

\end{document}